%% file: thesis.tex
\documentclass[a4paper,12pt,numbered,oneside,print,index,custommargin]{Classes/PhDThesisPSnPDF}

\input{Preamble/preamble}

\input{thesis-info}

\ifdefineAbstract
\fi

\begin{document}

\frontmatter

\begin{titlepage}
  \maketitle
\end{titlepage}

\include{Declaration/declaration}

\include{Acknowledgement/acknowledgement}
\include{Abstract/abstract}


\tableofcontents

\listoffigures


\printnomenclature

\mainmatter

\part{Background material}

\include{Chapter1/chapter1}
\include{Chapter2/chapter2}

\part{Code development work}

\include{Chapter3/chapter3}

\part{Research work}

\include{Chapter4/chapter4}
\include{Chapter5/chapter5}

\part{Conclusions, references and appendices}

\include{Chapter6/chapter6}


\begin{spacing}{0.9}



\bibliography{References/references.bib}{}
\bibliographystyle{utcaps}



\end{spacing}


\begin{appendices} 

\include{Appendix1/appendix1}
\include{Appendix2/appendix2}

\end{appendices}


\end{document}

%% file: Preamble/preamble.tex
\usepackage{geometry}
\ifsetCustomMargin
  \newgeometry{left=40mm,right=20mm,top=35mm,bottom=20mm}
  \setFancyHdr 
\fi

\ifsetCustomFont

\fi

\RequirePackage[labelsep=space,tableposition=top]{caption}

\usepackage{float}
\usepackage{epsfig}
\usepackage{amsmath}
\usepackage{amsfonts}
\usepackage{amssymb}
\usepackage{graphicx}
\usepackage{epstopdf}
\usepackage{morefloats}
\usepackage{url}
\usepackage{color,soul}
\usepackage{subfigure}
\usepackage{csquotes}
\usepackage{tcolorbox}

\usepackage{booktabs} 
\usepackage{multirow}

\usepackage{siunitx} 

\usepackage{setspace}
\ifuseCustomBib

\RequirePackage[backend=biber, style=numeric-comp, citestyle=numeric, sorting=nty, natbib=true]{biblatex}
\bibliography{References/references} 
\fi

\usepackage{cite} 

\newcommand{\Beq}{\begin{eqnarray}}
\newcommand{\Eeq}{\end{eqnarray}}
\newcommand{\eqn}[1]{Eqn. (\ref{#1})}

\newcommand{\grchombo}{\mathtt{GRChombo}}
\newcommand{\Lean}{\texttt{Lean}}

\newcommand{\tgamma}{\tilde{\gamma}}
\newcommand{\tGamma}{\tilde{\Gamma}}
\newcommand{\tA}{\tilde{A}}
\newcommand{\tD}{\tilde{D}}

\newcommand{\mpl}{M_{\mbox{\tiny Pl}}}

\newcommand{\onehalf}{\frac{1}{2}}
\def\lsim{\mathrel {\vcenter {\baselineskip 0pt \kern 0pt \hbox{$<$} \kern 0pt \hbox{$\sim$} }}}
\def\gsim{\mathrel {\vcenter {\baselineskip 0pt \kern 0pt \hbox{$>$} \kern 0pt \hbox{$\sim$} }}}

\usepackage[utf8]{inputenc}
\usepackage{nomencl}
\usepackage[intoc]{nomencl}
\makenomenclature



%% file: thesis-info.tex
\title{Scalar Fields in Numerical General Relativity: Inhomogeneous Inflation and Asymmetric Bubble Collapse}

\subtitle{3+1D code development and applications}

\author{Katy Clough}

\dept{Department of Physics}

\university{King's College London}
\crest{\includegraphics[width=0.2\textwidth]{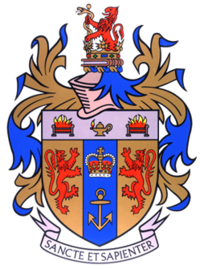}}

\supervisor{Dr Eugene Lim}

\degreetitle{Doctor of Philosophy}

\subject{LaTeX} \keywords{{LaTeX} {PhD Thesis} {Physics} {King's College London}}

%% file: Declaration/declaration.tex

\begin{declaration}

I hereby declare that except where specific reference is made to the work of 
others, the contents of this dissertation are original and have not been 
submitted in whole or in part for consideration for any other degree or 
qualification in this, or any other university. This dissertation is my own 
work and contains nothing which is the outcome of work done in collaboration 
with others, except as specified in the text and Acknowledgements. This 
thesis contains fewer than 100,000 words excluding the bibliography, footnotes, tables and equations.


\end{declaration}

%% file: Acknowledgement/acknowledgement.tex
\newgeometry{left=40mm,right=20mm,top=5mm,bottom=20mm}
\begin{acknowledgements}      
\vspace{-0.25cm}
I would like to sincerely thank my supervisor, Eugene Lim, for giving me the opportunity to pursue a career in research. I am grateful in particular for his support, encouragement and invaluable advice during the PhD. His contributions run throughout the work presented in this thesis, over and above his significant role as co-author on the papers presented in Chapters 3 to 6.

I am also grateful for the contributions of my $\grchombo$ collaborators, Hal Finkel, Pau Figueras, Markus Kunesch and Saran Tunyasuvunakool, for their work on the code development and testing, in particular in relation to the scaling and convergence tests presented in Chapter \ref{ch-GRChombo}. Moreover, I would like to thank them for their various useful insights and discussions, and for being a great team to work with - always willing to share ideas and knowledge. I look forward to continuing our collaboration in future, particularly on the new version of the code, for which Saran and Markus in particular deserve credit for their work alongside Intel.

Also in relation to the work presented in Chapter \ref{ch-GRChombo}, I would like to thank the $\Lean$ collaboration for allowing us to use their code as a basis for comparison, and especially Helvi Witek for helping with the setting up and running of the $\Lean$ simulation. I would like to thank Juha J{\"a}ykk{\"a}, James Briggs and Kacper Kornet at DAMTP for their amazing technical support.  The majority of the work in this thesis was undertaken on the COSMOS Shared Memory system at DAMTP, University of Cambridge operated on behalf of the STFC DiRAC HPC Facility. This equipment is funded by BIS National E-infrastructure capital grant ST/J005673/1 and STFC grants ST/H008586/1, ST/K00333X/1.The research also used resources of the Argonne Leadership Computing Facility, which is a DOE Office of Science User Facility, supported under Contract DE-AC02-06CH11357, and I benefitted greatly from attending their ATPESC 2015 course on supercomputing. I also used the ARCHER UK National Supercomputing Service (http://www.archer.ac.uk) for some simulations, and again, attended a number of their courses on High Performance Computing which I found invaluable. Part of the performance testing was performed on Louisiana State University's High Performance Computing facility.
\pagebreak
\newgeometry{left=40mm,right=20mm,top=35mm,bottom=20mm}
I would like to acknowledge my co-authors on the paper ``Robustness of Inflation to Inhomogeneous Initial Conditions'', presented in Chapter \ref{ch-Inflation}, in particular Raphael Flauger, for his deep knowledge of the topic of inflation, but also Brandon S. DiNunno, Willy Fischler and Sonia Paban. I am grateful to William East for sharing with us details of his simulations, and to Jonathan Braden, Hiranya Peiris, Matt Johnson, Robert Brandenberger, Adam Brown and Tom Giblin for useful conversations on this and related projects.

In Chapter \ref{ch-Conclusions} I briefly present work I was involved in for the paper ``Black Hole Formation from Axion Stars'', carried out with Thomas Helfer, David J. E. (Doddy) Marsh, Malcolm Fairbairn, and Ricardo Becerril. I acknowledge their contributions but in particular Thomas for doing most of the hard work on the simulations and Doddy for his encyclopaedic knowledge of axions.

I am grateful to Thomas Helfer and to James Cook for allowing me to train them in the art of GRChombo - I learnt as much as they did in the process. Thanks also to James for proof reading this work.

I am grateful to all the staff at King's College London who have supported me during my PhD, particularly Malcolm Fairbairn for his honest commentary on my presentation skills and materials, and to the administrative staff in Physics. Thank you to Lucy Ward for her tireless support and enthusiasm for outreach and women in science events, to the tutors who let me teach their problem classes and to the students who came and asked me tough questions. Thanks to Cardiff, Gottingen, Jena, Britgrav, YTF8, SCGSC, LCDM and Perimeter for inviting me to talk about my work, and to Helvi and Pau for letting me attend their Numerical GR lectures at Cambridge. 

I am grateful to my teachers and tutors throughout my education, all of whom have made a lasting impression on me, but in particular to my maths teacher Mrs Maule at Chesham High School and to my tutor Professor Kouvaritakis at Oxford.

Lastly, I am of course indebted to my partner and my family for their support.
\restoregeometry
\end{acknowledgements}
\restoregeometry

%% file: Abstract/abstract.tex

\newgeometry{left=40mm,right=20mm,top=1mm,bottom=20mm}
\begin{abstract}
\vspace{-0.5cm}
\nomenclature[z-pi]{GR}{General Relativity}
\nomenclature[z-pi]{NR}{Numerical Relativity}
\nomenclature[z-pi]{AMR}{Adaptive Mesh Refinement}
\nomenclature[z-pi]{MPI}{Message Passing Interface - parallel programming method}
Einstein's field equation of General Relativity (GR) has been known for over 100 years, yet it remains challenging to solve analytically in strongly relativistic regimes, particularly where there is a lack of a priori symmetry. Numerical Relativity (NR) - the evolution of the Einstein Equations using a computer - is now a relatively mature tool which enables such cases to be explored. In this thesis, a description is given of the development and application of a new Numerical Relativity code, $\grchombo$.

$\grchombo$ uses the standard BSSN formalism, incorporating full adaptive mesh refinement (AMR) and massive parallelism via the Message Passing Interface (MPI). The AMR capability permits the study of physics which has previously been computationally infeasible in a full $3+1$ setting. The functionality of the code is described, its performance characteristics are demonstrated, and it is shown that it can stably and accurately evolve standard spacetimes such as black hole mergers. 

We use $\grchombo$ to study the effects of inhomogeneous initial conditions on the robustness of small and large field inflationary models. We find that small field inflation can fail in the presence of subdominant scalar gradient energies, suggesting that it is much less robust than large field inflation. We show that increasing initial gradients will not form sufficiently massive inflation-ending black holes if the initial hypersurface is approximately flat. Finally, we consider the large field case with a varying extrinsic curvature $K$, and find that part of the spacetime remains inflationary if the spacetime is open, which confirms previous theoretical studies.

We investigate the critical behaviour which occurs in the collapse of both spherically symmetric and asymmetric scalar field bubbles. We use a minimally coupled scalar field subject to a ``double well'' interaction potential, with the bubble wall spanning the barrier between two degenerate minima. We find that the symmetric and asymmetric cases exhibit Type 2 critical behaviour with the critical index consistent with a value of $\gamma = 0.37$  for the dominant unstable mode. We do not see strong evidence of echoing in the solutions, which is probably due to being too far from the critical point to properly observe the critical solution.

We suggest areas for improvement and further study, and other applications.

\end{abstract}
\restoregeometry

%% file: Chapter1/chapter1.tex

\chapter{Introduction}  
\label{ch-Introduction}

\ifpdf
    \graphicspath{{Chapter1/Figs/Raster/}{Chapter1/Figs/PDF/}{Chapter1/Figs/}}
\else
    \graphicspath{{Chapter1/Figs/Vector/}{Chapter1/Figs/}}
\fi

This chapter provides an overview of the key background to the work in this thesis. The main themes are General Relativity (GR), its numerical formulation in Numerical Relativity (NR), and Scalar Fields (SF) coupled to gravity. These are considered in sections \ref{sec-GROverview}, \ref{sec-NROverview} and \ref{sec-SFOverview} respectively. 

The intention of this chapter is to provide an overview of the key motivations, intuitive principles and historical developments in each topic, in order to set the scene for the more technical detail given in Chapter \ref{ch-Technical}. 

In this thesis, we follow the indexing convention of \citep{ShapiroBook}. The signature is $(-+++)$. Low-counting Latin indices ($a,b,\dots$) are abstract tensor indices while Greek indices ($\mu,\nu,\dots$) denote spacetime component indices and run through $0,1,2,3$. Spatial component indices are labeled by high-counting Latin indices ($i,j,\dots$) which run through $1,2,3$.  Unless otherwise stated, we set Newton's gravitational constant $G=1$ and the speed of light $c=1$. Other symbols will be identified in the text, and are summarised in the nomenclature section at the start of the thesis. Where components are given we assume a coordinate basis unless otherwise specified, and the Einstein summation convention is used throughout. 

\nomenclature[a-pi]{$G$}{Newton's Gravitational Constant}
\nomenclature[a-pi]{$c$}{Speed of light in a vacuum}
\nomenclature[a-pi]{$a,b,\dots$}{Low counting Latin indices denote abstract tensor indices which run through $0,1,2,3$}
\nomenclature[g-pi]{$\mu, \nu, \dots$}{Greek indices denote spacetime component indices which run through $0,1,2,3$}
\nomenclature[a-pi]{$i,j,\dots$}{High counting Latin indices denote spatial component indices which run through $1,2,3$}

\nomenclature[z-pi]{SR}{Special Relativity}
\nomenclature[z-pi]{SF}{Scalar Field}
\nomenclature[z-pi]{BH}{Black Hole}
\nomenclature[z-pi]{EEP}{Einstein Equivalence Principle}
\nomenclature[z-pi]{EM}{Energy-Momentum (tensor)}
\nomenclature[z-pi]{SEP}{Strong Equivalence Principle}
\nomenclature[z-pi]{PDE}{Partial Differential Equation}
\nomenclature[z-pi]{ADM}{Arnowitt, Deser, Misner}
\nomenclature[z-pi]{BSSN}{Baumgarte, Shapiro, Shibata, Nakamura}
\nomenclature[z-pi]{FRW}{Friedman-Robertson-Walker(-Lemaitre)}
\nomenclature[z-pi]{ESA}{European Space Agency}
\nomenclature[z-pi]{EOM}{Equation of Motion}
\nomenclature[z-pi]{MHD}{Magnetohydrodynamics}
\nomenclature[z-pi]{QFT}{Quantum field theory}
\nomenclature[z-pi]{CMB}{Cosmic Microwave Background}
\nomenclature[z-pi]{VEV}{Vacuum Expectation Value}


\section{General Relativity}
\label{sec-GROverview}

After the theory of Special Relativity (SR) was proposed, and found to be consistent with observation, it became clear that Newtonian gravity could not be correct. For a force to act between two massive bodies simply due to their existence would require ``action at a distance'' - should one pop out of existence the other would immediately be freed from its orbit, requiring signals between the two to travel faster than the speed of light. Newton himself had objected to this idea, writing in correspondence in 1692 \citep{CohenBook}:
\begin{displayquote}
\textit{``It is inconceivable that inanimate matter should, without the mediation of something else, which is not material, operate upon, and affect other matter without mutual contact.''}
\end{displayquote}
However, as with any successful effective theory, the model of Newtonian gravity was too useful to be discounted on purely philosophical grounds, and thus the issue was largely ignored until SR prompted it to be revisited, and ultimately solved, by Einstein. 

Einstein's work on the problem was motivated by two key considerations. Firstly, the principle of general covariance, which states that the laws of physics must be the same for all observers, and secondly the principle of equivalence, that all objects fall with the same acceleration in a gravitational field regardless of mass. The path from these relatively simple tenets to General Relativity - a new, more accurate, theory of gravity - is described in sections \ref{sec-GRGeneralCovariance} and \ref{sec-GREquivalencePrinciple}. The consequences of the theory of GR are numerous, and revolutionise how we understand the Universe. Several will be discussed briefly in section \ref{sec-GRConsequences}.

To replace Newtonian gravity, the new theory had to relate the way that matter moved in a gravitational field, and the source of that field. The result obtained by Einstein was a radical new description of the Universe - gravity was no longer a force acting between two bodies, but an effect of matter curving the 4 dimensional spacetime around it, like placing a rock on an elastic sheet. As described succinctly by John Wheeler \citep{WheelerBook}:
\begin{displayquote}
\textit{``Matter tells space how to curve, space tells matter how to move.''}
\end{displayquote}
Or, in mathematical notation
\begin{equation}
G_{ab} \equiv R_{ab} + \onehalf R g_{ab} = 8 \pi T_{ab}  \label{eqn:EEM_intro} ,
\end{equation}
where the left hand side terms describe the curvature of the space, and the right hand side relates to the matter content. The components of this equation and its derivation will be discussed in the next chapter, in section \ref{sec-GRTheory}. This chapter aims to first introduce the main concepts in GR, without the distraction of the mathematical detail which is necessary for a complete description.

\nomenclature[a-pi]{$G_{ab}$}{The Einstein Curvature Tensor}
\nomenclature[a-pi]{$R_{ab}$}{The Ricci Tensor}
\nomenclature[a-pi]{$R$}{The Ricci Scalar}
\nomenclature[a-pi]{$g_{ab}$}{The 4 dimensional spacetime metric}
\nomenclature[a-pi]{$T_{ab}$}{The Energy Momentum (EM) Tensor or Stress Energy Tensor}

\subsection{General Covariance}
\label{sec-GRGeneralCovariance}

The principle of general covariance motivates the introduction of tensors as the key components of any physical law. Tensors are geometric objects which are invariant under a change of coordinates. Thus physical properties which are expressed in terms of tensors would be the same no matter what coordinates they are expressed in, for example, in cartesian or spherical coordinates. 

\begin{figure}
\begin{center}
\includegraphics[width=.8\textwidth]{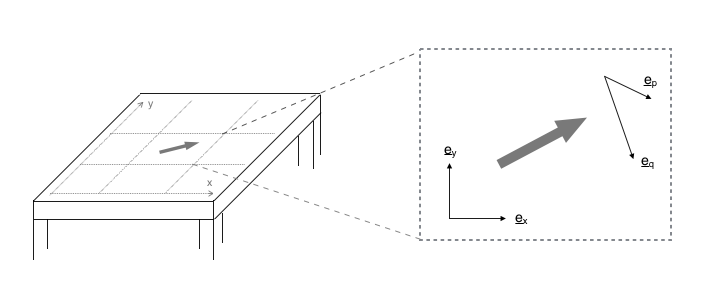}
\caption[Flat space vector]{If I draw an arrow on my desk, I am free to describe it in any coordinate basis I choose, but it will always have the same length. The arrow is a vector, which is a geometric invariant.}
\label{fig-FlatDesk}
\end{center}
\end{figure}

A very simple example is a vector, which is in fact a rank 1 tensor - if I draw an arrow on my desk, it has a certain length and direction (see figure \ref{fig-FlatDesk}). I might choose to describe this vector, $\vec{V}$ in terms of a cartesian coordinate basis $\vec{e}_\mu   = \begin{bmatrix} \vec{e}_x & \vec{e}_y \end{bmatrix}^T$ with the $x$ and $y$ basis vectors being unit vectors parallel to the horizontal and vertical directions. Alternatively I could describe it in terms of a randomly aligned pair of basis vectors labelled by $p$ and $q$ so that $\vec{e}_\mu   = \begin{bmatrix} \vec{e}_p & \vec{e}_q \end{bmatrix}^T$ which may or may not be orthogonal unit vectors, and may or may not be a coordinate basis. In each case the components of the vector $V^\mu$ in the basis will differ, but my vector is still physically the same - it has a fixed length which I could calculate in either basis and I should get the same result\footnote{Assuming that neither coordinate system is boosted relative to the other, since SR tells us that length is not in fact an invariant quantity, only spacetime length, but for the purposes of the example a boosted coordinate system seems an unlikely choice.}. 

Take a moment here to notice some notational conventions and distinguish the different objects involved. The vector $\vec{V}$ is the invariant thing - when I think of this object I am thinking of arrow on my desk as a physical thing, independent of any coordinate system. What an A-level maths student might think of as ``the vector'' are in fact its coordinates  $V^\mu$ in some basis where the $\mu$ labels each basis component and is related to a corresponding basis vector $\vec{e}_\mu$. Specifying these values is meaningless unless one also specifies the basis. The $\mu$ can thus take the values $x$ and $y$, or alternatively $p$ and $q$, but in any chosen basis should run through as many values as there are dimensions in the space under consideration, if the basis is to form a complete set. Here the number of basis components is two as the surface of my desk is two dimensional. The invariant vector itself is the product of the vector components and the basis, i.e. 

\begin{equation}
\vec{V} = V^\mu \vec{e}_\mu = V^x \vec{e}_x + V^y \vec{e}_y = V^p \vec{e}_p + V^q \vec{e}_q \label{eqn:vector} \, .
\end{equation}

Note that I may also write $\vec{V}$ as $V^a$, where the lower counting Latin index ($a,b,\dots$) indicates that I mean the tensor object rather than its components in some basis, $V^\mu$, for which Greek indices ($\mu,\nu,\dots$) are used. In this thesis, the components of a tensor $T_{\mu \nu}$ will often be discussed, since it is these numbers, in some assumed basis (usually cartesian or spherical polar), that are ultimately what we need to tell a computer to ask it to model the system. But we may also refer to abstract tensorial objects like the Energy-Momentum (EM) tensor $T_{ab}$, and it should be remembered that this object is invariant, although its components in an arbitrary basis will not be.

\nomenclature[a-pi]{$\vec{V}, V^a$}{An arbitrary vector}
\nomenclature[a-pi]{$V^\mu$}{The components of an arbitrary vector $\vec{V}$}

\begin{figure}
\begin{center}
\includegraphics[width=.8\textwidth]{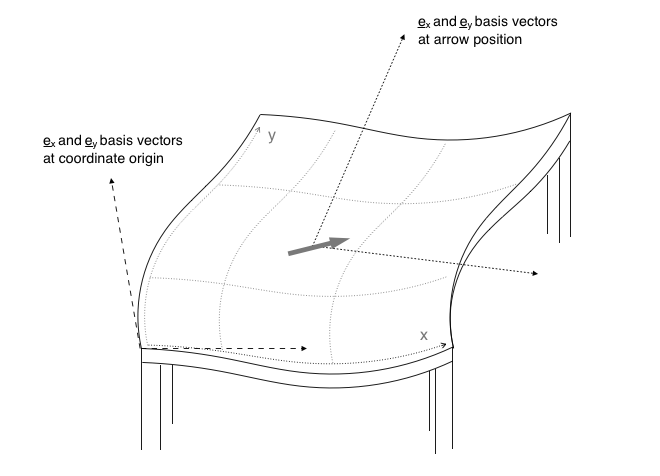}
\caption[Curved space vector]{If my desk is a curved surface, the directions of the basis vectors tangent to the coordinate lines at the bottom left hand corner are different to those where I draw my arrow. The local tangent frame changes at different points on a curved manifold.}
\label{fig-CurvedDesk}
\end{center}
\end{figure}

Notice here another point that seems rather trivial but is not - in figure \ref{fig-FlatDesk} I defined my basis vectors at the position of the arrow, rather than draw axes with an origin at the bottom left hand corner of the desk and define the $x$ and $y$ basis vectors as being parallel to those, as I might have been tempted to do. Why not? Well if my desk is not flat then I will see that the basis vectors I draw tangent to the surface at the bottom left hand corner won't obviously define the same directions (looking at the surface ``globally'') as those I would draw at my arrow, even though I have been careful to keep the $x$ and $y$ coordinate lines ``locally straight'' on the surface. See figure \ref{fig-CurvedDesk}.

This is the important notion that vectors can usually only be defined locally on a curved surface, and not globally. In fact this is not an easy notion to visualise - in particular there is an important distinction between intrinsic curvature (the surface is curved) and extrinsic curvature (the surface is embedded in a higher dimensional space), and we will seek to clarify this further in Chapter \ref{ch-Technical}. Here note that it is the intrinsic curvature of the desk surface which causes problems for comparing vectors at different positions. I could measure this curvature by drawing triangles on the desk and measuring the angles between the sides, which, if they do not sum to $\pi$ radians, tell me that the surface is not intrinsically flat, independently of being able to visualise it in a higher dimensional space.

My warped desk is a 2-dimensional example of a manifold, which can be thought of as a continuous and smooth surface, for which the number of coordinates required to uniquely define each point is the manifold dimension. Here ``smooth'' means locally Euclidean - that is, one can attach a flat plane to each point which is tangent to the surface there, matching both the value at the point and its first derivatives. This definition of a manifold may be easier to understand with a counterexample - if the edge of my desk is a very sharp right angle, this part of the surface would not be a manifold, because at the very corner point I have a discontinuity, to which I can't attach a tangent plane. Thus if we look at a small enough patch on a manifold, we can define vectors lying in this tangent plane and do calculations with them as if we were in flat space. But as we move away from that point the manifold may bend and change shape, meaning that the local tangent ``flat space'' I previously drew is no longer the same one for a neighbouring point. In fact, more than this, the very notion of being ``the same'' at different points is no longer an obvious concept and we will have to define it.

It turns out, as we will see in the next section, that spacetime is a 4-dimensional manifold, and that these ideas of local flatness and global curvature are fundamental to understanding the effects of gravity.

\subsection{The Equivalence Principle}
\label{sec-GREquivalencePrinciple}

The equivalence principle in its most basic form may be stated as the fact that inertial masses (as in the classic Newtonian relation $F=ma$) and gravitational masses (as in $F= GMm/r^2$) are equal. The consequence of this is that all objects fall at the same speed in a gravitational field, unlike in, say, an electric field, where their acceleration depends on their charge to (inertial) mass ratio. (This is of course also true in a gravitational field, it is simply that the gravitational ``charge'' is equal to the inertial mass and thus the ratio is always exactly one for all objects). Einstein rightly believed that this was not a coincidence, but an indication that we had missed something fundamental in our understanding of the laws of gravity. 

In a constant, uniform field, saying that all objects fall at the same speed implies that, in the freely falling frame, the gravitational force vanishes and the frame is an inertial one as in SR. In an inertial frame any object placed at rest in that frame will stay at rest, and clearly if I attach my coordinate system to one of the falling objects, then because they all fall at the same speed, all the objects will appear, as viewed in this coordinate system, to stay at rest. Thus in this frame, one does not need to take the gravitational force into account, and can calculate the motion of the objects relative to each other as if there were no external forces (as in SR). This is rather counterintuitive to humans on Earth because we are used to the Earth pushing up on us - it seems obvious that we ``feel'' gravity. But satellites experience roughly the same gravitational field as we do at the surface of the Earth, and astronauts in them feel nothing - they float about as if in deep space, because to be in orbit is essentially to be in freefall around the Earth. In their coordinate frame, attached to them as they fall, they perceive no gravitational force.

The Einstein Equivalence Principle (EEP) goes further than this basic statement to say that \emph{all physical laws} reduce to those of SR locally for objects in a freely falling frame, thus in such a frame one cannot ``detect'' gravity by \emph{any} local experiment. This is a stronger statement because it puts bounds on the ways in which other forces like electromagnetism and the strong and weak forces can couple to gravity - essentially it means that as far as these forces are concerned, any locally flat patch of space looks identical to another. The even stronger Strong Equivalence Principle (SEP) requires additionally that gravity behaves in the same way everywhere. It thus includes objects with strong gravitational self-interactions and rules out the possibility of a varying gravitational constant, $G$. The SEP applies to unmodified (Einstein) gravity with a minimally coupled scalar field, which is what is considered in this thesis, but for modified gravity theories, the SEP may be violated. For example, in Brans-Dicke gravity the gravitational constant is sourced by a scalar field which may vary in space and time. This variation would in theory be detectable at two separated points, even though each was locally flat, and this violates the SEP.

These statements about equivalence relate to regions which are small enough such that the gravitational field is constant, or ``locally flat''. However, in nature there is no such thing as a truly constant, uniform gravitational field. Gravitational fields are generally sourced by objects which are localised in space, and thus create radial fields. In a small enough region (say in a 1$\rm{m}^3$ box at the surface of the Earth, or the classic ``scientist in a falling lift'' scenario) the field will be \emph{approximately} uniform, but in reality \emph{any} movement away from a single point will result in an (albeit very small) change in the magnitude or direction of the field. So when we have a non point-like object, the gravitational forces can never be completely removed from all parts simultaneously by a coordinate choice, as each point is experiencing a different gravitational field, and thus requires a different choice of freely falling frame to cancel it out. 

So then, this suggestion of making the gravitational force ``disappear'' seems rather limited in its usefulness - if it is only exact at a single point and just a convenient approximation elsewhere, then we are back to approximating everything as SR in some small enough region, albeit we can now also do this in a falling lift and not just for rockets passing each other at constant velocities in outer space. We appear to understand things better, but this local picture doesn't, of itself, get us nearer our aim of relating gravitational effects and their sources.

The missing ingredient is the observation that variation in the gravitational field leads to the phenomenon of tidal forces. Standing on the Earth my head feels less gravity than my feet, since the gravitational force decreases as $1/r^2$ from the centre of the Earth, and so I am being stretched as if someone were pulling me in two directions. I don't notice this because I am not especially tall and so the difference is minute, but close to a black hole the effect of these tidal forces would be sufficient to pull me apart, and so I would be unwise to neglect them. To reduce the description of tidal forces to its simplest form - two particles at different points in a non uniform field, initially with the same velocity, will not maintain a constant separation, but will move apart or together, as if acted on by forces of different magnitude or direction. Tidal forces are a measurable physical phenomenon (they cause the tides in the sea, amongst other things), and so clearly cannot be removed by a coordinate choice.

If we now restate the original idea in more geometric language, we are saying that for (temporally and/or spatially) varying gravitational fields, then \emph{locally} one can find a coordinate basis that is flat in the sense of Minkowski-like; but this ``locally adapted'' frame changes (smoothly) from point to point, such that one cannot choose a global coordinate basis which applies at all points. Einstein's great insight was to realise that this is \emph{exactly} equivalent to the description of a curved manifold that we gave earlier - locally one can create a flat patch by an appropriate coordinate choice, but as we move away from that point this local ``flat space'' is no longer the appropriate one for a neighbouring point. There is no global coordinate system that is tangent to the whole space, as in our example earlier of the warped desk: spacetime is curved. 

In this picture tidal forces can be seen to be a manifestation of the curvature of the manifold, which cannot be entirely removed from the whole body in any chosen frame. But now the word force is actually misleading - there \emph{is no external gravitational force}, as can be seen from the fact that it can be removed at a single point by a convenient choice of coordinates. The so-called ``tidal forces'' that result in two separate objects moving apart are not true forces pulling them in opposite directions, but a consequence of their moving along geodesics (lines that are locally straight) in a curved spacetime. Even the word ``field'' is now somewhat inappropriate in its conventional context, and makes sense only if we think of the gravitational ``field'' as encoding the curvature of spacetime (which is indeed what we will do). 

\begin{figure}
\begin{center}
\includegraphics[width=.8\textwidth]{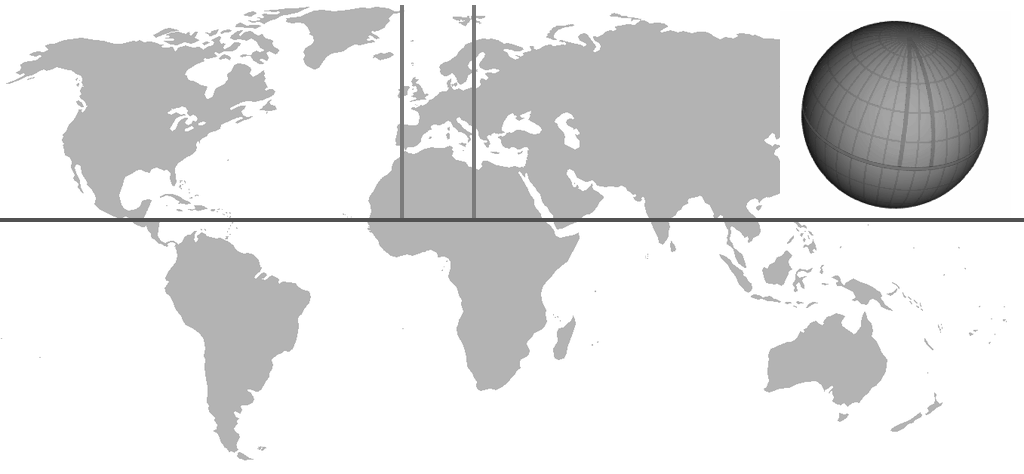}
\caption[The Earth is not flat, neither is spacetime]{In the flat space picture of the Earth, if two people start walking parallel to each other, at the equator, and travel in a straight line due North, they will maintain an equal separation. However, because the Earth is curved, they will in fact meet at the North pole as in the inset figure. If they believed that the Earth was flat, this would be rather surprising, and they might believe that they had been pulled towards each other by a mysterious force. This is analogous to the historic misunderstanding of gravity, which was for many years thought to be an attractive force, rather than a consequence of spacetime curvature. (Image from \citep{Wikimedia})}
\label{fig-CurvedEarth}
\end{center}
\end{figure}

This is analogous to what happens on the surface of the Earth if two people take initially parallel paths and both walk in a straight line, say due North. Because of the curvature of the Earth they will eventually meet, and if they believe the Earth to be flat as our ancestors did, they might erroneously conclude that they had been ``pulled together'' by some mysterious force. In fact their apparent ``attraction'' is purely a geometric effect of travelling on the surface of a sphere - a curved 2 dimensional manifold. See figure \ref{fig-CurvedEarth}.

This insight gives us the key we need to find the equation of motion for gravity. We know from Newtonian physics how matter gives rise to tidal forces which pull objects apart. If we can generate their observed effect - the way in which two separated objects move apart in the field - with a spacetime curvature instead, we can eliminate these fictitious forces from our equation altogether. We will have the desired relation to replace Newtonian gravity - a link between matter and spacetime curvature.

The mathematical derivation of \eqn{eqn:EEM_intro} requires some additional geometric ideas, not least the definition of the terms appearing on either side of the equation, which are not concise to state. Thus a more complete derivation is left to section \ref{sec-derivingEEgeo} of the following chapter. 

However, this approach begs the question - if one can already calculate the tidal forces on bodies with Newtonian gravity, why bother to replace them with spacetime curvature at all? The answer is that although the agreement between Newtonian Gravity and GR is (necessarily for consistency) very good at lower energies, there are other, unexpected effects which cannot be predicted from the force picture, which come into play when the spacetime curvature is high. Some of the consequences are quite revolutionary, as will be discussed in the following section.

\subsection{Consequences of GR}
\label{sec-GRConsequences}

\begin{figure}
\begin{center}
\includegraphics[width=.6\textwidth]{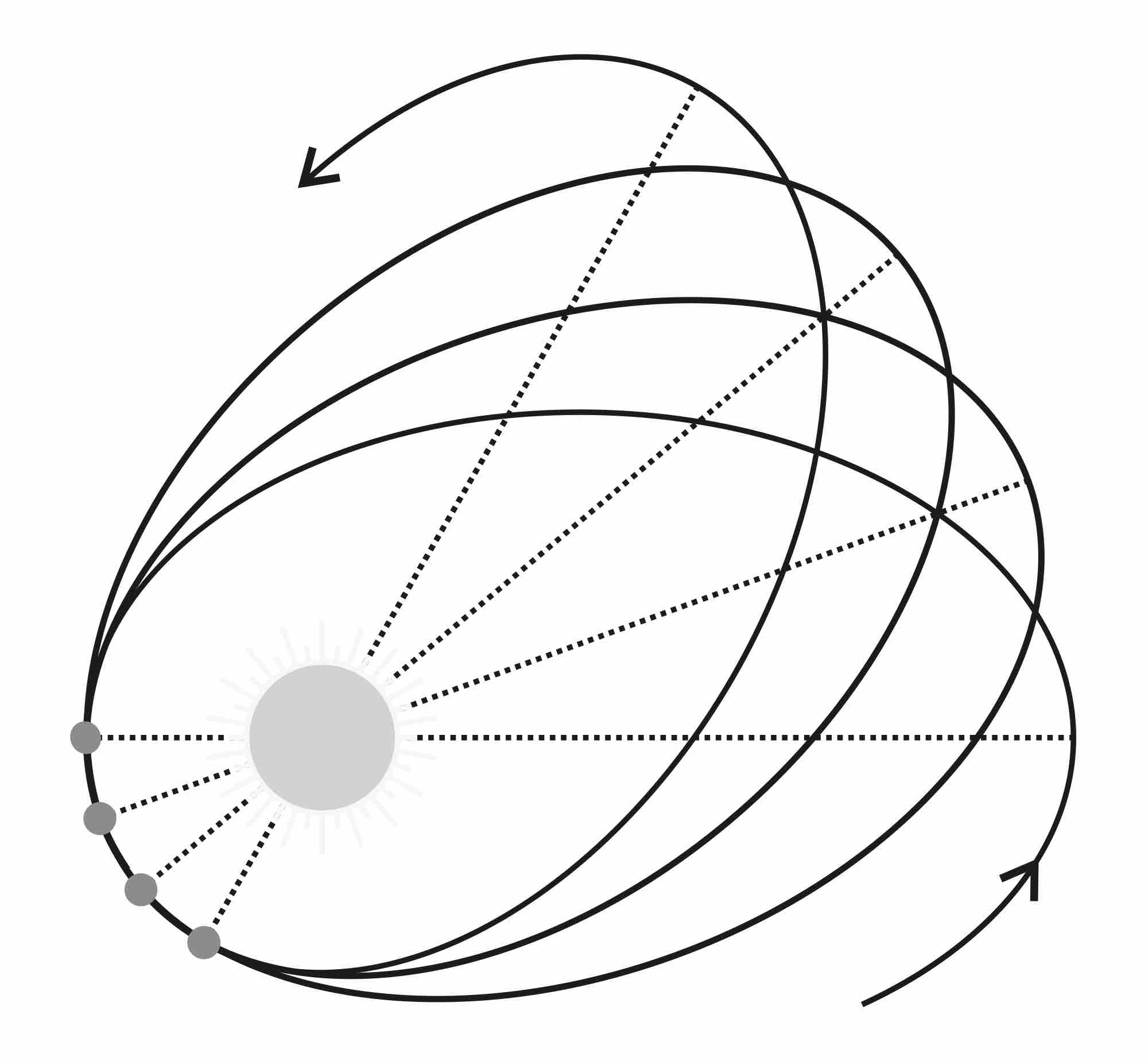}
\caption[Perihelion precession]{General relativity predicts that the perihelion direction rotates gradually in the orbital plane over progressive orbits. This effect does not appear in Newtonian Gravity and its agreement with the observed orbit of Mercury was one of the early successes of the theory. (Image from \citep{Wikimedia})}
\label{fig-perihelion}
\end{center}
\end{figure}

The effects of GR on our understanding of the universe are profound. At the lowest level, corrections are found to Newtonian gravity, and these corrections are the basis for some of the earliest tests of GR. A good example is the precession of the perihelion (the direction of closest pass) in the orbit of Mercury. In Newtonian gravity an isolated star and planet system would maintain a constant perihelion direction over the course of many orbits, but the inclusion of relativistic terms results in its direction gradually rotating in the orbital plane (see figure \ref{fig-perihelion}). Other classic tests include the bending of light from stars around the Sun (which would not occur in Newtonian gravitation as photons are massless\footnote{Although one can regard the photon as having a mass in terms of its angular frequency $\omega$, $m = \hbar \omega/c^2$, one will not get the correct deflection for a massless particle using the Newtonian result.}), and the measurement of gravitational redshift, firstly in the Pound-Rebka experiment \citep{PoundRebka1959} in 1959, and nowadays on a daily basis by anyone using GPS. 

\nomenclature[a-pi]{$\hbar$}{Reduced Planck's constant}

\begin{figure}
\begin{center}
\includegraphics[width=.6\textwidth]{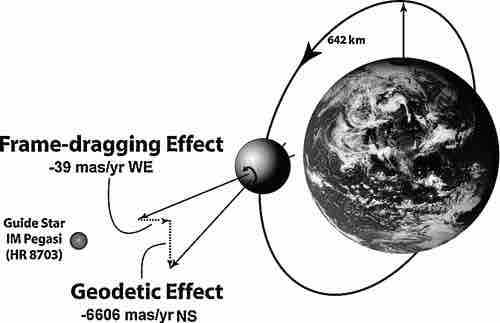}
\caption[Gyroscopic tests]{The geodetic effect results in the direction of a gyroscope appearing to precess on each orbit of the Earth. This is a consequence of the space around the Earth being curved by the mass energy and so occurs for an orbit in any direction, such that the gyroscope appears to turn in the plane of the orbit. The image also shows the effect of frame dragging, which results in the direction of a gyroscope appearing to turn in the direction of rotation of the Earth. If a polar orbit is chosen, as in Gravity Probe B, the two effects will be perpendicular. (Image from \citep{GravityProbeB})}
\label{fig-gyroscope}
\end{center}
\end{figure}

More modern tests include gravitational lensing of distant objects (see \citep{Bartelmann:2010fz} for a review), and satellite tests to observe the geodetic effect (also called de Sitter precession) and frame dragging (specifically Lense-Thirring precession). The former, the geodetic effect, results in the direction of a gyroscope appearing to precess as it orbits the Earth, due to the curvature of the space around the Earth resulting from its mass. The latter, frame dragging, occurs because the Earth is spinning, which causes the gyroscopic direction to be ``dragged'' round in the direction of rotation of the Earth. Figure \ref{fig-gyroscope} illustrates the recent Gravity Probe B experiment which tested these effects \citep{GravityProbeB}.

These corrections to Newtonian gravity are inferred from ``solutions'' to the equations of GR, which may be found in given circumstances. In this context, a solution is a description of the spacetime curvature resulting from a given matter distribution - \emph{matter tells space how to curve}. Where the situation has some high level of symmetry, and where simplifying assumptions may be made, it is possible to find analytic expressions for the spacetime curvature and its variation over time. From these solutions, the motion of a small test mass (which it is assumed does not materially affect the overall curvature) can be inferred - \emph{space tells matter how to move}.

One of the most well known solutions is the Schwarzschild metric \citep{Schwarzschild:1916uq} which describes the curvature of space outside a point mass, typically a black hole although it may also be applied outside extended bodies like the Earth (it is from this solution that the geodetic effect is calculated). It may seem rather remarkable that in the vacuum around a mass like the Sun, the space will be affected just by its presence, but it is exactly this solution which resolves the paradox of action at a distance which prompted the discovery of GR - the curvature of spacetime is the mediator of the gravitational effects between two separated bodies. Since disturbances in the curvature cannot travel faster than the speed of light, no gravitational signal can propagate between two points in spacetime faster than this limit, and causality is assured\footnote{Modulo the construction of spacetimes with closed time-like curves, e.g. wormholes, see \citep{MTWormholes}.}. Consideration of masses with angular momentum leads to the Kerr solution \citep{Kerr1963}, from which frame dragging can be deduced, and including electric charge gives the Reissner-Nordström metric \citep{Reissner1916}. The solution for a black hole which is both charged and rotating is the Kerr-Newman metric \citep{KerrNewman}.

These vacuum solutions give us new insights into potential phenomena around black holes, but also contain singularities - points at which spacetime becomes infinitely curved. The breakdown in our understanding at these points highlights the fact that, although GR is a far more accurate theory of gravity than the Newtonian one, it must still be an effective low energy theory - one requires a unified theory of gravity and the Standard Model at higher energies. That is, one expects that new physics might prevent the collapse of matter to an infinite density around the Planck scale, just as electron and neutron degeneracy pressures prevent gravitational collapse in white dwarfs and neutron stars respectively. However, such effects are well beyond the energy scales which we can currently probe, and in addition, the Cosmic Censorship conjecture asserts that singularities will always be enclosed by an event horizon, from which information about their nature cannot be extracted (although this remains, as the name implies, a conjecture, and considers only classical effects). 

\begin{figure}
\begin{center}
\includegraphics[width=.6\textwidth]{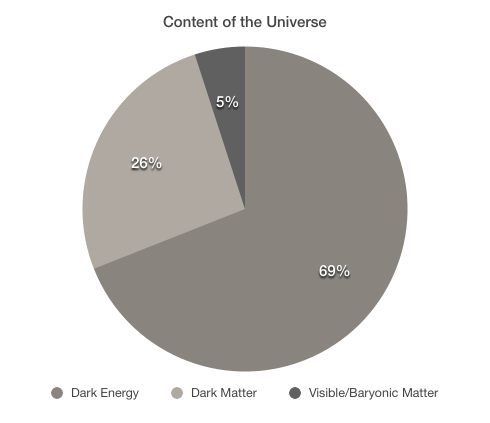}
\caption[Energy Content of the Universe]{Observations of the Universe allow us to constrain its energy and matter content as above. This leads to the somewhat surprising result that much of the matter and energy in the Universe appears to be invisible to us. (Data from 2015 Planck results \citep{PlanckResults2015})}
\label{fig-ContentUniverse}
\end{center}
\end{figure}

Another interesting ``solution'' in GR is found by applying the Einstein equation to our Universe as a whole. The resulting Friedmann-Robertson-Walker-Lemaitre (henceforth FRW, as is conventional) solution for a homogeneous and isotropic universe provides the basis of modern cosmology, as will be discussed in section \ref{sec-SFCosmology} below, and in section \ref{sec-Cosmology} of the following chapter. Here, simply note that GR gives us the ability to predict the future evolution of the Universe on large scales, given a knowledge of its energy and matter content. Turning this around, one obtains a possibly more useful result - observations of the evolution of the Universe allow us to constrain its content, and in doing so one is led to the realisation that much of the matter and energy content of the Universe is unaccounted for by visible matter - see figure \ref{fig-ContentUniverse} - the so-called problems of Dark Energy and Dark Matter.

\begin{figure}
\begin{center}
\includegraphics[width=.95\textwidth]{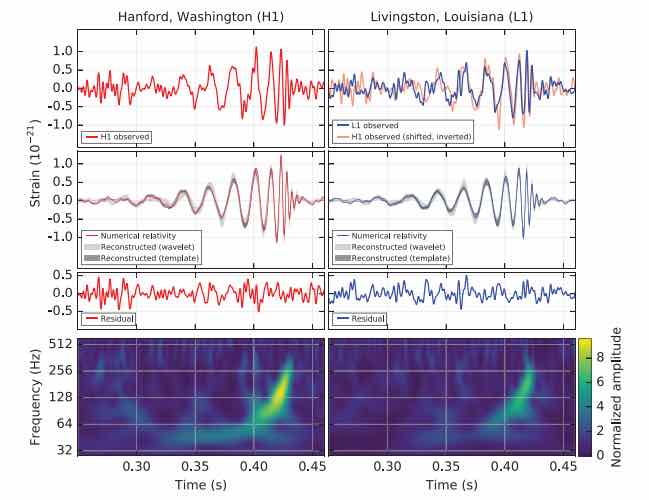}
\caption[LIGO Gravitational Wave signal]{GW150914, the gravitational wave signal detected by Advanced LIGO in 2015. Note that the solid lines show a numerical relativity waveform for the inferred binary black hole system, which is an excellent match to the measured data. Templates produced by Numerical Relativity are essential for understanding the late inspiral and merger phase of the signal. (Image from \citep{Abbott:2016blz})}
\label{fig-LIGO}
\end{center}
\end{figure}

Finally, and perhaps most timely at the moment of writing this thesis, the theory of GR predicts the existence of propagating waves in spacetime - gravitational waves. Such waves are emitted by the relative motion of masses, in particular, as a result of a quadrupole moment in the mass distribution. Gravitational waves emanating from a binary black hole collision approximately 1.4 billion light years away were measured for the first time on 14 September 2015 by the two Advanced LIGO detectors in Hanford and Livingston \citep{Abbott:2016blz}, see figure \ref{fig-LIGO}.  A network of ground based detectors is being established to further study this new area of observational cosmology. In the longer term, the European Space Agency (ESA) has designated the space-based LISA detector an L3 launch slot (expected launch date around 2034), and this seems to be on track following the LISA Pathfinder spacecraft's thus far successful test mission this year. As well as providing further confirmation of the accuracy of the theory of GR, the discovery of gravitational waves has the potential to revolutionise our understanding of the Universe, as it is an entirely new source of information about its content and history. An understanding of gravity and its effects is vital for studying the data gathered, and a key part of this effort will come from Numerical Relativity, which will be discussed in the next section.

\section{Numerical Relativity} 
\label{sec-NROverview}

Almost a hundred years after Einstein wrote down the equations of General Relativity \citep{Einstein1916}, solutions of the Einstein equation remain notoriously difficult to find beyond those which exhibit significant symmetries. Even for these highly symmetric solutions, basic questions remain unanswered. A famous example is the question of the non-perturbative stability of the Kerr solution -- more than 50 years after its discovery, it is not known whether the exterior Kerr solution is stable. The main difficulty of solving the Einstein equation is its non-linearity, which defies perturbative approaches. 

One of the main approaches in the hunt for solutions is the use of numerical methods. In Numerical Relativity (NR) the 4-dimensional Einstein equation \eqn{eqn:EEM_intro} is formulated as a 3+1 dimensional Cauchy problem, where the Cauchy initial data, specified on some 3-dimensional spatial hyperslice, is evolved forward in time. An alternative approach, the Characteristic formulation, is not considered in this thesis, but further details can be found in the review by Winicour \citep{Winicour}.

\subsection{NR as a Cauchy problem}
\label{NRCauchy}

\eqn{eqn:EEM_intro} is an inherently 4-dimensional equation. Each of the tensors it contains are geometric objects which exist on a 4-dimensional manifold and the coordinate system within the manifold may be specified arbitrarily. There is thus (in the general case) no natural foliation of the coordinates one chooses into space and time, as what one calls ``time'' will depend on the observer, and their position and velocity within the spacetime.

However, as humans our brains are not well adapted to visualise a 4-dimensional space and we naturally find it more easy to visualise spatial surfaces being evolved over some chosen time-like coordinate. As long as one is careful with the interpretation of the results which are obtained, as far as possible drawing conclusions in a coordinate independent way, this is a useful tool for understanding gravitational solutions. Moreover, it provides a means by which to answer the question ``what happens next?'' which is often of interest for a given scenario.

\begin{figure}
\begin{center}
\includegraphics[width=.95\textwidth]{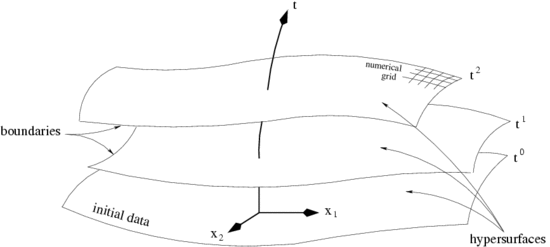}
\vspace{10mm}
\caption[Numerical slicing of a spacetime]{In NR, the metric and matter data on an initial spatial hypersurface is evolved along a local time-like coordinate according to the Einstein equation, to give data on the next hypersurface. The process is then repeated. Here the spatial slices are represented as 2-dimensional for clarity, whereas the data evolved will be 3-dimensional for a full spacetime. (Image from \citep{MPA:SliceNR})}
\label{fig-NRSlices}
\end{center}
\end{figure}

In NR we thus decompose our spacetime into a 3-dimensional spatial slice, and a time-like direction ``off'' the surface - see figure \ref{fig-NRSlices}. Such a decomposition allows us to specify constraint satisfying initial data on some (3-dimensional) Cauchy surface, which may then be evolved forward in discrete steps along the time coordinate.

For example, our initial data may be two black holes boosted in opposite directions so as to give a binary inspiral like the one seen by Advanced LIGO. The initial data would describe the curvature of the spacetime around the black holes, and its derivative with respect to time (see section \ref{sec-ADMTheory} for a more exact description). This is analogous to specifying the initial position and velocity of a particle, which will then be evolved subject to some second order equation of motion (EOM).  

Einstein's equation \eqn{eqn:EEM_intro} is what provides the EOM for the spacetime curvature under gravity. In its 4-dimensional tensor form, it constrains the relationship between the curvature and its derivatives on the 4-dimensional spacetime manifold. It can thus provide, once expanded out in some coordinates which delineate space and time, a set of nonlinear, coupled second order partial differential equations (PDEs) which relate the derivatives in space, the derivatives in time, and the matter content present. These can be rearranged to give us the time derivatives of the curvature as a function of the spatial derivatives and matter content, thus allowing us to generate the future evolution of the curvature at each point from the initial data. 

Note that in a black hole evolution, one does not evolve the central singularity of the black hole in which the mass is contained, which will be excised, or a clever choice of coordinates used to avoid it. In other systems, initial data for the matter field and its time-like derivatives must also be specified, along with an EOM for how that matter type evolves in a curved spacetime. This will be discussed below and in the next chapter for a scalar field matter source. 

The problem of solving a Cauchy problem for a system of coupled PDEs from an initial data set is a classic numerical problem, used in various other fields such as fluid dynamics. There are a number of subtleties and challenges which arise in the specific case of gravity, which will be explored further in the following chapter, but in principle there is no difference between evolving a fluid flow and evolving a pair of black holes, each set of variables simply obeys a different set of PDEs.

A more detailed description of the theory behind the formulation of the Cauchy problem is given in section \ref{sec-ADMTheory} of the next chapter.

\subsection{Key historical developments in NR}
\label{NRHistoricalDevelopments}

Numerical methods have been used to solve the Einstein equation for many years, but the past decade has seen a culmination of theoretical and technical developments, leading to tremendous advances. 

Three key milestones are worth mentioning. Firstly, the development of the ADM formulation of the Einstein equations in 1962 gave a natural decomposition into a 3+1 form suitable for use in a Cauchy problem as described above (see section \ref{sec-ADMTheory}). Originally formulated by Arnowitt, Deser and Misner from a field-theoretic perspective \citep{Arnowitt:1962hi}, as a Hamiltonian formulation for use in quantum gravity, the form now used in NR and referred to as the ``standard ADM decomposition'' more closely resembles the reformulation by York in 1979 \citep{York1979}. This form is mathematically different\footnote{The evolution system for preserving the constraints is well posed for York, whereas in the original ADM formulation it is not, although both are only weakly hyperbolic in terms of the evolution equations, see \citep{AlcubierreBook}.}, but should give the same results for real physical systems. 

Secondly, the discovery that the ADM decomposition was not numerically stable (see section \ref{sec-NRStability}), and its reformulation in a more stable form by Baumgarte, Shapiro, Shibata and Nakamura\footnote{Oohara and Kojima were co-authors of the original paper with Nakamura in 1987, but unfortunately are not usually included in the abbreviation, although some texts use BSSNOK to recognise their contribution.} (the ``BSSN'' form \citep{Nakamura:1987zz,Shibata:1995we,Baumgarte:1998te}), enabled long term stable evolutions of strongly gravitating spacetimes.

The final breakthrough was the development in 2005 of suitable gauge choices for evolving realistic astrophysical scenarios such as neutron stars, core collapse, and the inspiral merger of two black holes \citep{Pretorius:2005gq,Baker:2005vv,Campanelli:2005dd}. The use of Generalised Harmonic Coordinates (GHC) with explicit excision \citep{Pretorius:2004jg}, and ``moving puncture" gauge excision, enabled the study of spacetimes containing moving singularities. This is discussed further in section \ref{sec-InitialConditionsGauge}.

The other driver of developments in NR is an explosion in the availability of large and powerful supercomputing clusters and the maturity of parallel processing technology such as the Message Passing Interface (MPI) and OpenMP \citep{MPIwebsite, OpenMPwebsite}, which open up new computational approaches to solving the Einstein equation.

We anticipate that the development of NR will continue to accelerate, especially given the recent discovery of gravitational waves at Advanced LIGO described above. Beyond searching for gravitational waves and black holes, NR is now beginning to find uses in the investigation of other areas of fundamental physics. For example, standard GR codes are now being adapted to study modified gravity \citep{Berti:2015itd}, cosmology \citep{Wainwright:2014pta,Johnson:2011wt} and even string theory motivated scenarios  \citep{Cardoso:2012qm,Chesler:2013lia,Cardoso:2014uka,Choptuik:2015mma}. In particular, there is an increasing focus on solving GR coupled to matter equations in the strong-field regime: cosmic string evolution with GR, realistic black hole systems with accretion disks, non-perturbative systems in the early universe, etc. Since it is often difficult to have an intuitive picture of the entire evolution ahead of time, the code must be able to automatically adapt to ensure that all regions of interest remain adequately resolved. This nascent, but growing, interest in using NR as a mature scientific tool to explore other broad areas of physics was a key motivation of the $\grchombo$ code development, and the research work described in this thesis demonstrates its suitability for solving these types of problems. 

\subsection{Existing numerical codes and AMR}
\label{NRExistingCodes}

In the NR community, the requirement for varying resolution is largely met through a moving-box mesh refinement scheme. This type of setup consists of hierarchies of boxes nested around some specified centres, and the workflow typically requires the user to specify the exact size of these boxes beforehand. These boxes are then moved around, either along a pre-specified trajectory guided by prior estimates, or by automatically tracking certain quantities or features in the solution as it evolves. Boxes which come within a certain distance of each other may also be allowed to merge. A number of moving-box mesh refinement codes have been made public over the recent years, many of which are built on top of the well-known $\mathtt{CACTUS}$ framework \citep{Goodale2002a,Loffler:2011ay}. One such implementation is the McLachlan/Kranc code \citep{Brown:2008sb,Kranc:web}, which uses finite difference discretisation and the Baumgarte-Shapiro-Shibata-Nakamura (BSSN) evolution scheme  \citep{Baumgarte:1998te,Shibata:1995we}. Similarly, the \texttt{LEAN} code \citep{Sperhake:2006cy,Zilhao:2010sr}, which uses the \texttt{CACTUS} framework, and \texttt{BAM} and AMSS-NCKU \citep{Marronetti:2007ya, PhysRevD.82.024005} also implement the BSSN formulation of the Einstein equations.  There is also $\mathtt{GRHydro}$ which implements general-relativistic magnetohydrodynamics (MHD) for the Einstein Toolkit \citep{EinsteinToolkit:web}, building yet another layer of physics on top of evolution codes such as McLachlan/Kranc. There are also non-$\mathtt{CACTUS}$ codes such as $\mathtt{SPeC}$ \citep{Pfeiffer:2002wt} and \texttt{bamps} \citep{Hilditch:2015aba}, which implement the generalised harmonic formulation of the Einstein equations using a pseudospectral method. In addition to these public codes, there is a plethora of closed-source codes.

The moving-box mesh refinement technique has found great success in astrophysically motivated problems such as two-body collision/inspiral. Outside of this realm, however, the setup can quickly become impractical, especially where one expects new length scales of interest to emerge dynamically over the course of the evolution. This can occur generically in highly nonlinear regimes, either by interaction between GR and various matter models, or by gravitational self-interaction itself which can exhibit complicated unstable behaviour in higher dimensions. In such situations, it is necessary to develop a code which has the flexibility to create refinement regions of arbitrary shapes and sizes, anywhere in the computational domain as may be required. This can be achieved by using a fully adaptive mesh refinement (AMR) technique, whose feature is generally characterised by the ability to monitor a chosen quantity at each time step and insert higher resolution sub-regions where this quantity fails to lie within some chosen bounds. Of course, the efficacy of such codes depend crucially on a sensible choice of these criteria, however when implemented correctly they can be an extremely powerful tool. The advantage here is twofold: AMR ensures that small emergent features remain well-resolved at all times, but also that only those regions which require this extra resolution get refined, thus allowing more problems to fit within a given memory footprint. 

\begin{figure}
\begin{center}
\includegraphics[width=.8\textwidth]{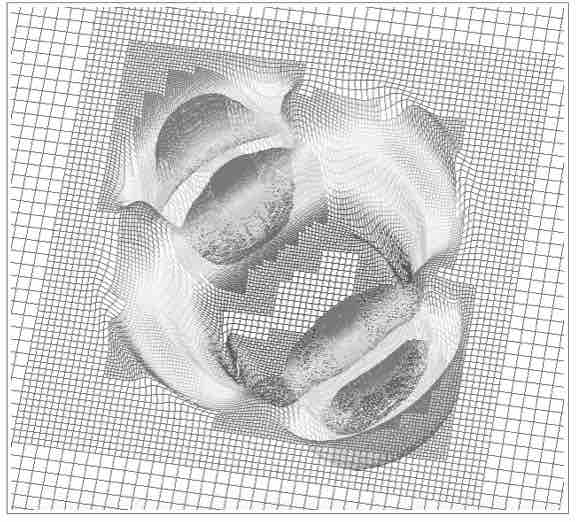}
\caption[Adaptive Mesh Refinement]{This image from a $\grchombo$ simulation shows a 2-dimensional slice through the 3-dimensional spacetime during an evolution from Chapter 5 of an asymmetric scalar field bubble collapse. The elevation at each point is equal to the value of the scalar field. The AMR mesh is overlaid, from which it can be seen that the refinement is increased in regions of strong gradients in the field. The remeshing is signficantly more flexible than the moving-box refinement used by most NR codes.}
\label{fig-AMR}
\end{center}
\end{figure}

In this thesis we will describe the development of a new code for Numerical GR called $\grchombo$ with full AMR. A detailed description of the code, its AMR implementation and details of the code tests are provided in Chapter \ref{ch-GRChombo}. An illustration of AMR in $\grchombo$ is shown in figure \ref{fig-AMR}. To the best of our knowledge, \texttt{PAMR/AMRD} \citep{PAMR} and \texttt{HAD} \citep{Neilsen:2007ua} are the only two codes with full adaptive mesh refinement (AMR) capabilities in numerical GR, but we understand that these are significantly less flexible in their refinement ability than the code we have developed. 

\section{Scalar fields with gravity}
\label{sec-SFOverview}

\begin{figure}
\begin{center}
\includegraphics[width=.6\textwidth]{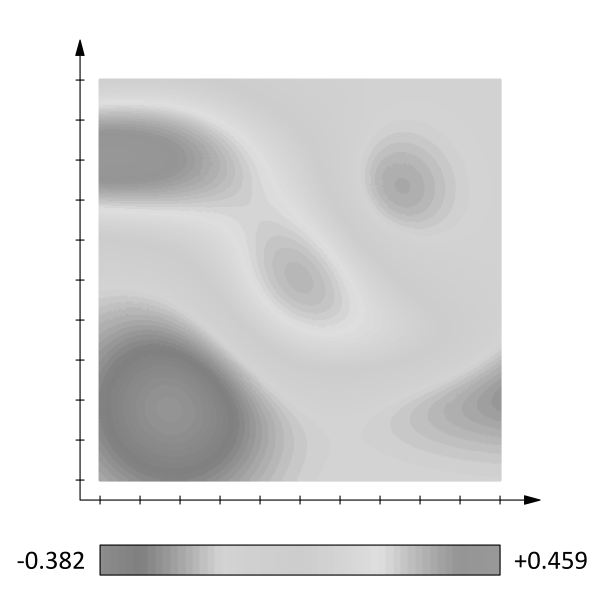}
\caption[A scalar field]{A scalar field is described by a single value at each point in space. (Image from \citep{Wikimedia})}
\label{fig-ScalarField}
\end{center}
\end{figure}

A scalar field is a simple idea often introduced in elementary physics by thinking about a temperature field in a room. The field is a scalar in the sense that it has a value at each point in space which can be described by a single real number, unlike, say, a vector field which requires a magnitude and direction to be fully specified. One expects the field to vary continuously across the space and it is possible to plot its variation in any chosen direction. See figure \ref{fig-ScalarField} for an illustration.

However, temperature is not a \emph{fundamental} scalar field, it is simply a macroscopic property of space at each point, determined by other factors such as the proximity of heat sources. When a physicist talks about fields (scalar or otherwise), they usually have in mind something more abstract - a fundamental field of nature, which takes a value at each point in space and may couple to other fields. In quantum field theory (QFT), particles like electrons are localised fluctuations in these fundamental fields, and particle collisions create new particles because they transfer energy to, and thus excite fluctuations on, other coupled fields. Scalar fields are called spin zero fields because they are invariant under a Lorentz transformation (they transform under the trivial (0,0) representation of the Lorentz group).

In this section we consider some examples of scalar fields and their applications, in particular the two applications considered in this thesis - critical collapse and cosmology.

\subsection{Scalar fields and scalar potentials}
\label{sec-SFPotentials}

The Higgs field is currently believed to be the only truly fundamental scalar field which has been observed in nature, \citep{EnglertBrout1964, Higgs1964, Kibble1964}. However, it is possible that other fundamental scalar fields exist which were active in the early universe, but now lie dormant as the average energy density has decreased to such an extent that there is no longer sufficient energy to excite them. A candidate for such a field is the inflaton, which plays a key role in the theory of inflation, as discussed further in section \ref{sec-SFCosmology} below.

Scalar fields are also useful in effective theories, where they may describe the low energy behaviour of more fundamental degrees of freedom. For example, the Landau-Ginzberg model \citep{Ginzburg:1950sr}, which describes the dynamics of ``Cooper pairs'' (pairs of electrons with opposite spins) in conventional superconductivity, is equivalent to (and in fact preceded) the Albelian Higgs model, with the Cooper pairs being treated as a single scalar particle. Similarly in particle physics, pi mesons (``pions'') are described at low energies as a scalar particle, despite being composite particles made up of two quarks.

Finally, scalar fields often provide a simple toy model for understanding the behaviour of more general fields in more complicated scenarios, such as those in which they are coupled to strong gravity, and also for unusual gravity effects, such as Critical Collapse, introduced in section \ref{sec-SFCriticalCollapse} below.

\begin{figure}
\begin{center}
\includegraphics[width=.8\textwidth]{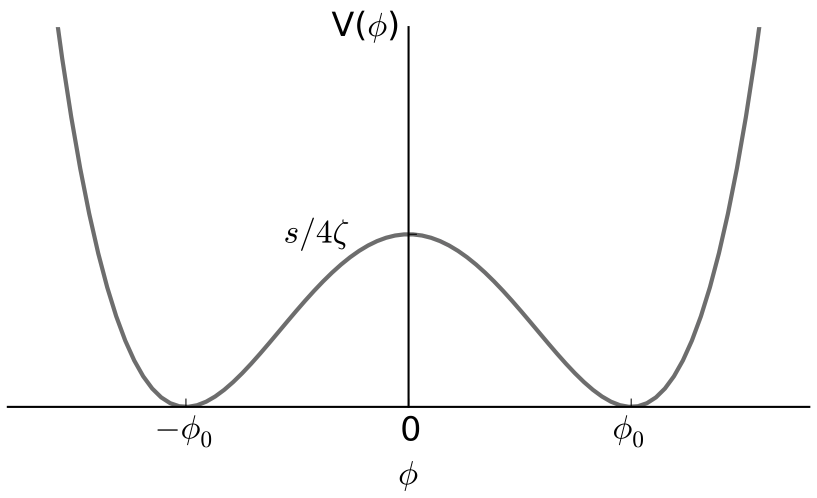}
\caption[A potential function]{A $\phi^4$ potential for a scalar field $\phi$. Each possible value of $\phi$ has an associated potential value (an energy density). The form of the potential affects its evolution, and, in the absence of strong gradients in space and time, the field will tend to fall to the local minimum in the potential, just like a ball in a hilly landscape.}
\label{fig-PotentialSF}
\end{center}
\end{figure}

The equation of motion for a scalar field $\phi$ in flat space, subject to a scalar field potential $V(\phi)$ is the Klein Gordon equation, which can be written as
\begin{equation}
- \frac{\partial^2 \phi}{\partial t^2} + \frac{\partial^2 \phi}{\partial x^2} + \frac{\partial^2 \phi}{\partial y^2} + \frac{\partial^2 \phi}{\partial z^2} = \frac{dV(\phi)}{d \phi} . \label{eqn:KGFlat}
\end{equation} 
The term $V(\phi)$ (with the exception of any terms in $\phi$ or $\phi^2$) results in a non linear self interaction of the field. That is, for a non trivial potential $V(\phi)$, two plane waves in the field will not simply superpose but will interact in a non trivial way. The form of $V(\phi)$ can be thought of as a property of the field - in the case where $V(\phi) = \onehalf m^2 \phi^2$ then $m$ can be identified with the ``mass'' of the field. In more complicated forms it still determines how the field propagates, but in a more involved manner. The key point is that the field has a tendency to want to fall to the minima of the potential, and then stay there unless excited. It can thus have a strong effect on how the field evolves. The shape of the potential for a field must be assumed, or derived from some higher energy theory in the case where the field is only an effective description. Multi minima potentials are thought to arise in the low energy effective theories of several string theories, but one would need an exact model to be able to derive their form. An example of a potential is shown in figure \ref{fig-PotentialSF}.

\nomenclature[g-pi]{$\phi$}{Scalar field}
\nomenclature[a-pi]{$V(\phi)$}{Scalar field potential}
\nomenclature[a-pi]{$m$}{The field mass, in an $m^2\phi^2$ potential}

In both the cases of fundamental scalar fields mentioned - the Higgs field and the inflaton field - the shape of the potential is essential in determining its behaviour and properties. One must take care to distinguish the motion of the field in the potential from the motion of the field in physical space. When we consider the motion of the field in the potential, we are considering only a single point and its field value, and looking at the corresponding value and slope of the potential at that point. The evolution of that point in physical space will be determined by \eqn{eqn:KGFlat}, which combines both its tendency to ``roll downhill'' in potential space, and the effect of its spatial gradients in physical space, which tend to pull it into a flatter spatial configuration.

In this thesis and in the code we have developed, the behaviour described is entirely classical. We consider only classical scalar fields and classical effects, and not quantum ones, although we know that all fields are fundamentally described by QFT.  In effect, the field value being evolved is the expectation value of the field operator, and the approach assumes that the quantum field is in an approximately coherent state. For example, we cannot model the quantum tunnelling between minima which may result in the bubble solutions described in Chapter \ref{ch-CriticalCollapse}, although we can take the tunnelling solutions as an initial condition and evolve forward classically. Equally, we cannot model the propagation of individual particles - our modelling of the field as a purely classical one is only valid in the limit where occupation number in the underlying field is high, and/or the wavelength of the fluctuations in the classical field are much larger than the compton wavelength of the quanta of the field. In post-inflationary cosmology this is usually the case - quantum effects are almost always negligible in comparison to the effects of gravity which dominate over larger scales. In smaller scale problems, such as axion stars, or during inflation where quantum fluctuations are ``blown up'' to larger scales, one must be more careful to consider whether quantum effects are relevant.

\subsection{Scalar fields in cosmology}
\label{sec-SFCosmology}

\begin{figure}
\begin{center}
\includegraphics[width=.6\textwidth]{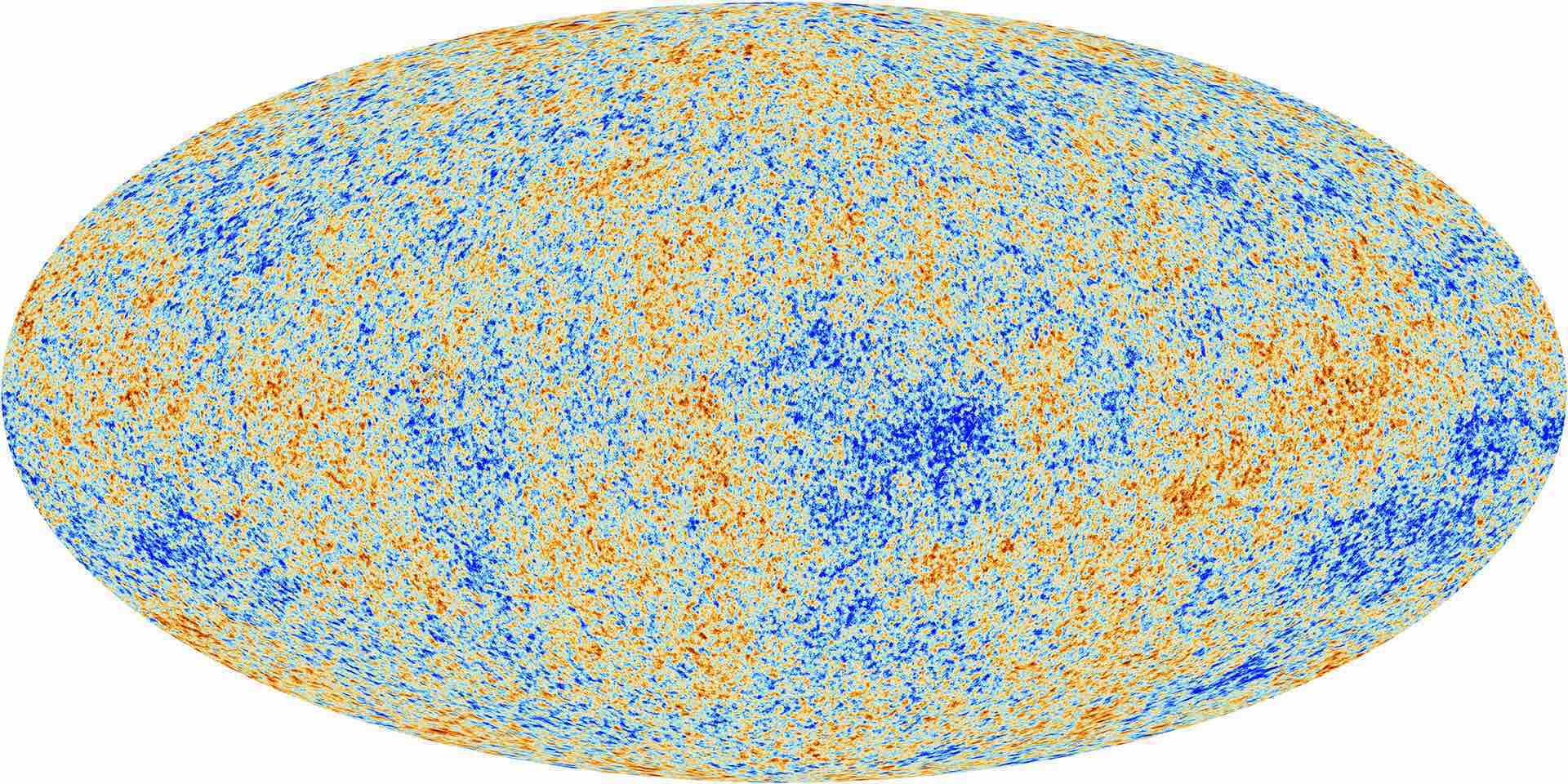}
\caption[Cosmic Microwave Background radiation]{The image shows the temperature anisotropies of the Cosmic Microwave Background (CMB) radiation as observed by Planck. The CMB is the oldest observable light in our Universe, emitted from the last scattering surface. The temperature fluctuations correspond to  the density variations which provided the seeds of future structure - stars and galaxies - in the Universe. (Image from ESA/Planck Collaboration)}
\label{fig-CMB}
\end{center}
\end{figure}

As discussed above, one can obtain an analytic solution to Einstein's equation for the universe as a whole if one assumes a space which is homogeneous and isotropic, and filled with some kind of fluid matter. The result is an expanding space which proves to be a good description of our universe on larger scales, if a certain matter content is assumed - the FRW spacetime. In particular, the model is consistent with observations of the Cosmic Microwave Background (CMB) radiation, which is light emitted from the last scattering surface, after recombination of the hot plasma into neutral atoms. The CMB data from the Planck and WMAP satellites, see figure \ref{fig-CMB}, provides an enormous amount of information and has led to an age of ``precision cosmology''. 

However, whilst the FRW model and cosmological data explain many things, they also raise a number of questions, one of which is, why does the universe look so similar in all directions? If the Universe is simply ``rewound'' using the FRW model, it is clear that opposite sides of the observable universe could not have been in causal contact when the CMB light was emitted. Thus, assuming they started out with some random configuration (which is what physicists tend to assume), they should look very different from each other now. This is not the case, which implies that the model is incomplete - casual contact must have occurred at some point\footnote{Causal contact tends to smooth out differences, as regions in contact equilibrate over time - think of putting two tanks of water at different temperatures in contact, side by side - after some time all the water will be at a constant temperature everywhere.}. 

\begin{figure}
\begin{center}
\includegraphics[width=.8\textwidth]{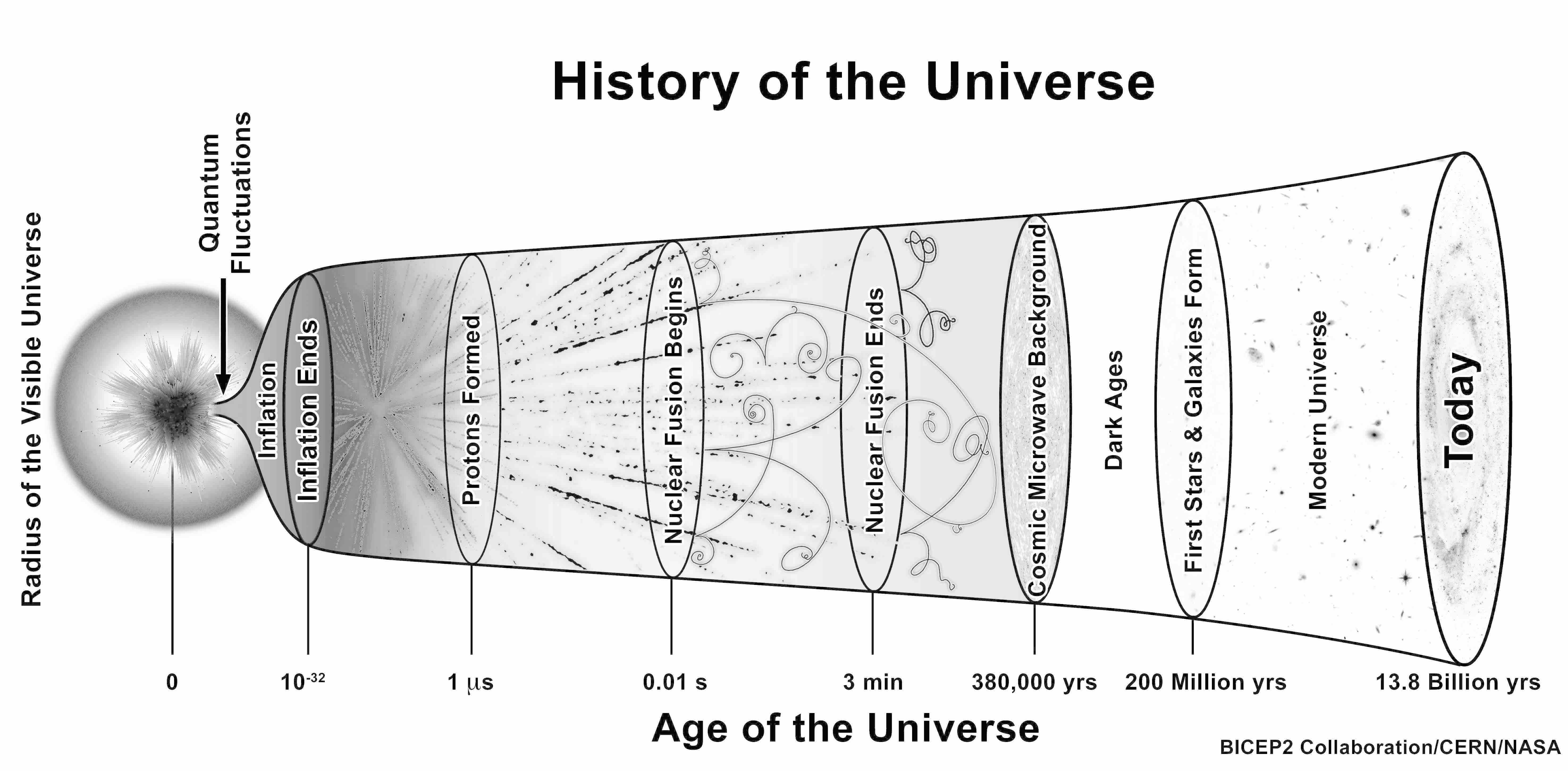}
\caption[History of the Universe]{The image shows our current understanding of the timeline for the evolution of the Universe. Inflation is an early phase in which the Universe undergoes a period of accelerated expansion. (Image from \citep{BICEPwebsite})}
\label{fig-HistoryUniverse}
\end{center}
\end{figure}

A solution to this problem is inflation, first proposed by Guth and Starobinsky, and later updated by Linde, and, independently, Albrecht and Steinhardt \citep{Guth:1980zm,Linde:1981mu,Albrecht:1982wi,Starobinsky:1980te}. The theory also usefully explains the scarcity of magnetic monopoles and why the universe is flat on large scales. Inflation is a period of superluminal expansion in the early universe, which would allow distant regions to have been causally connected in the past. An illustration of the proposed history of the Universe, with a period of inflation at the beginning, is shown in figure \ref{fig-HistoryUniverse}. One possible source of such an expansion is a scalar field subject to a particular form of scalar potential. This theoretical scalar field is commonly referred to as the inflaton, and there are many possible models proposed for its behaviour.

Such inflationary models are well studied in the homogenous case, and in the perturbative regime. However, they are not well studied in cases where there are large variations in the initial conditions, such as large fluctuations in the value of the scalar field throughout space. If one wants to explain how random fluctuations can be eliminated via inflation, one should show that one can \emph{start inflation} with a truly random configuration in all variables, and still achieve the same homogeneous result. Otherwise we are really back to square one, as we now need to explain how to obtain a homogenous starting point for inflation to begin with.

Further technical details of FRW cosmology and inflation relevant to the current work are given in the next chapter in section \ref{sec-Cosmology}. In Chapter \ref{ch-Inflation} of this thesis, we complete a study of a class of inhomogeneous initial conditions, and their effects on inflation, considering the robustness of different inflationary models to perturbations in the field, and to non uniform initial expansion.

\subsection{Critical collapse of scalar fields}
\label{sec-SFCriticalCollapse}

\begin{figure}
\begin{center}
\includegraphics[width=.8\textwidth]{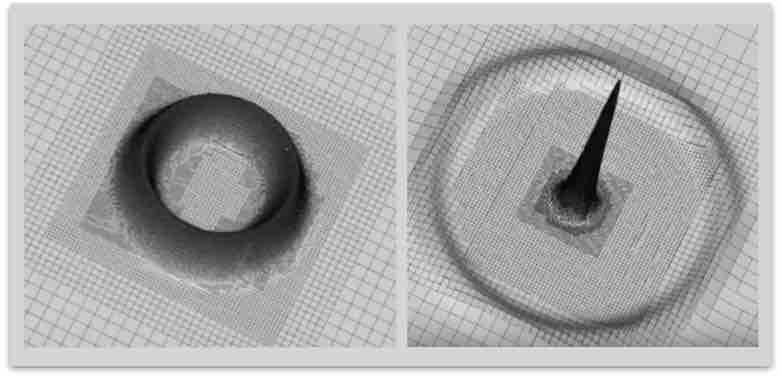}
\caption[Critical Collapse]{An illustration of a scalar field critical collapse. The image shows a 2D slice through the 3D spatial hypersurface, and the elevation is the value of the field at each point. The initial perturbation splits into an ingoing and outgoing spherical wave. If there is sufficient energy in the collapse, a black hole will form.}
\label{fig-CriticalCollapse}
\end{center}
\end{figure}

In a 4-dimensional spacetime, for any one parameter, $p$, family of initial configurations of a scalar field, the end state will be either a black hole or the dispersal of the field to infinity. The transition between these two end states occurs at a value of the parameter $p^*$, at which the critical solution exists. An illustration of a critical collapse is shown in figure \ref{fig-CriticalCollapse}, in which is a gaussian bump in a spherical shell (which appears as a ring in a 2D slice) collapses inwards. The parameter $p$ could be the initial height of the bump, or its radius. When $p$ is small the bump will collapse inwards and then disperse. As $p$ is increased, we are adding more energy into the gradient in the walls, and eventually we will have added a sufficient amount that on collapse a black hole will form. The value of the parameter at this point is $p^*$. This was almost exactly the procedure followed by Choptuik in his 1992 study \citep{Choptuik:1992jv}, and whilst this result may seem rather obvious, his studies revealed other behaviour near this critical point which was not.

Firstly, in a spherically symmetric collapse, the mass $M$ of any black hole that is formed close to the critical point follows the relation 
\begin{equation}
M \propto (p - p^*)^{\gamma_S},
\end{equation}
where the scaling constant $\gamma_S$ is universal in the sense that it does not depend on the choice of family of initial data - $p$ may be the initial height, width or any other scale which may be varied in the initial data. This phenomenon of universality implies that one can tune a black hole mass to zero, in theory creating a naked singularity in breach of the cosmic censorship conjecture. 

\nomenclature[a-pi]{$M$}{The mass of a black hole or compact object}
\nomenclature[a-pi]{$p$}{The critical collapse parameter, with critical value $p^*$}
\nomenclature[g-pi]{$\gamma_S$}{in critical collapse, the scaling exponent}
\nomenclature[g-pi]{$\Delta_S$}{in critical collapse, the scale-echoing constant}

The other key phenomenon observed is that of self-similarity in the solutions, or ``scale-echoing''. Close to the critical point, and in the strong field region, the fields are subject to a scaling relation in which, as the time nears the critical time, the same field profile is seen but on a smaller spatial scale. This scale-echoing may be either continuous or discrete, but the factors leading a system to either case are not well understood.  

Whilst spherically symmetric configurations have been well-studied analytically and numerically, axisymmetric and fully asymmetric configurations are much less well understood due to the high resolutions required to resolve the scale echoing. 

Further technical details of critical collapse are given in section \ref{sec-CriticalCollapse} of the next chapter. Chapter \ref{ch-CriticalCollapse} of this thesis presents work on the critical collapse of non spherically symmetric scalar field ``bubbles'' - solutions which interpolate between two minima in a $\phi^4$ potential. 

%% file: Chapter2/chapter2.tex

\chapter{Technical background}  
\label{ch-Technical}

\ifpdf
    \graphicspath{{Chapter2/Figs/Raster/}{Chapter2/Figs/PDF/}{Chapter2/Figs/}}
\else
    \graphicspath{{Chapter2/Figs/Vector/}{Chapter2/Figs/}}
\fi

In this chapter the key topics covered by the thesis are explored in more technical detail. We follow the theoretical steps in formulating a numerical evolution, and the background to the specific problems studied. Discussion of the implementation aspects of the numerical evolution are left to the following chapter in which the code which was developed is described. As in Chapter \ref{ch-Introduction}, we divide this chapter into three sections, GR, NR and Scalar Fields. 
\begin{itemize}
\item Section \ref{sec-GRTheory} concerns GR generally, and aims to summarise the Einstein Equation, its key geometric components and their physical interpretation from a geometric and a Lagrangian perspective.
\item Section \ref{sec-NRTheory} explains the key issues encountered in the numerical formulation of GR as a $3+1$D Cauchy problem which can be implemented and solved on a computer, including the ADM decomposition, numerical stability and gauge issues.
\item Section \ref{sec-SFTheory} discusses scalar fields coupled to gravity, and the specific problems of Inflation and Critical Collapse for which the research presented in Chapters \ref{ch-Inflation} and \ref{ch-CriticalCollapse} was undertaken.
\end{itemize}

\nomenclature[z-pi]{GHC}{Generalised Harmonic Coordinates}
\nomenclature[z-pi]{CTT}{Conformal Transverse Traceless}
\nomenclature[z-pi]{CTS}{Conformal Thin Sandwich}
\nomenclature[z-pi]{DM}{Dark Matter}
\nomenclature[z-pi]{DE}{Dark Energy}
\nomenclature[z-pi]{SEC}{Strong Energy Condition}
\nomenclature[z-pi]{CSS}{Continuous Self Symmetry}
\nomenclature[z-pi]{DSS}{Discrete Self Symmetry}
\nomenclature[z-pi]{CS}{Critical Surface}
\nomenclature[z-pi]{CP}{Critical Point}
\nomenclature[z-pi]{KE}{Kinetic Energy}

\nomenclature[g-pi]{$\epsilon_{ijk}$}{the three dimensional Levi-Civita symbol}
\nomenclature[g-pi]{$\delta^i_j$}{the Kronecker delta}
\nomenclature[g-pi]{$\eta_{ab}$}{the Minkowski metric}


\section{GR - key theoretical concepts} 
\label{sec-GRTheory}

In this section we aim to summarise the formulation of the Einstein equation, and highlight the key concepts which will be important in the numerical formulation. This is not intended to be a complete treatment of the subject of GR, and the reader is referred to a standard textbook on GR for further detail. In particular, the books by Schutz \citep{SchutzGR} and Carroll \citep{CarrollBook} give detailed and thorough introductions to the subject, whilst Wald \citep{wald1984general} is the key reference for more advanced topics, or as a concise reference.

In this section all references to the metric and its derived objects refer to the 4-dimensional versions, for example $R= {}^{(4)}R$. We will always assume a coordinate basis, which means that the basis vectors are defined as the tangents to coordinate curves. The result is that such basis vectors commute and the Christoffel symbols are symmetric\footnote{Schutz gives a good description of non coordinate bases in both his books \citep{SchutzGeo}, \citep{SchutzGR}, in particular there is a useful example in the latter which shows that the often-used unit vectors in (flat space) polar coordinates are not in fact a coordinate basis, which has consequences for tensor calculus.}. 

Note that we will include any cosmological constant contribution to the Einstein equation in the stress-energy tensor, rather than stating it separately, which effectively means that it is treated as a fluid which violates the strong energy condition (``SEC''). This corresponds to the treatment in Chapter 4, in which the inflaton scalar field sources the cosmological constant for inflation.

\subsection{Geometric preliminaries}
\label{sec-Geometry}

\subsubsection{Manifolds and metrics}

As stated in Chapter \ref{ch-Introduction} , an n-dimensional manifold may be thought of as a smooth and continuous surface. More explicitly, it is a set which at each point is homeomorphic to an n-dimensional Euclidean space, and may be continuously parameterised (locally at least) by some coordinates that can be mapped to the reals $\mathbb{R}^n$. 

\nomenclature[x-pi]{$\mathbb{R}^n$}{the real numbers}
\nomenclature[g-pi]{$\lambda$}{the affine parameter of a curve}
\nomenclature[g-pi]{$\tilde{d}\phi$}{an arbitrary one-form}

The differentiability of the manifold with reference to the local coordinates means that vectors can be defined as tangents to local curves, with components in some basis $d x^\mu / d \lambda$ where $\lambda$ parameterises the curve, and one-forms can be defined as linear, real valued functions of these vectors. There is a duality in the definition such that we can equally define a one form $\tilde{d}\phi$ as the geometric object with components $\partial \phi / \partial x^\mu$ in some basis (i.e. the gradient of a scalar function), and a vector as a linear, real valued function of the one form. The vector takes the one form (or vice versa) into the derivative of the scalar function along the curve to which it is tangent, ie 
\begin{equation}
\tilde{d}\phi(\vec{V}) = \vec{V}(\tilde{d}\phi) =  \frac{\partial \phi}{\partial x^\mu} \frac{d x^\mu}{d \lambda} = \frac{d\phi}{d\lambda} \, .
\end{equation}
The spacetime of GR is a pseudo-Riemannian manifold\footnote{The pseudo in pseudo-Riemannian means that the metric is not positive definite, ie $g_{ab} V^a V^b \ngtr 0$ $\forall ~ \vec{V}$, which is obviously very important physically as it is due to the minus sign associated with the time direction, but does not make a big difference to our discussion here of geometric properties.}, meaning that in addition to the above manifold coordinate structure, one has specified a metric, $g_{ab}$, which is a rank 2 tensor, at each point. This is an additional piece of information which defines the local distance $ds$ on the manifold, when an infintesimal (vector) step $\vec{dx}$ is taken
\begin{equation}
ds^2 = g_{ab} dx^a dx^b  \label{eqn:metric} \, .
\end{equation}
It also serves to define a one-to-one mapping between vectors and one-forms, such that the one form dual $\tilde{V}$ to the vector $\vec{V}$ is defined as
\begin{equation}
V_a = g_{ab} V^b  \label{eqn:dualspace} \, .
\end{equation}
In our Universe, a key feature of the spacetime manifold is that the metric has three positive eigenvalues and one negative eigenvalue, such that its signature is $3 - 1 = +2$, and that the metric is symmetric. 

This distinction between the coordinate labelling of the manifold and the physical distances is a very important point in GR, and becomes even more relevant when working with simulations in NR. In SR the metric is a constant everywhere in spacetime and equal in a cartesian basis to
\begin{equation}
g_{\mu \nu} = \eta_{\mu \nu}  \equiv \rm{diag}(-1,1,1,1) \label{eqn:SRmetric} \, .
\end{equation}
This means that distances are determined by the Pythagorean rules of flat space (ignoring the complications of the minus sign) and our coordinates will be linked directly to physical distances as measured by the observer in that frame. However, in GR this is no longer the case - the metric varies from place to place and the coordinates $x^\mu$ which we impose are simply an arbitrary labelling, embodying the gauge freedom which is exactly the principle of general covariance. Taken in isolation, the coordinates tell us simply how the spacetime is connected, so that $x=2$ is somewhere ``between'' $x=1$ and $x=3$, but the actual distances between the points are not necessarily equal to $1$. We require knowledge of the metric to understand the physical quantities - proper distances, times and volumes - which would be measured by an observer, according to \eqn{eqn:metric}.
 
Equivalently, the metric of GR is a geometric object which takes a value at each point on the 4 dimensional manifold. Expressed in some basis, it is a set of 10 quantities (it is symmetric), and is the fundamental object which is used to describe the curvature of the spacetime manifold. 

\subsubsection{Curvature and the Einstein Equation}

As was stated in Chapter \ref{ch-Introduction}, the interplay between matter and curvature is summarised by Einstein's field equation, an inherently local equation relating the Einstein curvature tensor $G_{ab}$ to the Energy-Momentum (EM) tensor $T_{ab}$ at each and every point in the spacetime

\begin{equation}
G_{ab} \equiv R_{ab} + \onehalf R g_{ab} = 8 \pi T_{ab}  \label{eqn:EEM} \, .
\end{equation}
The left hand side, $G_{ab}$, encodes the curvature, which is completely determined by specifying the metric across the spacetime (terms like $R_{ab}$ just being a shorthand to represent some convenient combinations of the metric and its derivatives, which will be defined below).

On the right hand side, the EM tensor $T_{ab}$ is usually defined in words in its raised component form $T^{\mu\nu}$, as ``the flux of four-momentum $p^\mu$ across a surface of constant $x^\nu$”. Its form depends on the type of matter - for example, for a perfect fluid with energy density $\rho$ and pressure $P$ the components measured by an observer with 4-velocity $U^\mu$ are
\begin{equation}
T^{\mu \nu} = (\rho + P) U^\mu U^\nu + P g^{\mu \nu} \label{eqn:EMTFluid} \, .
\end{equation}

\nomenclature[g-pi]{$\rho$}{energy density}
\nomenclature[a-pi]{$P$}{pressure (of a fluid, say)}
\nomenclature[a-pi]{$U^a$}{4-velocity (of a fluid, say)}
\nomenclature[a-pi]{$p^a$}{4-momentum (of a fluid, say)}

Returning to the curvature part, we define a particularly useful combination of the metric and its gradients, the Riemann curvature tensor, as
\begin{equation}
R^a_{\ bcd} = \partial_c \Gamma^a_{bd} - \partial_d \Gamma^a_{bc} + \Gamma^a_{ec} \Gamma^e_{bd} - \Gamma^a_{ed} \Gamma^e_{bc} \, , \label{eqn:RiemannTensorU} \, ,
\end{equation}
where we have implicitly chosen the torsion-free ``Levi-Civita'' or ``metric'' connection in our definition, as is usual in GR. The Christoffel symbols (which are not tensors) are the components of the Levi-Civita connection in some basis, and can be expressed in terms of derivatives of the metric as
\begin{equation}
\Gamma^\rho_{\mu\nu} = \onehalf g^{\rho\sigma} \left(  \partial_\mu g_{\sigma \nu} +  \partial_\nu g_{\mu \sigma} -  \partial_\sigma g_{\mu \nu}  \right)  \label{eqn:Christoffels} \, .
\end{equation}
In effect, the Christoffel symbols describe how the basis vectors change from place to place on the manifold. If we choose a locally flat inertial frame, in which the Christoffel symbols (but not their derivatives) vanish, the components of the Riemann tensor can be written in a lowered form as
\begin{equation}
R_{\alpha \beta \gamma \delta} = \onehalf (\partial_\beta \partial_\gamma g_{\alpha \delta} - \partial_\beta \partial_\delta g_{\alpha \gamma} + \partial_\alpha \partial_\delta g_{\beta \gamma} - \partial_\alpha \partial_\gamma g_{\beta \delta} ) \, ,  \label{eqn:RiemannTensorL} 
\end{equation}
which makes explicit their dependence on the second derivatives of the metric, and the many symmetries (which must hold in all bases, as they can be expressed as valid tensor equations such as $R_{abcd} = - R_{bacd}$, see for example Schutz \citep{SchutzGR}).

\nomenclature[a-pi]{$R^a_{~bcd}$}{Riemann curvature tensor}
\nomenclature[g-pi]{$\Gamma^\rho_{\mu \nu}$}{4-dimensional Christoffel symbols}
\nomenclature[g-pi]{$\nabla_a$}{4-dimensional covariant derivative}

\begin{figure}
\begin{center}
\includegraphics[width=.4\textwidth]{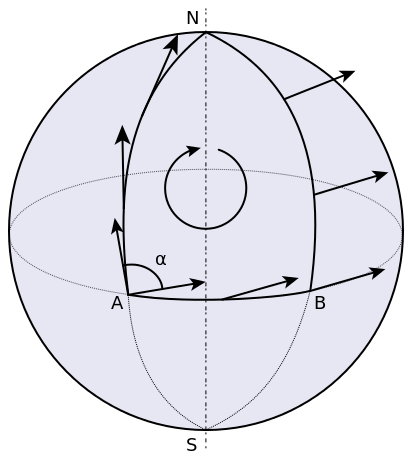}
\caption[Parallel Transport]{The Riemann Tensor describes the change in direction of a vector after it is parallel transported around a closed loop in the manifold. One can see that when a vector is moved around a loop on a curved surface, this change is non zero. Image from \citep{Wikimedia}.}
\label{fig-ParallelTransport}
\end{center}
\end{figure}

It can be shown that the Riemann tensor has two interpretations. Firstly, it defines the change in direction of a vector as it is parallel transported around a closed curve (see figure \ref{fig-ParallelTransport}). Explicitly, the change in the vector component $\delta V^\alpha$ will be 
\begin{equation}
\delta V^\alpha = R^\alpha_{~ \beta \gamma \delta} dx^\gamma dx^\delta V^\beta  \label{eqn:RiemannPT} \, .
\end{equation}
Equivalently, the Riemann tensor is the commutator of the covariant derivative acting on a vector, ie
\begin{equation}
\nabla_c \nabla_d V^a - \nabla_d \nabla_c V^a = R^a_{~ bcd} V^b  \label{eqn:RiemannCommutator} \, .
\end{equation}
Note that whilst covariant derivatives of scalars commute, on a curved manifold covariant derivatives of vectors do not.

The quantities appearing in the Einstein Equation, the Ricci tensor $R_{ab}$ and its trace, the Ricci scalar
\begin{equation}
R = Tr(R_{ab}) = g^{ab} R_{ab} = R^a_a  \label{eqn:RicciScalar} \, ,
\end{equation}
are defined by the contraction of the Riemann tensor\footnote{Why this contraction between the first and third indices, rather than others? One can show that any other contraction is either equal to zero or $\pm R_{ab}$ due to the symmetries of the Riemann tensor, so it is in effect the only one possible. We will also see in \ref{sec-derivingEEgeo} why this contraction is relevant in relation to tidal forces.}
\begin{equation}
R_{ab} = R^c_{\  acb}  \label{eqn:RicciTensor} \, .
\end{equation}
The result of these relations is that the curvature term appearing on the left hand side of the Einstein field equation is a (quite complex) non linear combination of the metric and its first and second derivatives with respect to space and time. 

Note that if all components of the Riemann tensor are zero, the space is flat. The same is not true of the Ricci tensor or Ricci scalar, which may be zero in a curved space, leading to non trivial solutions, even when $T_{ab} = 0$, the so called ``vacuum solutions''.

In the next section, we use these geometric ideas to motivate the Einstein equations, as was described qualitatively in Chapter \ref{ch-Introduction}, by relating the separation of neighbouring particles due to tidal forces to movement on a curved manifold. 

\subsection{The Einstein Equation from geometric principles}
\label{sec-derivingEEgeo}

At the end of section \ref{sec-GREquivalencePrinciple}, we stated the following:
\begin{displayquote}
\textit{We know from Newtonian physics how matter gives rise to tidal forces which pull objects apart. If we can generate their observed effect - the way in which two separated objects move apart in the field - with a spacetime curvature instead, we can eliminate these fictitious forces from our equation altogether. We will have the desired relation to replace Newtonian gravity - a link between matter and spacetime curvature.}
\end{displayquote}
Here we proceed with this interpretation, having developed the machinery we need in the previous section to describe the effects of curvature, in particular, the Riemann tensor. 

Consider a single particle at a position $x_1$ falling freely in a gravitational field, with four-velocity $\vec{U}$. For a Newtonian gravitational potential $\Psi$ the acceleration arises from the potential gradient,
\begin{equation}
\frac{d\vec{U}}{dt} = - \vec{\nabla} \Psi  \label{eqn:dvdtNWT} \, .
\end{equation}
The equivalent statement in GR is the geodesic equation, which can be written (with proper time $\tau$ as the affine parameter)
\begin{equation}
\frac{dU^\alpha}{d\tau} = - \Gamma^\alpha_{\beta \gamma} \frac{dx^\beta}{d\tau} \frac{dx^\gamma}{d\tau} \label{eqn:dvdtGR} \, .
\end{equation}
Thus we can see that the Christoffel symbols act as $\vec{\nabla} \Psi$. The statement that we can find a local frame in which the gravitational force disappears is equivalent to the statement that we can find a local frame in which the Christoffel symbols are zero. 

\nomenclature[g-pi]{$\Psi$}{the Newtonian gravitational potential}
\nomenclature[g-pi]{$\tau$}{proper time (in cosmology, as measured by a comoving observer)}

\begin{figure}
\begin{center}
\includegraphics[width=.6\textwidth]{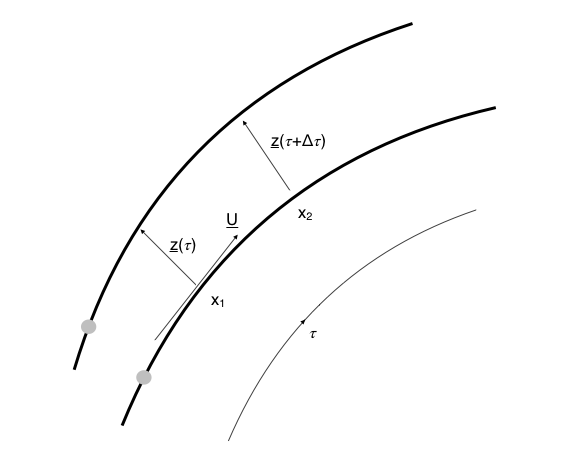}
\caption[Geodesic Devation]{Two particles are separated by $\vec{z}$ as they fall freely (along geodesics) in a gravitational field. Due to the curvature of spacetime their separation will change as they advance along the curves.}
\label{fig-GeodesicDeviation}
\end{center}
\end{figure}

If we introduce a second particle at $x_2$, which is slightly separated from the first but also falling freely, and define the separation between two point particles $\vec{z}$ as $z^\alpha = x_2^\alpha - x_1^\alpha$ (see figure \ref{fig-GeodesicDeviation}), then the tidal acceleration is given by Newton as
\begin{equation}
\frac{d^2\vec{z}}{dt^2} = - (\vec{z} . \vec{\nabla}) \vec{\nabla} \Psi  \label{eqn:tidalaccelerationN} \, .
\end{equation}
Considering, equivalently, the motion of particles in a curved space, which we assume would follow geodesics, one can show (see for example Schutz \citep{SchutzGR}) that the equation of geodesic deviation is
\begin{equation}
\frac{d^2z^\alpha}{d\tau^2} = - z^\beta R^\alpha_{~  \mu\beta\nu} U^\mu U^\nu  \label{eqn:tidalaccelerationGR} \, .
\end{equation}
Comparing these two we make the connection\footnote{Although we are cheating a bit since the $\nabla$ in the Newtonian case is the 3 dimensional spatial gradient and not a four dimensional quantity. We should really show that the time components do not contribute in some chosen frame and then generalise from a tensor equation.} that
\begin{equation}
z^\beta \nabla_\beta \nabla^\alpha \Psi \leftrightarrow  z^\beta R^\alpha_{\  \mu \beta \nu} U^\mu U^\nu  \label{eqn:equateacceleration} \, ,
\end{equation}
and hence
\begin{equation}
\nabla^2 \Psi \leftrightarrow R^\alpha_{\  \mu \alpha \nu} U^\mu U^\nu  \label{eqn:nabla2psi} \, .
\end{equation}
The Newtonian potential is sourced by the mass density $\rho$ according to the Poisson equation
\begin{equation}
\nabla^2 \Psi = 4 \pi \rho  \label{eqn:newtonFE} \, .
\end{equation}
We already have our ``GR'' version of the left hand side in \eqn{eqn:nabla2psi}. For the right hand side, given the definition of the EM tensor, the energy density measured by an observer moving along the geodesic is:
\begin{equation}
\rho = T_{\mu \nu} U^\mu U^\nu \label{eqn:GRrho} \, .
\end{equation}
Combining these results and requiring that they are true for all $\vec{U}$ gives us
\begin{equation}
R^\alpha_{~  \mu \alpha \nu} = R_{\mu \nu} = 4 \pi T_{\mu \nu}  \label{eqn:GRSimple} \, .
\end{equation}
This is clearly close to \eqn{eqn:EEM} but not quite right, as we are missing a factor of 2 and the Ricci scalar contribution. However, at this point note that the inability to make tidal forces disappear in any frame is directly connected to having a non zero Riemann tensor, as expected.

The problem we have is that physically we know that the EM tensor on the right hand side $T^{ab}$  is divergenceless, and thus so should the Ricci Tensor be. It happens that this puts big restrictions on what form $R_{ab}$ can take. To solve this problem, $R_{ab}$ is replaced by the Einstein Tensor $G_{ab} \equiv R_{ab} - \onehalf R g_{ab}$, for which $G^{ab}$ is divergenceless as an identity (see for example Schutz \citep{SchutzGR}). The factor of two then comes in so as to recover the correct Newtonian limit. This divergenceless property of the Einstein Tensor is very important, and gives rise to the Bianchi Identities
\begin{equation}
\nabla_a G^{ab} = 0  \label{eqn:BianchiIdentities} \, .
\end{equation}

\subsection{The Einstein Equation from action principles}
\label{sec-derivingEEaction}

The form of the Einstein field equation \eqn{eqn:EEM} can be derived in several ways. The fact that there are many consistent ways in which it can be reached is part of the elegance of the theory. 

\nomenclature[a-pi]{$S_G$}{the Einstein-Hilbert action}
\nomenclature[a-pi]{$\mathcal{L}$}{Langrangian density}
\nomenclature[a-pi]{$\mathcal{L}_G$}{Langrangian density for the Einstein-Hilbert action}
\nomenclature[a-pi]{$\mathcal{L}_M$}{Langrangian density for a matter field}
\nomenclature[a-pi]{$\mathcal{L}_{SF}$}{Langrangian density for a scalar field}
\nomenclature[a-pi]{$g$}{determinant of the four dimensional spacetime metric}
\nomenclature[a-pi]{$\mathcal{M}$}{a manifold}

An alternative to the geometric approach is the minimisation of the Einstein-Hilbert action
\begin{equation}
S_G[g^{ab}] = \int d^4 x \sqrt{-g} R \, . \label{eqn:Hilbertaction}
\end{equation}
where $g$ is the determinant of the four dimensional spacetime metric, and its (negative) square root encodes the dependence of the volume element on the metric. The action $S_G$ can be considered as a map from a certain field configuration (of $g_{ab}$) on a manifold $\mathcal{M}$ into the real numbers $\mathbb{R}$. The integrand is the Lagrangian density for GR
\begin{equation}
\mathcal{L}_G =  \sqrt{-g} R \, . \label{eqn:LagrangianG}
\end{equation}
which excluding the volume factor $\sqrt{-g}$ is simply the Ricci scalar $R$. Since this is the only non trivial scalar one can obtain from contractions of the Riemann tensor, it is the obvious choice for the scalar Lagrangian.

Taking the functional derivative of this action with respect to the inverse of the metric (and assuming zero surface terms\footnote{This will be correct if the change in the metric $\delta g^{ab}$ and its derivatives go to zero at infinity}) gives
\begin{equation}
\frac{\delta S[g^{ab}]}{\delta{g^{ab}}} = \sqrt{-g} (R_{ab} -\frac{1}{2} R g_{ab} ) \, . \label{eqn:actionderiv}
\end{equation}
we see that minimisation of the action leads directly to the (vacuum) Einstein field equation:
\begin{equation}
R_{ab} -\frac{1}{2} R g_{ab} = 0 \, . \label{eqn:EEV}
\end{equation}

Including an energy-momentum source provides an alternative definition of the EM tensor in terms of the minimisation of a matter action. One defines a new Lagrangian density as
\begin{equation}
\mathcal{L} = \mathcal{L}_G + C_M \mathcal{L}_M \, . \label{eqn:Lagrangian}
\end{equation}
where $C_M$ is a constant which depends on the energy momentum source, being $16\pi$ for scalar field matter. The minimisation of the combined action means that the field equation gains an extra term, recovering \eqn{eqn:EEM} if the EM tensor is defined to be
\begin{equation}
T_{ab} = - \frac{C_M}{8 \pi \sqrt{-g}} \frac{\delta S_M[g^{ab}]}{\delta{g^{ab}}} . \label{eqn:EMTdef} \, .
\end{equation}
which can also be written in terms of the Lagrangian density as
\begin{equation}
T_{ab} = - \frac{C_M}{8 \pi \sqrt{-g}} \frac{\partial \mathcal{L}_M[g^{ab}]}{\partial{g^{ab}}} \, . \label{eqn:EMTdef2}\ .
\end{equation}
Now the requirement that the matter action is diffeomorphism invariant leads to the requirement (see for example Wald \citep{wald1984general}) that for a matter field which satisfies the field equations, the EM tensor is divergenceless, that is
\begin{equation}
\nabla^a T_{ab} = 0 . \label{eqn:ConsTab} \, ,
\end{equation}
which is consistent with the expected conservation of energy and momentum from its physical definition above in terms of fluxes across a surface.

For example, for a minimally coupled scalar field, with a simple kinetic term, the Lagrangian density is
\begin{equation}
\mathcal{L}_{SF} = - \onehalf \sqrt{-g} \left( g^{ab} \nabla_a \phi \nabla_b \phi + 2 V(\phi) \right) \label{eqn:LagrangianSF} \, .
\end{equation}
One can verify that \eqn{eqn:EMTdef2} then leads to the EM tensor  
\begin{align}
T_{ab} = \nabla_a \phi \nabla_b \phi - \onehalf g_{ab} (\nabla_c \phi \, \nabla^c \phi + 2V) \, .  \label{eqn:EMTensorSF}
\end{align}

In some ways this derivation of the Einstein equation is more elegant than the geometric approach, because $R$ is the obvious choice for the scalar to play the role of the Lagrangian, and we don't have to do a last minute switch from $R_{ab}$ to $G_{ab}$. However, a geometric understanding is probably more important in the field of NR, and is closer to the original derivation followed by Einstein. We present both here because scalar fields are often expressed in the language of Lagrangians and it is thus valuable to connect the two approaches in the context of this work. We will continue to make this connection in the following section when we decompose the metric in the 3+1 formalism using both a geometric and Lagrangian approach.


\section{NR - key theoretical concepts}
\label{sec-NRTheory}

In this section we describe the key issues encountered in the numerical formulation of GR as a $3+1$D Cauchy problem which can be implemented and solved on a computer. As discussed in Chapter \ref{ch-Introduction}, when one wishes to solve the Einstein equation numerically, the usual scenario is that one knows or postulates some initial condition on a spatial hypersurface, and wants to find out ``what happens next'', that is, one wishes to evolve the slice forward in time. This in principle a tractable problem - if one knows the metric on a hyperslice and its derivatives as one moves ``off'' the slice, that should be enough to populate the rest of spacetime, using the Einstein field equations. 

One must define what is meant by the spatial hypersurface. In GR, there is no preferred time-like direction and, crucially, no global concept of time. This makes the problem of solving the Einstein equation numerically substantially different from normal Cauchy problems. The data on the initial 3 dimensional spatial hyperslice is evolved forward along a local time coordinate, with each point corresponding to an ``observer'' who moves through the spacetime, rather than any fixed spatial point. The freedom to choose the path of these observers, the so-called ``gauge choice'', is discussed in section \ref{sec-NRStability}.

There exists a ``natural'' decomposition of the Einstein equations which is well motivated from both the Lagrangian and geometric approaches - the ADM (Arnowitt Deser Misner) decomposition \citep{Arnowitt:1962hi}. As we have mentioned, the original decomposition by Arnowitt et al was reformulated by York \citep{York1979}, and this is the one which we describe here. 

As the York distinction implies, several formulations are possible. The different formulations must agree for physical data (otherwise they will not describe gravity as we observe it), but they may have different global mathematical properties, and thus behave differently as one moves off the constraint surface (i.e. into regions of non-physical data). This has important consequences for numerical stability, and is discussed further in section \ref{sec-NRStability}. In this section we will also introduce the formulation used in the work presented in this thesis - the BSSN formalism - and explain why it has desirable properties. 

\subsection{ADM decomposition}
\label{sec-ADMTheory}

In this section, the ADM decomposition is derived both as a geometric problem, and from variational principles of the Einstein-Hilbert action. 

\subsubsection{Spacetime slicing and kinematics}

\begin{figure}
\begin{center}
\includegraphics[width=.95\textwidth]{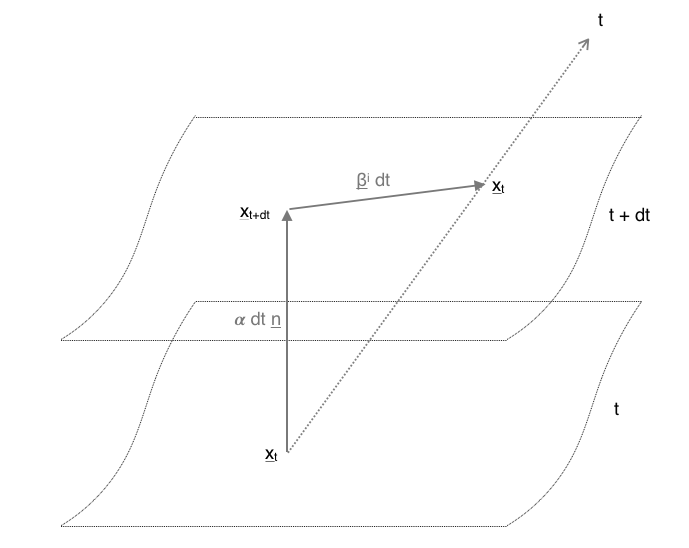}
\caption[$3+1$D Foliation of spacetime]{Foliation of a 4 dimensional spacetime into a 3 dimensional ``spatial'' hyperslice, and a ``time-like'' normal to that slice. The gauge variables - the lapse $\alpha$ and shift $\beta^i$ - are also illustrated. In this picture space is represented as a two dimensional surface, whereas in full GR each spatial slice is a 3 dimensional volume.}
\label{fig-Foliation}
\end{center}
\end{figure}

Consider the foliation of 4-dimensional spacetime into a 3-dimensional ``spatial'' hyperslice, and a ``timelike'' normal to that slice, as illustrated in figure \ref{fig-Foliation}. It is assumed that the spacetime is \emph{globally hyperboloidal}, that is, that it can be foliated into level sets of a universal time function $t$ which are distinct and cover the whole spacetime.

The spatial coordinates $x^i(t)$ label the points on the spatial hypersurface at some coordinate time $t$. Within this slice, the proper distance $dl$ is determined by a 3-dimensional spatial metric $\gamma_{ij}$ according to
\begin{equation}
dl^2 = \gamma_{ij} dx^i dx^j \, .
\end{equation}
The normal direction to the hyperslice at each point is given by the unit vector $\vec{n}$, which is the 4-velocity of the normal observers. Travelling along this direction, the distance in proper time $\tau$ to the slice at $t+dt$ is given by:
\begin{equation}
d\tau = \alpha dt \label{eqn:LapseDefinition} \, .
\end{equation}
Here $\alpha$ is the lapse function, which takes a value at each point on the slice. The lapse encodes our freedom to slice the time-like evolution as we choose - it is a gauge variable. A value of $\alpha$ of less than one, for example, indicates that coordinate time runs slower than proper time at this point, but this should make no difference to the physical results we obtain in this basis.

\nomenclature[g-pi]{$\alpha$}{the lapse}
\nomenclature[g-pi]{$\beta^i$}{the (spatial components of the) shift vector}
\nomenclature[g-pi]{$\gamma_{ij}$}{the 3-dimensional spatial metric (in the adapted basis)}
\nomenclature[a-pi]{$n^a$}{the unit normal vector to the spatial slice, also the 4-velocity of the normal observers}

As we move onto the next slice, we may use the equivalent spatial coordinate freedom to relabel the coordinates on our hyperslice. This relabelling is parameterised by the shift vector $\beta^i$. It may not immediately be clear why we would want to do this - surely it is simpler to leave the points at fixed locations in space? Unfortunately this is not possible in the general case. Firstly, it turns out that the freedom to move our coordinates dynamically on each slice can improve the stability of our numerical evolutions, in particular in black hole spacetimes. Secondly, it is important to understand that each coordinate on the spatial slice does not correspond (necessarily) to a fixed point in space, but rather a \emph{particular observer} moving through the spacetime. The observer labelled by $x^i$ can move with reference to a fixed point from slice to slice, \emph{even with a zero shift vector}. We will return to these points below and in section \ref{sec-InitialConditionsGauge} on gauge choices. For now simply note that the shift vector moves the coordinates according to the following relation
\begin{equation}
x^i(t+dt) = x^i(t) - \beta^i dt \label{eqn:ShiftDefinition} \, ,
\end{equation}
where the notation is rather confusing but should be read as \emph{``the observer moving with 4-velocity $n^a$ who, on the slice at $t$, was labelled with the coordinates $x^i(t)$ is labelled on the timeslice at $t+dt$ by the coordinates $x^i(t+dt)$''}. 

Using simple addition of vectors, we can see that the 4-dimensional spacetime distance $ds$ is given by
\begin{equation}
ds^2 = (-\alpha^2 + \beta^i \beta_i) dt^2 + 2 \beta_i dx^i dt + \gamma_{ij} dx^i dx^j \label{eqn:ADMMetric} \, .
\end{equation}

\nomenclature[a-pi]{$ds$}{the spacetime distance}
\nomenclature[a-pi]{$dl$}{the distance within the spatial hypersurface}

Notice that we have, without justifying it, introduced a coordinate system which is adapted to the slicing - the $\vec{e}_0$ basis vector is tangent to the lines of constant $x^i$ (along the $t$ coordinate line), and the $\vec{e}_i$ basis vectors are tangent to the slice. This is a natural choice, and will make things simpler as it will mean a lot of the components of our geometric objects reduce to zero. For example, in this basis, the unit normal vector has the components in raised and lowered forms of
\begin{equation}
n^\mu = (1/\alpha, - \beta^i/\alpha)  \quad n_\mu = (- \alpha, 0, 0, 0) \, ,
\end{equation}
from which we can see that it is normalised and timelike such that $n^\mu n_\mu = -1$. In this basis objects living in the spatial slice can have their indices raised and lowered with the spatial metric $\gamma_{ij}$. We will refer to this coordinate choice as ``the adapted basis''.

\nomenclature[a-pi]{$P^a_b$}{the projection operator for the spatial hyperslice}
\nomenclature[g-pi]{$\gamma_{ab}$}{the spatial metric (in a general basis)}
\nomenclature[g-pi]{$\beta^a$}{the shift vector (in a general basis)}

However, it is important to be aware that we can define all our quantities independently of this coordinate system, as we will now do. Knowing the unit normal vector to the slice $\vec{n}$, we can define the projection operator which projects indices onto the spatial hypersurfaces as
\begin{equation}
P^a_b \equiv \delta^a_b + n^a n_b \, .
\end{equation}
Applying this to the metric gives the (4-dimensional) metric induced on the spatial slice as
\begin{equation}
\gamma_{ab} = P^c_a P^d_b g_{cd} = g_{ab} + n_a n_b \, ,
\end{equation}
from which we can see that the projection operator is in fact the spatial metric, that is
\begin{equation}
P^b_a = \gamma^b_a \, .
\end{equation}
The lapse function is defined as 
\begin{equation}
\alpha = ( - \vec{\nabla}t ~.~ \vec{\nabla}t )^{-\onehalf} \, ,
\end{equation}
and the shift vector $\beta^a$ is defined such that it is orthogonal to $\vec{n}$ by construction, with $\beta^0 = 0$ and
\begin{equation}
\beta^i = - \alpha ( \vec{n} ~.~ \vec{\nabla}x^i )\, ,
\end{equation}
so that the time vector has components
\begin{equation}
t^a = \alpha n^a + \beta^a \ .
\end{equation}
Note that in a general basis $t^a \neq \nabla^a t$, but $t^a \nabla_a t = 1$ in all bases. We can also see that $\beta^a$ is the projection of $\vec{t}$ onto the spatial hypersurface
\begin{equation}
P^a_b t^b = (\delta^a_b + n^a n_b)(\alpha n^b + \beta^b) = \beta^a \, .
\end{equation}
We won't have much need to use more general bases, and work mainly in the adapted basis, in which case we can frequently ignore the time-like components of the geometric objects which are intrinsic to the spatial slice. The key point to take away is simply that these quantities are geometric objects which exist independently, and are not defined by, the obvious coordinate choice aligned to the slices. As a consequence, we can only ignore the time-like components when we make this particular choice, and not in the general case.

\subsubsection{More kinematics - the extrinsic curvature}

\nomenclature[a-pi]{$K_{ab}$}{the extrinsic curvature tensor}
\nomenclature[a-pi]{$K_{ij}$}{the extrinsic curvature (in the adapted basis)}
\nomenclature[x-pi]{$\pounds_{\vec{V}}$}{the Lie derivative along the vector field $\vec{V}$}
\nomenclature[a-pi]{$D_a$}{the covariant derivative defined with respect to the spatial metric $\gamma_{ab}$}

To fully specify our decomposed spacetime, we must also define an object called the extrinsic curvature, $K_{ab}$, which describes how the spatial hypersurface is embedded in the 4-dimensional spacetime. 

\begin{figure}
\begin{center}
\includegraphics[width=.95\textwidth]{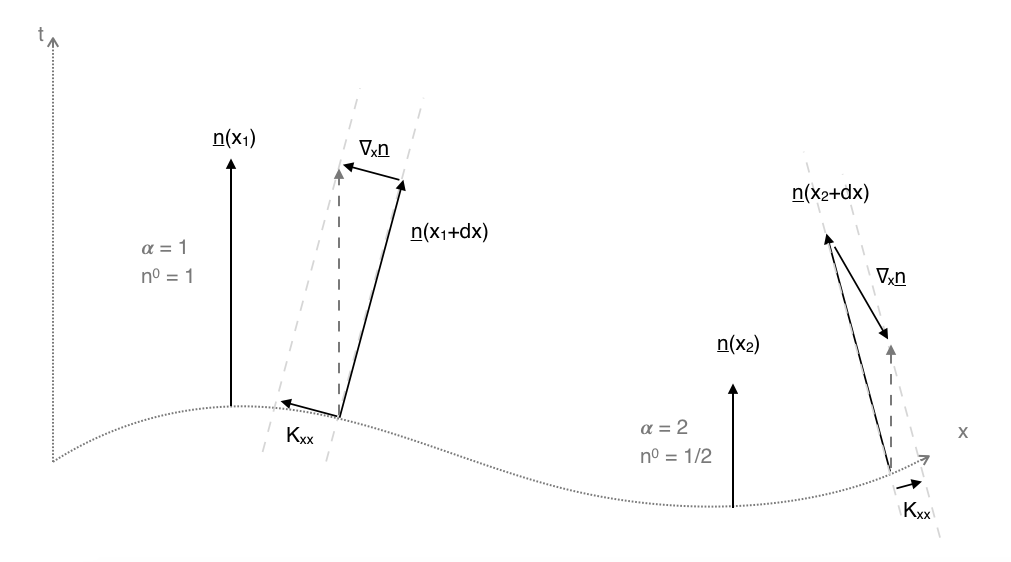}
\caption[Extrinsic Curvature]{In 1+1D the extrinsic curvature in the adapted slicing has only one non trivial component, $K_{xx}$. The figure illustrates how its value relates to the change in the normal vector as it is parallel transported along the slice. In the vicinity of $x_1$ the lapse is constant in space, and so the covariant derivative of the normal vector naturally has all its components tangent to the slice. At another point $x_2$ the lapse is not constant and so the normal vector at $x_2+dx$ is a different length (which has been exaggerated for effect, it should vary smoothly and thus be only infintessimally different in length). In this case it is necessary to project the covariant derivative into the slice to find the value of $K_{xx}$. Note that for simplicity in the figure the shift is equal to zero, so that the time-like direction is parallel to the normal vector, which is not true in the general case.}
\label{fig-ExtrinsicCurvature}
\end{center}
\end{figure}

The notion of extrinsic curvature is in some ways more intuitive that the notion of intrinsic curvature. Consider a cylinder in 3-dimensional space - the intrinsic curvature of the 2-dimensional surface is zero - it is flat in the sense that the parallel transport of a vector around a loop on the surface does not lead to a change in its direction. However, our human brains consider this surface to be curved, which it is \emph{in the 3-dimensions in which it is embedded} - it has an \emph{extrinsic curvature}. 
This extrinsic curvature can be defined in two equivalent ways. Firstly, it can be defined as the change in the direction of the normal vector under parallel transport a small distance away along the surface, see figure \ref{fig-ExtrinsicCurvature}. 
Geometrically, this definition corresponds to
\begin{equation}
K_{ab} = - P_a^c \nabla_c n_b = - (\nabla_a n_b + n_a n^c \nabla_c n_b) \, . \label{eqn:Kabdefinition1}
\end{equation}
Note that the quantity we might naively specify, $\nabla_a n_b$, is projected into the spatial slice (there is also a minus sign which is just a matter of convention). This is to remove the effect of the lapse, which is not intrinsic to the slice, but appears in the normalisation of the normal vector. By projecting into the slice we ensure that $K_{ab}$ is symmetric and intrinsic to the slice. In effect, $K_{ab}$ is defined as if all normal observers followed the geodesic congruence with $\alpha=1$, which means that it is a gauge independent quantity.
Equivalently, the extrinsic curvature may be defined as the Lie derivative of the metric along the normal direction, i.e.
\begin{equation}
K_{ab} = - \onehalf \pounds_{\vec{n}} \gamma_{ab} \, .  \label{eqn:Kabdefinition2}
\end{equation}
That this is equivalent to \eqn{eqn:Kabdefinition1} can be shown by expanding out the Lie derivative (a short discussion of Lie derivatives and the derivation of this result is given in the Appendix \ref{sec-appendix1B}). If we now choose the adapted basis, the time components of the extrinsic curvature can be ignored - they are zero in the raised form, and although non-zero in the lowered form, all their information will effectively be contained in the spatial components. These components are then
\begin{equation}
K_{ij} = - \frac{1}{2 \alpha}(\partial_t \gamma_{ij} - D_i \beta_j - D_j \beta_i) \, . \label{eqn:Kij}
\end{equation}
Where $D_i$ is the covariant derivative defined with respect to the spatial metric $\gamma_{ij}$\footnote{This is equivalent to the projection of the covariant derivative into the spatial slice $D_a = P^b_a \nabla_b$ for the derivative of a scalar or a purely spatial tensor, but when acting on a general four tensor, one must also project the indices of the tensor itself into the spatial slice.}.

Contracting \eqn{eqn:Kabdefinition1} with the metric it can be seen that its trace is equal to the divergence of the normal lines
\begin{equation}
K =  g^{ab} K_{ab} = - \nabla_c n^c \, , \label{eqn:KtraceVolume}
\end{equation}
where the second term vanishes because $\vec{n}$ is unitary and so its gradient is orthogonal to it. This means that it corresponds to the changing volume element of the normal observers. We will see that in the special case of an isotropic and homogeneous Universe, with geodesic observers, $K$ is related to the Hubble constant as $K = - 3H$. Negative $K$ thus corresponds to an expanding space, and positive $K$ to a collapsing one\footnote{This is true for the definition of $K$ used here, but an opposite sign convention for $K_{ab}$ is possible and used by some authors. In addition, as will be discussed later, in more general cases the volume growth may be a gauge effect, rather than due to the physical expansion of the space.}.

\subsubsection{Constraints and dynamics}

In previous sections we have discussed only the kinematics derived from a 3+1 slicing of the metric. In this section we introduce the dynamics and physical constraints imposed on the metric by the Einstein equation. 

The method to be followed consists in projecting the Einstein equation both onto the spatial surface that we have constructed, and normal to it. In fact there are three options - either both indices can be projected into the spatial hypersurface, both normal to it, or (since the tensors are symmetric), either one of the indices can be projected into the slice and the other one out of it. 

We start from two well-known relations which are simply (but lengthily) derived from the geometric slicing described above. Firstly, the Gauss-Codazzi equation
\begin{equation}
P^e_a P^f_b P^g_c P^h_d ~ {}^{(4)}R_{efgh} = {}^{(3)}R_{abcd} + K_{ac} K_{bd} - K_{ad} K_{bc} \, , \label{eqn:GaussCodazzi}
\end{equation}
and secondly the Codazzi-Mainardi equation
\begin{equation}
P^e_a P^f_b P^g_c n^h ~ {}^{(4)}R_{efgh} = D_b K_{ac} - D_a K_{bc} \, .  \label{eqn:CodazziMainardi}
\end{equation}
Contracting both sides of \eqn{eqn:GaussCodazzi} twice with the metric $g^{ac} g^{bd}$ gives
\begin{equation}
n^a n^b G_{ab} =  {}^{(3)}R + K^2-K_{ab}K^{ab} \, , \label{eqn:ContractedGC}
\end{equation}
from which, using the Einstein equation to replace $G_{ab}$ with $T_{ab}$, and using the adapted basis, we obtain 
\begin{equation}
\mathcal{H} = {}^{(3)}R + K^2-K_{ij}K^{ij}-16\pi \rho = 0 \, . \label{eqn:HamiltonianConst}
\end{equation}
where $\rho \equiv n^a n^b T_{ab}$ is the energy density measured by a normal observer. This relation is the \emph{Hamiltonian constraint}. It involves no time derivatives and is independent of the gauge parameters\footnote{When we start specifying scalar field data on the initial slice, the gauge parameters will appear in the constraint equation. This is a consequence of the fact that the time derivatives of the scalar field are usually specified with reference to coordinate time and not for the (gauge independent) normal geodesic observer. Their appearance in the constraints is then to \emph{remove} their effect from the gauge dependent quantities, rather than because the constraints depend on the gauge.} $\alpha$ and $\beta^i$. It is not, therefore, related to the evolution of the quantities but their relation within a slice. It tells us that we are not free to specify any data we like for the metric and the energy density - the data must satisfy this relation or it will not satisfy the Einstein Equation. This is quite clear when you think about it physically - if I were completely free to choose all my quantities, I could put a very large mass in the centre of my space, and insist that the spacetime around it was completely flat. This is clearly not a valid physical scenario, and we see that it is indeed ruled out by \eqn{eqn:HamiltonianConst}.

\nomenclature[a-pi]{$\mathcal{H}$}{the Hamiltonian constraint}
\nomenclature[a-pi]{$\mathcal{M}^i$}{the momentum constraints}
\nomenclature[a-pi]{$S^i$}{the momentum density as measured by normal observers}
\nomenclature[a-pi]{$S_{ij}$}{the spatial part of the energy momentum tensor in the adapted basis}
\nomenclature[a-pi]{$S$}{the trace of $S_{ij}$, i.e. $\gamma^{ij} S_{ij}$}
\nomenclature[a-pi]{$K$}{the trace of $K_{ab}$, i.e. $g^{ab} K_{ab}$, also $\gamma^{ij}K_{ij}$ in the adapted basis}
\nomenclature[a-pi]{$R_{ij}$}{the 3-dimensional Ricci tensor (in the adapted basis)}

The same contraction of \eqn{eqn:CodazziMainardi} gives the projection of the Einstein equation with one index in and one out of the slice
\begin{equation}
P^{ab} n^c G_{bc} = D_b (\gamma^{ab} K - K^{ab}) \, ,
\end{equation}
from which, again using the Einstein equation to eliminate $G_{ab}$, we obtain the \emph{momentum constraints} in the adapted basis
\begin{equation}
\mathcal{M}^i = D_j (\gamma^{ij} K - K^{ij}) - 8\pi S^i \, , \label{eqn:MomentumConst}
\end{equation}
where $S^i \equiv - \gamma^{i \mu} n^\nu T_{\mu \nu}$ is the momentum density as measured by normal observers. Again these (three) relations must be satisfied by the data on the each slice if it is to represent a true ``physical'' spacetime. However, as with the Hamiltonian constraint, it gives us no data about how the quantities should evolve in time, save that these relations should continue to be satisfied.

The four constraints reduce the number of degrees of freedom from ten (the symmetric components of $g_{\mu \nu}$) to six. To obtain the remainder, we require the projection of both indices into the slice. Starting with \eqn{eqn:GaussCodazzi} again and contracting with the metric $g^{ac}$ we have
\begin{equation}
P^f_b P^h_d \left( {}^{(4)}R_{fh} + n^e n^g ~ {}^{(4)}R_{efgh} \right) = {}^{(3)}R_{bd} + K K_{bd} - K_{bc} K^c_d \, .
\end{equation}
One can also show that the last non trivial projection of the 4D Riemann tensor is
\begin{equation}
P^f_b P^h_d \left( n^e n^g ~ {}^{(4)}R_{efgh} \right) =  \pounds_{\vec{n}} K_{bd} - K_{bc} K^c_d + \frac{1}{\alpha} D_b D_d \alpha \, .
\end{equation}
Before we can equate these two (purely geometric) relations, we need to eliminate the term $P^f_b P^h_d {}^{(4)}R_{fh}$. In deriving the constraints, we have eliminated similar terms by expressing them in terms of $G_{ab}$ and then making a substitution for the EM tensor, thereby introducing the ``physics'' of GR. Whilst we can do the same here, we choose instead to replace $R_{ab}$ directly using an alternative form of the Einstein Equation
\begin{equation}
R_{ab} = 8 \pi (T_{ab} - \onehalf g_{ab} T) \, .
\end{equation}
where $T \equiv T^a_a$. This step is the key difference between the York and original ADM formulations. If we used the Einstein tensor instead, we would add a term proportional to the Hamiltonian constraint to the evolution equation derived, which for physical data is zero and so the two are the same. Combining these results and expressing them in the adapted basis gives the evolution equation for $K_{ij}$ as  
\begin{multline}
\partial_t K_{ij} = \beta^k \partial_k K_{ij} + K_{ki} \partial_j \beta^k + K_{kj} \partial_i \beta^k - D_i D_j \alpha \\ + \alpha \left({}^{(3)}R_{ij} + K K_{ij} - 2 K_{ik} K^k_j \right) + 4 \pi \alpha \left( \gamma_{ij}(S - \rho) - 2 S_{ij} \right) \, ,  \label{eqn:dtKij}
\end{multline}
where $S_{ij} \equiv \gamma_{i \mu}\,\gamma_{j \nu} T^{\mu \nu}$ and $S \equiv S^i_i$. Combining this with the definition of $K_{ij}$ \eqn{eqn:Kij} above, which can be rearranged to give
\begin{equation}
\partial_t \gamma_{ij} = - 2\alpha K_{ij} + D_i \beta_j + D_j \beta_i \, , \label{eqn:dtgammaij}
\end{equation}
gives a full set of evolution equations for the spatial metric and the extrinsic curvature. Note that since the Einstein equation involved second derivatives of the metric with respect to time, we have effectively performed the usual trick of decomposing a second order differential equation into two first order ones, by first defining $K_{ij}$ to be (loosely speaking) the first time derivative of the metric as in \eqn{eqn:dtgammaij}, and then giving the time evolution in terms of $K_{ij}$. Thus one should not strictly see \eqn{eqn:dtgammaij} as an evolution equation for the metric (since it is not derived from the Einstein equations in any way) but rather as the definition of $K_{ij}$. This is equivalent to how in Newtonian mechanics $F=m ~ dv/dt$ is the evolution equation whereas $v = dx/dt$ is just the definition of the velocity, although together they allow one to derive the overall (second order) evolution of position $x$ from two first order equations.

A couple of notes to close this section - firstly, one can show that if the constraint equations are satisfied on the initial slice, the evolution equations \eqn{eqn:dtKij} and \eqn{eqn:dtgammaij} will preserve them on future slices, due primarily to the effect of the Bianchi identities in \eqn{eqn:BianchiIdentities}. Secondly, we consider again the ten degrees of freedom in the full metric. We have already said that four of these are removed by the constraints. A further four of the six dynamical degrees of freedom represent the freedom to choose one's gauge variables in space and time. That leaves two physical degrees of freedom of the gravitational field, which corresponds to the two polarisations of gravitational waves.

A summary of the key equations derived in this section is given in Appendix \ref{sec-appendix1C} for reference.

\subsubsection{ADM decomposition - Lagrangian formulation}

Arnowitt, Deser and Misner's original paper derived the ADM decomposition as a minimisation of the classical action \eqn{eqn:Hilbertaction}, with the 4 dimensional Ricci scalar ${}^{(4)}R$ decomposed into its components in the slice using
\begin{equation}
R = 2 \left( n^a n^b G_{ab} - n^a n^b R_{ab} \right) \ .
\end{equation}
The first term is the contracted Gauss-Codazzi relation in \eqn{eqn:ContractedGC} above, and the second can be shown using the definition of the Riemann tensor in \eqn{eqn:RiemannCommutator} to be
\begin{equation}
K^2 - K_{\mu \nu} K^{\mu \nu} + \textrm{a total divergence} \ .
\end{equation}
The problem was originally formulated as a Hamiltonian problem with the conjugate momenta
\begin{equation}
\pi^{ij} = \frac{\partial \mathcal{L}}{\partial \dot{\gamma}_{ij}} = - \sqrt{\gamma}\left( K^{ij} - \gamma^{ij} K \right)
\end{equation} 
in place of the extrinsic curvature, but for consistency with the above work we retain $K_{ij}$ here, which is just a change of variable. Recognising that the volume element $\sqrt{-g} ~ d^4x = \alpha \sqrt{\gamma} ~ d^4x$, where $\gamma$ is the determinant of the 3 dimensional spatial metric, gives the action as
\begin{equation}
S[\gamma^{ij}, \alpha, \beta] = \int d^4 x ~ \alpha \sqrt{-\gamma} \left( {}^{(3)}R + K_{ij} K^{ij} - K^2 \right). \label{eqn:Hilbertaction_AMR}\,
\end{equation}

\nomenclature[g-pi]{$\pi^{ij}$}{the conjugate momenta of the spatial metric fields}
\nomenclature[g-pi]{$\gamma$}{the determinant of the 3 dimensional spatial metric}

Taking the functional derivative of this action and minimising it with respect to each of the components of the decomposed 4-metric in turn, the lapse $\alpha$, shift $\beta^i$ and spatial metric $\gamma_{ij}$, gives the Hamiltonian constraint, the momentum constraints, and the evolution equation for $K_{ij}$ respectively. For example, considering minimisation with respect to the lapse, and ignoring surface terms
\begin{equation}
\frac{S[\gamma^{ij}, \alpha, \beta]}{\delta{\alpha(y)}} = \int d^4 x \sqrt{-\gamma} \left({}^{(3)}R - K_{ij} K^{ij} + K^2 \right) \delta(x-y)  = 0 \,
\end{equation}
we obtain the Hamiltonian constraint as per \eqn{eqn:HamiltonianConst}.

Thus we see how the ``gauge choice'' variables dictate that each slice must be correctly embedded in the higher dimensional spacetime such that the constraints are satisfied, with, as one might expect, the time-like lapse $\alpha$ giving rise to the equation of energy conservation, and the spatial gauge variable $\beta^i$ giving rise to momentum conservation. The true ``equation of motion'' for the metric then comes from the minimisation of the action with respect to $\gamma_{ij}$, which gives rise to an equation similar to \eqn{eqn:dtKij} (differing only by the addition of a multiple of the Hamiltonian constraint) for the evolution of $K_{ij}$.

The equivalence of the field theoretic approach with the geometric approach is part of the beauty of the theory of gravity, but it is also useful. By formulating in terms of an action, one may study modified gravity theories derived from new actions, with symmetries motivated by other physical ideas. Since one may derive new equations of motion from such a new action, one can in theory evolve these numerically and compare their results to standard Einstein gravity, thus exploring higher energy deviations from the accepted model. For example, $f(R)$ gravity in which the $R$ is replaced by some function of $R$ in the Einstein-Hilbert action \citep{amendola2010dark}, or $f({}^{(3)}R)$ in Horava Lifshitz gravity \citep{Horava:2009uw}, in which the 4-dimensional diffeomorphism is broken. Probing these modified gravity models is one of the key aims of ESA's Euclid mission \citep{Amendola:2016saw}, which will map large scale structure of galaxies and galaxy clusters across a significant portion of the sky.

Although one can formulate an equation of motion in $3+1$ dimensions for modified gravity models in NR by following the prescription above, it turns out that ensuring the numerical stability and well-posedness of the equations which are obtained is not trivial. This issue, in the context of standard Einstein gravity, is considered in the following section.

\subsection{Numerical stability}
\label{sec-NRStability}

The $3+1$D ADM decomposition of the Einstein Equation presented above and summarised in Appendix \ref{sec-appendix1C} is already, in theory, in a form suitable for evolution on a computer. However, one finds that it results in large instabilities developing during the simulation. This can be shown to be due to the equations being \emph{weakly hyperbolic} rather than \emph{strongly hyperbolic}, which means they are not \emph{well-posed}, such that certain modes may grow without bound. In the first part of this section we will define and explain these terms, and describe the key points that lead to numerical problems in the ADM formalism. 

Many numerical relativity codes implement the so called BSSN form of the Einstein equation \citep{Nakamura:1987zz,Shibata:1995we,Baumgarte:1998te}. This admits a strongly hyperbolic formulation of the Einstein equation, and together with the ``$1+\log$'' slicing \citep{Bona:1994dr} and the ``gamma-driver'' gauge conditions \citep{Alcubierre:2002kk}, has allowed the stable simulation of dynamical spacetimes of interest, including black hole binaries. We will present this formulation in the second part of this section, and describe how its well-posedness is achieved.

More recently, other refined formulations of the Einstein equation based on the Z4 system \citep{Bona:2003fj,Gundlach:2005eh} have been proposed, most notably the Z4c formulation \citep{Bernuzzi:2009ex} and the CCZ4 formulation \citep{Alic:2011gg}. In the Z4 system, both the Hamiltonian and the momentum constraint are promoted to dynamical variables and hence constraint violating modes can propagate and eventually exit the computational domain, which can result in a more stable evolution. In this thesis we use only the BSSN formulation, which was found to be sufficiently stable for our purposes in the research presented.

The other main approach in NR is Generalised Harmonic Coordinates (GHC), which takes a completely different approach and evolves the full spacetime metric $g_{ab}$. This has been used with much success by groups including Pretorius \citep{Pretorius:2004jg}, but again it is not used for this work and so we do not consider it further here.

Note that going forward, all references to the metric and its derived objects refer to the 3-dimensional versions, for example $R={}^{(3)}R$, unless otherwise specified. The dimension will be specified where there is potential for confusion, but it should be clear from the indexing convention (Roman indices for 3-dimensional objects and Greek indices for 4-dimensional ones). 

\subsubsection{Well-posed and hyperbolic formulations}

\nomenclature[a-pi]{$\bf{u}$}{the state vector (an ordered list of the evolution variables, rather than a geometric vector)}
\nomenclature[a-pi]{$\bf{w}$}{the vector of eigenfunctions $w$ (an ordered list, rather than a geometric vector)}
\nomenclature[a-pi]{$M^i$}{the characteristic matrix for the $i$th spatial direction}
\nomenclature[a-pi]{$P_s$}{the principle symbol matrix}
\nomenclature[a-pi]{$R_s$}{the matrix of eigenvectors of $P_s$}
\nomenclature[a-pi]{$H_s$}{the symmetrising matrix of $P_s$}
\nomenclature[g-pi]{$\Lambda_s$}{the matrix of eigenvalues of $P_s$}
\nomenclature[a-pi]{$s({\bf{u}})$}{source terms in the evolution system for $\bf{u}$}
\nomenclature[a-pi]{$A_i$}{spatial derivative of the lapse, used in stability analysis of section \ref{sec-NRStability}}
\nomenclature[a-pi]{$d_{ijk}$}{spatial derivative of $\gamma_{ij}$, used in stability analysis of section \ref{sec-NRStability}}
\nomenclature[a-pi]{$X^k_{ij}$}{combination of evolution variables, used in stability analysis of section \ref{sec-NRStability}}

If a system of PDEs is \emph{well posed}, this means that a small change in initial data results in a small change in the solution. This can be expressed by the condition
\begin{equation}
\Vert {\bf{u}}(t,x) \Vert \leq C_1 e^{C_2 t} \Vert {\bf{u}}(0,x) \Vert
\end{equation}
with the constants $C_1$ and $C_2$ independent of the initial data. This only requires a less than (or equal to) exponential growth in the initial conditions, which is not in itself fantastic - such growth may still cause problems within a simulation. However, it is certainly a necessary (if not sufficient) condition for good numerical behaviour.

We will now describe how a system which is \emph{strongly hyperbolic} can be shown to be well-posed as a result. We consider a system of the form
\begin{equation}
\partial_t {\bf{u}} + M^i d_i {\bf{u}} = s({\bf{u}}) \, ,
\end{equation}
where $\bf{u}$ is not a vector in the geometric sense, but an ordered list of the evolution variables, $K_{ij}, \alpha$ etc. and so we denote it in bold face. We will ignore the effect of the source term, setting $s({\bf{u}})=0$, as it does not play a role in our discussions here (but note that it may indeed affect the analysis if it is non linear in the variables). The $i$ spans the spatial dimensions and thus there is one \emph{characteristic matrix} $M^i$ for each direction. If one considers an arbitrary unit vector $\vec{s} = (s_x, s_y, s_z) $ one can construct the \emph{principle symbol} matrix $P_s(\vec{s}) = M^i s_i$. If this has real eigenvalues and a complete set of eigenvectors for all $s_i$, the system is \emph{strongly hyperbolic}\footnote{If the former condition, real eigenvalues, is met but not the latter, a complete set of eigenvectors, it is only \emph{weakly hyperbolic}.}. In this case one can always find a symmetric, positive definite matrix $H_s = (R_s^{-1})^T R_s^{-1} $  which ``symmetrises'' $P_s$ such that
\begin{equation}
H_s P_s - P_s^T H_s^T = H_s P_s - P_s^T H_s = 0 \, ,
\end{equation}
where $R_s$ is the matrix of eigenvectors of $P_s$, and $\Lambda_s$ the diagonal matrix of eigenvalues such that
\begin{equation}
P_s R_s = \Lambda_s R_s \, .
\end{equation}
We can then use $H_s$ to construct an ``Energy-Norm'' of the initial condition vector $\bf{u}$ and its adjunct ${\bf{u}}^\dagger$
\begin{equation}
\Vert {\bf{u}} \Vert^2 = {\bf{u}}^\dagger H_s {\bf{u}} \, .
\end{equation}
Using a Fourier mode ${\bf{u}}(x,t) = \tilde{{\bf{u}}}(t) e^{ik \vec{x} . \vec{s}}$ as the initial condition we can see that
\begin{equation}
\partial_t \Vert {\bf{u}} \Vert^2 = \partial_t({\bf{u}}^\dagger H_s {\bf{u}}) = ik \tilde{{\bf{u}}}^T (P_s^T H_s - H_s P_s)\tilde{{\bf{u}}} =  0 \, ,
\end{equation}
meaning that the energy norm is constant in time, and thus the system is well posed as required. If we reduce our system to derivatives of a single direction $x$, (which we can do with our tensor variables since they have no preferred direction), we can also define eigenfunctions $\bf{w}= R_s^{-1}\bf{u}$ such that
\begin{equation}
\partial_t {\bf{w}} + \Lambda_s d_x {\bf{w}} = 0 \, ,
\end{equation}
so that the evolution equations for the eigenfunctions decouple, and propagate with speeds equal to the eigenvalues of the principle symbol. In this case we are effectively only considering the matrix $M^x$. Often the eigenfunctions can be found by inspection, rather than explicit calculation of the eigenvectors.

We will now sketch the key steps in analysing the ADM system of PDEs according to this approach. For a more complete treatment, on which this discussion is based, see Chapter 5 of Alcubierre \citep{AlcubierreBook}. 

Firstly, in order to take all the equations into first order form, we define the following new variables:
\begin{equation}
A_i = \partial_i (\ln \alpha) \quad d_{ijk} = \onehalf \partial_i \gamma_{jk} \, ,
\end{equation}
The lapse is considered to be a dynamically varying quantity, subject to the evolution
\begin{equation}
\partial_0 \alpha = -\alpha^2 Q \, ,
\end{equation}
where $Q$ is some function of the other variables, and $\partial_0$ is shorthand for $\partial_t - \beta^i \partial_i$. The shift is considered to be an a priori function of space and time and thus it and its derivatives are treated as source terms. We consider only the ``principle parts'', that is, we keep only the highest order derivatives present, which dominate the behaviour, whilst the remaining terms are considered source terms and set to zero. This means that we ignore the evolution equations for the lower order quantities $\alpha$ and $\gamma_{ij}$. Our equations reduce to those for the 27 independent variables
\begin{align}
&\partial_0 A_i \simeq - \alpha \partial_i Q  \, , \\
&\partial_0 d_{ijk} \simeq - \alpha \partial_i K_{jk}  \, , \\
&\partial_0 K_{ij} \simeq - \alpha \partial_k X^k_{ij}  \, . \label{eqn:HyperbolicADM0}
\end{align}
up to the principle part, where $X^k_{ij} = d^k_{ij} + \delta^k_{(i} \left( a_{j)} + d_{j)m}^m - 2 d^m_{mj)} \right)$. Considering only one spatial derivative direction $x$, many of the equations immediately decouple and thus some eigenfunctions can be found by inspection, for example, the eigenfunctions
\begin{align}
&w = A_i \quad i \neq x \quad \rm{with} ~ \lambda = -\beta^x \, , \\
&w = d_{ijk} \quad i \neq x \quad \rm{with} ~ \lambda = -\beta^x  \, .  \label{eqn:HyperbolicADM1}
\end{align}
Additionally by considering how $X^i_{jk}$ evolves, one can show that another set of eigenfunctions is
\begin{equation}
w = \sqrt{\gamma_{xx}} K_{ij} \mp X^x_{ij} \quad i,j \neq x \quad \rm{with} ~ \lambda = -\beta^x \pm \alpha \sqrt{\gamma_{xx}}  \, .
\end{equation}
However, in trying to extend this to the directions involving $x$, one finds a problem. We can see from \eqn{eqn:HyperbolicADM0} that $K_{xi}$ evolves as a function of the derivatives of $X^x_{xi}$, but since
\begin{equation}
\partial_0 X^x_{xi} \simeq \alpha \gamma^{xj} \partial_x K_{ij} \quad i,j \neq x \, , \label{eqn:HyperbolicADM2}
\end{equation}
it is clear that the time derivative of $X^x_{xi}$ is independent of $K_{xi}$. This means that this subsystem cannot be symmetrised, and that $K_{xi}$ can grow in an unbounded way unless $X^x_{xi}=0$. The ADM equations are thus only weakly hyperbolic. This can be fixed by assuming that the momentum constraint is satisfied, in which case \eqn{eqn:HyperbolicADM2} becomes 
\begin{equation}
\partial_0 X^x_{xi} \simeq \alpha \gamma^{xx} \partial_x K_{ix} \quad i \neq x \, ,
\end{equation}
for which the two quantities recouple, allowing them to be symmetrized, with the same eigenfunction structure as \eqn{eqn:HyperbolicADM2}.

We therefore find that the ADM decomposition will be strongly hyperbolic \emph{only if} the momentum constraint is satisfied at all times. In a numerical simulation this cannot be guaranteed, which is what leads to the instabilities in the formalism. In order to remove this dependence, we need to alter the structure of the characteristic matrix in such a way as to ensure hyperbolicity regardless of the constraint violation. We want to do this without altering the physical behaviour of the system, and fortunately the freedom to add and subtract arbitrary multiples of the constraints (which should be zero) gives us the ability to do exactly that. Another possibility is found in promoting certain quantities to dynamical variables as is considered in the next section in the BSSN formalism. As we have mentioned, other formalisms exist, but we refer the reader to the standard NR texts \citep{AlcubierreBook, ShapiroBook} for further details. 

Note that a second condition which is required to achieve hyperbolicity in the ADM formalisms is that the lapse cannot be chosen to be an a priori function of space and time as was done for the shift - it must evolve as a dynamical variable\footnote{Alternatively the desensitised lapse $\alpha/\sqrt{\gamma}$ can be specified as a function of space and time, but we do not take this approach in this work. One can see that in any case the slicing conditions we use results in a similar behaviour to specifying a constant desensitised lapse, with the lapse becoming smaller in regions in which the normal observer volume is shrinking.}. Some possible dynamic conditions will be considered in \ref{sec-InitialConditionsGauge}.

\subsubsection{BSSN}

The BSSN system is derived from the ADM system with the following key steps:

\begin{enumerate}
\item{\emph{Introduce the conformal connection coefficients $\tilde{\Gamma}^i$}: These three auxiliary variables are promoted to dynamical evolution variables. They allow us to adjust the form of the characteristic matrix and thus change the hyperbolicity of the system.}
\item{\emph{Replace certain combinations of variables with multiples of the constraints}: In several of the evolution equations, multiples of the constraints are added, again these change the characteristic matrix and thus improve stability.}
\item{\emph{Decompose ADM variables into conformal versions}: Improving the hyperbolicity is a necessary but not sufficient condition for well behaved numerics. Another key feature of BSSN is the conformal decomposition of variables, which has been found in practise to improve stability.}
\end{enumerate}

Starting with the last point, we decompose the induced metric as $\gamma_{ij}=\frac{1}{\chi^2}\,\tilde\gamma_{ij}$ so that $\det\tilde\gamma_{ij}=1$ and $\chi = \left(\det\gamma_{ij}\right)^{-\frac{1}{6}}$. As suggested in \citep{Campanelli:2005dd}, we use the inverse of the conformal factor $\psi$ which is specified in most texts, since this results in a conformal factor which goes to zero at a black hole singularity, rather than a $1/r$ singularity\footnote{Note that many texts also use a $\chi$ which is equivalent to our $\chi^2$. One should thus take care with conventions when comparing results.}. The components of the conformal metric thus remain O(1) and encode the directional stretching of space, whilst the conformal factor gives the overall scale of the metric. For weak gravity cases in a conformally flat space, $\chi$ is approximately related to the Newtonian gravitational potential $ \Psi$ by
\begin{equation}
\chi = \sqrt{\frac{1}{1-2  \Psi}},
\end{equation}
and a value of $\chi$ less than $1$ can thus be loosely thought of as corresponding to a gravitational ``well'', and at flat space infinity, $\chi$ tends to 1.

\nomenclature[g-pi]{$\chi$}{the conformal factor of the spatial metric in BSSN, defined as $\chi = \left(\det\gamma_{ij}\right)^{-\frac{1}{6}}$}
\nomenclature[a-pi]{$\tilde{A}_{ij}$}{the traceless part of the extrinsic curvature in the BSSN decomposition}
\nomenclature[g-pi]{$\tilde{\gamma}_{ij}$}{the conformal metric in the BSSN decomposition}
\nomenclature[g-pi]{$\tilde{\Gamma}^i$}{the conformal connection coefficients in the BSSN decomposition}
\nomenclature[g-pi]{$\Gamma^i_{jk}$}{the three dimensional Christoffel symbols associated with the metric $\gamma_{ij}$}
\nomenclature[g-pi]{$\tilde{\Gamma}^i_{jk}$}{the three dimensional Christoffel symbols associated with the conformal metric $\tilde{\gamma}_{ij}$}

Similarly, the extrinsic curvature is decomposed into its trace, $K=\gamma^{ij}\,K_{ij}$, and its traceless part so that
\begin{equation}
K_{ij}=\frac{1}{\chi^2}\left(\tilde A_{ij} + \frac{1}{3}\,K\,\tilde\gamma_{ij}\right)\,,
\end{equation}
with $\tilde\gamma^{ij}\,\tilde A_{ij}=0$. It is not actually clear why such a decomposition should improve numerical stability, but in practise it has been found to do so where the tracelessness of $\tilde{A}_{ij}$ is enforced at each step.

One can then find the evolution equations for the decomposed variables directly from the ADM versions, taking care to account for the tensor density nature of $ \chi, \, \tilde{\gamma}_{ij}$ and $\tilde{A}_{ij}$ when expanding the Lie derivatives (further details of this are given in Appendix \ref{sec-appendix1LieShift}). This gives
\begin{align}
&\partial_t\chi=\frac{1}{3}\,\alpha\,\chi\, K - \frac{1}{3}\,\chi \,\partial_k \beta^k + \beta^k\,\partial_k \chi\,, \label{eqn:dtchi}\\
&\partial_t\tilde\gamma_{ij} =-2\,\alpha\, \tA_{ij}+\tgamma_{ik}\,\partial_j\beta^k+\tgamma_{jk}\,\partial_i\beta^k-\frac{2}{3}\,\tgamma_{ij}\,\partial_k\beta^k +\beta^k\,\partial_k \tgamma_{ij}\,, \label{eqn:dttgamma} \\
&\partial_t K = -\gamma^{ij}D_i D_j \alpha + \alpha\left(\tilde{A}_{ij} \tilde{A}^{ij} + \frac{1}{3} K^2 \right) + \beta^i\partial_iK + 4\pi\,\alpha(\rho + S), \label{eqn:dtK} \\
&\partial_t\tilde A_{ij} = \chi^2\left[-D_iD_j \alpha + \alpha\left( R_{ij} - 8\pi\,\alpha \,S_{ij}\right)\right]^\textrm{TF} + \alpha (K \tA_{ij} - 2 \tA_{il}\,\tA^l{}_j) \nonumber \\
&\hspace{1.3cm} + \tA_{ik}\, \partial_j\beta^k + \tA_{jk}\,\partial_i\beta^k-\frac{2}{3}\,\tA_{ij}\,\partial_k\beta^k+\beta^k\,\partial_k \tA_{ij}\,, \label{eqn:dtAij}
\end{align}
where in the equation for $\partial_t K$ the Ricci scalar $R$ has been eliminated by an addition of (minus $\alpha$ times) the Hamiltonian constraint $\mathcal{H}$. Here $D_i$ is the metric compatible covariant derivative with respect to the physical metric $\gamma_{ij}$ and $[\ldots]^\textrm{TF}$ denotes the trace free part of the expression inside the parenthesis. 

We now introduce the conformal connection functions $\tilde\Gamma^i=\tilde\gamma^{jk}\,\tilde\Gamma^i_{~jk}$ where $\tilde\Gamma^i_{~jk}$ are the Christoffel symbols associated to the conformal metric $\tilde\gamma_{ij}$,
\begin{equation}
\tilde \Gamma^i_{~jk} = \frac{1}{2}\,\tilde\gamma^{il}\left(\partial_j\tilde\gamma_{kl} + \partial_k\tilde\gamma_{jl} - \partial_l\tilde\gamma_{jk}\right) \, ,
\end{equation}
which gives the dynamical variables for the BSSN system as
\begin{equation} 
\{ \chi, \, \tilde{\gamma}_{ij}, \, K, \, \tilde{A}_{ij}, \, \tilde{\Gamma}^i \, \} \, .
\end{equation}

In order to write down the evolution equations for $\tilde{\Gamma}^i$, we simply use their definition and the evolution equation for $\gamma_{ij}$, \eqn{eqn:dtgammaij}, to find
\begin{equation}
\partial_t \tilde \Gamma^i= \partial_j(\pounds_{\vec{\beta}} \gamma^{ij}) - 2 (\alpha \partial_j \tilde{A}^{ij} + \tilde{A}^{ij} \partial_j \alpha) \, .
\end{equation}
A crucial step is that we now eliminate the divergence term $\partial_j \tilde{A}^{ij}$ using the momentum constraint expanded in terms of the conformal variables. We also have to account for $\tilde{\Gamma}^i$ not being a tensor (and not even a tensor density as it is composed of Christoffel symbols) when expanding the Lie derivative. We therefore obtain a number of additional terms including second derivatives of the shift, in addition to the terms relating the the Lie derivative of a tensor density (see Appendix \ref{sec-appendix1LieShift}). The resulting evolution equation is
\begin{align}
&\partial_t \tilde \Gamma^i=-2\,\tA^{ij}\,\partial_j \alpha +2\,\alpha\left(\tilde\Gamma^i_{jk}\,\tA^{jk}-\frac{2}{3}\,\tilde\gamma^{ij}\partial_j K - 3\,\tA^{ij}\frac{\partial_j \chi}{\chi}\right) \nonumber \\
&\hspace{1.3cm} +\beta^k\partial_k \tilde\Gamma^{i} +\tilde\gamma^{jk}\partial_j\partial_k \beta^i +\frac{1}{3}\,\tilde\gamma^{ij}\partial_j \partial_k\beta^k \nonumber \\
&\hspace{1.3cm} + \frac{2}{3}\,\tilde\Gamma^i\,\partial_k \beta^k -\tilde\Gamma^k\partial_k \beta^i - 16\pi\,\alpha\,\tilde\gamma^{ij}\,S_j \, .\label{eqn:dtgamma1}
\end{align} 

\nomenclature[a-pi]{$\tilde{R}_{ij}$}{the part of the Ricci tensor related to the conformal metric}
\nomenclature[a-pi]{$R^{\chi}_{ij}$}{the part of the Ricci tensor related to the conformal factor}
\nomenclature[a-pi]{$\tilde D_i$}{the metric compatible covariant derivative with respect to the conformal metric $\tilde\gamma_{ij}$}

We can now use the derivatives of the evolved quantity $\tilde{\Gamma}^i$ in calculating the three-dimensional Ricci tensor, $R_{ij}$ used in the evolution equation for $\tilde{A}_{ij}$. The Ricci tensor is split as
\begin{equation}
R_{ij} = \tilde R_{ij} + R^\chi_{ij}\,,
\end{equation}
where
\begin{equation}
\tilde R_{ij} = -\frac{1}{2}\tgamma^{lm}\partial_m\partial_l\tgamma_{ij}+\tGamma^k\tGamma_{(ij)k}+\tgamma^{lm}(2\tGamma^k_{l(i}\tGamma_{j)km}+\tGamma^k_{im}\tGamma_{klj}) \label{eqn:conformalR}
\end{equation}
and
\begin{equation}
R^{\chi}_{ij}=\frac{1}{\chi}(\tD_i\tD_j\chi + \tgamma_{ij}\tD^l\tD_l \chi)-\frac{2}{\chi^2}\tgamma_{ij}\tD^l\chi \tD_l\chi. \label{eqn:Rchi}
\end{equation}
where $\tilde D_i$ is the metric compatible covariant derivative with respect to the conformal metric $\tilde\gamma_{ij}$. Note that the three-dimensional Ricci Scalar is then $R = \gamma^{ij}R_{ij}$. We could of course have simply calculated the Ricci Tensor as before, by calculating the Christoffel symbols and their derivatives from the metric derivatives on the grid. In fact it is usual in simulations to continue to use the data for $\tilde{\gamma}_{ij}$ on the slice to calculate $\tilde{\Gamma}^i$. However, using the evolved $\tilde{\Gamma}^i$ in the derivative terms $\partial_j \tilde{\Gamma}^i$ makes the equations ``more hyperbolic'' in the sense that the second derivatives of $\gamma_{ij}$ in the evolution equation for $\tilde{A}_{ij}$ (which are found in the Ricci tensor), now reduce to the scalar Laplace operator $\tilde\gamma^{kl}\partial_k \partial_l \tilde\gamma_{ij}$, with all other terms being rewritten in terms of first derivatives of $\tilde{\Gamma}^i$ \footnote{Note that it is also possible to add terms to the evolution equations proportional to the difference between the evolved $\tGamma^i$ and that calculated from the derivatives of the metric on the slice, in order to stabilise the evolution, as in \citep{Yo:2002bm}, but we do not use this ``constraint damping'' method in the work presented here.}. More formally, an analysis of the eigenfunctions, first done by Sarbach et al \citep{Sarbach:2002bt}, shows that in this form the equations are indeed strongly hyperbolic, and thus well posed.

\noindent The BSSN equations are summarised in Appendix \ref{sec-appendix1D} for reference.

\subsection{Initial conditions, gauge choice and interpretation of results}
\label{sec-InitialConditionsGauge}

Having formulated well-posed evolution equations which are suitable for numerical implementation, there are still some open points remaining before we can perform a time evolution. In particular, we need to:
\begin{enumerate}
\item{\emph{Specify initial data}: Given a coordinate grid on some spatial hyperslice, one must specify data at each point for the metric $\gamma_{ij}$, the extrinsic curvature $K_{ij}$ and stress energy components $S_i$, $S_{ij}$ and $\rho$. As we have established this is not ``free data'' in the sense that it must satisfy the Hamiltonian and momentum constraints.}
\item{\emph{Choose a gauge}: The lapse and shift are free parameters at the start, so we need to specify their initial values and how they will evolve with coordinate time. It turns out that although all gauge choices should (in theory) give the same physical result, the choice is important for achieving long term evolutions of spacetimes. In fact, gauge choice is often one of the most difficult problems in achieving a stable simulation.}
\item{\emph{Interpret the results}: Having evolved the initial data using the BSSN formulation discussed above, in the chosen gauge, one must interpret the data obtained in a gauge independent way. For example, one may wish to find event horizons or extract gravitational wave signals.} 
\end{enumerate}
These points will be considered briefly in this section. This is quite a limited overview of what are individually very large topics in their own right. The intention is to explain the methods which will be used in this thesis to the level required to understand the research undertaken, highlighting the key ideas and giving references for further details.

\subsubsection{Initial data}

Specifying the initial data amounts to specifying the 6 components of the spatial metric and the 6 components of the extrinsic curvature at each point on the initial hypersurface, given an initial matter configuration. This data must satisfy the Hamiltonian and momentum constraints, which, in the most general case, represent a set of four coupled, elliptic PDEs. 

Clearly the constraints can only remove 4 degrees of freedom from the initial data - the remaining 8 must be chosen according to physical principles or knowledge about the system, which is a non-trivial problem. A common choice is to choose the metric to be conformally flat (thus removing 5 degrees of freedom) and to impose some condition on the extrinsic curvature to remove the remaining three. For example, the extrinsic curvature can be decomposed into the product of two three-vectors, one of which can be set to zero. The problem then reduces to (the still highly difficult problem of) solving for the conformal factor $\chi$ and the remaining three-vector component of the extrinsic curvature, using the four constraint equations.

We will not consider the most general case here and simply note that the two main methods are the conformal transverse-traceless (CTT) or York-Lichnerowicz decomposition, see \citep{Cook:2000vr} for a review, and the conformal thin sandwich (CTS) decomposition which is also due to York \citep{York:1998hy}. These methods provide some guidance on how to choose values for the many unset degrees of freedom, but in the absence of significant symmetries many of the choices are arbitrary. In particular, doing things like imposing conformal flatness and setting components to zero can be shown in some cases to be ``unnatural'', in the sense that setting the problem up in this way results in some spurious gravitational wave (GW) emission at the start, before the simulation ``settles down''. 

Note that it is not that the starting point is \emph{unphysical}, as in our previous example of setting up a large mass in a completely flat space - the chosen data can satisfy the constraints exactly. It is more that the data chosen would not naturally ``spring into being'' in that configuration, so that we have artificially distorted the spacetime compared to the ``natural'' configuration we are most likely looking for. A good example is the binary black hole collision used in the convergence test performed in Chapter \ref{ch-GRChombo}. Two black holes are set up, stationary (with $K_{ij}=0$), at a fixed separation, from which they fall in by gravitational attraction. Clearly, the two bodies would not appear out of nowhere in this stationary state - they ought to have some inward velocity as a result of having (presumably) fallen in from being stationary at a large initial separation. As a result of this ``unnatural'' start, a burst of GW is produced initially before the main collision signal. In this case the problem could clearly be reduced by starting them much further apart (although this might be too computationally expensive), but in most cases there is no clear physical interpretation which would guide you to a better choice. Whilst one might consider it acceptable to have some junk GW data at the start of a simulation, it introduces the additional problem of calibrating the mass and angular momentum of the objects being evolved. That is, the configuration that the simulation ``settles into'' will not have the same mass as the initial data, as some energy is lost in the GW content. In high accuracy simulations of binary black hole mergers, this can be a potentially significant source of error, but in the research presented here it does not generate significant problems.

Returning to the methods used in this thesis, in testing the code, we use several analytic results which satisfy the constraints, such as black hole and wave data. These specific examples will be presented in Chapter \ref{ch-GRChombo}.

In our critical collapse simulations in Chapter \ref{ch-CriticalCollapse}, we choose the initial conditions such that the metric is conformally flat at a moment of time symmetry, i.e. where $K_{ij}=0$. In such a scenario, where the momentum flux $S^i$ is also zero, the momentum constraint is trivially satisfied. Choosing an initial field configuration for the field $\phi$, it is possible to solve the Hamiltonian constraint for the conformal factor $\chi$, the only remaining degree of freedom. This may be done numerically, and we take the (rather inefficient) method of relaxing $\chi$ over an initial period, until the Hamiltonian constraint is sufficiently satisfied and converges, according to 
\begin{equation}
\partial_t \chi = C_R \mathcal{H} \ ,
\end{equation}
where $C_R$ is some user defined constant which effectively sets the relaxation speed. The same approach was taken in the inflationary scenarios in Chapter \ref{ch-Inflation}, except that we have a non zero initial expansion rate $K$ which means that it is no longer time symmetric, although $\tilde{A}_{ij}=0$. The effect of this in solving the initial conditions (especially in the case of periodic (spatial) boundary conditions considered) will be discussed further in that chapter.  

Part of specifying the initial conditions involves specifying the boundary conditions for the edges of the numerical grid. In our simulations we use either periodic boundary conditions or asymptotically flat spacetimes with radiative (Sommerfeld) boundary conditions \citep{Alcubierre:2002kk}. These will be discussed further \ref{sec-BC}.

The final part of specifying the initial conditions lies in specifying the gauge conditions. As was mentioned previously, this is independent of specifying the data on the initial slice as the constraints do not depend on the gauge variables. However, because the time derivatives of matter fields are often specified with respect to coordinate time rather than proper time, the lapse and shift will (indirectly) affect the initial data and satisfaction of the constraints in these cases. For example, saying that $\partial_t \phi = 3$ means something different depending on the values of the lapse and shift. If instead the conjugate momenta of the field is specified, the physical situation would be invariant under a change of choice of the lapse and shift, but in practise it is more common to specify the coordinate time derivatives, so one must check that the implementation is consistent with what is intended. Gauge conditions are considered further in the sections below, split out into the choice of lapse and shift respectively.

\subsubsection{Gauge choice - lapse}

As we discussed above, the lapse determined the relation between coordinate time and proper time according to \eqn{eqn:LapseDefinition}. This is a local definition at each point on the slice, therefore observers at different locations (recall that each coordinate point represents an observer rather than a location in space) can have different values of the lapse, and thus travel at different rates of proper time. 

The normal observers will have some proper acceleration $\vec{a}$, and we can calculate this as
\begin{equation}
a^\mu = n^\nu \nabla_\nu n^\mu \,
\end{equation}
Expanding this relation in terms of its timelike and spacelike coordinates, and expressing the 4-dimensional Christoffel symbols in terms of 3-dimensional quantities gives
 \begin{equation}
a_0 = \beta^k \partial_k \ln(\alpha) \quad a_i = \partial_i \ln(\alpha) \, \label{eqn:ProperAcceleration}
\end{equation}
so we can see that a spatially varying lapse results in acceleration of the normal observers. 

\nomenclature[a-pi]{$\vec{a}$}{the acceleration of the normal observers}

We have said in section \ref{sec-NRStability} above that the lapse cannot be an a priori function of space and time, but must evolve dynamically. However, it is nevertheless instructive to consider the simplest possible choice of lapse, $\alpha=1$ everywhere, which seems like it ought to simplify things quite a lot. Combined with a zero shift $\beta^i=0$, this corresponds to the coordinate observers following \emph{geodesics}, since they have zero acceleration. In the case of a flat and static spacetime, this will correspond to the observers staying at a fixed location in space, which would indeed be very simple. Unfortunately, in most cases of interest, we are considering some matter distribution, and in this space geodesics will tend to focus on areas of high density. Even if those areas of overdensity subsequently disperse, the geodesic observers would continue with a constant velocity to the site to which they were previously attracted, with the result that eventually all the coordinate points will converge on the same physical point. In the most extreme case, a black hole spacetime, a grid set up with this slicing will simply fall into the event horizon (in a time $t=\pi M$ for the observers initially at the horizon), eventually causing the code to crash when the spacetime volume of each observer is too tiny to resolve. Even if we could continue to resolve the smaller and smaller volumes, our finite grid will quickly shrink until it covers only a very small region of space within the black hole, making it useless for observing external behaviour.

The focussing of observers is related to the evolution of the trace of the extrinsic curvature $K$. We have shown that this is related to the rate of growth of volume elements of the normal observer according to \eqn{eqn:KtraceVolume}, so that a positive $K$ represents a collapse of the volume elements. Consider the evolution equation for $K$ in the case of a constant unitary lapse and zero shift
\begin{equation}
\partial_t K =  K_{ij} K^{ij} + 4\pi (\rho + S) \ .
\end{equation}
This is positive definite assuming that the strong energy condition holds, leading to an ever-growing $K$, and thus ever-collapsing coordinates. This makes sense when one considers that gravity is always attractive for normal matter, so geodesic observers will always be focussed. 

A solution to this focussing of observers is the \emph{maximal slicing} condition, which preserves $K=0$ and $\partial_t K=0$ on all slices. This necessitates that the following condition is satisfied on each slice
\begin{equation}
D^2\alpha = \alpha[K_{ij}K^{ij}+4\pi(\rho+S)] .
\end{equation}
However, since this condition needs to be integrated on each timeslice, it is not well suited to a dynamic, parallelised evolution (which prefers conditions based only on local quantities, rather than global ones), and would be costly to perform, especially on a 3D AMR grid such at that used by $\grchombo$.

\begin{figure}
\begin{center}
\includegraphics[width=.8\textwidth]{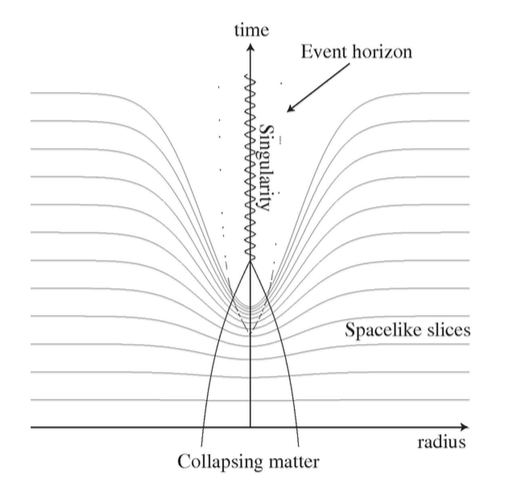}
\caption[The lapse condition - slice stretching and singularity avoidance]{The figure shows the effect of reducing the lapse in areas of high curvature, as occurs in the alpha-driver condition. Slowing the passage of proper time near the singularity creates an acceleration of the normal observers away from the point with infinite curvature, improving long term stability. However, the slices are ``stretched'' because of this, which can lead to numerical instabilities. Figure taken from Alcubierre \citep{AlcubierreBook}.}
\label{fig-SliceStretching}
\end{center}
\end{figure}

Instead we vary the lapse dynamically according to a generalised hyperbolic slicing condition, often referred to as a Bono-Masso type slicing condition \citep{Bona:1994dr}, which is designed to be ``singularity avoiding''. This is based on local quantities and derivatives at each point and thus is well suited to our implementation. The basic idea is to reduce the lapse in regions of high curvature, which tends to create an outward acceleration per \eqn{eqn:ProperAcceleration}, and in effect slows the passage of the normal observers to the point of focussing, see figure \ref{fig-SliceStretching}. The so-called \emph{alpha-driver} condition is
\begin{equation}
\partial_t \alpha = -\mu_{\alpha_1}\alpha^{\mu_{\alpha_2}}K + \mu_{\alpha_3}\beta^i\partial_i \alpha. \label{eqn:alphadriver}\,
\end{equation}
for which the commonly used $1+\log$ slicing applicable for black hole inspirals corresponds to $\mu_{\alpha_1}=2$, $\mu_{\alpha_2}=1$ and $\mu_{\alpha_3}=1$.  The optimal coefficients in this relation are in general physics dependent and we will describe in Chapter \ref{ch-CriticalCollapse} the intuition developed for these coefficients, which represented a significant part of stabilising the evolutions in the presence of matter.

\nomenclature[g-pi]{$\mu_{\alpha_i}$}{parameters in the alpha-driver lapse condition, $i=1,2,3$}

One problem with this slicing condition is that it leads to \emph{slice stretching}, in which the metric suffers shear as a result of the differing passage of time of the normal observers. This problem can be mitigated by the use of a dynamical shift condition, which we will discuss next.

\subsubsection{Gauge choice - shift}

As we found when considering the lapse, dynamical gauges are an essential element in making NR simulations stable in the presence of singularities. 

\begin{figure}
\begin{center}
\includegraphics[width=.6\textwidth]{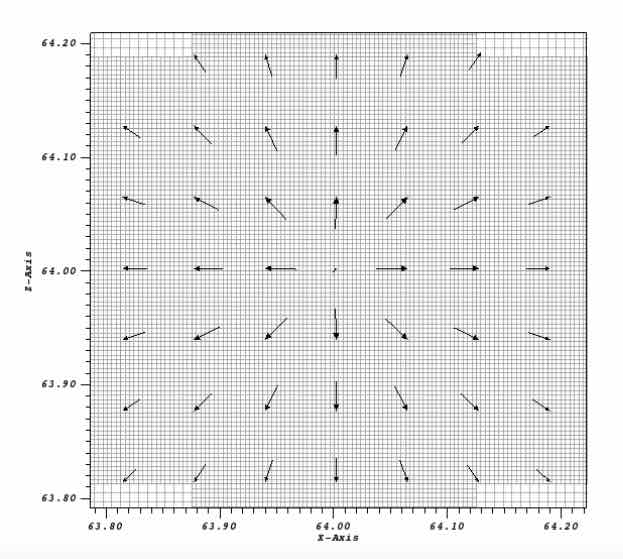}
\caption[Shift Vector]{A plot of the shift vector around a black hole using the gamma-driver condition, in a 2D slice through the centre. We see that the shift vector tends to move the observers away from the central singularity, which reduces slice-stretching and prevents them from falling into the singularity.}
\label{fig-ShiftVector}
\end{center}
\end{figure}

We have seen that we can impose a slicing condition for the lapse which slows the passage of the normal observers towards singularities. However, the lapse will never fall exactly to zero (when it does this tends to cause numerical issues), and so the central grid points will continue to infall, albeit very slowly. However, we can use our freedom to relabel spatial points to ``shift'' the observers back away from the singularity. The shift condition imposed (which will be specified below) thus tends to point away from a central singularity, see figure \ref{fig-ShiftVector}, to (approximately) maintain the positions of the observers in space relative to the singularity.
In addition, a non zero shift aims to reduce the ``slice-stretching'' caused by the alpha-driver lapse condition. This result is achieved by prescribing a condition for minimal distortion. A strict criteria would require $\tilde{\Gamma}^i=0$ and $\partial_t \Gamma^i = 0$, which necessitates the solution of a coupled set of elliptic equations on each slice, but, as with the lapse condition, we can achieve this approximately by implementing a condition which aims to instead drive $\tilde{\Gamma}^i$ to zero dynamically. This is the so-called \emph{gamma-driver} condition \citep{Alcubierre:2002kk},
\begin{equation}
\partial_t \beta^i  = \eta_{\beta_1} B^i \ , \quad \partial_t B^i = \mu_{\beta_1}\alpha^{\mu_{\beta_2}}\partial_t \tilde{\Gamma}^i-\eta_{\beta_2} B^i \ , \label{eqn:gammadriver}
\end{equation}
where $B^i$ is an auxiliary vector field, while $\eta_{\beta_1}$, $\eta_{\beta_2}$, $\mu_{\beta_1}$ and $\mu_{\beta_2}$ are input parameters. The usual hyperbolic gamma-driver condition uses the parameters $\eta_{\beta_1}=3/4$, $\eta_{\beta_2}=1$, $\mu_{\beta_1}=1$ and $\mu_{\beta_2}=0$. One may also include parameters that allow one to turn on standard advection terms in \eqn{eqn:gammadriver}, but we have not found these to be of particular use in our work so far. Again, this is a local condition which makes implementation simple.

\nomenclature[g-pi]{$\mu_{\beta_i}$}{parameters in the gamma-driver shift condition, $i=1,2$}
\nomenclature[g-pi]{$\eta_{\beta_i}$}{parameters in the gamma-driver shift condition, $i=1,2$}
\nomenclature[a-pi]{$B^i$}{An auxilliary vector field used in the gamma-driver shift condition, related to the time derivative of the shift}

The so-called \emph{moving punctures method} \citep{Campanelli:2005dd,Baker:2005vv}, is a combination of the $1+\log$ slicing for $\alpha$ and gamma-driver for $\beta^i$. As we mentioned in Chapter \ref{ch-Introduction}, the development of this gauge choice was one of the key steps in the development of the field of NR. It is strongly singularity avoiding, and is the standard choice for black hole spacetimes. Evolving a black hole in this gauge results in the ``trumpet'' solution, where the central points asymptote to a finite distance from the singularity \citep{Hannam:2008sg}. In this way we never resolve the singularity, and are able to achieve long term stable evolutions of the spacetime around it. One might worry about taking derivatives across this singular point, even within the event horizon, but in practise this does not cause issues, and the evolution is not spoiled by artefacts at the puncture \citep{Hannam:2006vv}.

\subsubsection{Interpretation of results}

In this thesis we have used several method to analyse our results. Often simply viewing the evolution of the variables in coordinate time (and taking account of the gauge issues discussed above) is sufficient to draw conclusions for our purposes. Methods which apply more generally, and which will be introduced in more detail in Chapter \ref{ch-GRChombo} on the code implementation and testing, include:

\begin{enumerate}
\item{\emph{Apparent Horizon Finder}: In most of the work we have used a spherically symmetric apparent horizon finder to identify black hole formation and quantify the mass of the black hole formed.}
\item{\emph{Mass, Angular Momentum and Momentum - ADM quantities}: One can extract data from the asymtotically flat regions of the spacetime regarding the mass, angular momentum and linear momentum of the spacetime. This was used in our testing phase but not extensively in our other work.}
\item{\emph{Gravitational Wave extraction}: Our convergence testing used the Newman-Penrose method for extracting gravitational waveforms. This has not been used in the other research presented here. Following the results of LIGO, extracting waveforms will clearly become a focus of much interest.} 
\end{enumerate}

Again these individually represent substantial topics, and we will give only a brief discussion of how they are used for our purposes in the following chapter, along with relevant references for further details. 


\section{Scalar fields with gravity} 
\label{sec-SFTheory}

In this section we summarise the key points regarding the addition of matter to the decomposed equations. We then discuss in more detail the two applications of scalar fields which are considered in this thesis - cosmology and critical collapse. 

\subsection{Scalar matter with minimal coupling} 
\label{sec-SF}

As we noted in Chapter \ref{ch-Introduction}, the equation of motion for a scalar field in flat space is the Klein Gordon equation, which can be written as
\begin{equation}
\eta^{\mu \nu}\partial_\mu \partial_\nu \phi = \frac{dV(\phi)}{d \phi} , \label{eqn:KGFlat2}
\end{equation} 
where $\eta^{\mu \nu}$ is the Minkowski metric of flat space and $V(\phi)$ is the scalar potential.

The equivalence principle (specifically the EEP) motivates the idea of minimal coupling, which is a method of modifying the flat space Lagrangian (or equivalently equation of motion) of some matter content for curved space. In this prescription, partial derivatives are replaced with covariant ones and any terms in the Minkowski metric are replaced by the full metric in curved spacetime $g_{\mu\nu}$. Thus the Klein-Gordon equation in curved space becomes
\begin{equation}
g^{\mu \nu}\nabla_\mu \nabla_\nu \phi = \frac{dV(\phi)}{d \phi} . \label{eqn:EOMKGCurved}
\end{equation} 
One can see that if the coordinates are chosen such that we are in a freely falling frame, locally the metric will be flat and the equation will reduce to the form in \eqn{eqn:KGFlat2}; thus satisfying the EEP. Note, however, that this is not the only form that we could have chosen that would satisfy the EEP, although it is the simplest. Minimal coupling will be assumed throughout this thesis where scalar fields are coupled to gravity. Note that $\nabla_\mu \phi = \partial_\mu \phi$ for a scalar field $\phi$.

In the Lagrangian picture, we have included a single minimally coupled scalar field $\phi$ as matter content
\begin{equation}
{L}_{SF} = \frac{1}{2}\nabla_\mu \phi\nabla^{\mu} \phi + V(\phi) \label{eqn:scalaraction},
\end{equation}
leading to the second order equation of motion \eqn{eqn:EOMKGCurved}, which, as is usual, we decompose into two first order equations using the variables $\phi$ and $\Pi_M$, with
\begin{equation}
\Pi_M \equiv \frac{1}{\alpha}(\partial_t \phi -\beta^i\partial_i \phi) \label{eqn:PiM}.
\end{equation}
We note that our $\Pi_M$ is the negative of $\Pi$ in some references, e.g. \citep{ShapiroBook}, and thus the negative of the conjugate momentum of the field. \eqn{eqn:EOMKGCurved} may then be decomposed into the following evolution equations in the adapted basis
\begin{equation}
\partial_t \phi = \alpha \Pi_M +\beta^i\partial_i \phi \label{eqn:dtphi}
\end{equation}
and
\begin{equation}
\partial_t \Pi_M=\beta^i\partial_i \Pi_M +\gamma^{ij}(\alpha\partial_j\partial_i \phi + \partial_j \phi\partial_i \alpha)+\alpha\left(K\Pi_M-\gamma^{ij}\Gamma^k_{ij}\partial_k \phi+\frac{dV}{d\phi}\right). \label{eqn:dtphiM} 
\end{equation} 
Again, \eqn{eqn:dtphi} is actually just the definition of $\Pi_M$. The true EOM for the field is given by \eqn{eqn:dtphiM}. 

\nomenclature[g-pi]{$\Pi_M$}{(minus) the conjugate momentum of the scalar field $\phi$}

We will also require the energy momentum tensor of the scalar field for calculating the matter components of the EM tensor in \eqn{eq:Mattereqns}. As was found in \eqn{eqn:EMTensorSF} above, the components are
\begin{equation}
T_{\mu\nu} = \nabla_\mu \phi \nabla_\nu \phi - \onehalf g_{\mu \nu} (\nabla_\rho \phi \, \nabla^\rho \phi + 2V) \ .
\end{equation}

The addition of a scalar field to the BSSN equations allows us to explore a range of effects involving gravity and fields. We now introduce the two key applications which are explored in this thesis - cosmology and critical collapse.

\subsection{Early universe cosmology}
\label{sec-Cosmology}

The Einstein equation can be used to provide insight into some of the biggest questions in physics. In particular, since gravitational effects dominate on large scales, it can be applied to the observable universe, to better understand the history of its expansion, and its future trajectory. Combined with an abundance of high accuracy data from large scale observational experiments (such as the Planck satellite) this has led to the development of a very successful model, $\Lambda$CDM, to explain the current composition of the universe.

Whilst the model is highly accurate and consistent with the measurements taken to date, many questions about the exact nature of the components remain unanswered. In addition, when the model is ``rewound'' to the start of time, significant inconsistencies are revealed, for which the theory of inflation is proposed as a solution. The nature of inflation is not well understood and although the basic principle fits well with available data, it is difficult to propose specific tests which would confirm or exclude it, or constrain the possible models. The most common model is ``slow-roll'' inflation in which a scalar field, called the ``inflaton'', drives the expansion.

In this section we summarise the key ideas in cosmology which are relevant to the work in this thesis. Further details can be found in Baumann's cosmology notes \citep{BaumannNotes} as a comprehensive starting point, or Weinberg \citep{WeinbergBook} for a complete treatment.

Note that in this section we do not set $G=1$ but follow the convention in \citep{WeinbergBook} and replace it with (non-reduced) Planck units $\mpl = \sqrt{\hbar c/G} = 2.17 \times 10^{-8} ~ \rm{kg}$, with $\hbar=c=1$, which is standard practise in cosmology\footnote{It is also common to use the reduced Planck mass $\mpl^{reduced} = \sqrt{\hbar c/8\pi G}$ which eliminates some factors of $8 \pi$ in the equations. However, since in our GR work we tend to keep the $8\pi$'s explicit, we also keep them here.}. We take the same approach when presenting our work in Chapter \ref{ch-Inflation}, although our numerical code $\grchombo$ always works in geometric units, which must then be scaled accordingly. A brief note on the conversion between these units is given in Appendix \ref{sec-appendix1A}. As in the previous sections, any cosmological constant is treated as being a component of the EM tensor, rather than being stated separately.

\subsubsection{Cosmology - kinematics and dynamics}
\label{sec-Cosmo-cos}

\nomenclature[a-pi]{$k_c$}{in cosmology, the curvature parameter for space}

Assuming a homogenous and isotropic universe, the metric can be written in terms of the coordinates $x^i$ as follows
\begin{equation}
ds^2 = a^2(t) \left( - dt^2 +  \gamma_{ij} dx^i dx^j \right) \, ,  \label{eqn:FRWmetric}
\end{equation}
where in radial polar coordinates
\begin{equation}
\gamma_{ij} dx^i dx^j = \frac{dr^2}{1 - \kappa_c r^2} + r^2 d\Omega^2 \, .
\end{equation}
Due to the scaling symmetry of the metric, it is possible to choose which of $r$, $a$ and $\kappa_c$ are dimensionful, and normalise them relative to some spatial scale $R_0$. Here we have used $\kappa_c = k_c / R_0^2$, so that it is a measure of the curvature of the surface with dimensions $(\rm{length})^{-2}$ (note that it may take any real value). The scale factor $a(t)$ is then dimensionless, and $r$ has units of length. $k_c$ is the curvature parameter which is normalised such that
\begin{align}
&k_c = -1 ~ \rm{for ~ negative ~ spatial ~ curvature} \, , \\
&k_c = +1 ~ \rm{for ~ positive ~ spatial ~ curvature} \, , \\
&k_c = 0 ~ \rm{for ~ spatial ~ flatness} \, .
\end{align}

\begin{figure}
\begin{center}
\includegraphics[width=.55\textwidth]{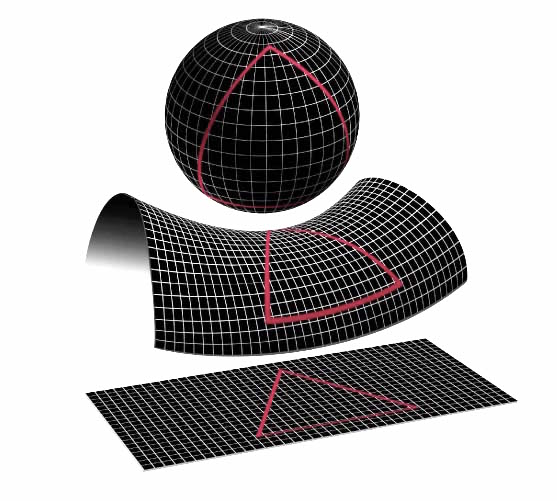}
\caption[Spatial Curvature]{The three cases of spatial curvature are illustrated. From top to bottom; Spherical/positive curvature, $k_c=+1$, areas are greater than they would be in flat space. Hyperboloidal/negative curvature, $k_c=-1$, areas are smaller than they would be in flat space. Flat space, $k_c=0$. (Remember that the curvature of interest here is the intrinsic one, thus a cylinder depicted here would also have $k_c=0$, since a triangle drawn on its surface would have the same area as in the bottom case.) Figure from \citep{Wikimedia}.}
\label{fig-SpatialCurvature}
\end{center}
\end{figure}

\begin{figure}
\begin{center}
\includegraphics[width=.8\textwidth]{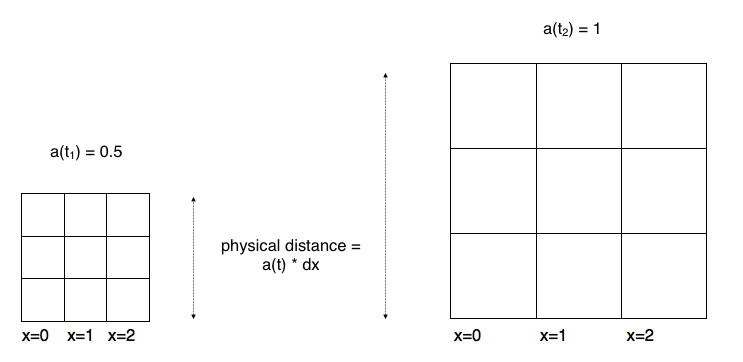}
\caption[Comoving Distance]{The coordinates $x$ are the \emph{comoving coordinates}, and do not represent physical distances. Physical distance at any time $t$ is given by $dl = a(t) dx$.}
\label{fig-ComovingDistance}
\end{center}
\end{figure}

\begin{figure}
\begin{center}
\includegraphics[width=.8\textwidth]{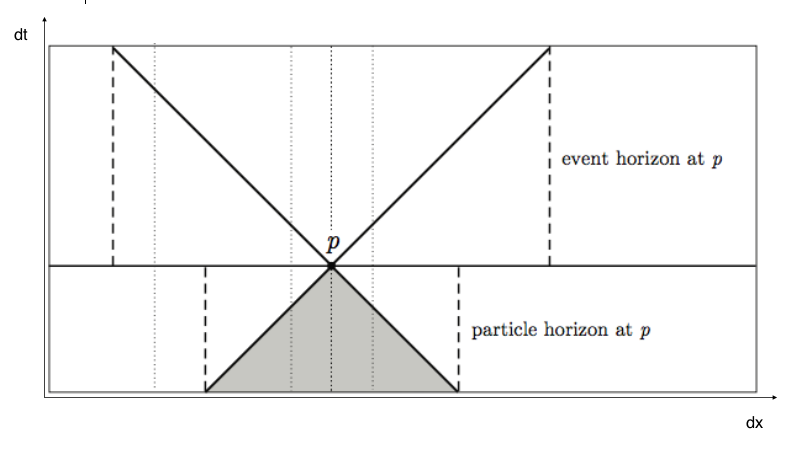}
\caption[Horizons]{In conformal time $t$ and comoving spatial coordinates $x$, with zero curvature $k_c=0$, light propagates along straight lines, which makes the past and future light cones which define the particle and event horizons easy to draw. Figure adapted from \citep{BaumannNotes}.}
\label{fig-Horizons}
\end{center}
\end{figure}

Note the following:
\begin{enumerate}
\item{\emph{No centre of the Universe}: The origin of the spatial coordinates is not a special point - choosing a different centre would give the same line element.}
\item{\emph{Constant spatial curvature}: Homogeneity and isotropy do not force the spatial metric to be flat, but they do impose a constant (intrinsic) curvature everywhere on the hyperslice, with the sign determined by $k_c$. The three possible cases are illustrated in figure \ref{fig-SpatialCurvature}. }
\item{\emph{Flat space $\neq$ flat spacetime}: The case of $k_c=0$ corresponds to a flat spatial metric, but note that this is still not necessarily a flat spacetime. The spatial slice is \emph{intrinsically flat}, but if the spacetime is expanding it has an \emph{extrinsic curvature}. It may therefore still have a non-zero curvature in 4 dimensions.}
\item{\emph{Coordinates}: The spatial coordinates $x, y, z$ are the \emph{comoving coordinates}, and as in NR they are just labels for specific points, rather than physical distances. This is illustrated in figure \ref{fig-ComovingDistance}. The time coordinate $t$ is the \emph{conformal time}, from which we can recover the change in physical proper time experienced by a comoving observer as $d\tau = a(t) dt $\footnote{Note that the use of $t$ and $\tau$ is the opposite convention to many standard Cosmology texts. The aim of using them in this way is to make a connection with the NR work described above. In cosmology conformal time $t$ is the ``unphysical'' coordinate time whereas the time $\tau$ is the proper time for a comoving observer, so it seems more consistent with the other material presented here to use them in this way.}.}
\item{\emph{A global time}: The homogeneity and isotropy of the spatial slices allows us to define a global time coordinate, the proper time measured by comoving observers. Because all points on the spatial slice are effectively the same, all comoving observers will measure the same proper (and conformal) time.}
\end{enumerate}
From now on we will consider the case of $k_c=0$ which is what we observe currently in the Universe (modulo some very small number). One can show that for a standard cosmology (made up of material which obeys the Strong Energy Condition (SEC)), the Universe must have been \emph{even flatter} in the past. This is called the ``flatness problem''. It provides a key motivation for having a period of inflation, as during such a period the spatial curvature would tend to reduce. However, we will focus in this section on the second key motivation - spatial homogeneity, which is a stronger constraint on the amount of inflation required.
In conformal time $t$ and comoving spatial coordinates $x$, light propagates along null geodesics with $ds^2 = 0$. This makes the past and future light cones (which define the particle and event horizons respectively) diagonal lines on an $x-t$ plot, as illustrated in figure \ref{fig-Horizons}.
Since the scale factor is allowed to be a function of time, one defines the Hubble parameter $H$ as
\begin{equation}
H = \frac{\dot{a}}{a} \, ,
\end{equation}
where the dot indicates a derivative with respect to physical time $\tau$. One uses the rescaling symmetry of the metric to set the current day value of the scale factor $a_0 = 1$, and observations then show that $H_0 \approx ~ 67 \rm{km~s^{-1}~Mpc^{-1}}$, where the true units are clearly time in accordance with its definition ($a$ is dimensionless), but the rather odd ones given refer to its historic definition according to the approximate relation
\begin{equation}
H = z/d \, ,
\end{equation}
for nearby objects, with $z$ redshift (in $\rm{km~s^{-1}}$) and $d$ distance (in $\rm{Mpc}$). 

\nomenclature[a-pi]{$H$}{in cosmology, the Hubble parameter}
\nomenclature[a-pi]{$a$}{in cosmology, the scale factor}
\nomenclature[a-pi]{$z$}{in cosmology, redshift}

To find the dynamics of the scale factor, one must use Einstein's equation. The requirement for the universe to be isotropic and homogeneous force the EM tensor to have the form per \eqn{eqn:EMTFluid} of
\begin{equation}
T^{\mu \nu} = (\rho + P) U^\mu U^\nu + P g^{\mu \nu} \, ,
\end{equation}
where $P$ is the pressure and $\rho$ is the energy density of the fluid as measured by a comoving observer, and $U^\mu$ is the 4-velocity of the observer with respect to the comoving frame. (Note that again we are treating any cosmological constant as contributing to the EM tensor, rather than adding a separate component on either the curvature or matter sides of the Einstein Equation.) Using the fact that the EM tensor is divergenceless per \eqn{eqn:ConsTab}
\begin{equation}
\nabla^\mu T_{\mu \nu} = 0 \, ,
\end{equation}
one obtains the continuity equation
\begin{equation}
\dot{\rho} + 3 H (\rho + P) = 0 \, , \label{eqn:ContinuityRho}
\end{equation}
from which one can obtain the scaling of the energy density for different types of matter as defined by their equation of state
\begin{equation}
w_i = P/ \rho \, , \label{eqn:EOS}
\end{equation}
as
\begin{equation}
\rho \propto a^{-3(1+w_i)}  \, .  \label{eqn:rhoEOS}
\end{equation}
From the Einstein equations themselves, we can relate the curvature to the energy content, which gives us the two ``Friedmann equations'', which are (again ignoring the curvature contribution, so $k_c=0$)
\begin{equation}
H^2 = \frac{8\pi \rho} {3 \mpl^2}  \, , \label{eqn:FRW1}
\end{equation}
and
\begin{equation}
\frac{\ddot{a}}{a} = - \frac{4 \pi (\rho+3P)} {3 \mpl^2}  \, . \label{eqn:FRW2}
\end{equation}
From \eqn{eqn:FRW1} one can define the critical density of the Universe, which is the total energy density of the Universe with zero spatial curvature $k_c$ (so the total energy density of the Universe as we currently observe it, since according to observations it is flat)
\begin{equation}
\rho_{crit} = \frac{3 \mpl^2 H_0^2}{8\pi} \, .
\end{equation}
We then express the energy density of each component of the universe as a dimensionless density parameter, representing the fraction it contributes to the total current energy density
\begin{equation}
\Omega_i = \frac{\rho_i}{\rho_{crit}} \, .
\end{equation}
In a universe dominated by a single component, combining \eqn{eqn:rhoEOS} and \eqn{eqn:FRW1} allows us to work out the evolution of the scale factor by integrating the equation
\begin{equation}
\dot{a} \propto a^{-\onehalf(1+ 3 w_i)}  \, . \label{eqn:EvolveSF}
\end{equation}
From this equation one can see that the value $w_i = -1/3$ is a critical value, which results physically from the fact that matter which obeys the SEC has a value of $w_i > -1/3$.

\nomenclature[g-pi]{$\rho_{crit}$}{in cosmology, the critical density of the universe for $k_c=0$}
\nomenclature[a-pi]{$w_i$}{in cosmology, the state parameter for a component $i$ defined by $w_i = P/ \rho$}
\nomenclature[g-pi]{$\Omega_i$}{in cosmology, the dimensionless density parameter for a component $i$}
\nomenclature[g-pi]{$\Lambda$}{in cosmology, the cosmological constant}

The table in figure \ref{table-Content} summarises the key types of matter that are currently components of our Universe, and the dependency of the energy density and scale factor in a universe in which they dominate. 
\begin{figure}
\begin{center}
\noindent \begin{tabular}{lp{5.0cm}SSSSS} \toprule
    {Type} & {Description} & {$w_i$} & {$\Omega_i$} & {$\rho(a)$} & {$a(\tau)$}  & {$a(t)$} \\ \midrule
    {Radiation} & {Gas of relativistic particles, energy density dominated by KE, e.g. photons, neutrinos} & {$1/3$} & {0.0001} & {$a^{-4}$} & {$\tau^{1/2}$} & {$t$} \\  \midrule
    {Matter} & {Gas of non-relativistic particles, pressureless, e.g. dark matter, baryons}  & {$0$} & {0.32} & {$a^{-3}$} & {$\tau^{2/3}$} & {$t^2$} \\  \midrule
    {Dark Energy} & {Vacuum Energy / Cosmological constant} & {$-1$} & {0.68} & {$a^0$} & {$e^{H \tau}$} & {$-1/t$} \\ \bottomrule
\end{tabular}
\caption[Summary of components of the Universe]{The table summarises the current known components of the Universe and their characteristics as introduced in this section.}
\label{table-Content}
\end{center}
\end{figure}
\noindent Knowing the current contributions of each component allows us to ``rewind'' the evolution of the scale factor in the Universe, according to
\begin{equation}
H^2(a) = H_0^2 \left[\frac{\Omega_{r,0}}{a^4} + \frac{\Omega_{m,0}}{a^3} + \Omega_\Lambda  \right]  \, , \label{eqn:FRW3}
\end{equation}
where the $0$ subscript denotes the present day values, and $a_0 = 1$ as is conventional\footnote{However, note that in Chapter 4 we will set $a=1$ at the start of inflation rather than at the current time, for numerical convenience.}. When we do so we find something rather surprising - \emph{there is not enough conformal time}. How much is enough? Consideration of the temperature change since the CMB was emitted gives the scale factor at that time as $a_{CMB} \approx 10^{-28}$. As we explained in Chapter \ref{ch-Introduction}, we observe the Universe to be homogeneous on extremely large scales in the CMB, and we assume that this is a result of thermalisation - the points being able to exchange signals prior to the light being emitted. The alternative explanation, that the Universe sprang into being in a completely homogeneous state is considered unnatural\footnote{What constitutes a ``natural'' initial state for the universe can be more a question of philosophy than physics. If we had a better understanding of what happened at higher energies, for example, a quantum theory of gravity, we would be better placed to comment on what is ``natural'' in this context. However, it is generally true in physics that randomness is more natural than a very ordered state.}.

\begin{figure}
\begin{center}
\includegraphics[width=.8\textwidth]{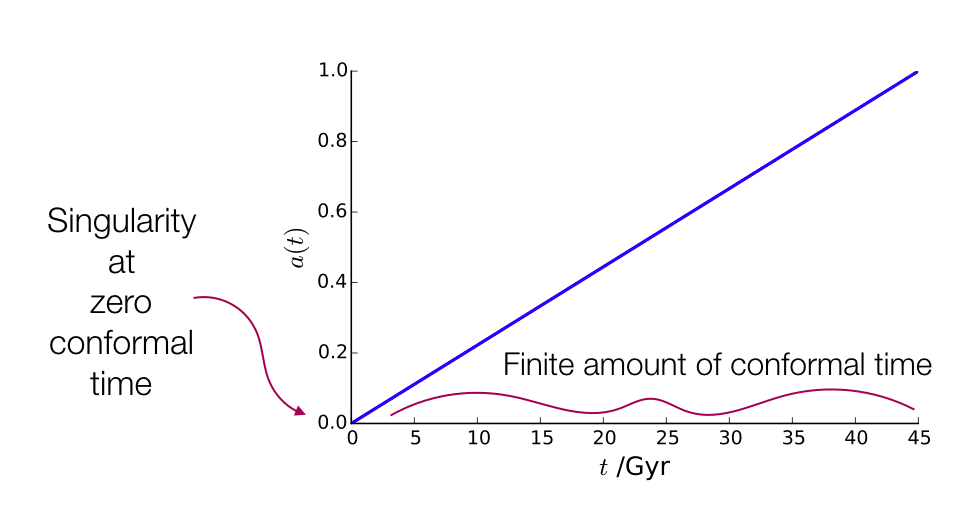}
\caption[Conformal Time in a Standard Cosmology]{Here we have illustrated the evolution of the scale factor as if it was dominated by radiation throughout the history of the Universe, but the result is essentially the same if we include the later transitions to matter and Dark energy domination. There is a finite amount of conformal time after the Big Bang singularity at which $a=0$ and before the CMB was emitted at $a_{CMB} \approx 10^{-28}$. This is insufficient to explain the homogeneity of the Universe on large scales, which led to the theory of inflation being proposed.}
\label{fig-StandardCT}
\end{center}
\end{figure}

We have said that in conformal time coordinates, the past light cones define the particle horizon - the separation of points that can have exchanged a signal at some point in the past. This then represents the locus of points that can have been in thermal contact. We therefore want the particle horizon of a distant point on the sky to have covered every other point in the observable sky \emph{at the time the CMB was emitted}. So we need as much conformal time \emph{before} the CMB was emitted (at $a_{CMB} \approx 10^{-28}$) as has passed since. That is a lot of conformal time, and it turns out to be far more than we have, assuming the matter and radiation have scaled as expected. This is illustrated in figure \ref{fig-StandardCT}, where we see that rewinding the universe leads to a singularity (the scale factor going to zero), in a finite amount of conformal time. 
The proposed solution to this so-called ``Horizon problem'' is considered in the next section: Inflation.

\subsubsection{Inflation and slow-roll models}
\label{sec-Cosmo-inf}

\nomenclature[a-pi]{$X_{ph}$}{in cosmology, the particle horizon}

As we have said, in conformal time and comoving coordinates light rays follow null geodesics that are straight lines, such that
\begin{equation}
\Delta x = \Delta t  \, .
\end{equation}
We can thus write the particle horizon $X_{ph}$ as 
\begin{equation}
X_{ph} = \int^{\tau_f}_{\tau_i} \frac{d\tau}{a(\tau)} \, , 
\end{equation}
or in terms of the scale factor
\begin{equation}
X_{ph} = \int^{a_{fin}}_{a_{init}} \frac{1}{a H} \frac{da}{a}  \equiv t_f - t_i  \, . \label{eqn:ParticleHorizon}
\end{equation}
The comoving Hubble radius is defined to be $(aH)^{-1}$. For a single component fluid then from \eqn{eqn:EvolveSF} this is  (again taking $a_0=1$)
\begin{equation}
(aH)^{-1} = H_0^{-1} a^{\onehalf (1+3w_i)} \, \label{eqn:HHorizon}
\end{equation}
and we can see by integrating \eqn{eqn:ParticleHorizon} that the initial conformal time
\begin{equation}
t_{init} = \frac{2 H_0^{-1}}{1+3w} a_{init}^{\onehalf(1+3w_i)} \, 
\end{equation}
is zero when $a_{init} \rightarrow 0$ for $w_i >-1/3$, as was illustrated in figure \ref{fig-StandardCT} for radiation domination. The particle horizon is then determined by the final value of conformal time, and is of order of the comoving Hubble radius, ie $X_{ph} = t_{fin} \sim (aH)^{-1}$. This gives a finite amount of conformal time since the initial singularity, which is insufficient to explain the homogeneity observed in the CMB on the largest scales. 

\begin{figure}
\begin{center}
\includegraphics[width=.8\textwidth]{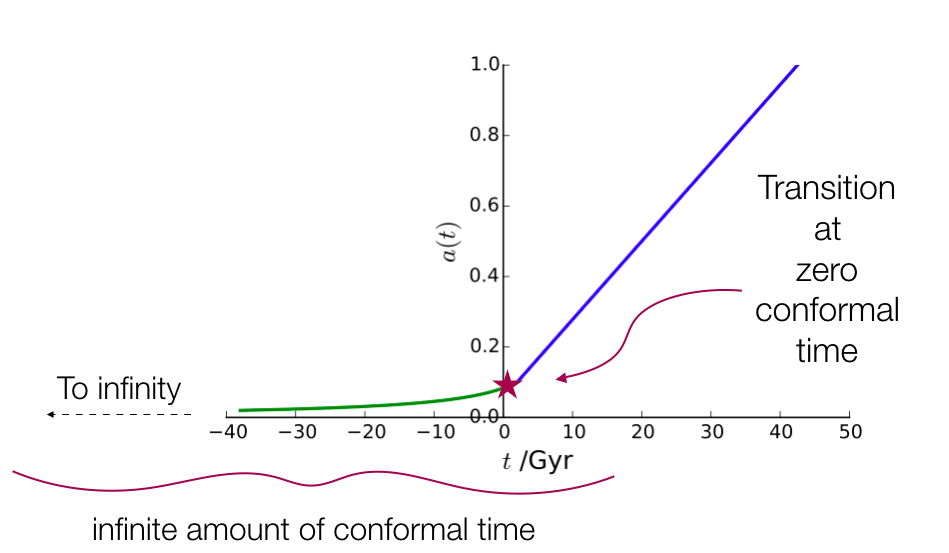}
\caption[Conformal Time with Inflation]{Here we have illustrated the evolution of the scale factor as if it was dominated by radiation throughout the late history of the Universe, but with a transition to an SEC violating fluid around $t=0$. There is now a potentially infinite amount of conformal time after the Big Bang singularity at which $a=0$, and before the CMB is emitted, giving distant parts of the Universe time to thermalise. Note that we have exaggerated the vertical scale of the transition point for clarity - the scale factor at transition would be less than $a_{CMB} \approx 10^{-28}$, whereas in the illustration it is shown as about $a=0.1$.}
\label{fig-InflationCT}
\end{center}
\end{figure}

However, for a fluid that violates the SEC $a_{init} \rightarrow -\infty$ as $w<-1/3$, as illustrated in figure \ref{fig-InflationCT}. There is now an (in principle) infinite amount of conformal time before the singularity is reached\footnote{Note that the additional conformal time is additional \emph{coordinate} time, and does not necessarily correspond to a large amount of additional proper time being experienced by a comoving observer.}, and so plenty of time for the Universe to have thermalised before the CMB was formed. Such a period, dominated by an SEC violating fluid, solves the horizon problem. By considering the amount of expansion which has occurred since the CMB was emitted, one finds that the minimum amount of inflation required to ensure thermal contact between the whole sky is roughly 60 $e$-folds, where the number of $e$-folds $\mathcal{N}$ satisfies
\begin{equation}
d\mathcal{N} = d(\ln a) = H d \tau \, .
\end{equation}

\begin{figure}
\begin{center}
\includegraphics[width=.6\textwidth]{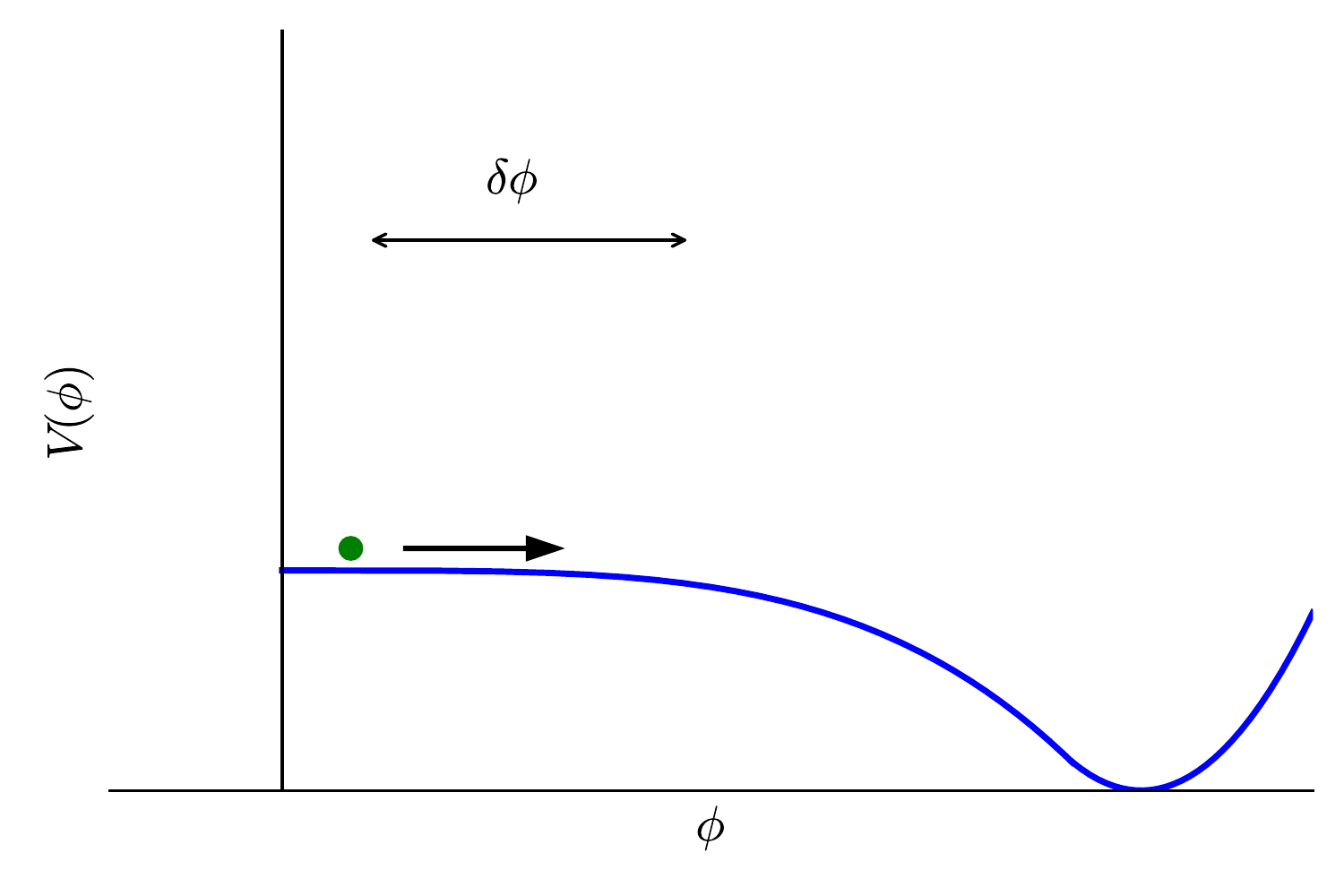}
\caption[Slow roll potential]{A typical \emph{slow-roll} potential is illustrated. The inflaton field $\phi$ sources the non zero energy density which causes inflation for a period as it rolls along a plateau in the potential, before falling to the minimum at which the value of $V(\phi)$ is zero, at which point inflation ends.}
\label{fig-SlowRoll}
\end{center}
\end{figure}

We can express the behaviour in the inflationary period in several equivalent ways, each of which can be taken as the basic definition of inflation and used to derive the other characteristics:
\begin{enumerate}
\item{\emph{A shrinking Hubble sphere}: We can see from \eqn{eqn:HHorizon} that, for $w_i < -1/3$, $(aH)^{-1}$ is smaller at the end of inflation that at the start.}
\item{\emph{A period of accelerated expansion}: A shrinking Hubble horizon implies that
\begin{equation}
\frac{d (aH)^{-1}}{d\tau}  = - \frac{\ddot{a}}{(\dot{a})^2} < 0 \, ,
\end{equation}
which corresponds to a period of positive acceleration $\ddot{a}$ in the expansion.}
\item{\emph{Roughly constant H}: A shrinking Hubble horizon also implies that
\begin{equation}
\frac{d (aH)^{-1}}{d\tau}  = \frac{1}{a} (\epsilon-1) < 0 \, ,
\end{equation}
where the Hubble slow roll parameter is defined as $\epsilon \equiv - \dot{H}/H^2$. Thus we require $\epsilon < 1$ and small $\dot{H}$ for the expansion.}
\item{\emph{De-Sitter like expansion}: For perfect inflation $\epsilon=0$ and the Hubble parameter $H$ is constant, which corresponds to a de-Sitter expansion, with $w=-1$, as for a cosmological constant.}
\item{\emph{Negative pressure}: An SEC violating fluid obeys $w= \rho/P < -1/3$ and thus has a negative pressure as the energy density measured by any observer should be positive.}
\item{\emph{Constant density}: From \eqn{eqn:ContinuityRho} one can show that
\begin{equation}
\frac{d \ln \rho}{d\ln a}  = 2 \epsilon \, ,
\end{equation}
so that small $\epsilon$ means $\rho$ is approximately constant.}
\end{enumerate}
Inflation is commonly thought of as a period of de Sitter expansion, but it cannot be sourced by a simple cosmological constant, because ultimately \emph{inflation ends}. A true cosmological constant would have continued to dominate the expansion to the present day, so this means that at some point the cosmological constant would have had to ``switch off'', which seems unnatural. The most commonly proposed solution to this problem is slow-roll inflation in which a scalar field $\phi$ sources the non zero energy density for a period as it rolls along a plateau in the potential, before falling to a different part at which the value of $V(\phi)$ is zero. A typical potential $V(\phi)$  for this ``inflaton'' field is shown in figure \ref{fig-SlowRoll}.
As illustrated in this figure, it is common to think of the inflaton as ``rolling down a hill''. If space is homogeneous then the value of the inflaton field is the same everywhere and the picture expresses how this universal value changes over time. However, if the space is \emph{inhomogeneous}, then the value of $\phi$ can differ at all points in space and each point will need its own picture, which in isolation is an incomplete description as it does not include the effects of spatial gradients in the field, which we know play a role per the Klein-Gordon equation \eqn{eqn:EOMKGCurved}.

In fact we ought to expect to need to add variation into the field - we propose inflation to explain the homogeneity of the universe as observed at the emission of the CMB, but there is no reason to suppose that it was homogeneous \emph{before inflation happened}. In fact, it is perverse to do so, since if the universe \emph{was} homogeneous prior to inflation then inflation is unnecessary - homogeneity is already assured. 

In the remainder of this section, we will summarise the key components of standard slow-roll inflation assuming spatial homogeneity in the field. Our work in Chapter \ref{ch-Inflation} concerns the effects of removing this assumption. 

The requirements of slow roll are usually summarised by the two \emph{Hubble slow roll parameters} $\epsilon$ and $\eta$. We have already defined 
\begin{equation}
\epsilon \equiv - \frac{\dot{H}}{H^2} \, ,
\end{equation}
for which the requirement that $\epsilon \ll 1$ corresponds to the requirement for H to be approximately constant. The second parameter is defined as
\begin{equation}
\eta \equiv \frac{\dot{\epsilon}}{H\epsilon} \, ,
\end{equation}
for which the requirement that $\vert \eta \vert \ll1$ corresponds to the requirement for $\epsilon$ to be approximately constant, such that inflation persists for a sufficient number of $e$-folds.

\nomenclature[g-pi]{$\epsilon$}{in cosmology, the first Hubble slow roll parameter}
\nomenclature[g-pi]{$\eta$}{in cosmology, the second Hubble slow roll parameter}
\nomenclature[g-pi]{$\epsilon_V$}{in cosmology, the first potential slow roll parameter}
\nomenclature[g-pi]{$\eta_V$}{in cosmology, the second potential slow roll parameter}

For a scalar field the values of $\rho$ and $P$ are
\begin{equation}
\rho_{\phi} = \onehalf \dot{\phi}^2 + V(\phi) \, ,  \quad P_{\phi} = \onehalf \dot{\phi}^2 - V(\phi) \, ,
\end{equation}
so that, for an SEC violating fluid, $V(\phi)$ must dominate over the kinetic energy. Substituting these density and pressure expressions into the Friedmann equations, \eqn{eqn:FRW1} and \eqn{eqn:FRW2}, we can obtain the evolution equation for the field, which is equivalent to the Klein-Gordon equation
\begin{equation}
\ddot{\phi} + 3 H \dot{\phi} + \frac{d V(\phi)}{d\phi} = 0  \, ,
\end{equation}
where the Hubble constant acts as a friction and the potential gradient acts as a force, driving the motion down the potential. We can also deduce that the slow roll parameters, see \citep{Copeland:1993jj, Liddle:1992wi}, are
\begin{equation}
\epsilon = \frac{4\pi \dot{\phi}^2}{\mpl^2 H^2}  \, ,
\end{equation}
and
\begin{equation}
\eta = \frac{2 \ddot{\phi}}{H\dot{\phi}} - 2 \epsilon \, .
\end{equation}
In the case of a homogeneous field in slow roll we can make several simplifying assumptions that allow us to characterise the behaviour in terms of the potential and its gradients only. In particular, assuming that the potential dominates over the kinetic energy $V(\phi) \gg \dot{\phi}^2 /2$ means that the first Friedmann equation \eqn{eqn:FRW1} becomes
\begin{equation}
H^2 \approx \frac{8\pi V}{3 \mpl}  \, ,
\end{equation}
and assuming that $\ddot{\phi} \approx 0$ gives the Klein Gordon equation
\begin{equation}
3H\dot{\phi} \approx - \frac{dV}{d\phi}  \, ,
\end{equation}
for which the time derivative is
\begin{equation}
3\dot{H}\dot{\phi} + 3H\ddot{\phi} \approx - \frac{d^2V}{d\phi^2} \dot{\phi}  \, .
\end{equation}
One then defines the \emph{Potential slow-roll parameters}, per \citep{Liddle:1992wi}, as
\begin{equation}
\epsilon_V \equiv \frac{\mpl^2}{16 \pi} \left( \frac{V'}{V} \right) ^2 \approx \epsilon \, ,
\end{equation}
where the dash denotes a derivative with respect to $\phi$, and
\begin{equation}
\eta_V \equiv \frac{\mpl^2}{8\pi} \frac{\vert V'' \vert}{V} \approx 2\epsilon - \frac{\eta}{2} \, .
\end{equation}
These parameters must both be much smaller than 1 for sufficient inflation to proceed. One can also work out the number of $e$-folds $\mathcal{N}$ as
\begin{equation}
\mathcal{N} = \int^{\phi_E}_{\phi_I} \sqrt{\frac{4\pi}{\epsilon}} \frac{d\phi}{\mpl} \, ,
\end{equation}
where $\epsilon$ can be replaced approximately by $\epsilon_V$, and $\phi_I$ and $\phi_E$ are defined as the values at which $\epsilon_V=1$.

\nomenclature[a-pi]{$\mathcal{N}$}{in cosmology, the number of $e$-folds}

After the inflaton falls into the minimum of the potential, its oscillations are expected to generate the particles of the standard model during the \emph{reheating} period. This marks the commencement of the ``standard big bang'' era of cosmology. 

\subsubsection{Cosmology and the ADM decomposition}
\label{sec-ADMCosmo}

It can be useful to relate the FRW cosmological properties that have been discussed to the equivalent NR quantities. In simplified cases there is a direct correspondence, and even in more complex cases, where the correspondence is broken due to a lack of homogeneity, it can be useful to think in these terms to develop some physical intuition for what is happening in a simulation.  

Consider the FRW metric in \eqn{eqn:FRWmetric}
\begin{equation}
ds^2 = a^2(t) \left( - dt^2 +  \gamma_{ij} dx^i dx^j \right) \, ,
\end{equation}
with the time evolution quantified by
\begin{equation}
H = \frac{\dot{a}}{a} \, ,
\end{equation}
and compare it to the BSSN metric
\begin{equation}
ds^2=-\alpha^2\,dt^2+\frac{1}{\chi^2} \tilde{\gamma}_{ij} dx^i dx^j \, ,
\end{equation}
with time evolution quantified by 
\begin{equation}
\partial_t \gamma_{ij} = - 2\alpha K_{ij} \, ,
\end{equation}
where we have immediately set the shift to zero, since we know that physically it does not change the embedding of the spatial hyperslice, but simply represents a relabelling of the coordinates on subsequent slices. In an FRW cosmology this would just correspond to relabelling the positions of our comoving observers - a change of what we call $x$, $y$ and $z$, for which we have already said the origin is an arbitrary point.

Consider the spatial part of the metric $\gamma_{ij}$. Isotropy, combined with setting the curvature parameter to zero, $k_c = 0$, corresponds to conformal flatness $\tilde{\gamma}_{ij} = \delta_{ij}$ in NR language. In this case the conformal factor is related directly to the scale factor by $\chi = 1/a$. However, in our simulations, even when they are conformally flat, $\chi$ can vary on the spatial slice whereas $a$ is assumed to be constant in space. It can be useful to consider small areas of the simulation domain as patches of FRW spacetime that are locally spatially flat but have undergone different amounts of expansion, and thus have a different scale factor. However, once spatial isotropy is lost, and especially in a varying gauge, this intuition can quickly break down.

Now consider the time evolution of the metric. We can connect the trace of the extrinsic curvature with the Hubble parameter in a homogeneous and isotropic universe as $K = -3H$. Whereas $H$ is a constant in space, we may have a spatially varying $K$, and again it can then be helpful to think of locally FRW patches sewn together, which are expanding at different rates. Note that a negative extrinsic curvature corresponds to an expanding universe, as expected from our NR convention.

Finally, looking at the time evolution, we see that in the case where $\alpha = 1$ we recover proper time measured by the comoving observers (who are the normal observers of NR). If we choose our lapse to be equal to the scale factor, $\alpha = a$, then our time coordinates will correspond to conformal time coordinates. In particular, if we choose the dynamical lapse to evolve as
\begin{equation}
\partial_t \alpha = - 3 \alpha^2 K \, ,
\end{equation}
then in an FRW spacetime this corresponds to 
\begin{equation}
\partial_t \alpha = \alpha^2 H = \frac{\alpha^2}{a^2} \partial_t a \, ,
\end{equation}
so that if the lapse is initially the same as the scale factor, $\alpha = a$, the two will evolve in the same way and therefore the slicing will be that of conformal time. Whilst we do not use exactly this slicing in our work in Chapter \ref{ch-Inflation}, we use something similar, such that the lapse gets bigger approximately in line with the increase in the scale factor. For stability, it is acceptable for the lapse to grow at a rate slower than the scale factor, but not faster, which would break the Courant condition (see section \ref{sec-BRAMR}). 

\subsection{Critical collapse}
\label{sec-CriticalCollapse}

One of the most fascinating and as yet not fully understood aspects of general relativity is the appearance of critical phenomenon in gravitational collapse as first discovered by Choptuik \citep{Choptuik:1992jv}. A comprehensive review can be found in \citep{Gundlach:2007gc}. 

Briefly, if we have an initial configuration, such as a Gaussian shaped bubble of scalar field, and allow this to evolve under the action of gravity, the result will be either the formation of a black hole, or dispersal of the field to infinity depending on the ``strength'' of the initial data. Varying any one initial parameter $p$ of the configuration (such as the height of the bubble), one finds that there is a critical point $p^*$ at which the transition between the two end states occurs, and that the mass of the black hole created on the supercritical side follows the scaling relation
\begin{equation}
M \propto (p - p^*)^{\gamma_S} \, , \label{eqn:scalingrelation}
\end{equation}
where the scaling constant $\gamma_S$ is universal in the sense that it does not depend on the choice of family of initial data. For a massless scalar in a spherically symmetric collapse, $\gamma_S$ has been numerically determined to be around 0.37. This index does, however, depend on the type of matter considered.

\begin{figure}
\begin{center}
\includegraphics[width=.8\textwidth]{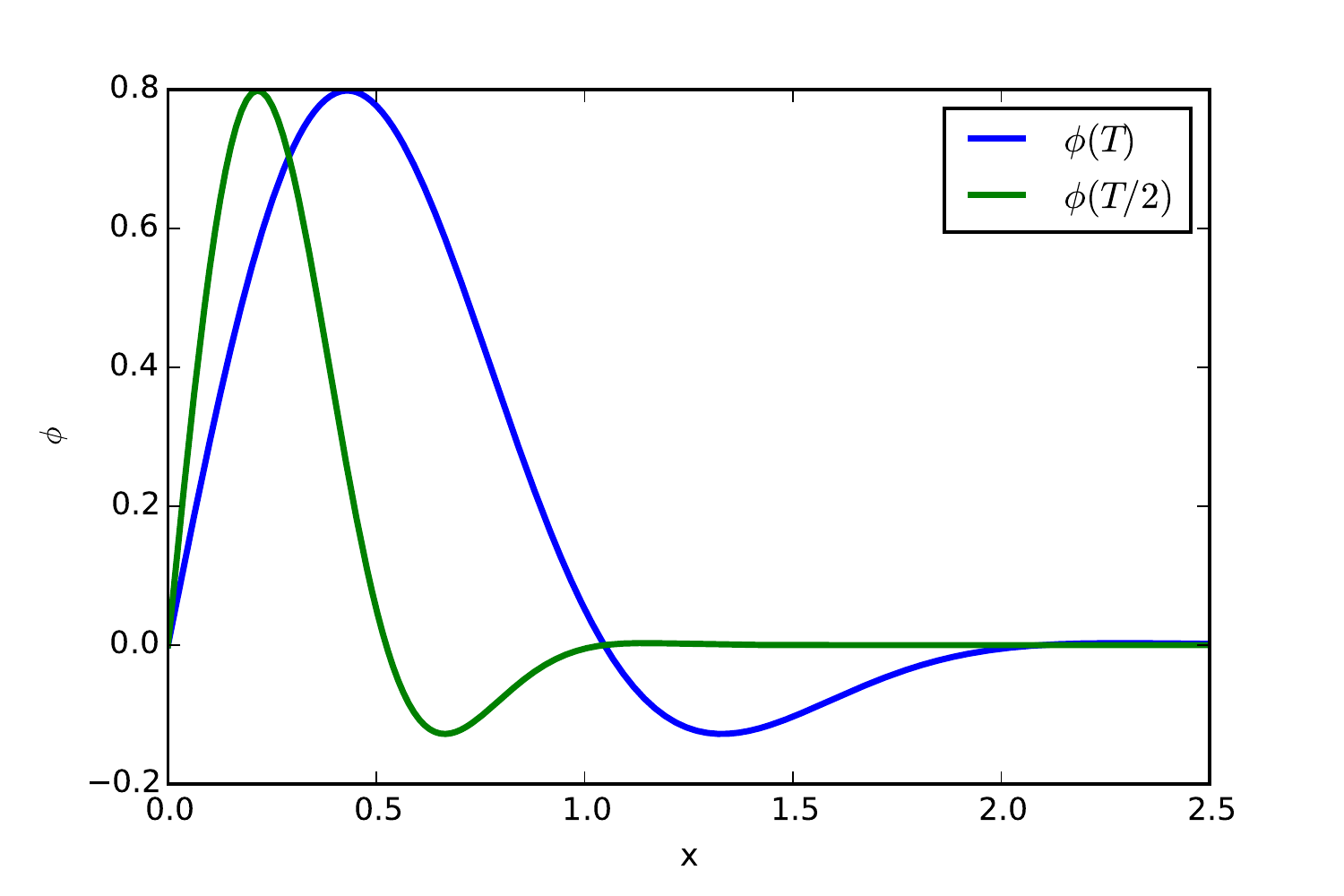}
\caption[Scale Echoing in Critical Collapse]{During a critical collapse, one sees that, as the time nears the critical time by a factor of $e^{-\Delta}$, the same field profile is seen but on a scale $e^{-\Delta}$ smaller. (Here, for example, $e^{-\Delta}=\onehalf$, where the value is just for illustration.)}
\label{fig-ScaleEchoing}
\end{center}
\end{figure}

The other key phenomenon which is observed is that of self-similarity in the solutions, or ``scale-echoing''. Close to the critical point, and in the strong field region, the value of any gauge independent field $\phi$ at a point $x$ and time $T$ exhibits the scaling relation
\begin{equation}
\phi (x,T) = \phi(e^{\Delta_S} x, e^{\Delta_S} T) \, , \label{eqn:echorelation}
\end{equation}
where $\Delta_S$ is a dimensionless constant with another numerically determined value of 3.44 for a massless scalar field in the spherical case. The time $T$ here is measured ``backwards'' - it is the difference between the critical time at which the formation of the black hole occurs and the current time, with time being the proper time measured by a central observer. What one sees is therefore that, as the time nears the critical time by a factor of $e^{\Delta_S}$, the same field profile is seen but on a scale $e^{\Delta_S}$ smaller, as illustrated in \ref{fig-ScaleEchoing}. 

\nomenclature[a-pi]{$t$}{co-ordinate time, conformal time in cosmology}
\nomenclature[a-pi]{$T$}{in critical collapse, proper time of a central observer, measured backwards from the critical time}

In this section we will summarise the key principles which underlie these characteristics, firstly explaining the origin of universality and how this leads to the scaling relation \eqn{eqn:scalingrelation}, and then discussing scale invariance and importance of gauge choice in looking for echoing solutions. 

The results presented here are only known to apply to the spherically symmetric case, which has been well studied. In Chapter \ref{ch-CriticalCollapse} we will detail the work which was undertaken to study scalar field bubbles in asymmetric configurations. This is expected to exhibit similar behaviour, but has been considerably more difficult to study due to the high levels of refinement required. 

\subsubsection{Universality and scaling}

One can consider the initial conditions of a GR spacetime, decomposed into the ADM quantities on some initial hypersurface, to be an infinite dimensional continuous dynamical system. Each point in the phase space is characterised by the configuration of the set of variables $\left\lbrace \gamma_{ij}, ~ K_{ij}, ~ \phi, ~\Pi_M \right\rbrace$ across the whole spatial slice, that must satisfy the constraint equations. Given some gauge choice of lapse and shift, the solution curves follow a trajectory in this phase space. 

For a massless scalar field, there are only two end points for an isolated system following collapse - formation of a black hole or dispersal of the field to infinity. Thus the phase space is divided into two halves - one for which all trajectories ultimately result in a black hole, the other in dispersal. A ``critical surface'' (CS) separates the two regions, forming a manifold with one less dimension that the full phase space. Since points on the surface go to neither of the two extremes, they will by definition stay within the surface if sufficiently finely tuned. One postulates that there is an attracting fixed point, or ``critical point'' (CP) somewhere in the CS and that it is an attractor of co-dimension one, i.e. there is a single growing mode which is not tangential to the CS.

\begin{figure}
\begin{center}
\includegraphics[width=.8\textwidth]{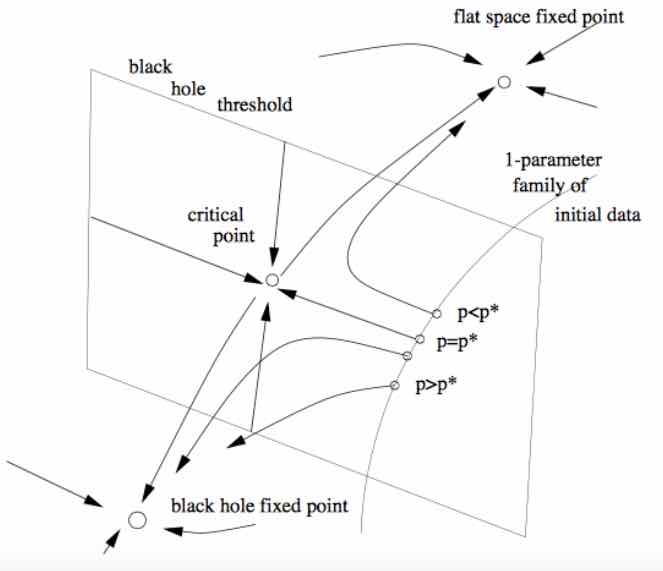}
\caption[Phase space picture of Critical Collapse]{The diagram illustrates how the phase space is separated into two regions by a critical surface. Initial data moves parallel to the surface towards the critical point, before diverging along the direction of the unstable mode. This is a very simplified diagram (the phase space is infinite dimensional), so one should take it as a heuristic picture rather than an exact representation. Diagram from \citep{Gundlach:2007gc}}
\label{fig-PhaseSpace}
\end{center}
\end{figure}

The result of this picture is that for initial data close to the critical surface, the evolution trajectory will be initially in the direction of the CP, moving parallel to the CS. As it nears the CP, it slows down, and then moves away in the direction of the unstable growing mode. This ``funnelling'' effect means that all initial data ends up following the same final path, differentiated only by how far they were initially from the CS. This is illustrated schematically in figure \ref{fig-PhaseSpace}. 

One can use this picture to derive the scaling relation from a simple dimensional analysis. This will be explained in the following section.

It is often found in dynamical systems analysis that critical points have additional symmetries. This appears to be the case in GR. In type II critical collapse, the critical spacetime is self-similar or scale invariant; i.e. if one assumes continuous self symmetry (CSS), there exists a ``homothetic'' vector field $\vec{\xi}$, which is one for which the Lie derivative of the metric satisfies
\begin{equation}
\pounds_{\vec{\xi}} ~ g_{ab} = 2 g_{ab}
\end{equation}
(see Appendix \ref{sec-appendix1B} for a discussion of Lie derivatives). In coordinates adapted to the symmetry 
\begin{equation}
x^\mu = (\sigma_S, ~ x^i) \, , \quad \xi = - \frac{\partial}{\partial \sigma_S} \, ,
\end{equation}
we can write
\begin{equation}
g_{\mu \nu}(\sigma_S,~ x^i) = e^{-2 \sigma_S} \tilde{g}_{\mu \nu} (x^i) \, .
\end{equation}
In discrete self similarity (DSS) the scaling is periodic in $\sigma_S$ with period $\Delta_S$ so that
\begin{equation}
 \tilde{g}_{\mu \nu} (\sigma_S, x^i)=  \tilde{g}_{\mu \nu} (\sigma_S + \Delta_S, x^i)  \, .
\end{equation}
Note that the proper time $T$ is generally related to the adapted coordinate $\sigma_S$ (which is dimensionless) as $\sigma_S \sim \ln{T/T_0}$, which is why in \eqn{eqn:echorelation} the $\Delta_S$ appears as a factor of $e^{\Delta_S}$. It is said that $\sigma_S$ is the logarithm of the spacetime scale. Thus the echoing is not spaced evenly in proper time unless expressed on a logarithmic scale. This is one factor which makes it difficult to resolve the many echos that occur over increasingly small timescales in a simulation.

\nomenclature[g-pi]{$\sigma_S$}{in critical collapse, the logarithm of the spacetime scale}

Another problem with observing the critical behaviour in simulations is that of gauge choice. One must choose a lapse and shift which adapts the slicing to the symmetries of the spacetime metric and the homothetic vector field. There is no obvious way to ensure this for general initial conditions, especially in the absence of spherical symmetry. It has been suggested by Smarr and York \citep{Smarr:1977uf} that maximal slicing and minimal strain should meet this requirement, but as we have explained above, these conditions, if enforced strictly, require the solution of elliptic equations on each slice, which is computationally demanding. One may hope that their approximate realisation as hyperbolic driver conditions would be sufficient, and there is evidence that this is the case from recent work by Akbarian et. al. \citep{Akbarian:2015oaa}.

\subsubsection{Deriving the scaling relation for type I and II critical collapse}

We first consider a type II critical collapse, which corresponds to the case where there is no mass scale in the problem, such as in the massless scalar field.

If we take some variable $Z$, which is a scale invariant variable such as $\tilde{g}_{\mu \nu}$, and rescaled matter variables $\tilde{\phi}$, then $Z(x)$ is an element of the phase space we have described up to a scale given by $\sigma_S$, which defines $Z(x,\sigma_S)$. 

\nomenclature[a-pi]{$Z$}{in critical collapse, a scale invariant variable}
\nomenclature[g-pi]{$\lambda_0$}{in critical collapse, the eigenvalue of the growing mode}

If we assume that the critical point has a CSS associated with it, then solutions in the phase space around it can be expanded to linear order in its perturbation modes as
\begin{equation}
Z(x, \sigma_S) \simeq  Z^*(x) + \sum^{\infty}_{i=0} C_i(p) e^{\lambda_i \sigma_S} Z_i (x)  \, ,
\end{equation}
where $C_i(p)$ are the perturbation amplitudes. These amplitudes depend on initial data only as a function of the distance from the CS, characterised by some parameter $p$, due to the funnelling effect of the solutions near to the CP. If the CP has only one growing mode, with a positive real $\lambda_i$, then in the limit of $\sigma_S \to \infty$ the other modes will vanish. Therefore in this limit
\begin{equation}
\lim_{\sigma_S \to \infty} Z(x, \sigma_S) \simeq  Z^*(x) + C_0(p) e^{\lambda_0 \sigma_S} Z_0(x)  \, .
\end{equation}
We can then expand the perturbation amplitude for this mode about the critical value of $p^*$
\begin{equation}
C_0(p) = C_0(p^*) +  \frac{dC_0}{dp} (p-p^*) ... \, .
\end{equation}
which, recognising that the $C_0(p^*)= 0$ as the perturbations are zero at the CS, gives
\begin{equation}
\lim_{\sigma_S \to \infty} Z(x, \sigma_S) \simeq  Z^*(x) +  \frac{dC_0}{dp} (p-p^*) e^{\lambda_0 \sigma_S} Z_0(x) \, .
\end{equation}
Then considering the solution at some $\bar{\sigma_S}$ defined by
\begin{equation}
\frac{dC_0}{dp} (p-p^*) e^{\lambda_0 \bar{\sigma_S}} = \epsilon \, ,
\end{equation}
where $\epsilon \ll 1$ so the linear approximation is still valid,
\begin{equation}
Z(x, \bar{\sigma_S}) \simeq  Z^*(x) + \epsilon Z_0(x) \, .
\end{equation}
This solution will be the same regardless of the value of $\epsilon$, apart from a scale given by $e^{-n\sigma_S}$ where n is the length (or mass) dimension of the variable $Z(x, \sigma_S)$. If the variable is a mass $n=1$ and the solution will scale as $e^{-\sigma_S}$, so
\begin{equation}
M \propto e^{-\sigma_S} \propto (p-p^*)^{1/\lambda_0} \, .
\end{equation}
This tells us that the critical exponent in  \eqn{eqn:scalingrelation} is related to the eigenvalue of the unstable mode of the critical solution as $\gamma_S = 1/\lambda_0$.

For DSS there is a small modification to the relation due to the period of the DSS, such that
\begin{equation}
\ln M = \gamma_S \ln (p-p^*) + C_D + f(\gamma_S \ln (p-p^*) + C_D) \, ,
\end{equation}
where $C_D$ is some constant, but $f(z)$ is a universal function with period $\Delta_S$.

In the case of Type I critical collapse, where a mass scale in the evolution equations is dynamically relevant (such as a scalar field with a large mass), the picture is similar. However, now the solution is not scale invariant but rather time invariant (or time periodic). Thus the critical solution has a finite mass and the universality in this context corresponds to the final BH mass near the critical threshold being independent of the initial data. The quantity that now scales with the distance to the CS is the lifetime of the intermediate state for which the solution is approximately critical, which scales as
\begin{equation}
T_p = (p-p^*)^{-\gamma_S} + C_I \, ,
\end{equation}
where $C_I$ is a data dependent constant.

%% file: Chapter3/chapter3.tex

\chapter{GRChombo - code development and testing}
\label{ch-GRChombo}

\ifpdf
    \graphicspath{{Chapter3/Figs/Raster/}{Chapter3/Figs/PDF/}{Chapter3/Figs/}}
\else
    \graphicspath{{Chapter3/Figs/Vector/}{Chapter3/Figs/}}
\fi

$\grchombo$ is a new multi-purpose numerical relativity code, which is built on top of the open source $\mathtt{Chombo}$ \citep{Chombo} framework. In this chapter, we will detail the capabilities of $\grchombo$ and illustrate how they expand the current field in numerical GR to permit new physics to be explored. The design methodology, scaling properties and performance of $\grchombo$ in a number of standard simulations are included. Videos of simulations using $\grchombo$ can be viewed via the website at www.grchombo.org. The work presented in this chapter is mainly derived from the paper ``GRChombo : Numerical Relativity with Adaptive Mesh Refinement'' \citep{Clough:2015sqa}.

\noindent The chapter is organised as follows:
\begin{itemize}
\item In section \ref{sec-Chombo} we describe the functionality of $\mathtt{Chombo}$ which is utilised by $\grchombo$, including the program structure, adaptive mesh refinement (AMR) methodology and load balancing.
\item In section \ref{sec-GRChombo} we describe the implementation of the code that we have built on top of $\mathtt{Chombo}$, including the finite differencing scheme, dissipation and equations of motion.
\item In section \ref{sec-Tests}, we present the results of standard tests, including the Apples with Apples tests \citep{Babiuc:2007}, black holes and black hole mergers, and critical collapse. We test the AMR capabilities of the code, its robustness to regridding errors, and its scaling and convergence properties.
\end{itemize}

\nomenclature[z-pi]{RK4}{Runge-Kutta 4th order}

\section{Features of Chombo}
\label{sec-Chombo}

$\mathtt{Chombo}$ is a set of tools developed by Lawrence Berkeley National Laboratory for implementing block-structured AMR in order to solve partial differential equations \citep{Chombo}. Some key features are:

\begin{itemize}
\item{\emph{C++ class structure}: $\mathtt{Chombo}$ is primarily written in the C++ language, using the class structure inherent in that language to separate the various processes. It is also possible to use a form of Fortran for the array operations, including the evolution equations.}
\item{\emph{Adaptive Mesh Refinement}: $\mathtt{Chombo}$ provides Berger-Oliger style \citep{bergeroliger,BergerColella} AMR with Berger-Rigoutsos \citep{BergerRigoutsis91} block-structured grid generation. Chombo supports full non-trivial mesh topology -- i.e. many-boxes-in-many-boxes. The user is required to specify regridding criteria.}
\item{\emph{MPI scalability}: $\mathtt{Chombo}$ contains parallel infrastructure which gives it the ability to scale efficiently to several thousand CPU-cores per run. It uses an inbuilt load balancing algorithm, with Morton ordering to map grid responsibility to neighbouring processors in order to optimise processor number scaling.}
\item{\emph{Standardised Output and Visualization}: $\mathtt{Chombo}$ uses the $\mathtt{HDF5}$ output format, which is supported by many popular visualization tools such as $\mathtt{VisIt}$. In addition, the output files can be used as input files if one chooses to continue a previously stopped run -- i.e. the output files are also checkpoint files.}
\end{itemize}

We detail some of the key features below. Note that there are many possibilities for configuring $\mathtt{Chombo}$, with regard to, for example, time stepping and block refinement; but here we focus on those features used by $\grchombo$.

\subsection{Chombo structure and classes} 
\label{sec-ChomboClasses}

$\mathtt{Chombo}$ uses the C++ language, the main purpose of which is to add ``object orientation'' to normal programming functionality. ``Classes'' are the central feature of object-oriented programming, and one can think of them as a somewhat complicated user defined type, like an integer, or double. Thus if one has a class Shape, one can define a Shape object in the same way as an integer:
\begin{verbatim}
int a = 7;
double b = 0.9;
Shape circle;
\end{verbatim}
However, a class in general has much more structure than a simple type like an integer. It may contain a number of variables or structures, and functions that set or operate on these members. By containing all of the information and operators in one structure, the interactions between different parts of the code are constrained - one should not (in theory) go in and amend the contents of an existing class - one should treat it as a ``black box'', which is designed to remain unmodified as new code is added. The new code must work within the constraints of the existing classes, using their access and member setting functions, or otherwise new classes must be written. 

In this context, another aspect of classes which is useful is their ability to inherit from an existing class. If, for example, one wants a class that does almost the same thing as ClassA, but contains an additional variable and a function on that variable, it is possible to make a class ClassB which inherits all the functionality of ClassA, but to which one can add additional structures and functions. A good easily readable reference for classes is \citep{TutorialsPoint}, or \citep{StroustrupBook}, which gives a more advanced treatment. 

The central class in $\mathtt{Chombo}$ is the AMR class. This is the class which operates the update process and all of the regridding and interlevel communications. Below this sits an AMRLevel class, which broadly defines what happens on each refinement level in a single update step. When writing a new physics code with $\mathtt{Chombo}$, the user writes a new class which inherits from AMRLevel, but which in addition implements user defined functions for the physics specific steps. For example, the AMRLevel class contains all the functionality to do a Runge-Kutta update, but the user must specify the equation of motion that gives the time derivatives of each variable $f$, $df/dt$, by writing a new version of the class member function evalRHS().

\begin{figure}
\begin{center}
\includegraphics[width=.8\textwidth]{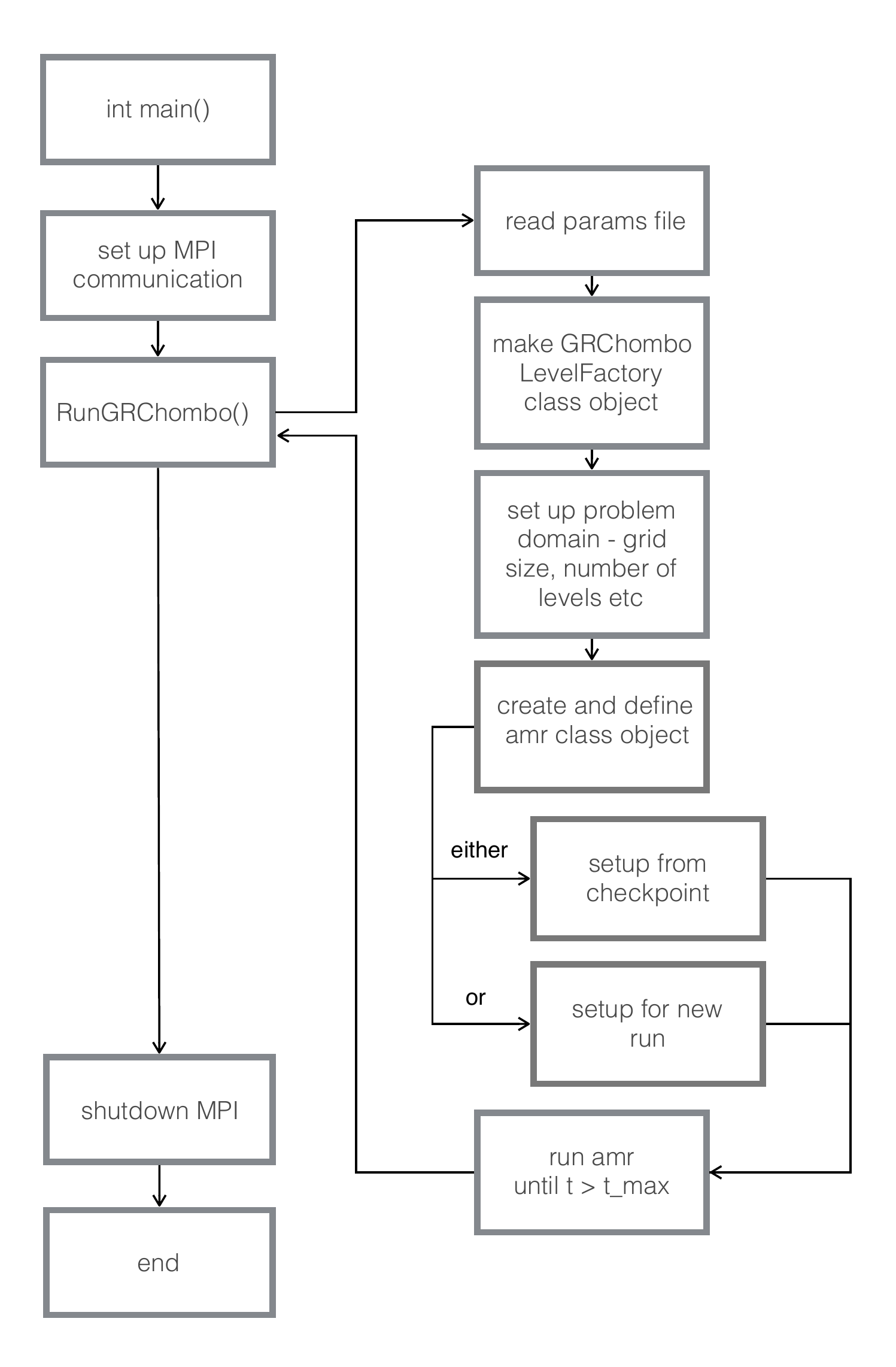}
\caption[Chombo structure diagram - program control flow]{The flow chart illustrates the program flow for setup and running.}
\label{fig-ChomboFlow1}
\end{center}
\end{figure}

The processes of $\mathtt{Chombo}$ are best described in a series of block diagrams. In figure \ref{fig-ChomboFlow1} we illustrate the overall program flow. We see that the main function simply sets up the MPI communication and calls a function ``RunGRChombo()''. That function reads in the parameters which will be used in the simulation, such as grid size, maximum number of refinement levels, time interval, etc, and sets up the basic structure of the problem domain. It creates a GRChomboClassFactory class object, which simply returns a pointer to the $\grchombo$ version of the AMRLevel class, GRChomboLevel. This information is then used to create an AMR class object, which is set up either to run from a checkpoint file, in which case the information from the checkpoint file is read in and used to set up an initial grid, or otherwise set up for a new run, in which case the initial data must be specified within the GRChomboLevel function InitialData(). 

\begin{figure}
\begin{center}
\includegraphics[width=.8\textwidth]{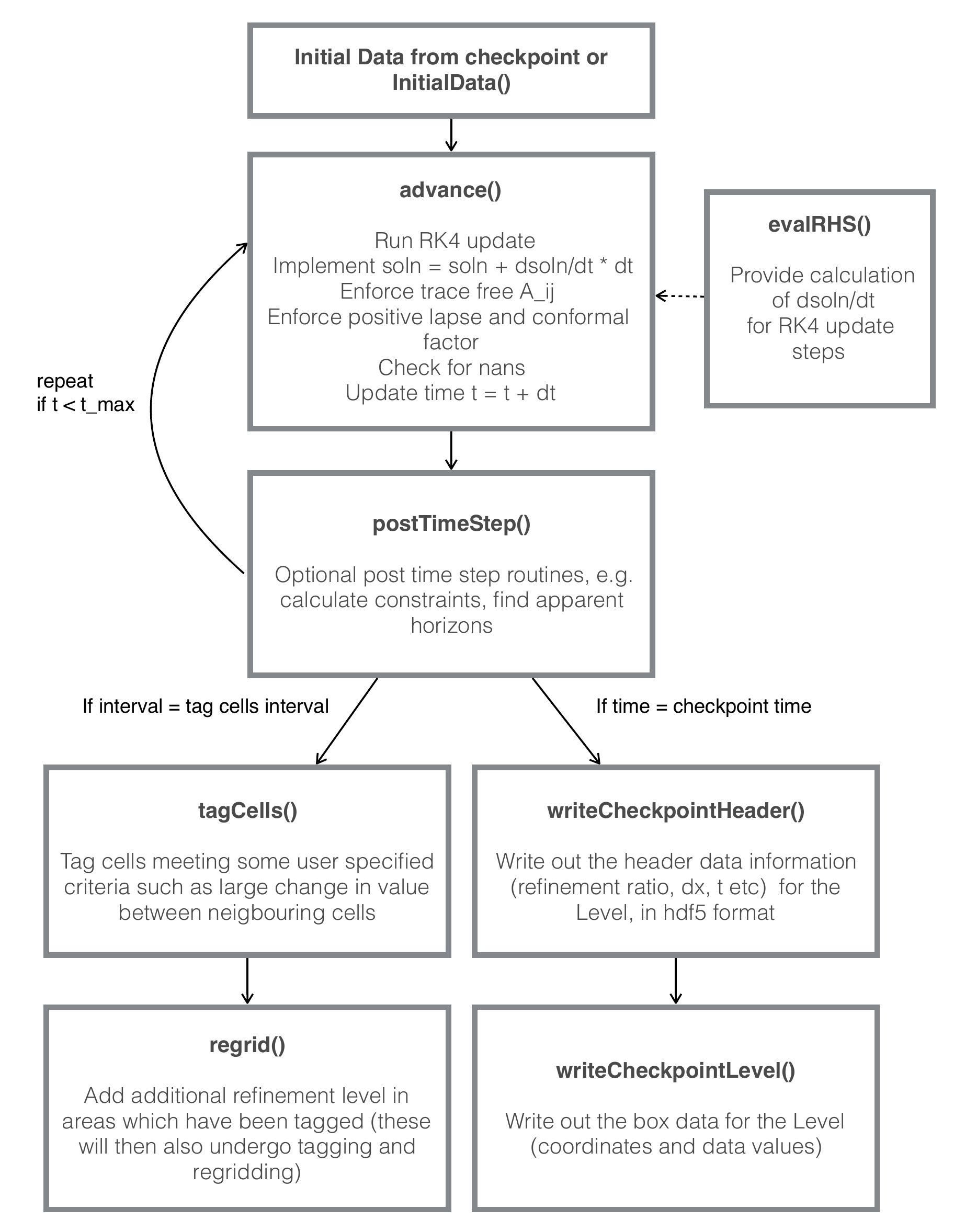}
\caption[Chombo structure diagram - evolution flow]{The flow chart illustrates the program flow for the evolution of a single AMR Level, highlighting the key functions in the GRChomboLevel class.}
\label{fig-ChomboFlow2}
\end{center}
\end{figure}

In figure \ref{fig-ChomboFlow2} we illustrate the flow which happens in each individual AMRLevel update. The key function is the advance() function. In $\grchombo$ we use a 4th order Runge-Kutta (RK4) update in time, but other options are available. The ``physics'' is contained in evalRHS(), the method used in the RK4 step to calculate the time derivatives of the physical variables - so this is what contains the BSSN equations and matter evolution equations. At user-defined intervals (which may differ on each Level), we check whether we want to add (or remove) additional resolution in different areas of the problem domain. This is done by tagging cells according to some refinement criteria and adding additional levels, if required, according to the algorithm described in section \ref{sec-BRAMR}. One may also write checkpoint files at multiples of the coarsest time step.

\begin{figure}
\begin{center}
\includegraphics[width=.8\textwidth]{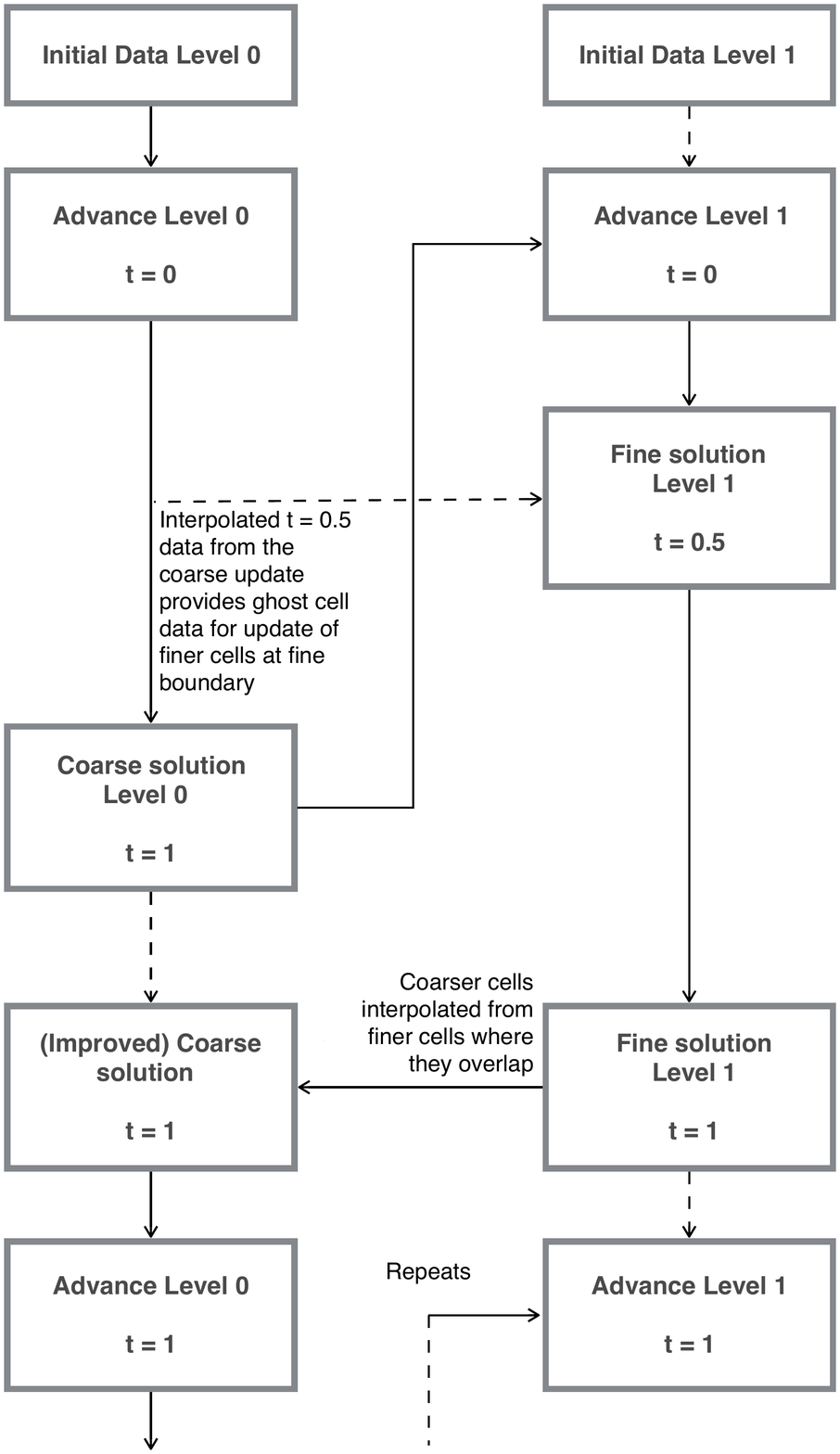}
\caption[Chombo structure diagram - interaction of levels]{The flow chart illustrates how a coarse and fine level interact during a single update step of the coarser level. For simplicity only two levels are shown.}
\label{fig-ChomboFlow3}
\end{center}
\end{figure}

In figure \ref{fig-ChomboFlow3} we illustrate the interaction between the coarser and finer levels, which is controlled by the AMR class and will be described in more detail in \ref{sec-BRAMR}. Note that the program flow takes each refinement level \emph{in serial}, starting with the \emph{coarsest level}. The parallelisation described in section \ref{sec-loadbalance} occurs entirely within a single level. Each refinement level is divided into a number of ``boxes'' (which may be disjoint) and it is these boxes which are divided between processors. As a result, at any one time the processors are working on boxes at a single level, and all processors must finish and synchronise before moving onto the next refinement level.

As shown in the figure, after a single coarse timestep, the finer level below takes a half step. To take the second step that will bring it in line with the coarser level, regions at the edge of the refined area need data with which to populate their ghost cells\footnote{Ghost cells are the outer boundary cells of the boxes, which must be exchanged between processors working in different regions.}. This is done by interpolating the data from the coarse time step. Once the finer level catches up with the coarser one, any coarse cells which are overlaid by better refinement are overwritten with the finer data (which should be more accurate). In the figure only two levels are shown, but in an actual simulation, each finer level will itself be a coarser level, and so the process will occur recursively down the hierarchy, until all levels are synchronised at the coarsest time step. In addition, each level may spawn additional finer levels at the end of a time step by regridding areas which have become poorly resolved. This is the main functionality of which $\grchombo$ takes advantage.

As will be discussed in Chapter \ref{ch-Conclusions}, we are currently in the process of rewriting $\grchombo$ to make it more modular and thus more adaptable. However, the broad functionality of $\mathtt{Chombo}$ described here remains the same.

\subsection{Berger-Rigoutsos block-structured AMR} 
\label{sec-BRAMR}

\begin{figure}
\begin{center}
\includegraphics[width=.8\textwidth]{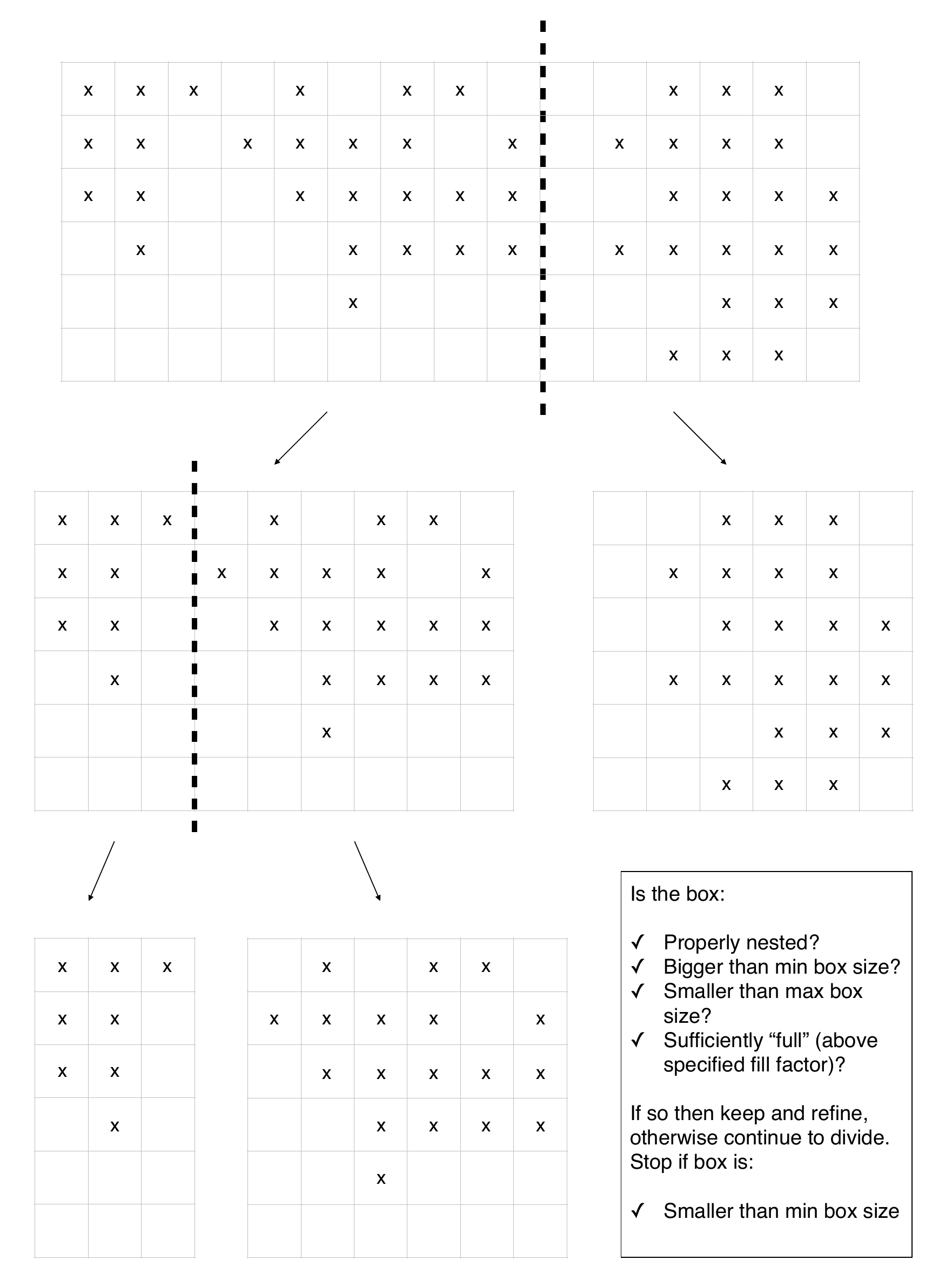}
\caption[Regridding algorithm in Chombo]{The figure illustrates the process for partitioning and selecting grids for refinement based on tagged cells (those marked with an x). The partitioning is shown schematically for 1 dimension, whereas the actual code divides the boxes in all three spatial directions.}
\label{fig-Regrid}
\end{center}
\end{figure}

$\grchombo$ uses {\tt Chombo}'s implementation of the Berger-Rigoutsos adaptive mesh refinement algorithm \citep{BergerRigoutsis91}, which is one of the standard block-structured AMR schemes.  Block-structured AMR regrids by overlaying variable size boxes, instead of remeshing on a cell-by-cell basis (the ``bottom-up'' approach). The main challenge is to find an efficient algorithm to \emph{partition} the cells which need regridding into rectangular ``blocks''. In this section, we will briefly discuss the algorithm. The basic idea is illustrated in figure \ref{fig-Regrid}.

For a given grid at some refinement level $l$, where $l=0$ is the base level and $l_{max}$ is some preset maximum refinement level, we first ``tag'' cells for which refinement is required. The refinement condition used by $\grchombo$ is discussed later in this section. The primary problem of AMR is to efficiently partition this grid into regions which require adaptive remeshing. In \emph{block-structured} AMR these regions are boxes in 3D or rectangles in 2D. \emph{Efficiency} is measured by the ratio of tagged over untagged cell points in the final partitions. 

In each partition, we compute the \emph{signatures} or \emph{traces} of the tagging function $f(x,y,z)$ of any given box
\begin{eqnarray}
X(x) &=& \int f(x,y,z) dy dz, \\
Y(y) &=& \int f(x,y,z) dx dz, \\
Z(z) &=& \int f(x,y,z) dy dx,
\end{eqnarray}
where $f(x,y,z) =1$ if a cell is tagged for refinement and $0$ otherwise. Given these traces, we can further compute the Laplacian of the traces $\partial^2_x X(x)$, $\partial^2_y Y(y)$ and $\partial^2_z Z(z)$. Given the Laplacians, the algorithm can search for all (if any) inflection points \emph{individually for each direction} -- i.e. the locations of zero crossings of the Laplacian. It can then pick the one whose $\delta (\partial^2_i X_i)$ is the greatest (corresponding to the line -- or plane in 3D -- separating the largest change in the Laplacian). This point then becomes the line of partition for this particular dimension. Roughly speaking, this line corresponds to an edge between tagged and untagged cells in the orthogonal directions of the signature. Furthermore, if there exists a point $x_i$ with zero signature $X_i(x_i)=0$ (i.e. no cells tagged along the plane orthogonal to the direction), then this ``hole'' is chosen to be the line of partition instead.

\nomenclature[a-pi]{$l$}{in $\grchombo$, the refinement level}

After a partitioning, we check whether or not each partition is \emph{efficient}, specifically whether it passes a user-specified threshold or \emph{fill factor}, $\epsilon_{ff} <1.0$, 
\begin{equation}
\frac{\mathrm{Tagged~Cells}}{\mathrm{Total~Cells}} > \epsilon_{ff}
\end{equation}
If this is true, then we check if this box is \emph{properly nested}\footnote{Properly nested means that (1) a $l+1$ level cell must be separated from an $l-1$ cell by at least a single $l$ level cell and (2) the physical region corresponding to a $l-1$ level cell must be completely filled by $l$ cells if it is refined, or it is completely unrefined (i.e. there cannot be ``half-refined'' coarse cells).} \citep{BergerColella, bergeroliger} and if so we accept this partition and the partitioning for this particular box stops. If not, then we continue to partition this box recursively until either all boxes are accepted or partitioning no longer can be achieved (either by the lack of any tagged cells or reaching a preset limit on the number of partitions). Furthermore, $\grchombo$ allows the user to set a maximum partition size, which if exceeded will force a partitioning of the box. This can be useful when trying to achieve good load balancing, as many smaller boxes can be shared more easily than several large ones.

\nomenclature[g-pi]{$\epsilon_{ff}$}{in regridding, the fill factor threshold}

Note that a higher value of $\epsilon$  means that the partitioning will be more aggressive, which will lead to a higher efficiency in terms of the final ratio of tagged to untagged cells -- generating more boxes in the process. However, this is not necessarily always computationally better as partitioning requires computational overhead, which depends on the number and topology of the processors. The ideal fill ratio is often a function of available processors, their topology and of course the physical problem in question.

A partitioned box is then \emph{refined}, i.e. its grids are split into a finer mesh using the (user definable) refinement ratio $n^l = \delta x^{l+1}/\delta x^l$, and the process continues recursively until we either have no more tagged cells, or when we reach a preset number of refinement levels $l_{max}$.

Finally we need to specify  a prescription for tagging which cells are required to be refined. $\grchombo$ tags a cell when any (set of) user selected fields $F \in {\bf{u}}$ pass a chosen threshold $\sigma_{tt}(F)$, which sets a limit on the $L^2$ norm of the change in the value of the field across that cell, i.e. 
\begin{equation}
f(x,y,z)=\left\{
  \begin{array}{cc}
   1 & \mathrm{if}~\sqrt{\sum\limits_{i=1}^{3}  (F({\bf x}+\delta x \, \hat{\bf x}_i) - F({\bf x}-\delta x \, \hat{\bf x}_i))^2} > \sigma_{tt}(F) \\
    \\
   0 & \mathrm{otherwise}.
    \end{array}
    \right. \label{eqn:tagging} 
\end{equation}

\nomenclature[g-pi]{$\sigma_{tt}$}{in regridding, the tagging threshold}

This condition can be augmented, for example by using estimated truncation errors as tagging conditions instead.

Partitioning can be done at every time-step for each refinement level, or the frequency may be defined by the user for each refinement level. The user may wish to select a lower frequency because it might be useful to not partition at every timestep for a given refinement level. Firstly, less frequent regridding saves computational overhead, and in addition it can be important to let numerical errors dissipate (e.g. via Kreiss-Oliger dissipation, see Sec. \ref{sec-kreissoliger}) before remeshing. Once a new hierarchy of partitions is determined, we interpolate via linear interpolation from coarse to fine mesh (higher order interpolation tends to overfit the data), and from fine to coarse mesh, which can introduce errors. We will see in the testing phase that this effectively reduces the convergence of the code from 4th order to 3rd order.

Since the finer mesh has a smaller Courant number, each mesh level's timestep is appropriately reduced via
\begin{equation}
\Delta t^{l+1} = \frac{\Delta t^l}{n^l}.
\end{equation}

$\grchombo$ follows the standard Berger-Collela AMR evolution algorithm \citep{BergerColella}, as was illustrated in figure \ref{fig-ChomboFlow3}. Starting from the coarsest mesh, it advances the coarse mesh 1 time step i.e. $t \rightarrow t+\Delta t^l$. Then it advances the next finest mesh $n^l$ times until the fine mesh ``catches up'' with the coarse mesh time. Once both coarse and fine mesh are at the same time $t$, $\grchombo$ synchronises them by interpolating from the fine cells to the coarse cells. 

Note that in a conservative system, this simple synchronisation is not conservative and requires proper \emph{refluxing} -- the coarse fluxes are replaced with a time-averaged fine mesh fluxes. This step is not implemented by $\grchombo$ as GR equations are not conservative, but would need to be considered if a fluid type matter was added.

\subsection{Load balancing} 
\label{sec-loadbalance}

$\grchombo$'s efficiency when running on a large number of distributed-memory nodes is highly dependent on efficient load balancing of the available computational work across those nodes. Load balancing seeks to avoid the situation where most of the nodes are waiting for some small subset of nodes to finish their computational work, and it does this by seeking to distribute the amount of work to be done per time step evenly among all of the nodes. This can be non-trivial when AMR boxes at many different locations are simultaneously being evolved across the system. In addition, even within a single node, multiple OpenMP threads might be running, and the per-node workload needs to be balanced amongst those threads.

For the inter-node load balancing, $\grchombo$ leverages {\tt Chombo}'s load balancing capabilities to distribute the AMR boxes among the available nodes. It does this by building a graph of the boxes to be distributed, adding edges between neighbouring and overlapping boxes. A bin packing / knapsack algorithm is used to balance the computational work among nodes, where the work is assumed to be proportional to the number of grid points, and then an exchange phase is used to minimise the communication cost. Because this load balancing procedure can be costly, we normally run it only every few time steps. In between runs of the load balancing procedure, new boxes generated by AMR refinement stay on the node which holds the parent box.

Within each node, the computational work is divided amongst the available OpenMP threads by iterating over the boxes to process using OpenMP's dynamic scheduling capability. This allows each thread to take the next available box from the queue of unprocessed boxes, instead of deciding ahead of time which boxes each thread will process. This is important because the boxes are varying in size. We generally divide even the coarsest level into multiple boxes so that it can be processed in parallel by multiple threads.

\section{Implementing GRChombo}
\label{sec-GRChombo}

$\grchombo$ is a physics engine built around $\mathtt{Chombo}$. $\grchombo$ solves the system of hyperbolic partial differential equations of the Einstein equation, with scalar matter content.

Below are the key features of $\grchombo$, which are built on top of the $\mathtt{Chombo}$ functionality described above.

\begin{itemize}
\item{\emph{BSSN formalism with moving puncture}: $\grchombo$ evolves the Einstein equation in the BSSN formalism with scalar matter. Singularities of black holes are managed using the moving puncture gauge conditions \citep{Campanelli:2005dd, Baker:2005vv}.}
\item{\emph{4th order discretisation in space and time}: We use 4th order spatial stencils combined with a 4th order Runge-Kutta time update. In section \ref{sec-Tests-convergence} we show that the convergence is approximately 4th order without regridding, but reduces to 3rd order convergence with regridding effects.}
\item{\emph{Kreiss-Oliger dissipation}: Kreiss-Oliger dissipation is used to control errors, from both truncation and the interpolation associated with regridding.}
\item{\emph{Boundary conditions}: In our work we either use radiative or periodic boundary conditions. In principle, one can implement other boundary condition methods in $\grchombo$. For many simulations, the AMR ability allows us to set the boundaries far enough away so that reflections do not affect the results during simulation time.}
\item{\emph{Initial Conditions}: As detailed in \ref{sec-InitialConditionsGauge}, for the work done to date we generally used simple analytic conditions, or relaxation of the Hamiltonian constraint in somewhat more general conditions. However, in principle any initial conditions can be read in, for example, where solutions to the constraints have been found numerically. Note that $\grchombo$ itself does not (currently) solve for the initial conditions.}
\item{\emph{Diagnostics}: The key diagnostics used in this work are apparent horizons, ADM mass and momenta, and gravitational waves.}
\end{itemize}

Additional details of these features are given in the sections below.

\subsection{Evolution equations} 
\label{sec-evoleqn}

$\grchombo$ evolves the BSSN equations described in section \ref{sec-NRStability} and summarised in Appendix \ref{sec-appendix1D}.

Note that in the actual evolution, the values of the three-vector $\Gamma^i$ are computed from the knowledge of the conformal metric, $\tilde\gamma_{ij}$, but for its spatial derivatives $\partial_i \Gamma^i$, the evolved $\Gamma_i$ is used. 

In addition, we hard code the condition $\alpha>0$ as is usual practice. For the algebraic constraints of BSSN, we do not enforce (by hand) the condition that the conformal metric has a determinant of one, but we do enforce after each RK4 step that $\tA_{ij}$ is traceless.

\subsection{Discretization and time-stepping} 
\label{sec-discretization}

We would like to evolve a set of fields in space, the state-vector
\begin{equation}
{\bf{u}}(x^i,t) = \{F_1,F_2,F_3,\dots\} 
\end{equation}
through time $t$ via the equations of motion
\begin{equation}
\frac{\partial {\bf{u}}}{\partial t} = {\cal F}(\bf{u}) ,
\end{equation}
where ${\cal F}$ is some operator on $\bf{u}$ which, in the case of the Einstein equation, is non-linear. 

In $\grchombo$, both the space and time coordinates are discretised. Evolution in time is achieved through time-stepping $t\rightarrow t+\Delta t$, where at each time step we compute the fluxes for each grid point individually. Time stepping is implemented using the standard 4th Order Runge-Kutta method, and hence, as usual, we only need to store the values of the state-vector at each time step.

The state vector $\bf{u}$ itself is discretised into a cell-centered grid. Spatial derivatives across grid points are computed using standard 4th order stencils for all spatial derivatives, except for advection terms which are implemented using an upwind stencil. The form of the stencils $\grchombo$ uses exactly follow equations (2.2) through (2.6) of \citep{Zlochower:2005bj}, which are, for the standard derivatives
\begin{align}
&\partial_x F_{i,j,k} =\frac{1}{12 dx} (F_{i-2,j,k} - 8F_{i-1,j,k} + 8F_{i+1,j,k} - F_{i+2,j,k}) \, , \\
&\partial_{xx} F_{i,j,k} =\frac{1}{12dx^2}(-F_{i+2,j,k} + 16F_{i+1,j,k} - 30F_{i,j,k} + 16F_{i-1,j,k} - F_{i-2,j,k}) \, , \\
&\partial_{xy} F_{i,j,k} = \frac{1}{144dxdy} ( F_{i-2,j-2,k} - 8F_{i-1,j-2,k} + 8F_{i+1,j-2,k} - F_{i+2,j-2,k}  \\
&\hspace{3cm}- 8(F_{i-2,j-1,k} - 8F_{i-1,j-1,k} + 8F_{i+1,j-1,k} - F_{i+2,j-1,k}) \\ 
&\hspace{3cm}+ 8(F_{i-2,j+1,k} - 8F_{i-1,j+1,k} + 8F_{i+1,j+1,k} - F_{i+2,j+1,k}) \\
&\hspace{3cm}- ~ (F_{i-2,j+2,k} - 8F_{i-1,j+2,k} + 8F_{i+1,j+2,k} - F_{i+2,j+2,k}) ) \, ,
\end{align}
whilst for the advection derivatives (of the form $\beta^i \partial_i F$) we use the upwinded stencils 
\begin{equation}
\partial_x F_{i,j,k} = \frac{1}{12dx} (-F_{i-3,j,k} + 6F_{i-2,j,k} - 18F_{i-1,j,k} + 10F_{i,j,k} + 3F_{i+1,j,k}) \, ,
\end{equation}
for $\beta^x < 0$ and
\begin{equation}
\partial_x F_{i,j,k} = \frac{1}{12dx} (F_{i+3,j,k} - 6F_{i+2,j,k} + 18F_{i+1,j,k} - 10F_{i,j,k} + 3F_{i-1,j,k}) \, ,
\end{equation}
for $\beta^x > 0$.

\subsection{Kreiss-Oliger dissipation} 
\label{sec-kreissoliger}

In a finite difference scheme, instabilities can arise from the appearance of high frequency spurious modes. Furthermore, regridding generates errors an order higher than the typical error of the evolution operator, hence it is crucial that we control these errors in an AMR code. 

The standard prescription is to implement some form of numerical dissipation to damp out these modes. $\grchombo$ implements $N=3$ Kreiss-Oliger \citep{TUS:TUS1547} dissipation. In this scheme, for all evolution variables in ${\bf u}$, the evolution equations are modified as follows for each spatial direction
\begin{multline}
\partial_t F_m \rightarrow \partial_t F_m + \\
 \frac{\sigma}{64 \Delta x}(F_{m+3}-6F_{m+2}+15F_{m+1}-20F_{m}+15F_{m-1}-6F_{m-2}+F_{m-3}),
\end{multline}
where $m\pm n$ labels the grid point, $n$ the total offset from $m$ in the spatial direction and $\sigma$ is an adjustable dissipation parameter usually of the order ${\cal O}(10^{-2})$. This 3rd order scheme is accurate as long as the integration order of the finite difference scheme is 5 or less (which it is in our implementation using 4th order Runge-Kutta).

\nomenclature[g-pi]{$\sigma$}{the Kreiss Oliger dissipation parameter}

\subsection{Boundary conditions} 
\label{sec-BC}

$\grchombo$ supports both periodic (in any direction) boundary conditions, as well as any particular boundary conditions the user may want to specify (such as Neumann or Dirichlet types). A particular popular type of boundary condition is the so-called Sommerfeld \citep{Alcubierre:2002kk} boundary condition, where out-going radiation is dissipated away. For any field $F$, we impose the condition at the boundary
\begin{equation}
\frac{\partial F}{\partial t} = -\frac{v x_i}{r} \frac{\partial F}{\partial x_i}  -v\frac{F-F_0}{r} \, ,
\end{equation}
where $r= \sqrt{x_1^2+x_2^2+x_3^2}$ is the radial distance from the center of the grid, $F_0$ is the desired space-time at the boundary (typically Minkowski space for asymptotically flat spacetimes) and $v$ the velocity of the ``radiation'', which is typically chosen to be 1.

It should be noted that, whilst these conditions work reasonably well in practice (in particular for harmonic oscillations of a massive field), they are not constraint preserving and affect the well-posedness of the system, and so their use is somewhat questionable. Where simulations are run for more than one light crossing time, allowing signals from the boundaries to propagate to the system under study, the effect of the boundary conditions on results may need to be considered.

\nomenclature[a-pi]{$r$}{radial coordinate distance from the centre of the computational grid}
\nomenclature[a-pi]{$x^i$}{coordinates of a point on the computational grid}

\subsection{Initial conditions} 
\label{sec-IC}

$\grchombo$ supports several ways of entering initial conditions. 
\begin{itemize}
\item{Direct equations -- Initial conditions which are described by known analytic equations, such as the Schwarzchild solution, can be entered directly in equation form.}
\item{Checkpointing -- The {\tt HDF5} format output files from $\grchombo$ double as checkpoint files. A run can simply be continued from any previous state as long as its {\tt HDF5} output file is available. }
\item{Entering from data file -- $\grchombo$ allows one to read in data from a file.}
\item{Relaxation -- $\grchombo$ has a rudimentary capability to solve for the conformal factor given some initial mass distribution, and assuming that the momentum constraint is satisfied by the other variables specified. Given a guess for $\chi$, $\grchombo$ relaxes it to the correct initial values using a dissipation term which is proportional to a user chosen dissipation coefficient times the Hamiltonian constraint.}
\end{itemize}

The initial conditions used in the code development are mostly analytic or approximate analytic solutions, and so are entered directly into the code.  In the critical collapse, a Mathematica numerical solution as a function of the radius is interpolated onto the initial grid. 

\subsection{Apparent horizon finder in spherical symmetry} 
\label{sec-AHFinder}

The presence of a black hole event horizon is gauge invariant. We use an apparent horizon finder which assumes spherical symmetry to identify marginally trapped surfaces on each spatial slice. Whilst these are local rather than global horizons, if we detect an apparent horizon on a time slice, the singularity theorems tell us that it must lie inside an event horizon (see, for example, section 7.1 of \citep{ShapiroBook}). Thus if we detect an apparent horizon we can infer that a black hole has formed, and the area of the apparent horizon provides a lower bound on the black hole mass. Note that the converse is not true -- the absence of an apparent horizon does not imply the absence of an event horizon.

\nomenclature[g-pi]{$\Theta$}{in the apparent horizon finder, expansion of the outgoing null geodesics}
\nomenclature[a-pi]{$s^i$}{outward pointing normal vector to a 2D surface within the spatial hypersurface}

One can find apparent horizons as the surface on which the expansion of the outgoing null geodesics is zero, that is
\begin{equation}
\Theta = \nabla_i s^i + K_{ij} s^i s^j - K = 0 \, ,
\end{equation}
where $s^i$ is the outward-pointing unit normal to the apparent horizon surface (which thus lies in the spatial slice, not to be confused with the normal vector to the slice itself which is time-like).

In spherical symmetry the outward pointing normal vector may be found at each point in the surface, assuming that the central point of the AH lies at the centre of the grid at $x^i=0$, as 
\begin{equation}
s^i = \frac{x^i}{r} \, ,
\end{equation}
where $x^i$ denotes the $x$, $y$ and $z$ coordinates of the point. The distance $r$ is the coordinate radius from the centre point $r^2 = x^2 + y^2 + z^2$. We can then calculate $\Theta$ at each point in the hyperslice as in \citep{Thornburg:2003sf}, as
\begin{equation}
\Theta =\frac{A}{D^{3/2}} + \frac{B}{D^{1/2}} + \frac{C}{D} - K ,
\end{equation}
where
\begin{align}
&A = - (\gamma^{ik} s_k)(\gamma^{jl} s_l) \partial_i s_j - \onehalf (\gamma^{ij} s_j)(s_k s_l \partial_i \gamma^{kl}) \, , \\
&B = (\partial_i \gamma^{ij}) s_j + \gamma^{ij} \partial_i s_j + g^{ij} s_j (\partial_i \ln \sqrt{\gamma}) \, , \\
&C = K^{ij} s_i s_j \, , \\
&D = \gamma^{ij} s_i s_j \, .
\end{align}

The mass $M$ of the apparent horizon is then found from its area $A$ as
\begin{equation}
M^2 = \frac{A}{16 \pi}  \, .
\end{equation}
By considering the point where the apparent horizon crosses the x axis, at a radius of $r$ from the centre point, M can be calculated in cartesian coordinates from the state values at that point as
\begin{equation}
M = (\gamma_{yy} \gamma_{zz})^{1/4} \frac{r}{2 \chi}  \, .
\end{equation}

One can extend the method to non spherical horizons but the normal vector in this case is not trivial to find as the surface may change shape at different points in a way that is not known a priori. One generally requires a trial surface which is then made to converge on the true horizon using Newton's method, as in Thornburg's fast apparent horizon finder \citep{Thornburg:2003sf}. In the work in this thesis, a spherical apparent horizon finder was sufficient, as even in asymmetric cases, we could wait for the solution to settle into an approximately spherical solution before measuring the mass. In future work we plan to extend the method to more general surfaces.

\subsection{Extracting ADM mass and momenta} 
\label{sec-ADMMJP}

A problem in GR is found in defining the total energy or momentum of a system. Whilst we have a local law for energy conservation, $\nabla_aT^{ab} = 0$, there is no global law of conservation of energy integrated over a finite volume. This is because $T_{ab}$ contains only the contributions from matter and not the energy of the gravitational field itself (which is in fact difficult to define). 

\nomenclature[a-pi]{$M_{ADM}$}{the ADM mass of a spacetime}
\nomenclature[a-pi]{$P^i_{ADM}$}{the ADM linear momentum of a spacetime}
\nomenclature[a-pi]{$J^i_{ADM}$}{the ADM angular momentum of a spacetime}

In the weak field case, in cartesian coordinates, one could define the energy, momentum and angular momentum in a slice, neglecting the gravitational contribution, as
\begin{equation}
M = \int \rho ~ dV \, , \quad P^i = \int S^i ~ dV \, , \quad J^i = \int \epsilon^{ijk} x_j S_k ~ dV \, .
\end{equation}
These can be rewritten in terms of the ADM variables using the Hamiltonian and Momentum constraints, assuming small $K_{ij}$, and converting to surface integrals using Gauss' theorem, to give
\begin{align}
&M_{ADM} = \frac{1}{16\pi} \oint_{S} (\delta^{ij} \partial_i \gamma_{jk} - \partial_k) s^k dS   \, , \\
&P^i_{ADM} = \frac{1}{8\pi} \oint_{S} (K^i_j - \delta^i_j K) s^k dS   \, , \\
&J^i_{ADM} = \frac{1}{8\pi} \oint_{S} \epsilon^{ijk} x_j (K_{kl} - \delta_{kl} K) s^l dS   \, ,
\end{align}
where $s^i$ is, as in the apparent horizon finder, the outward pointing unit normal 
\begin{equation}
s^i = \frac{x^i}{r} \, .
\end{equation}
In the strong field regime the volume expressions will no longer be valid because they do not contain the energy of the field. However, if the surface integrals are evaluated at infinity, in asymptotically flat space, they remain valid and thus we can define the ADM mass, linear momentum and angular momentum of an isolated system by the value of these expressions evaluated at spatial infinity. 

In practice, in our code, we evaluate these expressions as far out as is computationally feasible, given the finite size of the computational domain. 

\subsection{Gravitational wave detection} 
\label{sec-GWdetect}

In the section on convergence testing we extract a gravitational waveform from the code to test convergence. Here we will summarise briefly the method used to extract such a waveform in cartesian coordinates, following closely the description in section 6 of \citep{Cardoso:2014uka}. The reader is referred to the standard NR texts, in particular \citep{AlcubierreBook}, for a more full discussion. 

We use the Newman-Penrose formalism which is based on a tetrad of null vectors, $\vec{k}$, $\vec{l}$, $\vec{m}$, $\vec{\bar{m}}$ where the the first two are real vectors, and the latter two are part imaginary and complex conjugates of each other. In Cartesian coordinates these are found from a Gram-Schmidt orthonormalisation of the following spatial vectors
\begin{equation}
u^i = [x,y,z] \, , \quad v^i = [xz,yz,-x^2-y^2] \, , \quad w^i = \epsilon^i_{jk}u^j v^k \, , 
\end{equation}
which are used to construct the tetrad as
\begin{equation}
k^\alpha =  \frac{1}{\sqrt{2}}(n^\alpha + u^\alpha) \, , \quad l^\alpha =  \frac{1}{\sqrt{2}}(n^\alpha - u^\alpha) \, , \quad m^\alpha =  \frac{1}{\sqrt{2}}(v^\alpha + i w^\alpha) \, , 
\end{equation}
where $n^\alpha$ is the unit normal vector to the spatial slice as described in Chapter \ref{ch-Technical}, and the time components of $\vec{u}$, $\vec{v}$, and $\vec{w}$ are zero by construction.

\nomenclature[g-pi]{$\Psi_4$}{the Newman Penrose scalar}
\nomenclature[a-pi]{$C_{abcd}$}{the Weyl tensor}
\nomenclature[a-pi]{$E_{ij}$}{the electric part of the Weyl tensor}
\nomenclature[a-pi]{$B_{ij}$}{the magnetic part of the Weyl tensor}
\nomenclature[a-pi]{$h_+$}{in gravitational waves, the strain for the $+$ polarisation}
\nomenclature[a-pi]{$h_\times$}{in gravitational waves, the strain for the $\times$ polarisation}

The complex Weyl scalar $\Psi_4$ is defined as the projection of the Weyl tensor $C_{abcd}$ onto these vectors as follows
\begin{equation}
\Psi_4 \equiv C_{\alpha \beta \gamma \delta} l^\alpha \bar{m}^\beta l^\gamma \bar{m}^\delta  \, .
\end{equation}
Its physical significance is that it measures the outgoing radiation from a source. As its name implies, there are other Weyl scalars which come from different projections of the Weyl tensor (five in total, each complex, thus accounting for the 10 degrees of freedom in the Weyl tensor). However, for a plane wave spacetime, the tetrad can be chosen such that only $\Psi_4$ is non zero. The ``peeling theorem'' tells us that, at a large distance from an isolated source, the space looks locally like a plane wave. Therefore taken in asymptotically flat space, the quantity $\Psi_4$ contains all the relevant information about the gravitational wave.

The Weyl tensor is defined in 4 dimensions as
\begin{equation}
C_{abcd} = R_{abcd} - g_{a[c}R_{d]b} + g_{b[c}R_{d]a} + \frac{1}{3} g_{a[c}g_{d]b}R \, ,
\end{equation}
so we can calculate $\Psi_4$ directly from this, or alternatively by first calculating the electric and magnetic parts of the Weyl tensor in the adapted basis, with components
\begin{equation}
E_{ij} = R_{ij} - \gamma^{mn}(K_{ij}K_{mn} - K_{im}K_{jn}) \, , ~ \quad ~ B_{ij} = \gamma_{ik} \epsilon^{kmn} D_m K_{nj} \, .
\end{equation}
The expression for $\Psi_4$ is given in terms of these quantities as
\begin{multline}
\Psi_4 = - \onehalf \left( E_{mn}(v^m v^n - w^m w^n) - B_{mn}(v^m w^n + w^m v^n) \right)  \\ 
 + \onehalf \left( E_{mn}(v^m w^n - w^m v^n) + B_{mn}(w^m w^n + v^m v^n) \right) \, .
\end{multline}
This value of $\Psi_4$ can be extracted from a point in asymptotically flat space and used to construct the gravitational wave strains $h_+$ and $h_\times$  at that point by integrating the relation
\begin{equation}
\ddot{h}_+ - i \ddot{h}_\times = \Psi_4  \, .
\end{equation}
However, to reduce numerical error, it is common, rather than using a single point, to extract the multipole components of the $\Psi_4$ signal and quote these directly, as we have done when calculating the convergence. This effectively means that $\Psi_4$ is decomposed into its components in a basis of \emph{spin-weighted} spherical harmonics of spin weight $s=-2$, 
\begin{equation}
\Psi_4(t,\theta,\phi) = \sum_{l,m} \psi_{l,m}(t) Y_{\ell,m}^{-2} (\theta, \phi) \, ,
\end{equation}
where the components can be found by integrating the $\Psi_4$ values, multiplied by the (complex conjugate of the) spin weighted spherical harmonics, over the surface of a sphere at the radius of extraction, i.e.
\begin{equation}
\psi_{l,m}(t) = \int \Psi_4(t,\theta,\phi) \bar{Y}_{\ell,m}^{-2} (\theta, \phi) dS \, .
\end{equation}

\nomenclature[g-pi]{$\psi_{l,m}$}{the multipole components of the Weyl scalar}
\nomenclature[a-pi]{$Y_{\ell,m} (\theta, \phi) $}{the $(l,m)$ spherical harmonic}

\section{Testing GRChombo}
\label{sec-Tests}

We detail the results of the standard Apples with Apples tests \citep{Babiuc:2007} in Sec. \ref{sec-Apples} when turning off AMR and using fixed resolution grids. In Sec. \ref{sec-VacBH} we turn on the AMR abilities of the code and demonstrate that it can stably evolve spacetimes containing black-hole-type singularities. In Sec. \ref{sec-Tests-Chop} we demonstrate the ability of the code to evolve matter content by considering scalar fields with gravity, by recreating the results of the sub-critical and critical cases of Choptuik scalar field collapse detailed in \citep{AlcubierreBook}. In Sec. \ref{sec-Tests-convergence} we demonstrate convergence of the code and in Sec. \ref{sec-Tests-scaling} we discuss its scaling properties. In \ref{sec-Tests-performance} we simulate a head on collision of two black holes, and compare our code performance to an existing Numerical GR code. 

\subsection{Apples with Apples tests}
\label{sec-Apples}

In this section we describe the results of applying the code to the standard Apples with Apples tests in the paper \citep{Babiuc:2007}. Here we give a brief description of the key features of the tests, but the reader should refer to \citep{Babiuc:2007} for full specifications. Where we do not specify details, our treatment can be assumed to follow that of the standard tests. The AMR capabilities of the code are not utilised in these tests (which were designed for a uniform resolution) in order to make our results comparable with other codes. (We consider the effects of regridding on code performance in Section \ref{sec-VacBH}.)

\subsubsection{Robust stability test}
\label{sec-Apples-rob}

The robust stability test introduces small amounts of random noise to all of the evolution variables, in order to test the code's robustness against numerical errors. The test was conducted at resolutions of $\rho = 4$, $\rho=2$ and $\rho=1$, which correspond to grid spacings of 0.005, 0.01 and 0.02 respectively on a grid of width $1$. The amplitude of the noise is scaled as $1/\rho^2$. No dissipation was added in the test. 

\begin{figure}[H]
\begin{center}
\subfigure[Hamiltonian constraint]{
\includegraphics[height=.4\textwidth]{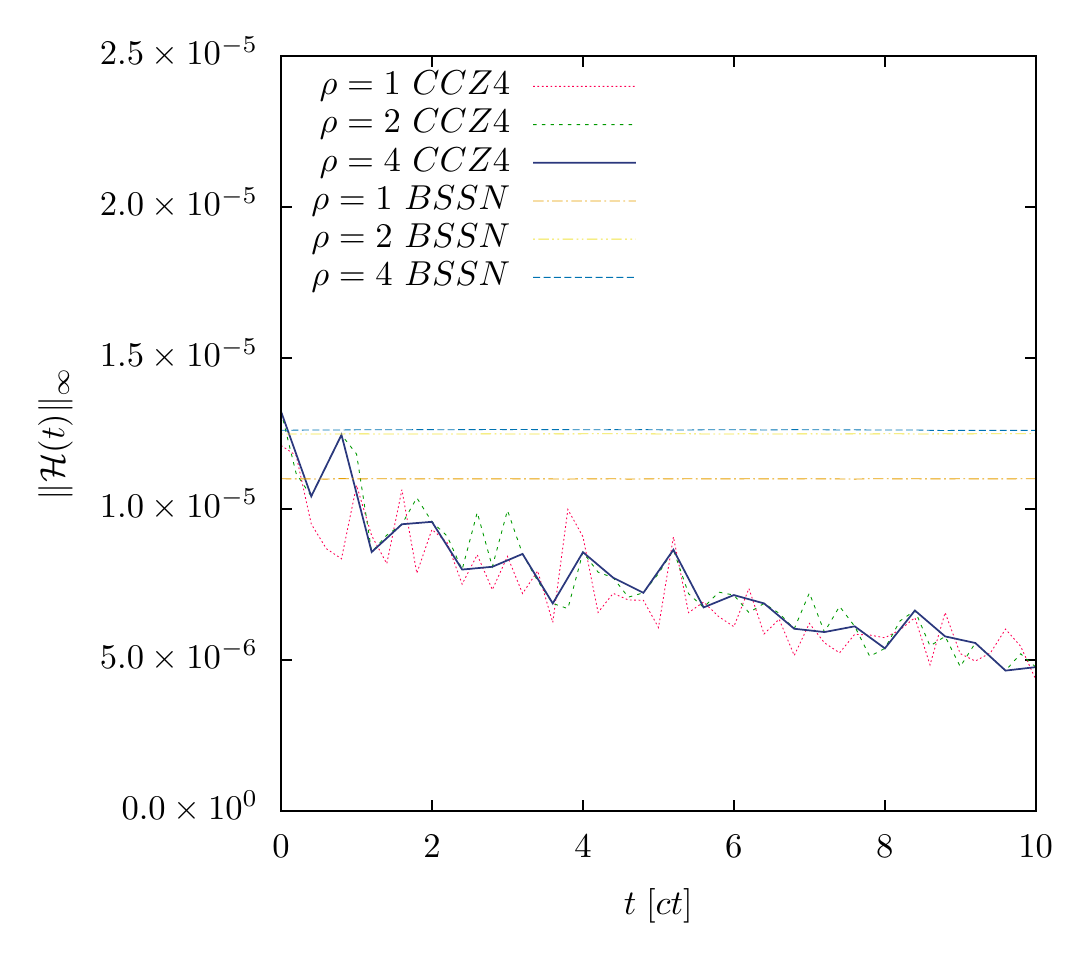}
}
\subfigure[Deviation in $\tilde \gamma_{xx}$]{
\includegraphics[height=.4\textwidth]{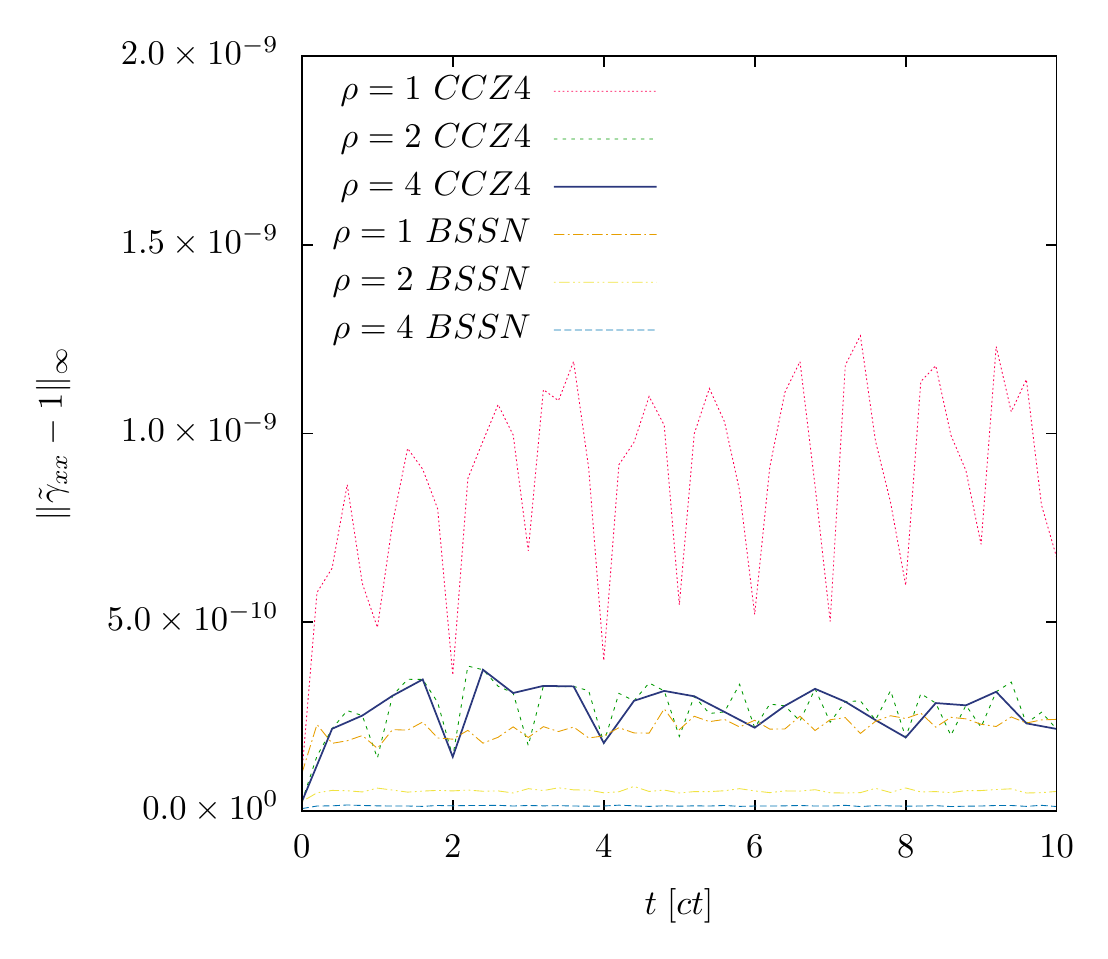}
}
\caption[Robust stability test]{Robust stability test for both the BSSN and the CCZ4 codes, with resolutions $\rho=1,2,4$ respectively. \textit{Left}: time evolution of the $L_{\infty}$ norm of the Hamiltonian constraint. \textit{Right}: deviation of $\tilde \gamma_{xx}$ from 1. Neither norm grows at an increasing rate with increasing resolution, and so the test is passed.
\label{fig-robust}}
\end{center}
\end{figure}

As shown in Figure \ref{fig-robust}, the error growth in the evolution variables did not increase with increasing grid resolution, and the Hamiltonian constraint $\mathcal{H}$ did not grow more for higher resolutions. Therefore, we conclude that the test is passed. 

\subsubsection{Linear wave test}
\label{sec-Apples-lin}

A wave of fixed amplitude is propagated across the grid in the $x$-direction with periodic boundary conditions, with
\begin{equation}
ds^2 =  -dt^2 +dx^2 + (1-H) dy^2 + (1-H) dz^2 \quad H = A \sin \left( 2\pi(x-t) \right) \, . 
\end{equation}

The amplitude $A$ is small enough that the non-linear terms are below numerical precision, such that the behaviour under the Einstein equation is approximately linear. The test measures the errors in magnitude and phase introduced by the code after 1000 crossing times. 

\begin{figure}[H]
\begin{center}
\subfigure[$g_{yy}$ at $T=1000$]{
\includegraphics[height=.35\textwidth]{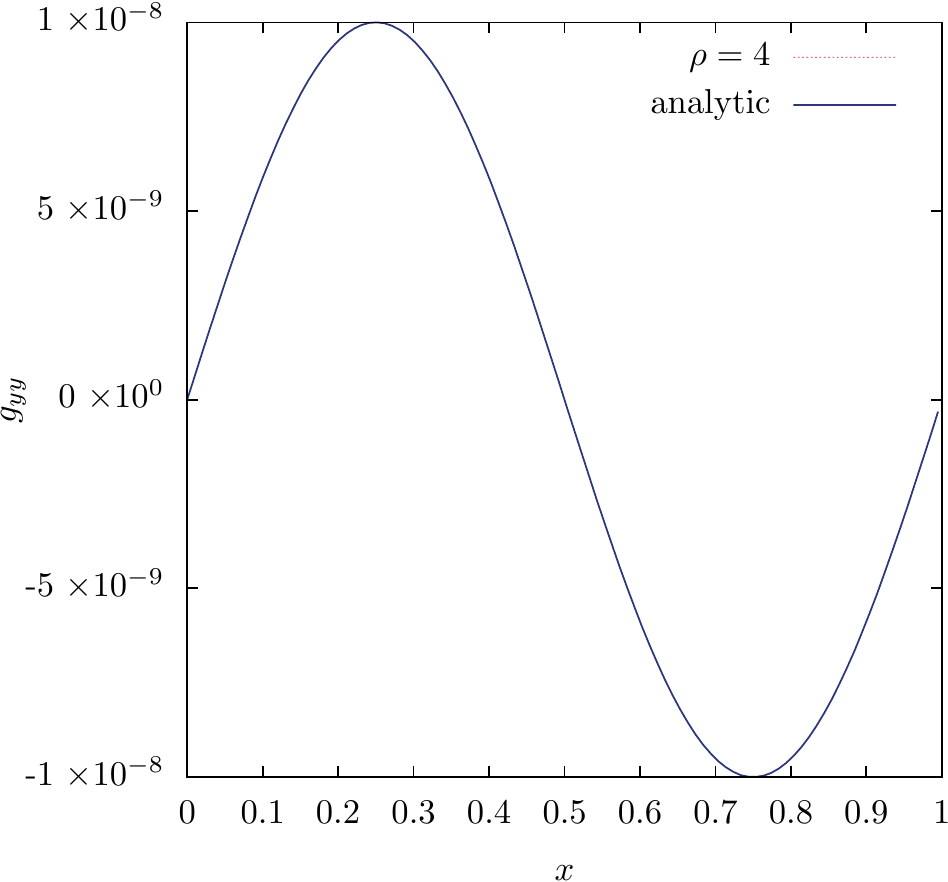}
}
\hspace{0.75cm}
\subfigure[Error in $g_{yy}$ at $T=1000$]{
\includegraphics[height=.35\textwidth]{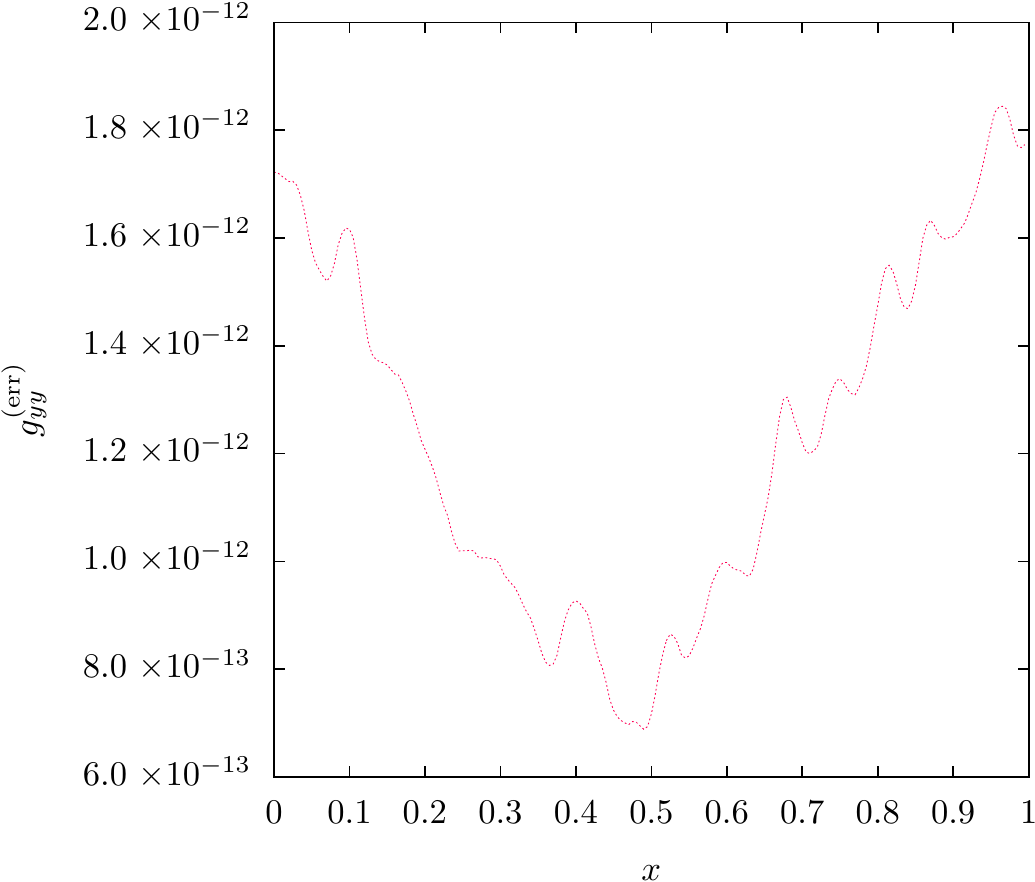}
}
\caption[Linear wave test]{Linear wave test. \textit{Left}: analytical solution and the evolved $g_{yy}$ component of the metric at $T=1000$ at resolution $\rho=4$, but the two are indistinguishable. \textit{Right}: absolute value of the error across the grid at $T=1000$, from which we can see more easily that some small errors in the magnitude and phase have been introduced.
\label{fig-linear}}
\end{center}
\end{figure}

As can be seen from Figure \ref{fig-linear}, this error is 4 orders of magnitude smaller than the signal and therefore negligible.

\subsubsection{Gauge wave tests}
\label{sec-Apples-gauge}

The gauge wave test requires the evolution of the metric
\begin{equation}
ds^2 =  (1-H)(-dt^2 + dx^2) + dy^2 + dz^2 \quad H = A \sin{2\pi(x-t)} \, . 
\end{equation}

\begin{figure}[H]
\begin{center}
\includegraphics[width=.55\textwidth]{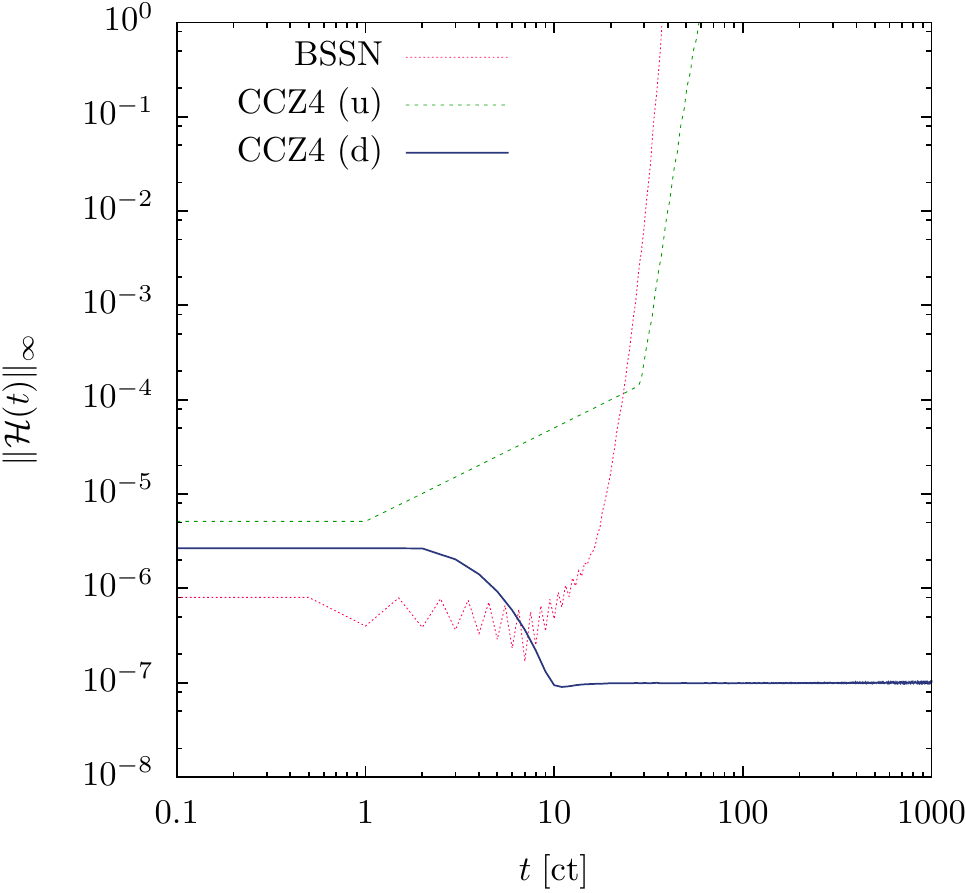}
\caption[Gauge wave test]{Gauge wave test. The increase in the $L_\infty$ norm of the Hamiltonian constraint means that the BSSN code only remains stable for less than 50 timesteps. Undamped (u) CCZ4, i.e. CCZ4 with $\kappa_1 = 0$, performs similarly. Damped (d) CCZ4 with $\kappa_1 = 1$ is stable for the full 1000 timesteps. The test was performed with initial amplitude of $A=0.1$, Kreiss-Oliger dissipation coefficient of $\sigma=0.1$ and a resolution of $\rho=4$. 
\label{fig-gauge}}
\end{center}
\end{figure}

The BSSN formulation is known to produce unsatisfactory results for the gauge wave tests. $\grchombo$ is no different in this respect. As can be seen in Figure \ref{fig-gauge}, it becomes unstable after around 50 crossing times, with the Hamiltonian constraint increasing exponentially, even for a relatively small initial amplitude of the gauge wave of $A = 0.1$. 

As was shown in \citep{Alic:2011gg} stability can be achieved by adding in the CCZ4 constraint damping terms. $\grchombo$ shows exactly this behaviour (figure \ref{fig-gauge}).

\subsubsection{Gowdy wave test}
\label{sec-Apples-gowdy}

The Gowdy wave evolves a strongly curved spacetime; an expanding vacuum universe containing a plane polarised gravitational wave propagating around a 3-torus. The metric is
\begin{equation}
ds^2 =  t^{-\onehalf} e^{\frac{\lambda}{2}} (-dt^2 + dz^2) + t e^P dx^2 + e^{-P} dy^2 ~ , 
\end{equation}
where $P$ is a function of $z$ and $t$ with Bessel functions $J_n$, in particular we use $P=J_0(2\pi t)\cos{2\pi z}$. $\lambda$ is also a function of $z$ and $t$ which may be expressed as a (rather more complex) product of Bessel functions, as given in eqn A.28 of \citep{Babiuc:2007}. In the forward time direction, $P$ decays to zero, and $\lambda$ undergoes linear growth due to the cosmological expansion.

In the expanding direction we use the slicing, $\partial_t \alpha = -\partial_t \sqrt{\gamma_{zz}}$. The collapsing direction is evolved starting at $t=t_0$ with harmonic slicing for the lapse and zero shift. A Kreiss-Oliger dissipation coefficient of $\sigma=0.05$ was used in both directions. 

\begin{figure}[H]
\begin{center}
\subfigure[BSSN Lapse $\alpha$ (collapsing)]{
\includegraphics[height=.4\textwidth]{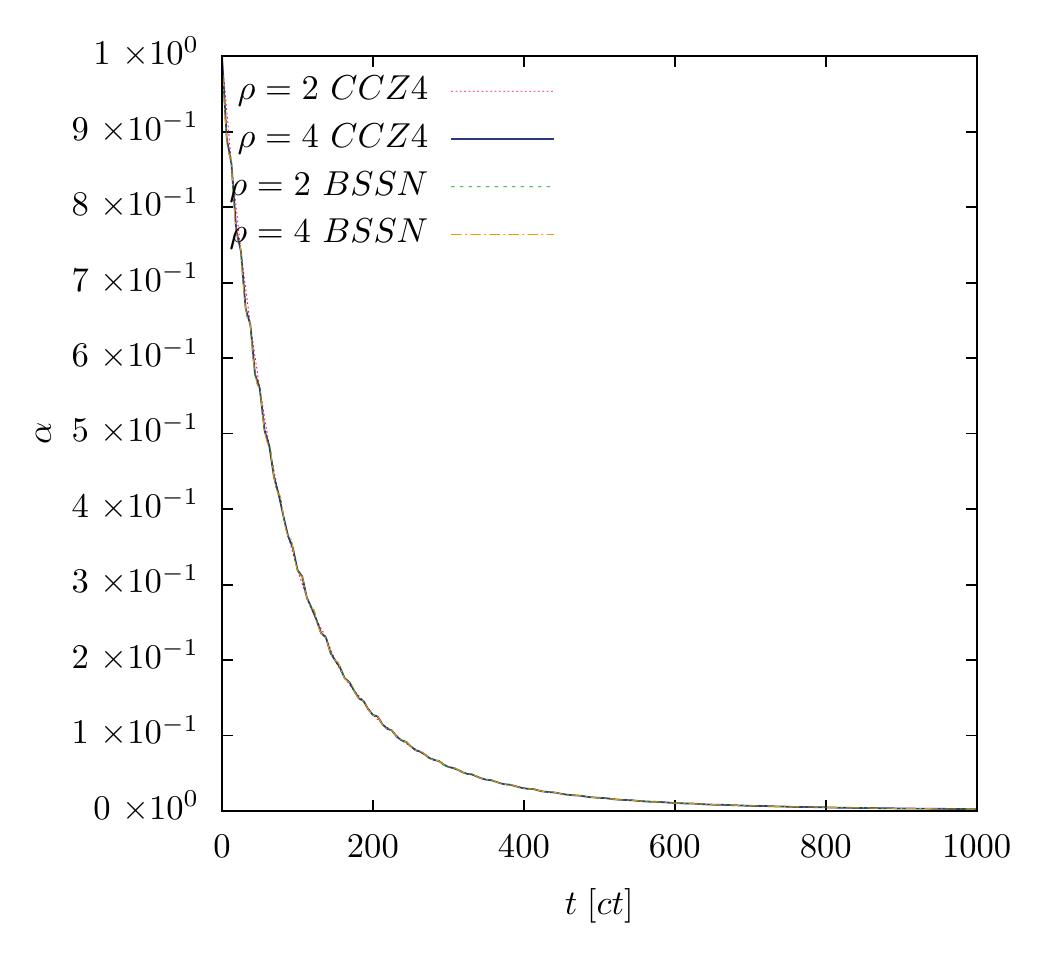}
}
\hspace{0.75cm}
\subfigure[Hamiltonian constraint (collapsing)]{
\includegraphics[height=.4\textwidth]{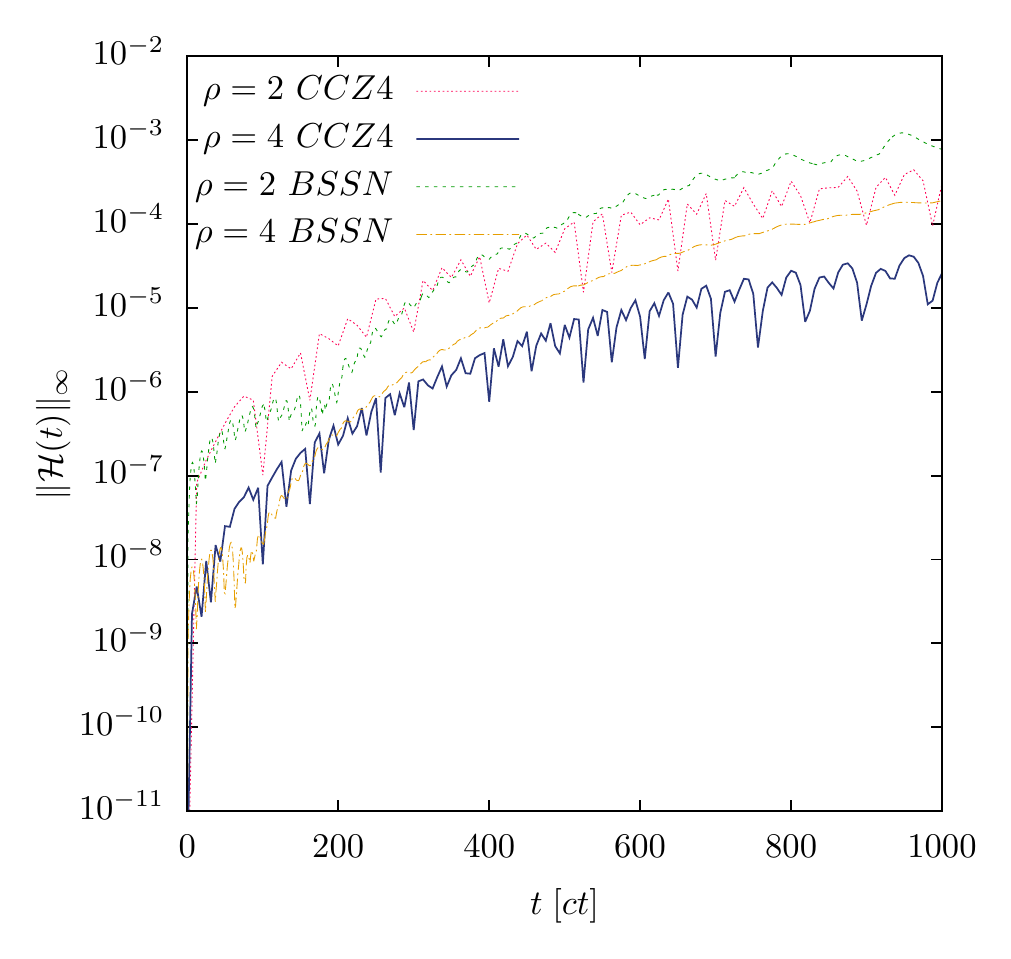}
}
\caption[Gowdy wave test, collapsing]{Gowdy wave test, collapsing. \textit{Left}: minimum value of the lapse $\alpha$ across the grid as the spacetime is evolved towards the singularity. As expected, the harmonic gauge causes the evolution to ``slow down'' as the singularity is approached. \textit{Right}: evolution of the  $L_\infty$ norm of the Hamiltonian constraint for two resolutions. The test reaches $T=1000$ crossing times without crashing.
\label{fig-gowdy-collapsing}}
\end{center}
\end{figure}

\begin{figure}[H]
\begin{center}
\subfigure[$K$ (expanding)]{
\includegraphics[height=.4\textwidth]{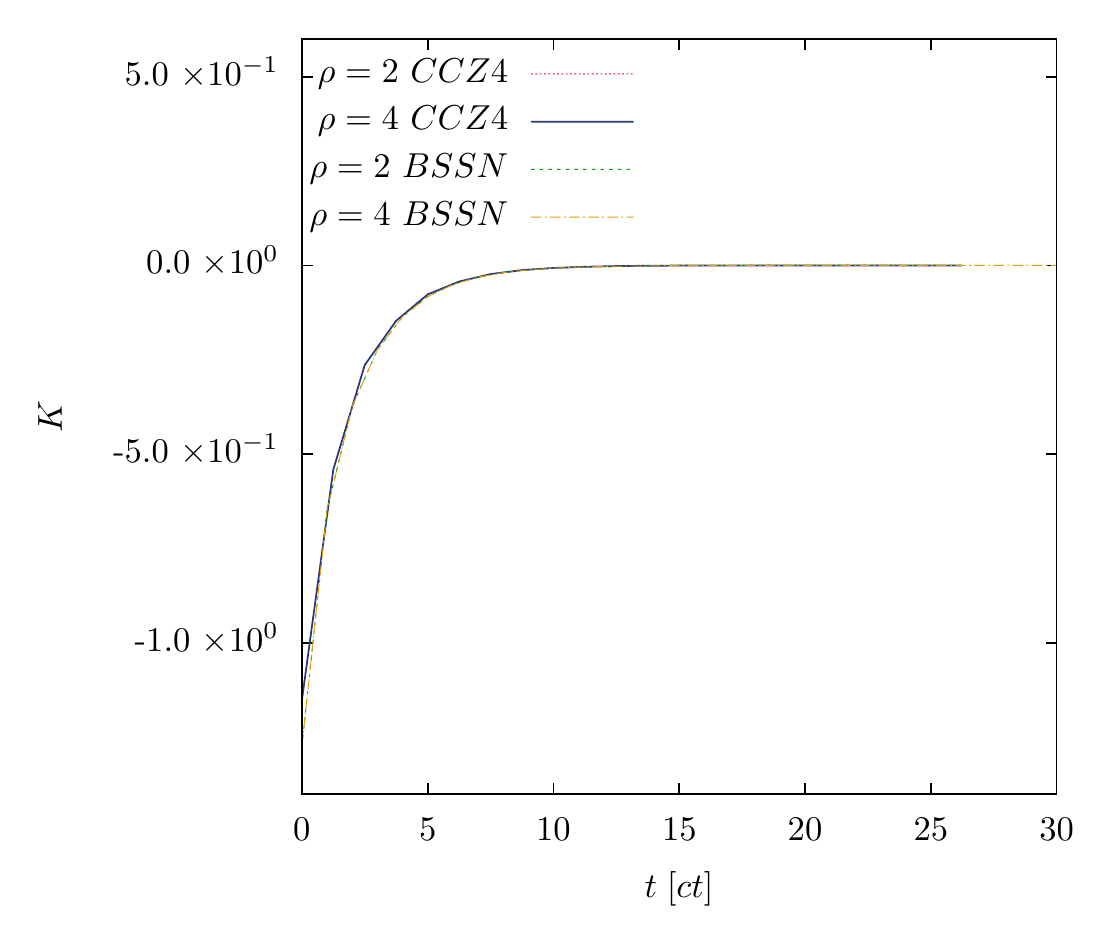}
}
\hspace{0.75cm}
\subfigure[Hamiltonian constraint (expanding)]{
\includegraphics[height=.41\textwidth]{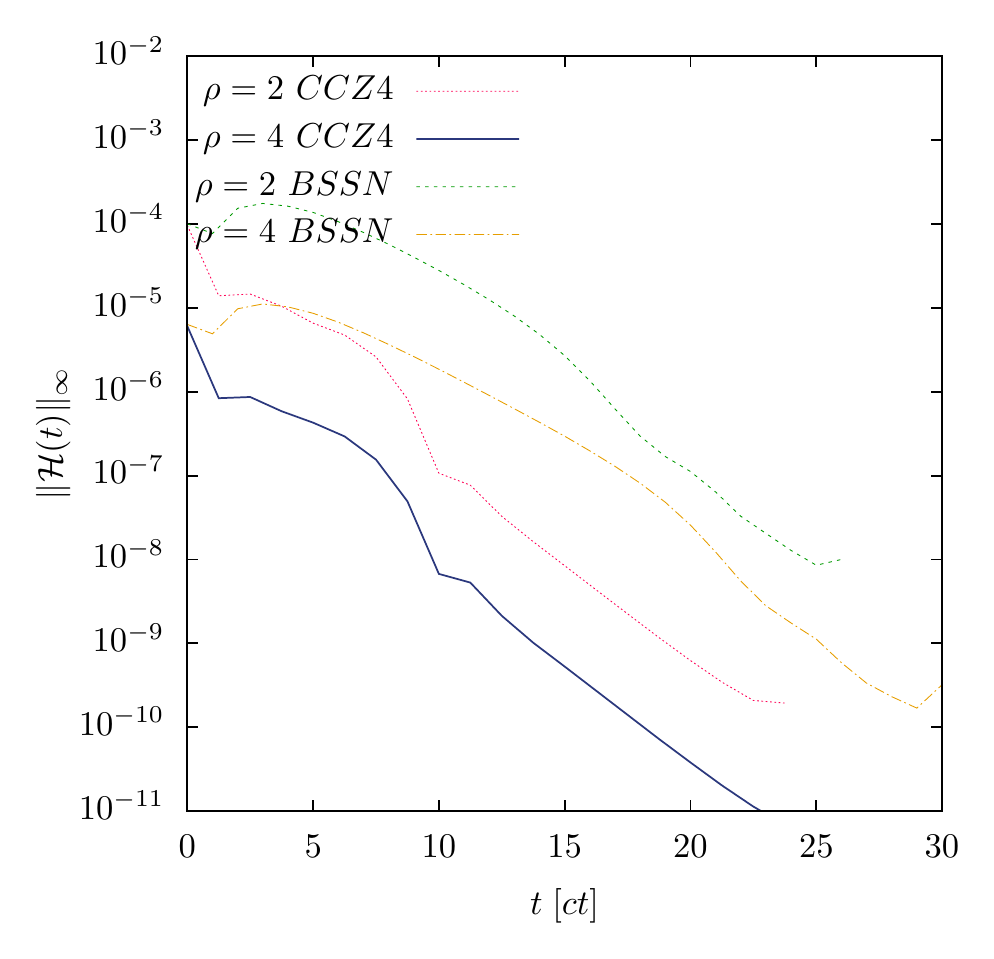}
}
\caption[Gowdy wave test, expanding]{Gowdy wave test, expanding. \textit{Left}: trace of the extrinsic curvature $K$ as the Gowdy wave spacetime is evolved in the collapsing direction. This correctly asymptotes to zero as the spacetime expands, but becomes unstable at around $t=30$. \textit{Right}: evolution of the  $L_\infty$ norm of the Hamiltonian constraint for two different resolutions.
\label{fig-gowdy-expanding}}
\end{center}
\end{figure}

The results for both the BSSN and CCZ4 codes in the collapsing direction are shown in Figure \ref{fig-gowdy-collapsing}, and in the expanding direction in Figure \ref{fig-gowdy-expanding}.

As is found in the Apples with Apples tests \citep{Babiuc:2007} for other simple BSSN codes, $\grchombo$ with BSSN and CCZ4 gives a less than satisfactory performance in this test in the expanding direction. The evolution is stable for approximately the first 30 crossing times, after which high frequency instabilities develop and cause code crash, due to the exponentially growing $\gamma_{zz}$ component. In \citep{Babiuc:2007} it was found that this behaviour of BSSN could be controlled with dissipation, but that long term accuracy was not achievable. 

\begin{figure}[H]
\begin{center}
\subfigure[Convergence (expanding)]{
\includegraphics[height=.4\textwidth]{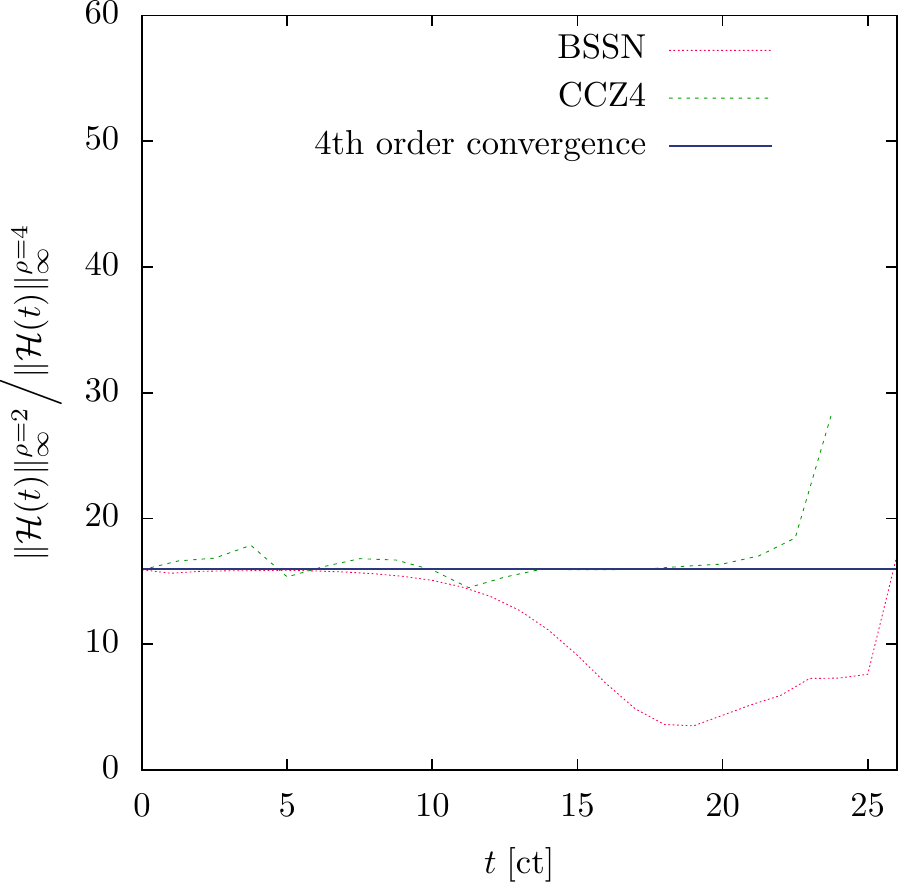}
}
\hspace{0.75cm}
\subfigure[Convergence (collapsing)]{
\includegraphics[height=.4\textwidth]{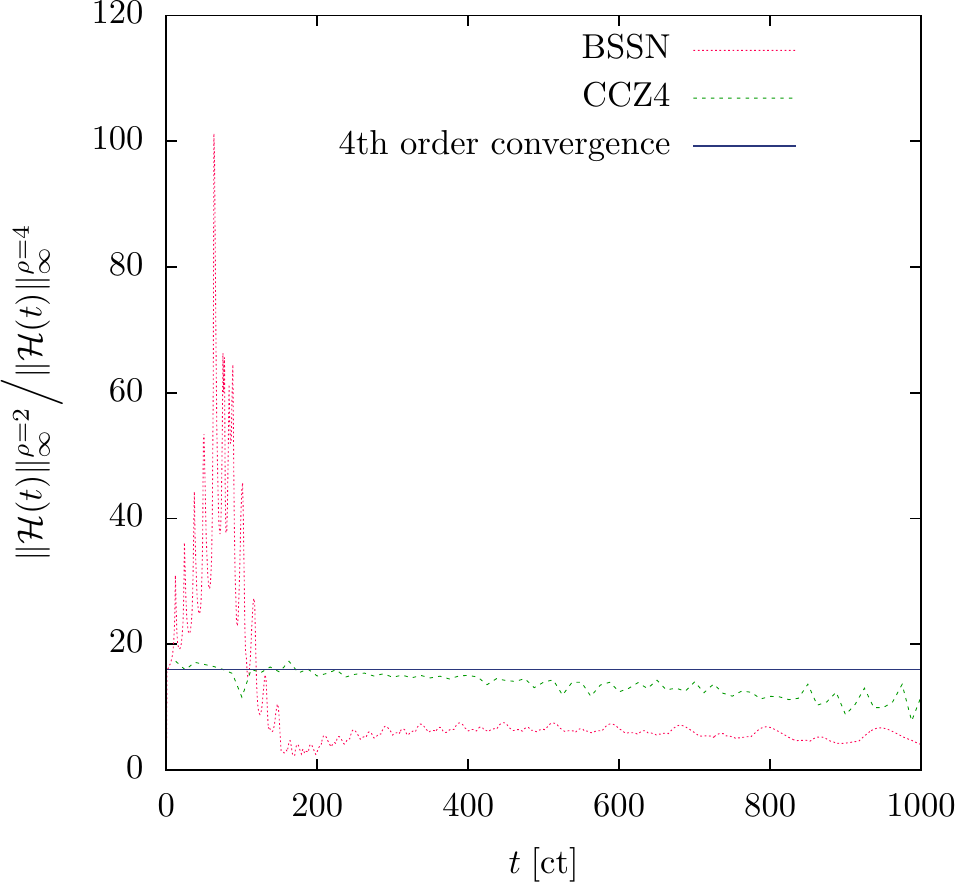}
}
\caption[Gowdy wave test, convergence]{Gowdy wave test, convergence. The ratio of the $L_\infty$ norm of the Hamiltonian constraint for the resolutions $\rho=4$ and $\rho=2$ is shown, for the expanding and collapsing directions for the BSSN and CCZ4 codes. A value of 16 indicates 4th order convergence, which is demonstrated by the codes initially, although lost at later times by BSSN. 
\label{fig-gowdy-convergence}}
\end{center}
\end{figure}

In the contracting direction the evolution is stable for the full 1000 crossing times and we were able to confirm the convergence of our code. As shown in Figure \ref{fig-gowdy-convergence}, both BSSN and CCZ4 exhibit 4\textsuperscript{th} order convergence initially. While convergence is never lost, the order is reduced at later times. This is similar to the behaviour found in \citep{Babiuc:2007} and \citep{Cao:2011fu}.

\subsection{Vacuum black hole spacetimes}
\label{sec-VacBH}

In this subsection we show that our code can stably evolve spacetimes containing black holes. 

All the simulations presented here used the BSSN formulation of the Einstein equations, along with the gamma-driver and alpha-driver gauge conditions. Adding CCZ4 constraint damping gives better performance for the Hamiltonian constraint, as would be expected, but the results are broadly similar and so are not presented here. Unless otherwise stated, we perform the simulations with up to 8 levels of refinement and  we based our tagging/regridding criterion, \eqn{eqn:tagging}, on the value of $\chi$. We emphasise that the purpose of this subsection is to demonstrate that we can stably evolve black hole spacetimes, but we are not interested in extracting gravitational wave data or in studying convergence; this will be done in the next subsection. 

Where we refer to taking an $L^2$ norm of the Hamiltonian constraint $\mathcal{H}$ in a test, this is calculated as follows (using the weighted variable sum function in VisIt):

\begin{equation}
|| \mathcal{H} ||_{2} = \sqrt{\sum_i m_i \mathcal{H}_i^2} , \label{eqn:L2norm}
\end{equation}
where $m_i=V_i/V_{tot}$, is the fraction of the total grid volume $V_{tot}$ occupied by the $i$th box. Where the grid contains a black hole, we excise the interior by setting $\mathcal{H}$ to zero within the region in which the lapse $\alpha$ is less than 0.3 (which is an approximate rule of thumb for the location of an event horizon for a black hole in the moving puncture gauge). The difference in the results is small, since the error norm is dominated by regridding errors at the boundaries between meshes. We also exclude the values on the outer boundaries of the grid, which can distort the results in cases where periodic boundaries are used. 

\subsubsection{Schwarzschild black hole}
\label{sec-VacBH-SS}

First we evolve a standard Schwarzschild black hole in isotropic gauge, with a conformally flat metric, the lapse initially set to one everywhere, and the conformal factor $\chi$ set to

\begin{equation}
\chi = \left(1+\frac{M}{2r} \right)^{-2} \, . \label{eqn:SCchi}
\end{equation}

\begin{figure}
\begin{center}
\includegraphics[width=.6\textwidth]{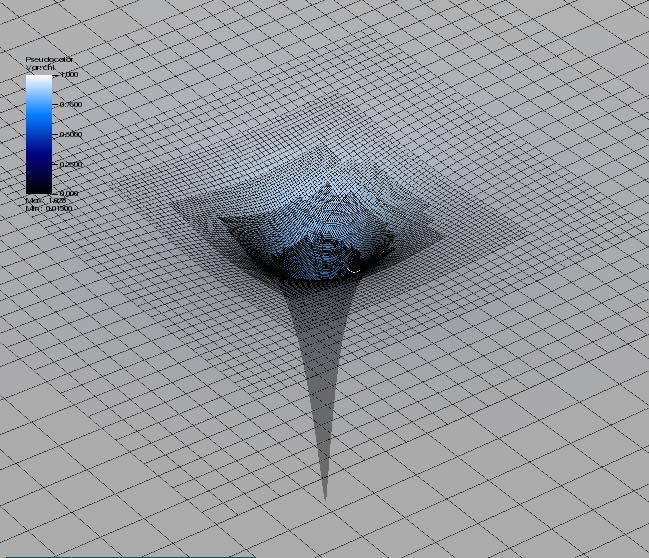}
\caption[Schwarzschild black hole, conformal factor]{The profile for $\chi$ through a slice perpendicular to the z axis is shown for a Schwarzschild black hole.
\label{fig-profile}}
\end{center}
\end{figure}

\begin{figure}[H]
\begin{center}
\subfigure[Hamiltonian constraint]{
\includegraphics[height=.37\textwidth]{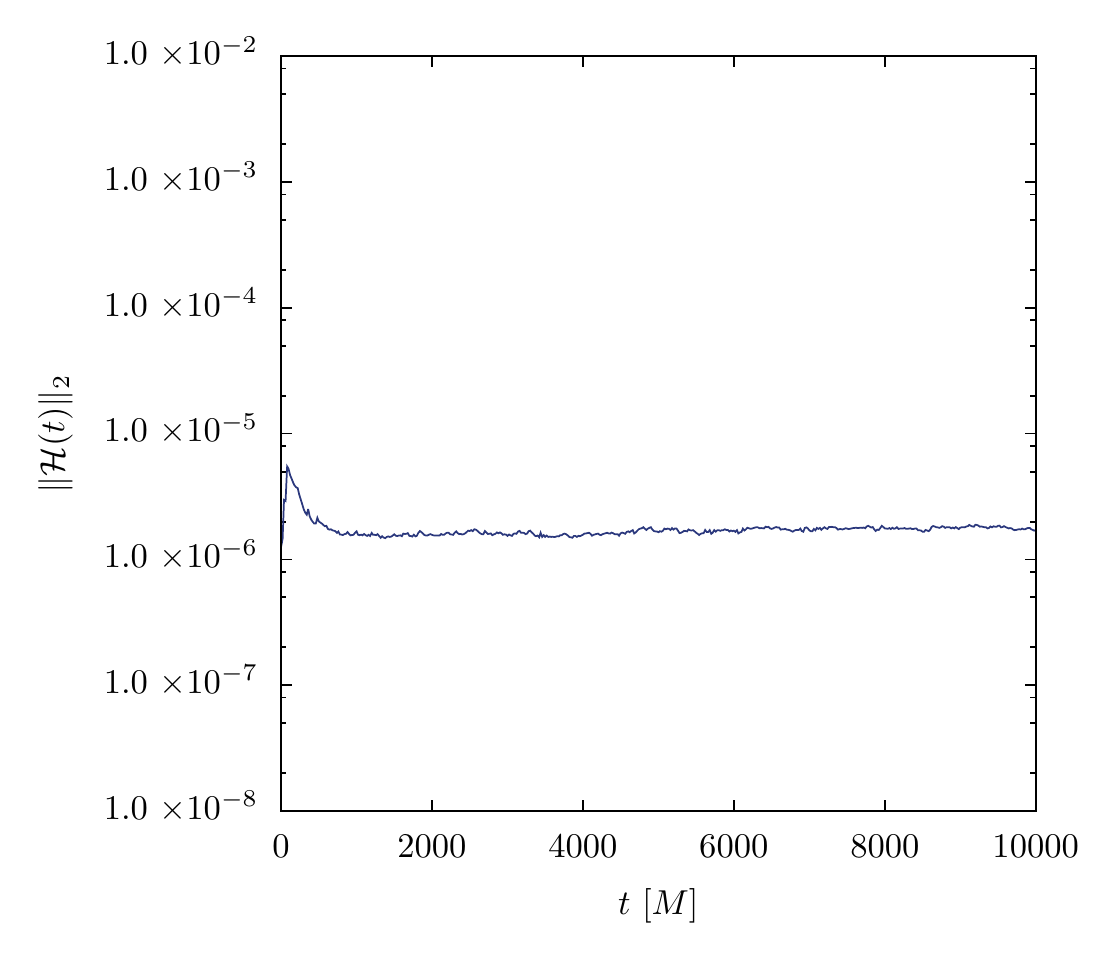}
}
\hspace{0.75cm}
\subfigure[ADM quantities]{
\includegraphics[height=.35\textwidth]{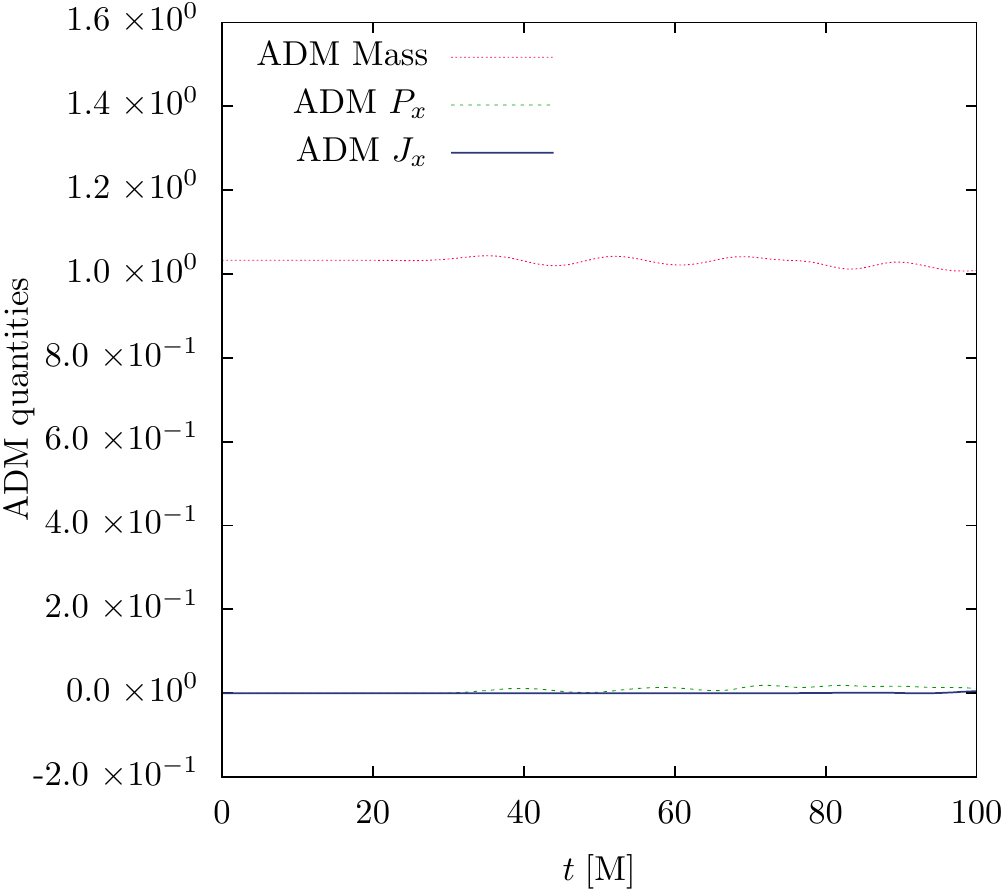}
}
\caption[Schwarzschild black hole tests]{Schwarzschild black hole simulations. \textit{Left:} Evolution of the $L^2$ norm of the Hamiltonian constraint up to $t=10000M$, showing long term stability.  \textit{Right:} ADM Mass, angular momentum and linear momentum (in the $x$ direction) during the initial stages of the evolution. These quantities remain approximately constant. 
\label{fig-SC}}
\end{center}
\end{figure}

In this simulation, we chose the outer boundary of the domain to be at $600M$ and the spatial resolution in the coarsest mesh as $10M$.  We impose Sommerfeld boundary conditions. The initial value of $\chi$ through a slice is shown in Figure \ref{fig-profile}. We see the expected ``collapse of the lapse'' at the singularity and the solution quickly stabilises into the ``trumpet'' puncture solution described in \citep{Hannam:2008sg}. We find an apparent horizon and are able to evolve the black hole stably and without code crash for well over $t = 10000M$ time steps as shown in Figure \ref{fig-SC} (left). In this figure we show the $L^2$ norm of the Hamiltonian constraint across the whole grid, and it remains bounded throughout the evolution.

We monitor the ADM mass of the black hole by integrating over a surface near the asymptotically flat boundary, as seen in Figure \ref{fig-SC} (right). We also monitor the angular momentum and linear momentum of the black hole, and find that these remain zero as expected, as shown in Figure \ref{fig-SC} (right). These simple ADM measures rely on asymptotic flatness at the surface over which they are integrated, and so are sensitive to errors introduced by reflections at the boundaries, initial transients from approximate gauge choice or if the black hole is moving nearer the boundary (as in the boosted case). They are therefore less reliable as the simulation progresses, and we use them simply to confirm that we are evolving the correct spacetime initially.

\subsubsection{Kerr black hole}
\label{sec-VacBH-Kerr}

In this sub-subsection we present the results of a simulation of the Kerr black hole spacetime in the quasi isotropic Kerr-Schild coordinates that were used in \citep{Brandt:1996si}, with the angular momentum parameter $a = J/M$ set to 0.2. The domain size was chosen to be $(320M)^3$ and the grid spacing in the coarsest level was $4M$. We impose periodic boundary conditions for simplicity, which limits the duration of the simulation due to boundary effects. 

\begin{figure}[H]
\begin{center}
\subfigure[Hamiltonian constraint]{
\includegraphics[height=.36\textwidth]{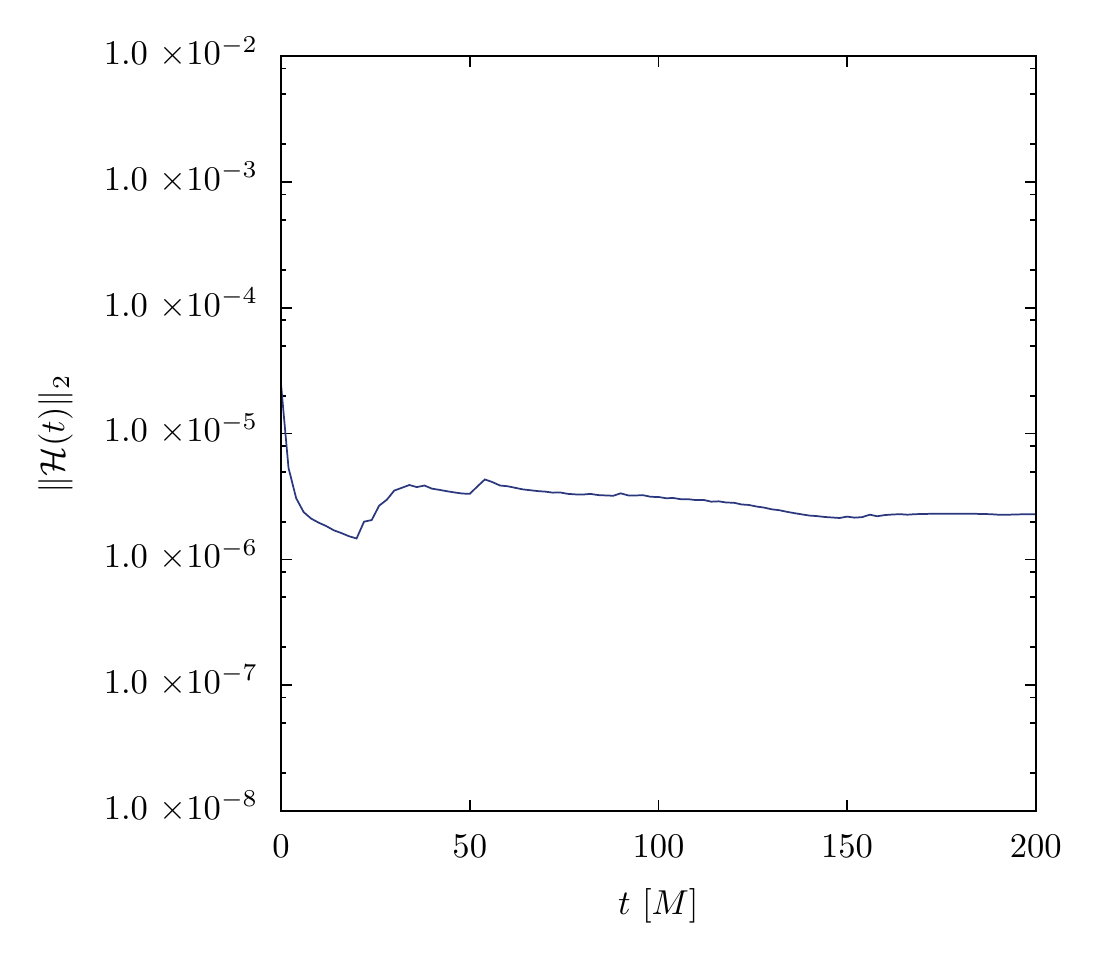}
}
\hspace{0.75cm}
\subfigure[ADM quantities]{
\includegraphics[height=.34\textwidth]{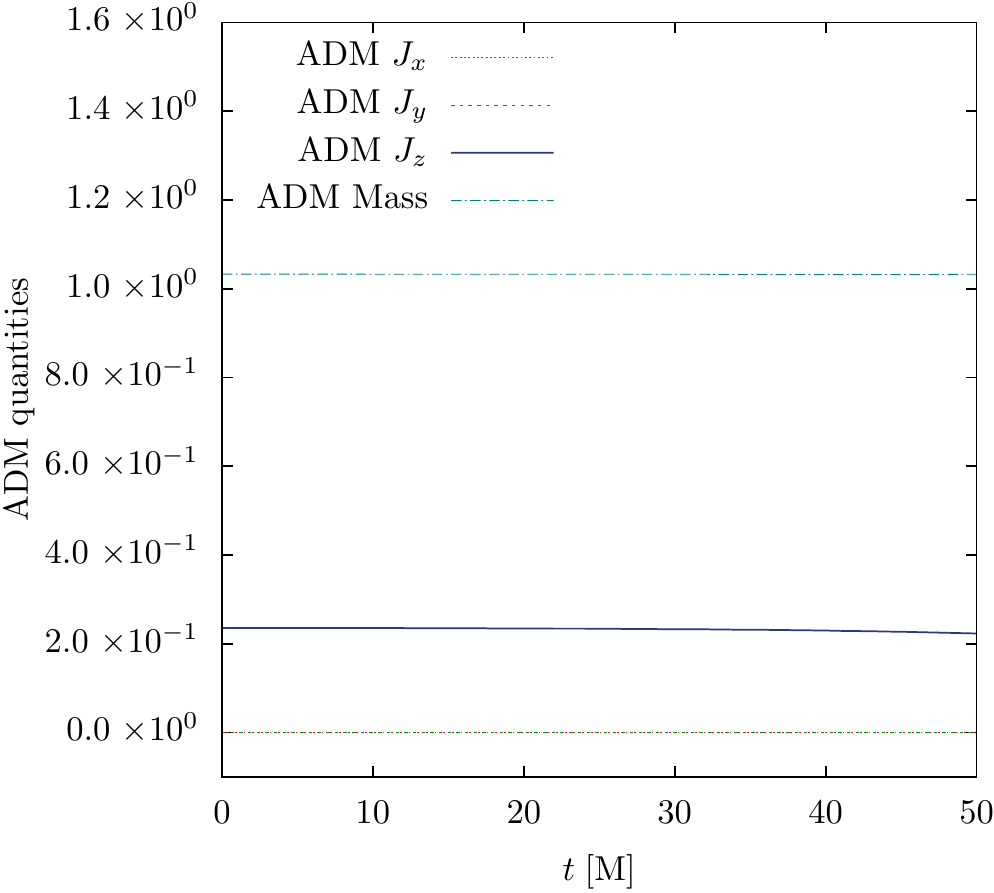}
}
\caption[Kerr black hole tests]{Kerr black hole simulations. \textit{Left}:  Evolution of the $L^2$ norm of the Hamiltonian constraint. \textit{Right}: Components of the angular momentum and mass of the Kerr black hole during the evolution. The ADM quantities remain constant. 
\label{fig-KE}}
\end{center}
\end{figure}

In Figure \ref{fig-KE} (left) we show the $L^2$ norm of the Hamiltonian constraint throughout the evolution. This plot shows that the amount of constraint violation remains stable during the simulation. In the right panel of Figure \ref{fig-KE} we display the ADM measures for the three components of the angular momenta and the mass. This Figure shows that these quantities remain (approximately) constant during the simulation.  

\subsubsection{Boosted black hole}
\label{sec-VacBH-Boost}

\begin{figure}[H]
\begin{center}
\subfigure[Initial position]{
\includegraphics[width=.8\textwidth]{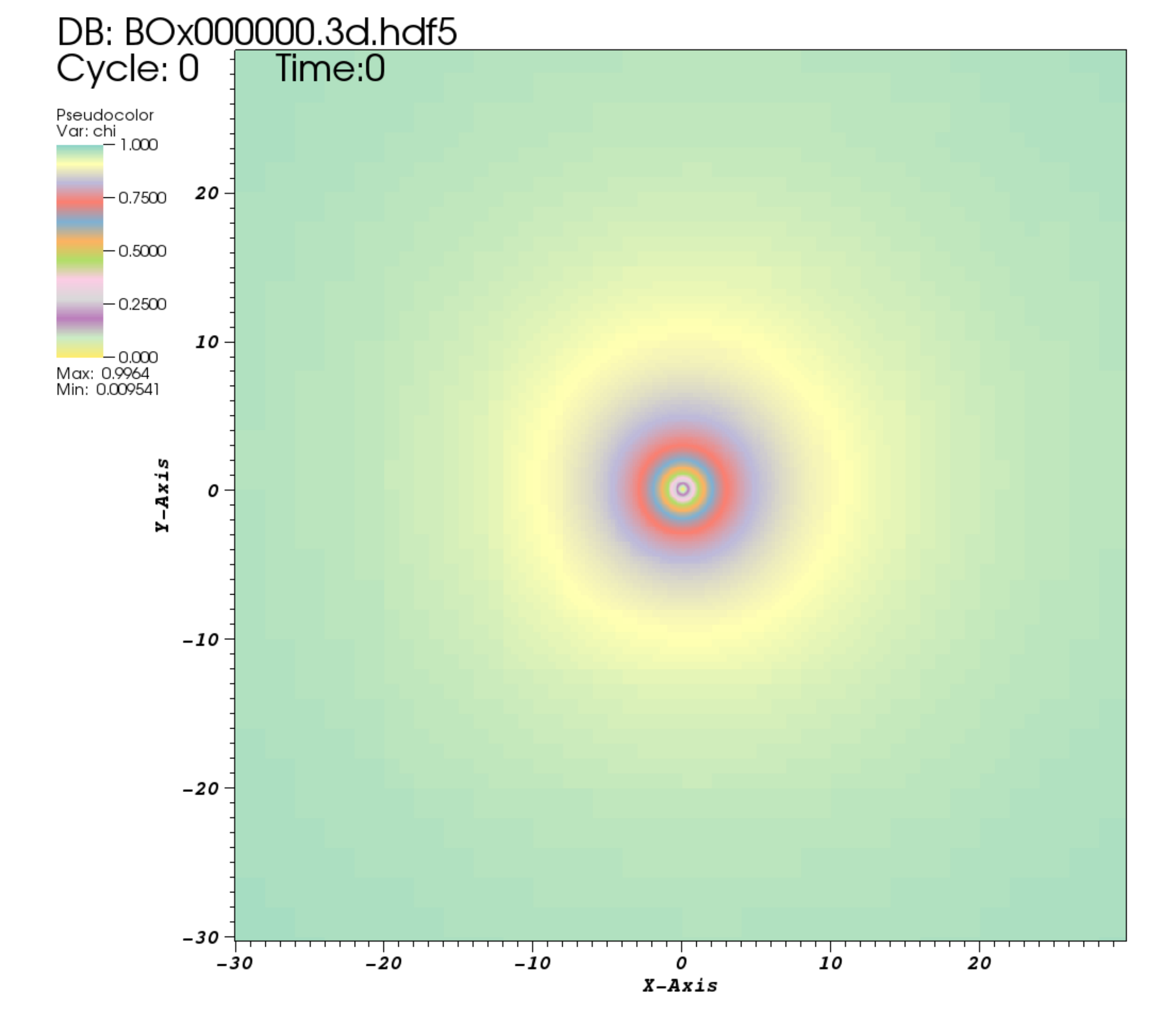}
\hspace{1.5cm}
}
\subfigure[Position at $t=100$]{
\includegraphics[width=.45\textwidth]{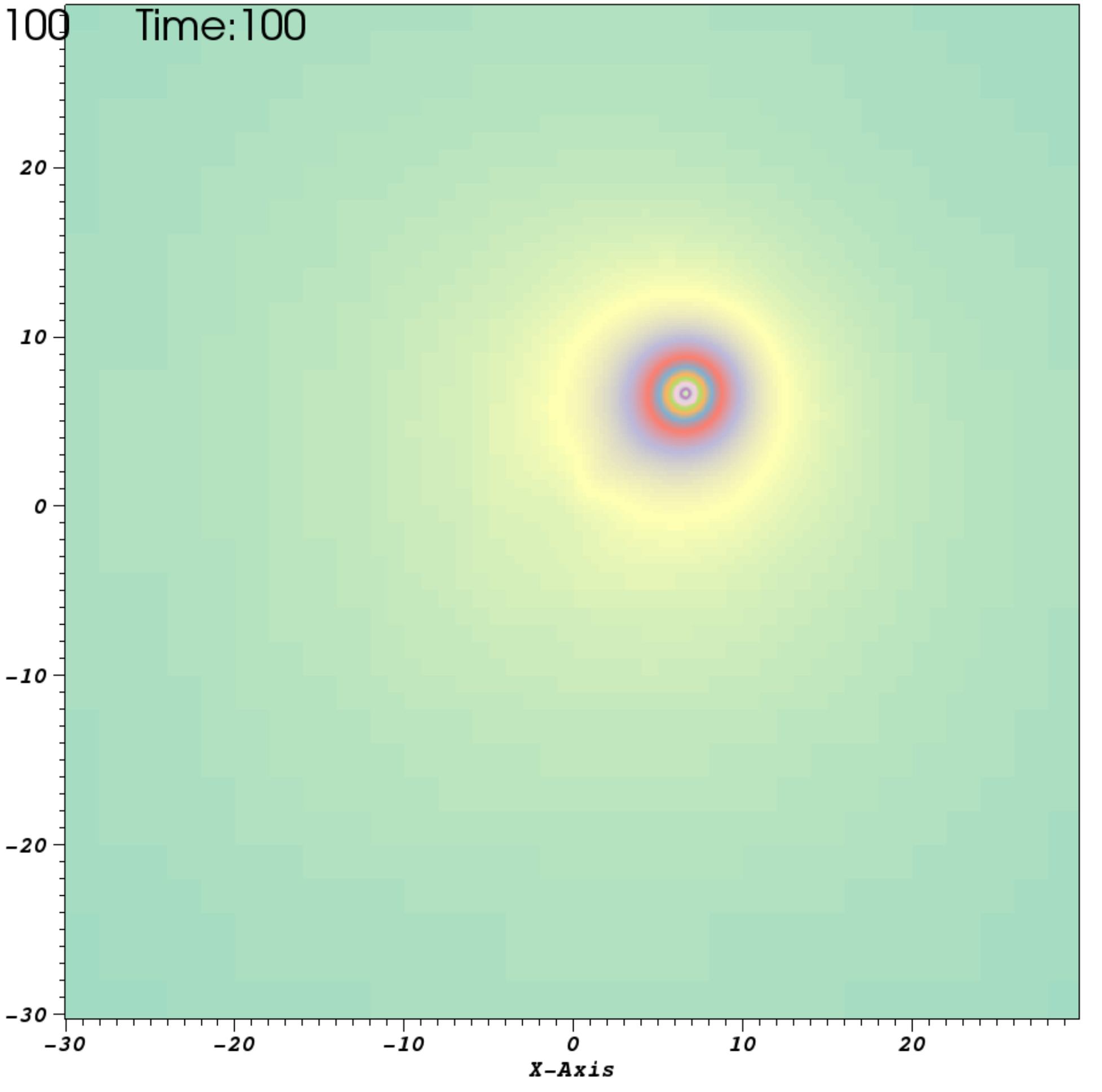}
}
\subfigure[Grid at $t=100$]{
\includegraphics[width=.45\textwidth]{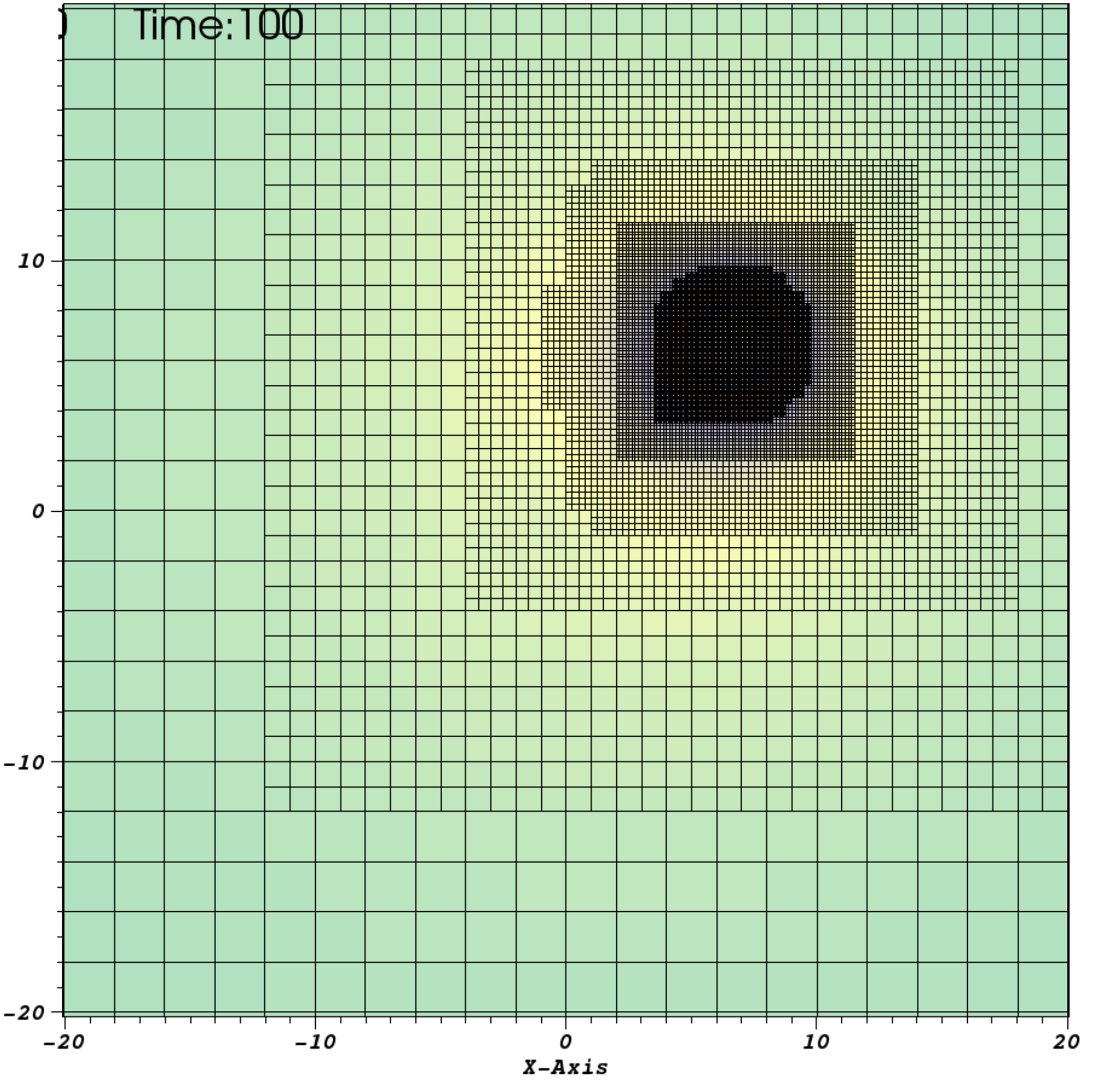}
}
\caption[Boosted black hole, moving]{The boosted black hole moves across the grid diagonally with initial momenta of $P_x = 0.02$,  $P_y = 0.02$ and $P_z = 0.0$, as expected, and the grid adapts to this movement, with the high resolution grids following the movement.
\label{fig-Boostmove}}
\end{center}
\end{figure}

In this sub-subsection we evolve a boosted black hole using the perturbative approximation from \citep{ShapiroBook}, in which the momentum constraint is solved by
\begin{equation}
\tilde{A}_{ij} = \frac{3 \chi^3}{2r^2} (p^i s^j + p^j s^i - (\delta^{ij} - s^i s^j) s_k p^k) \, ,
\end{equation}
where $p^i \ll 1.0$ are the initial momenta, and the Hamiltonian constraint is solved to first order in $p^2 = p^i p_i$ as
\begin{equation}
\chi = (1 + q + \frac{p^2}{M^2}(q_1 + \onehalf q_2 (3 \cos^2 \theta -1))^{-2} \, ,
\end{equation}
where
\begin{align}
&q = \onehalf \frac{M}{r} \, , \\
&q_1 = \frac{p}{8}(1+q)^{-5}(q^4 + 5q^3 + 10q^2 + 10q + 5) \, , \\
&q_2 = 0.05 (1+q)^{-5} q^2(84 q^5 + 378 q^4 + 658 q^3 + 539 q^2 + 192 q + 15) \nonumber \\
&\hspace{4cm}+ 4.2 q^3 \ln \frac{q}{1+q} \, .
\end{align}
The initial momenta are set to $p_x = 0.02$,  $p_y = 0.02$ and $p_z = 0.0$.  The domain size was chosen to be $(640)^3$, with spatial resolution in the coarsest grid of $4M$. We imposed periodic boundary conditions at the outer boundaries of the domain. The black hole moves across the grid diagonally as expected, as is seen in Figure \ref{fig-Boostmove}.

\begin{figure}[H]
\begin{center}
\subfigure[Hamiltonian constraint]{
\includegraphics[height=.37\textwidth]{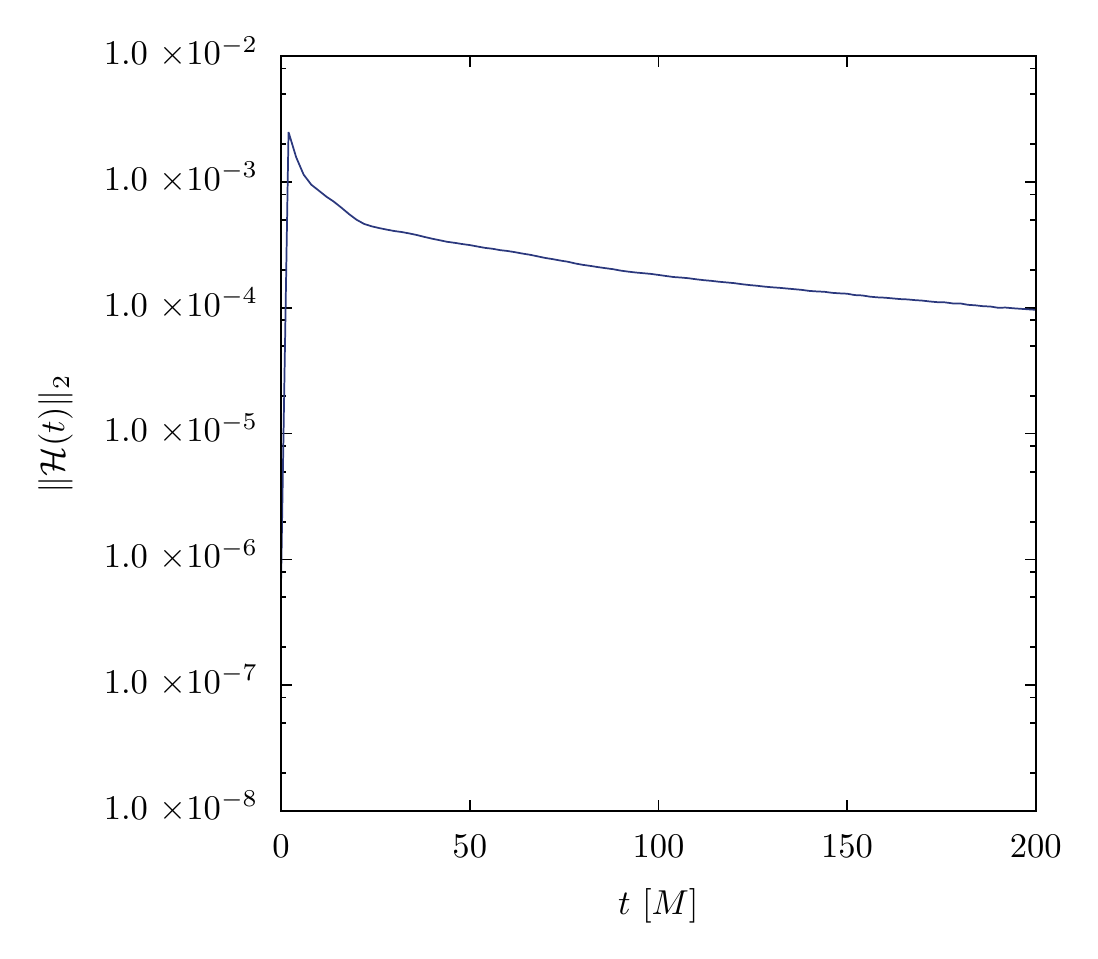}
}
\hspace{0.75cm}
\subfigure[ADM linear momentum]{
\includegraphics[height=.35\textwidth]{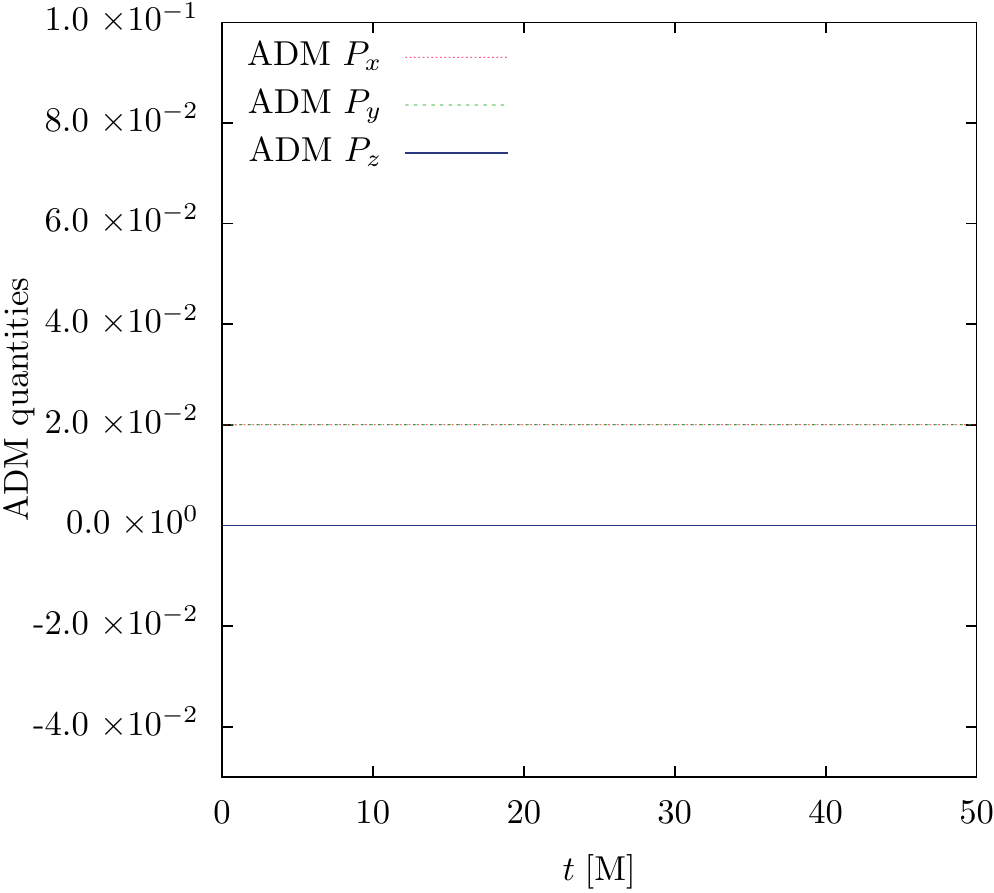}
}
\caption[Boosted black hole tests]{Boosted black hole simulations.  \textit{Left}: Evolution of the $L^2$ norm of the Hamiltonian constraint. \textit{Right}: Components of the ADM linear momentum during the evolution. They remain constant. Note that the lines for $P_x$ and $P_y$ overlap and so are difficult to distinguish.
\label{fig-BoostBH}}
\end{center}
\end{figure}

In the left panel of Figure \ref{fig-BoostBH} we show the $L^2$ norm of the Hamiltonian constraint across the domain as a function of time. This plot shows that the constraints remain bounded throughout the simulation. In the right panel of Figure \ref{fig-BoostBH} we display the components of the ADM linear momentum during the simulation. In the continuum limit they should be constant and in our simulation they are indeed approximately constant. 

\vspace{1cm}

\subsubsection{Binary inspiral}
\label{sec-VacBH-Binary}

In this sub-subsection we superpose the initial perturbative solution for two boosted black holes in \citep{ShapiroBook}, sufficiently separated, to simulate a binary inspiral merger. The domain size was $(200M)^3$ with a grid spacing in the coarsest level of $5M$. As in some of the previous tests, for simplicity we imposed periodic boundary conditions at the outer boundaries of the domain. 

\begin{figure}[H]
\begin{center}
\subfigure{
\includegraphics[width=.4\textwidth]{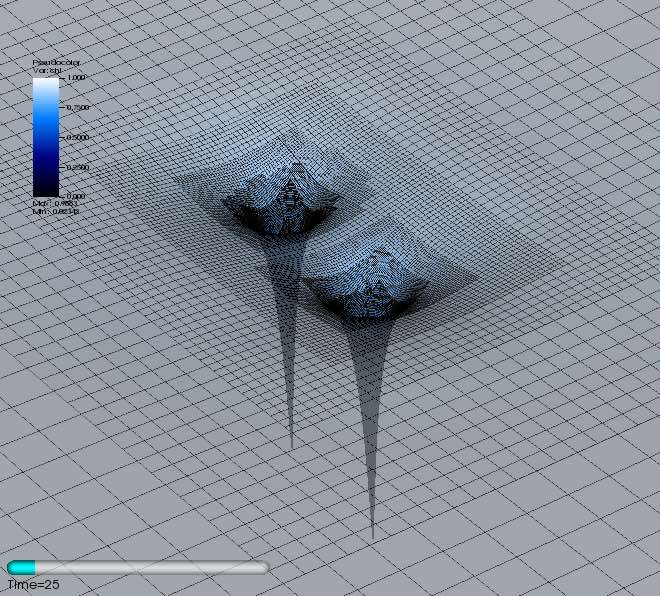}
}
\subfigure{
\includegraphics[width=.4\textwidth]{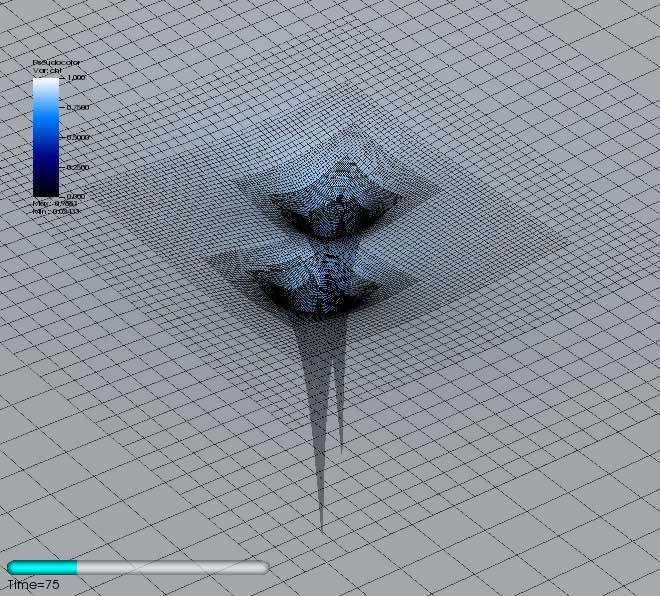}
}
\subfigure{
\includegraphics[width=.4\textwidth]{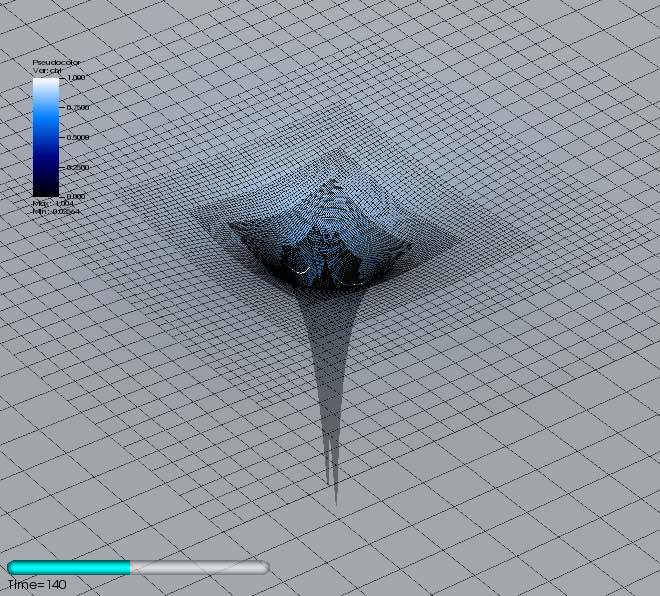}
}
\subfigure{
\includegraphics[width=.4\textwidth]{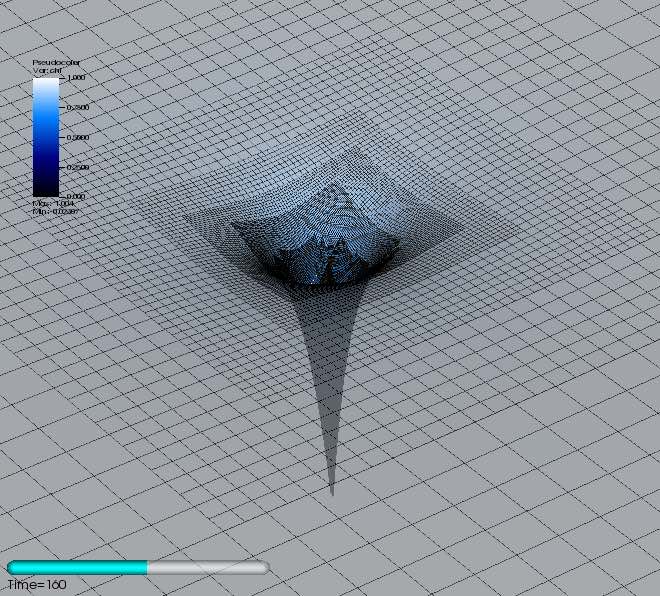}
}
\caption[Binary Black Hole merger]{Two black holes are evolved with $\grchombo$ in an inspiral merger. The final stages of the merger are shown. 
\label{fig-BiBH}}
\end{center}
\end{figure}

\begin{figure}[H]
\begin{center}
\includegraphics[height=0.45\textwidth]{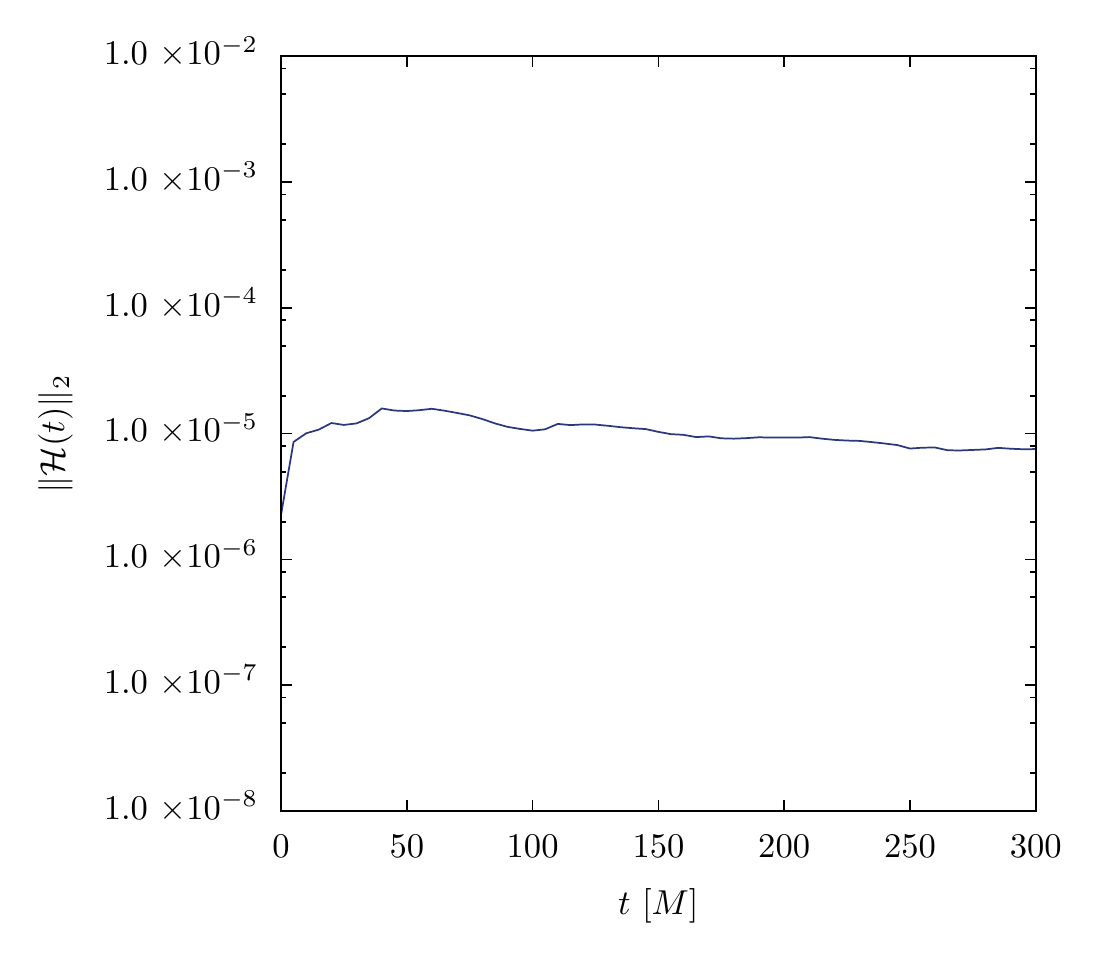}
\caption[Binary merger test]{$L^2$ norm of the Hamiltonian constraint across the whole domain for the binary black hole merger. The constraint violation remains bounded throughout the evolution, which includes the merger and ring-down phases.
\label{fig-Binary}}
\end{center}
\end{figure}

We are able to evolve the merger stably such that the two black holes merge to form one with a mass approximately equal to the sum of the two. The progression of the merger is shown in Figure \ref{fig-BiBH}. The time evolution of the $L^2$ norm of the Hamiltonian constraint across the grid is shown in Figure \ref{fig-Binary}. Again this remains stable throughout the simulation.



\subsection{Choptuik scalar field collapse}
\label{sec-Tests-Chop}

\begin{figure}[H]
\begin{center}
\vspace{1cm}
\subfigure{
\includegraphics[width=.46\textwidth]{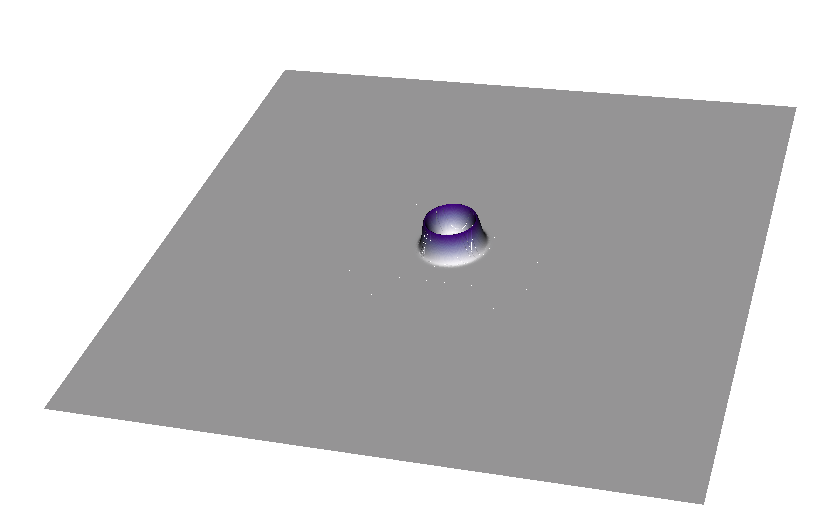}
}
\subfigure{
\includegraphics[width=.46\textwidth]{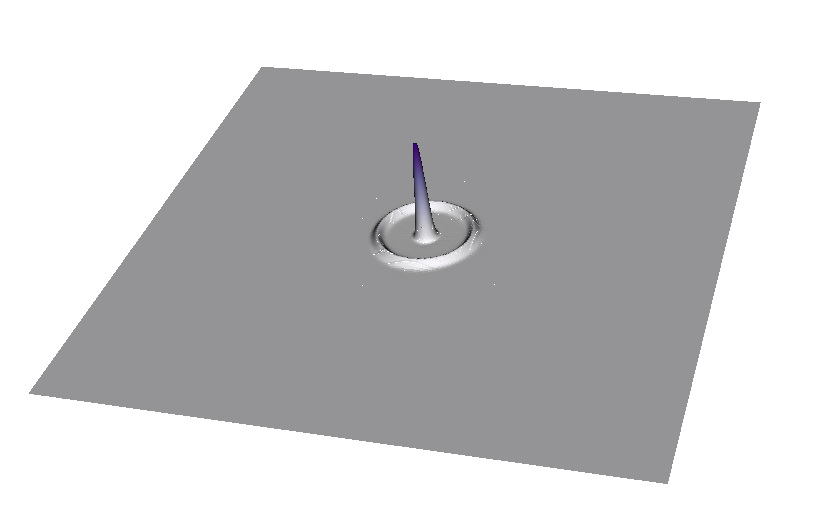}
}
\subfigure{
\includegraphics[width=.8\textwidth]{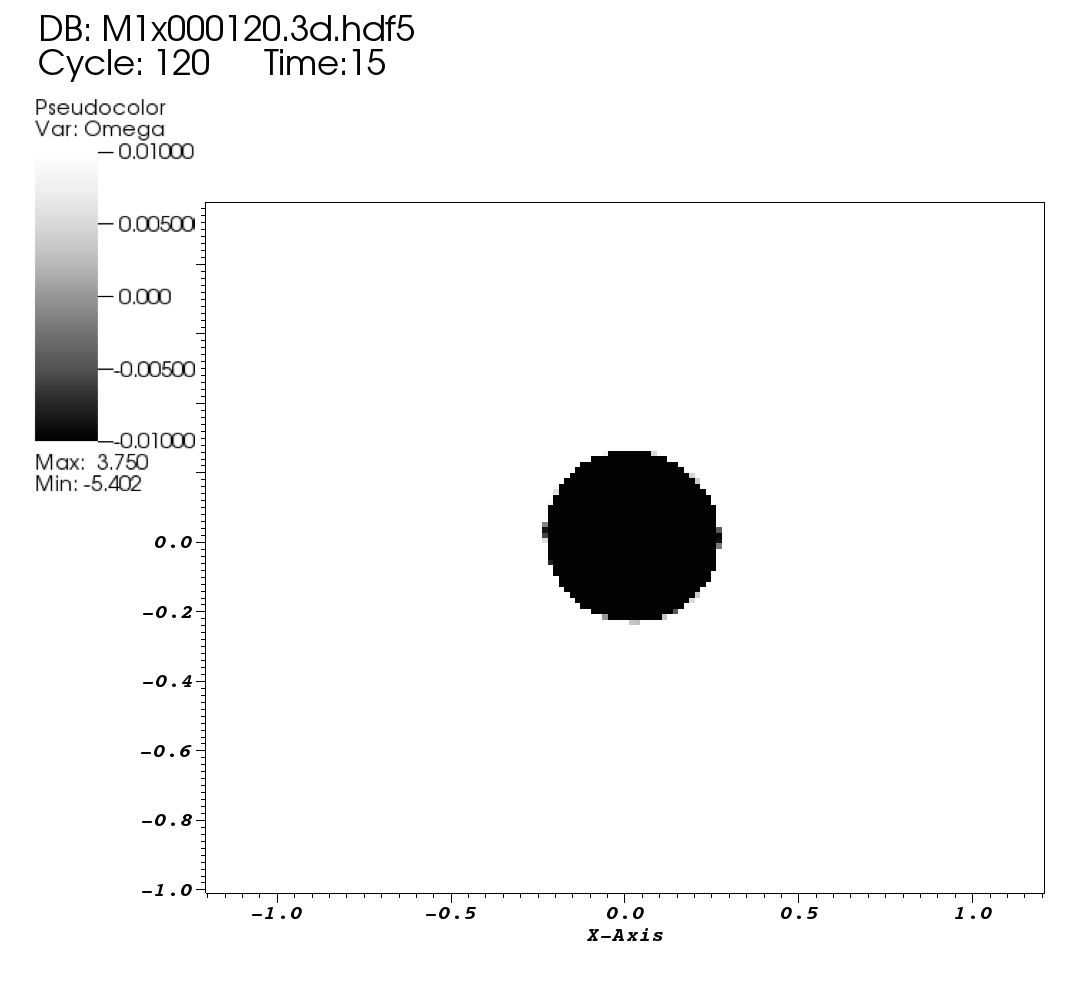} \hspace{1.5cm}
}
\vspace{1cm}
\caption[Choptuik Scalar Field collapse]{In Choptuik scalar field collapse, the initial specially symmetric configuration in the first figure (which shows the values on a slice perpendicular to the z axis) collapses, splitting into an ingoing and an outgoing wave as seen in the second image. If the amplitude of the initial perturbation is greater than a certain critical value, the ingoing wave will result in the formation of a black hole, as seen from the output of the apparent horizon finder in the third figure, which shows that an apparent horizon with a mass of about 0.25 has formed by $t=15$.
\label{fig-chop3D}}
\end{center}
\end{figure}

We now test the scalar field part of the code, by simulating the Choptuik scalar field collapse as described in \citep{AlcubierreBook} and illustrated in Figure \ref{fig-chop3D}. The referenced description is for a 1+1D simulation which is evolved using a partially constrained evolution. The lapse $\alpha$ and the single degree of freedom for the metric, $A$, are both solved for on each slice using ODEs obtained from the Hamiltonian constraint, and the polar areal slicing condition ($\partial_t \gamma_{\theta\theta}) = \partial_t \gamma_{\phi\phi} = 0 $. The only degrees of freedom which are truly evolved are those of the field variables, $\phi$, $\Psi$ and $\Pi$.

\begin{figure}[H]
\begin{center}
\subfigure[Subcritical profiles of $\phi$]{
\includegraphics[height=0.35\textwidth]{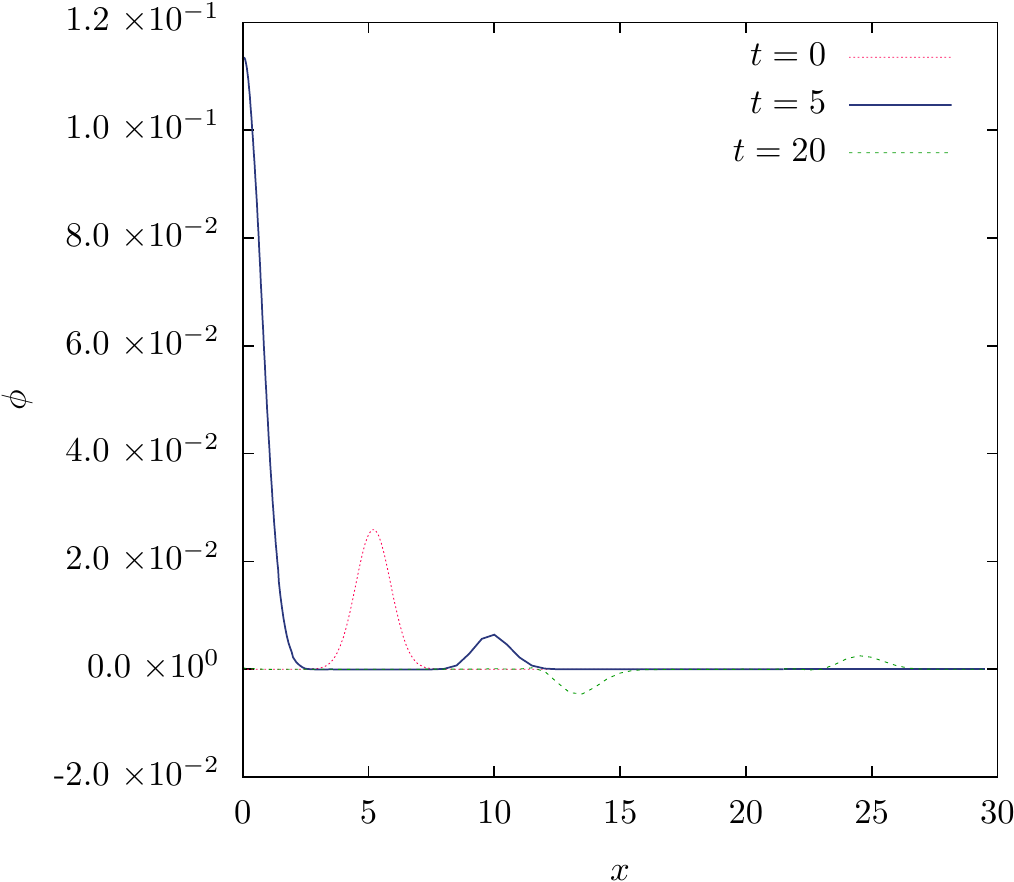}
}
\hspace{0.75cm}
\subfigure[Supercritical profiles of $\phi$]{
\includegraphics[height=0.35\textwidth]{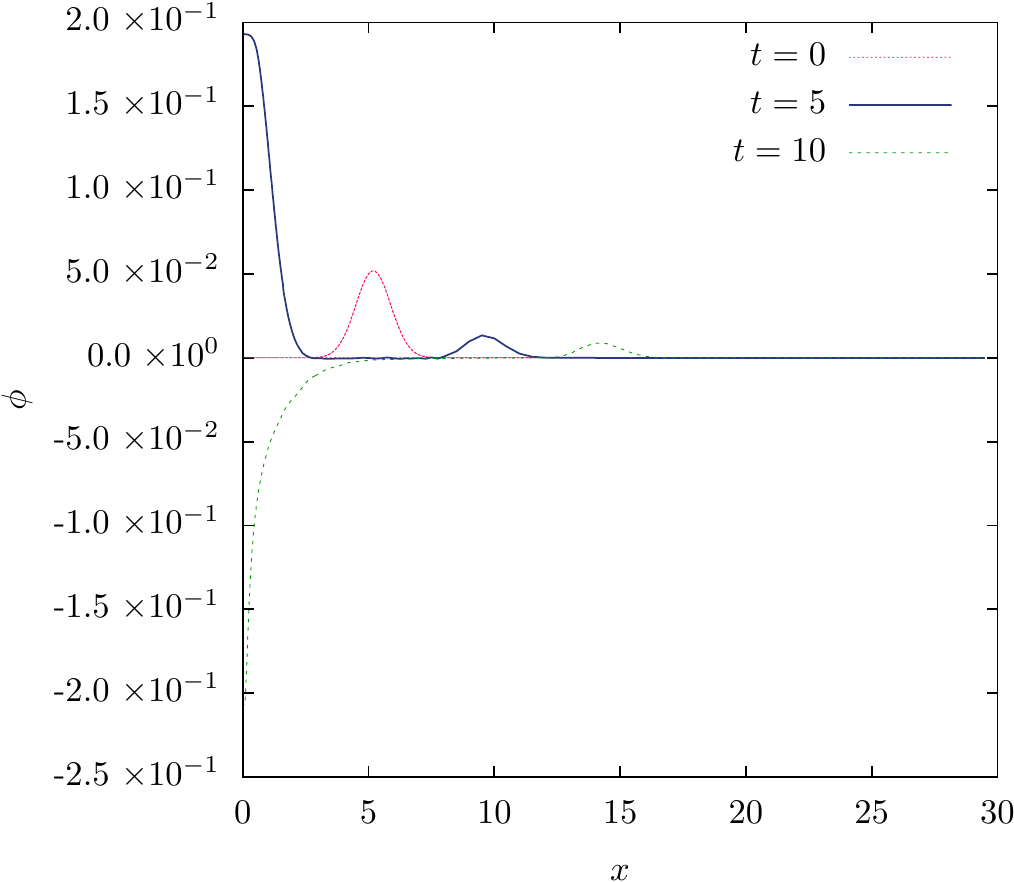}
}
\caption[Choptuik collapse test, scalar field]{Choptuik scalar field collapse. The profiles shown for the fields at $t=$ 0, 5 and 20 differ from those in \citep{AlcubierreBook} due to the different gauge conditions used. In the supercritical case we show the snapshot at $t=$ 10 rather than 20 as this is the point at which the evolution is frozen in the gauge choice in \citep{AlcubierreBook}. In the puncture gauge the evolution of the region within the event horizon continues and the result is that the large spike in the field effectively falls into the puncture, resulting in a zero field value at the centre of the coordinate grid. 
\label{fig-chopsnaps}}
\end{center}
\end{figure}

\begin{figure}[H]
\begin{center}
\subfigure[Subcritical lapse]{
\includegraphics[height=0.35\textwidth]{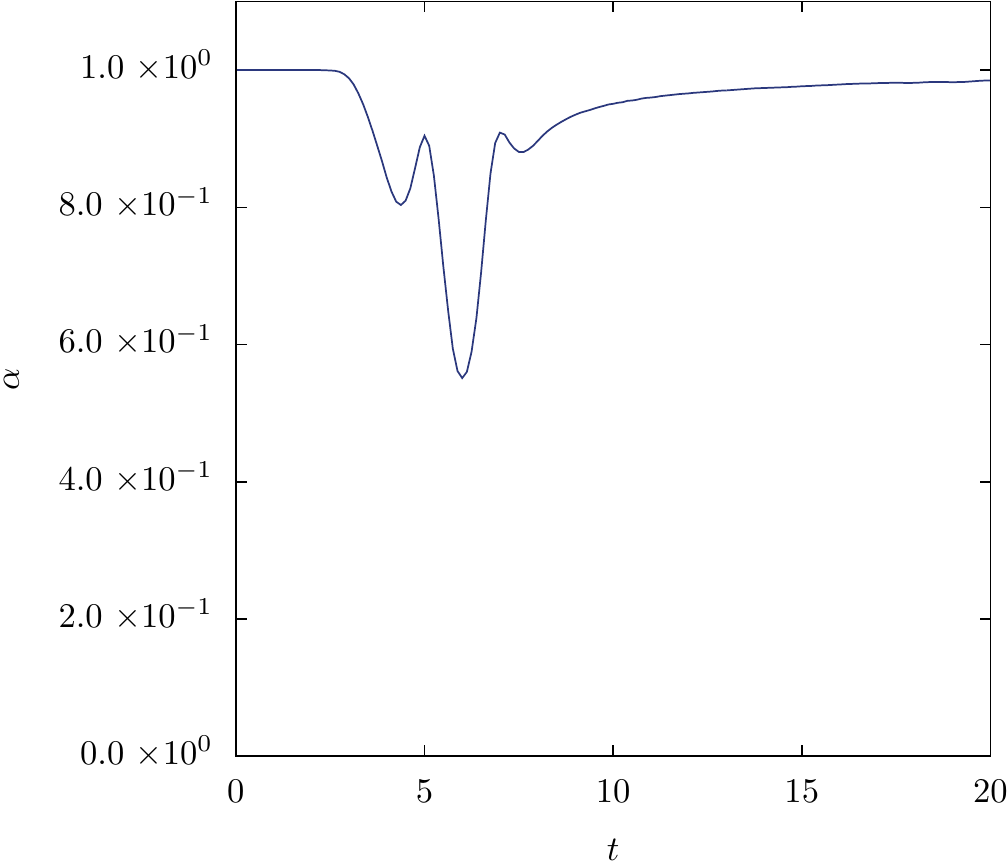}
}
\hspace{0.5cm}
\subfigure[Supercritical lapse]{
\includegraphics[height=0.35\textwidth]{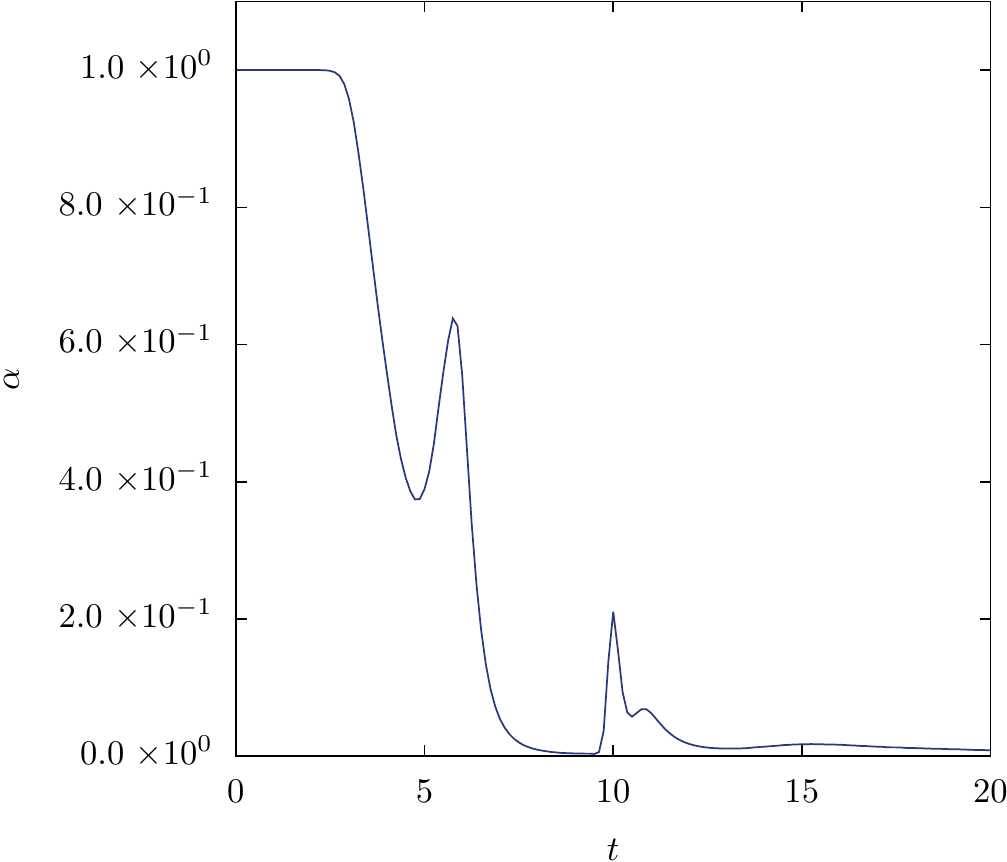}
}
\caption[Choptuik collapse test, lapse]{Choptuik scalar field collapse. The values of the lapse at the centre of the grid are given. It can be seen the the profiles are very similar to those obtained by Alcubierre in \citep{AlcubierreBook}, and that the one for the supercritical case shows the characteristic collapse of the lapse which is symptomatic of black hole formation.
\label{fig-chopalp}}
\end{center}
\end{figure}

\begin{figure}[H]
\begin{center}
\subfigure[Subcritical lapse]{
\includegraphics[height=0.65\textwidth]{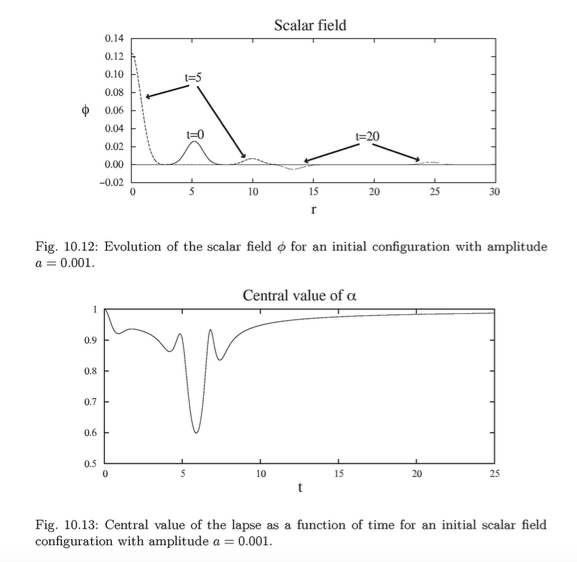}
}
\hspace{0.5cm}
\subfigure[Supercritical lapse]{
\includegraphics[height=0.65\textwidth]{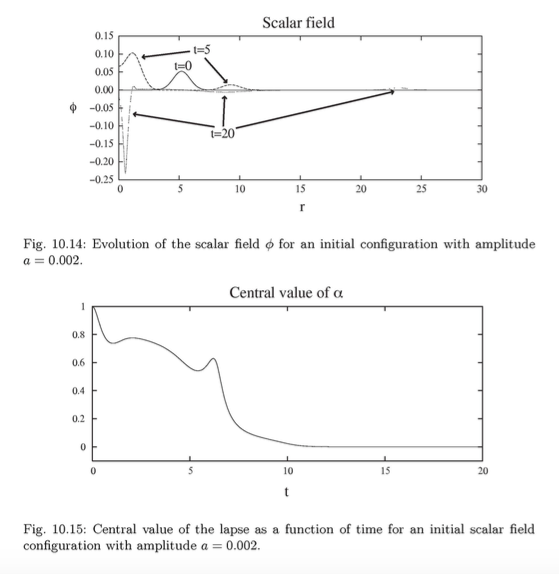}
}
\caption[Choptuik collapse test, Alcubierre results]{Choptuik scalar field collapse. The results obtained by Alcubierre in \citep{AlcubierreBook} are shown for comparison to our results.
\label{fig-Alcubierre}}
\end{center}
\end{figure}

Our evolution is carried out using the full $3+1$ BSSN equations, without assuming or adapting coordinates to spherical symmetry. We are able to replicate the results obtained in \citep{AlcubierreBook}, subject to some minor differences due to the fact that we evolve with the puncture gauge rather than according to the polar areal gauge, see figures \ref{fig-chopsnaps} and \ref{fig-chopalp}, and compare to \ref{fig-Alcubierre}.

We see that $\grchombo$ can accurately evolve the field profile in the presence of gravity, and copes with the collapse of the supercritical case into a singularity, without code crash. For the subcritical cases we see that the field disperses as expected. 


\subsection{Convergence test: head on collision of two black holes}
\label{sec-Tests-convergence}

In this subsection we simulate the head on collision of two black holes and analyse the convergence of the code. We set up Brill-Lindquist initial data \citep{Brill:1963yv} consisting of two static black holes of mass $0.5M$ with a separation of $10M$, located at the centre of the computational domain. We extract the gravitational wave signal (see figure \ref{fig-psi4} below). An initial burst of radiation is seen, which is a property of the superimposed initial data, prior to the main signal. Even though this set up could be simulated in axisymmetry, we have evolved the system without imposing any symmetry assumptions. So the results below correspond to a full $3+1$ simulation with \texttt{GRChombo}.

We performed runs at three different resolutions with 7 levels of refinement, each level having half the grid spacing as the previous one. The grid spacings were
\begin{itemize}
\item $0.03125M/4M$ for the low resolution run,
\item $0.02083M/2.66667M$ for the medium resolution run,
\item $0.01563M/2M$ for the high resolution run.
\end{itemize}
Here the numbers refer to the resolution on the finest/coarsest grids respectively. The outer boundary of the domain is located at $200M$ and we impose periodic boundary conditions for simplicity. This puts an upper bound on the time up to which we can evolve the system before boundary effects influence physical observables. 
\begin{figure}
\begin{center}
\subfigure{
\includegraphics[width=0.8\textwidth]{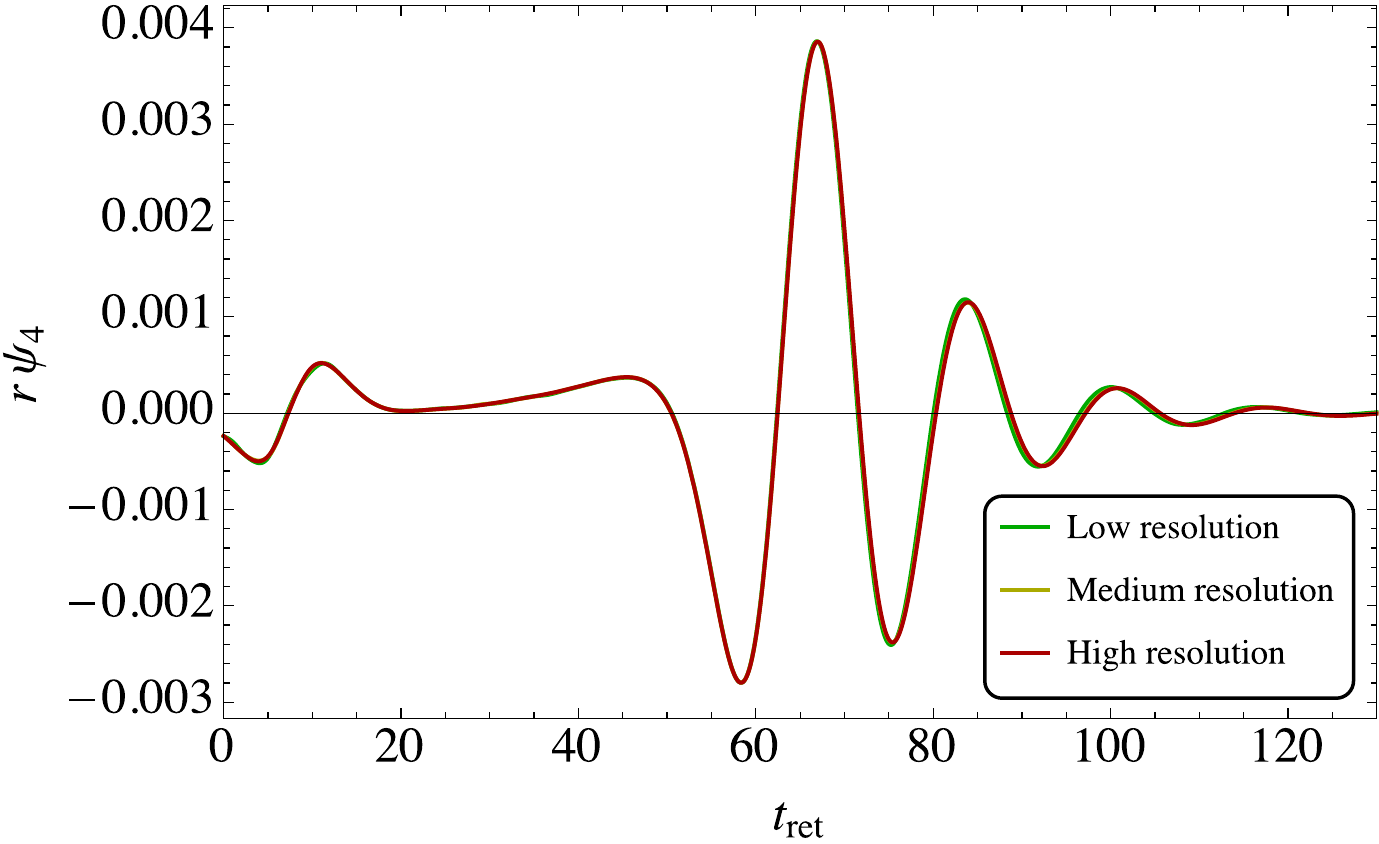}
}
\subfigure{
\includegraphics[width=0.75\textwidth]{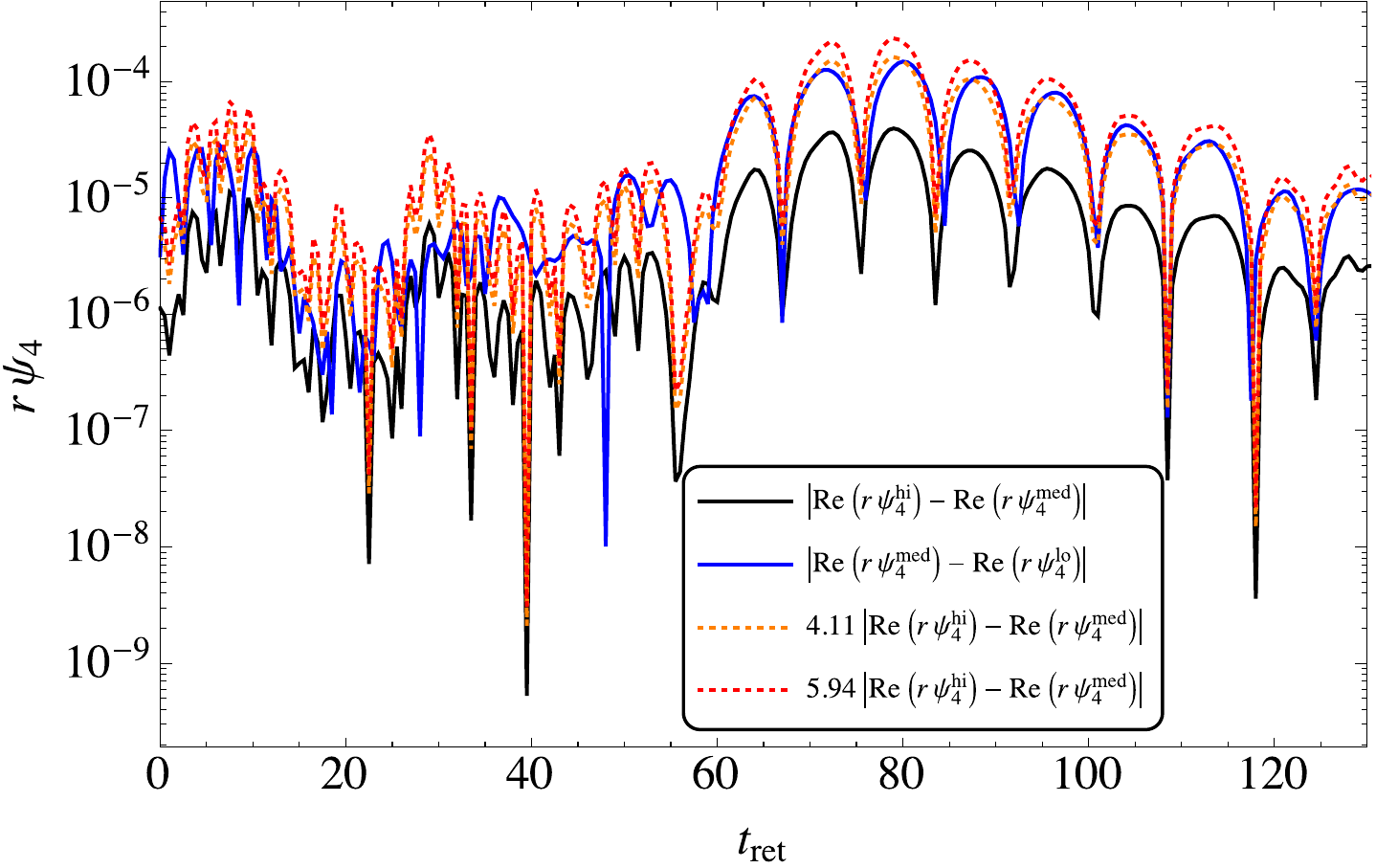}
}
\caption[Convergence test of head on collision]{Convergence test of head on collision: \textit{Top}: The real part of the $\ell=2$, $m=0$ mode of $r\Psi_4$ on the sphere of radius $R=60M$. \textit{Bottom}: Differences in the real part of the $\ell=2$, $m=0$ mode of $r\Psi_4$ between three different resolutions. We also show the data rescaled by a factor consistent with either third ($\times 4.11$) or fourth ($\times 5.64$) order convergence.
\label{fig-psi4}}
\end{center}
\end{figure}
In Figure \ref{fig-psi4} (top) we display the real part of the $\ell=2$, $m=0$ mode of $r\Psi_4$ extracted on a sphere of radius  $R=60M$ using 4th order interpolation. We use $320$ grid points\footnote{Something of order 64 grid points should in practise be sufficient.} in both the polar and azimuthal directions on the extraction sphere.  Following \citep{Loffler:2011ay}, we test convergence by comparing a physical quantity $\Psi$ at different resolutions. The convergence is of order $Q$ if for a set of grid spacings $h_1$, $h_2$, $h_3$, the differences between the numerically computed physical quantity $\Psi$ at successive resolutions satisfy
\begin{equation}
\frac{\Psi_{h_1} - \Psi_{h_2}}{\Psi_{h_2} - \Psi_{h_3}} = \frac{h_1^Q - h_2^Q}{h_2^Q - h_3^Q}\, ,
\end{equation}
ignoring higher order terms. Such terms may introduce errors in the convergence relation if the ratio between successive grids is not large, which in practise it is not (values of 1.2 are common, to avoid the highest resolution having a prohibitive computational expense). With the resolutions used in these runs, assuming $4^\textrm{th}$ order convergence the above factor is $\approx 5.953$, whilst assuming $3^\textrm{rd}$ order convergence the factor is $\approx 4.115$.

The gravitational wave content of the superimposed initial data is reflected in the non-zero initial signal. The collision of the two black holes takes place at $t\sim50$, so the signal before this collision time should be regarded as mostly unphysical (although some physical interaction - bremsstrahlung - can occur prior to the merger). As can be seen in the plot, the results for the two higher resolutions are indistinguishable on the scale employed here, whilst the lowest resolution shows a very slight drift towards later times, but is still in very good agreement. The bottom plot in Figure \ref{fig-psi4} shows the absolute value of the difference between $r\Psi_4$ computed at low and medium resolution (solid blue), medium and high resolution (solid black), and this latter curve scaled up by the convergence factor assuming $3^\textrm{rd}$ (dotted orange) and $4^\textrm{th}$ (dotted red) order convergence. This plot shows that in the highly dynamical stages of the evolution, when there is a lot of regridding and the boxes move around the domain, the convergence is closer to $3^\textrm{rd}$ order. On the other hand, when the system has nearly settled, and hence the boxes do not move much, the convergence order is somewhat closer to $4$. We can explain this loss of convergence due to regridding because in the interpolation used in \texttt{GRChombo} only the values of the functions are matched across levels, not their derivatives. 

\subsection{MPI scaling properties}
\label{sec-Tests-scaling}

We now turn to the performance aspects of $\grchombo$. Here we perform a number of scaling tests to show that our code can exploit the parallelism offered in modern supercomputers to a reasonable extent. Whilst {\tt Chombo} does have the capability to partially utilise threads through hybrid OpenMP routines, we will limit our attention to pure MPI mode in these tests, as we have found that this gives significantly smaller run-to-run performance variations.

\begin{figure}[htp]
\begin{center}
\includegraphics[height=8cm]{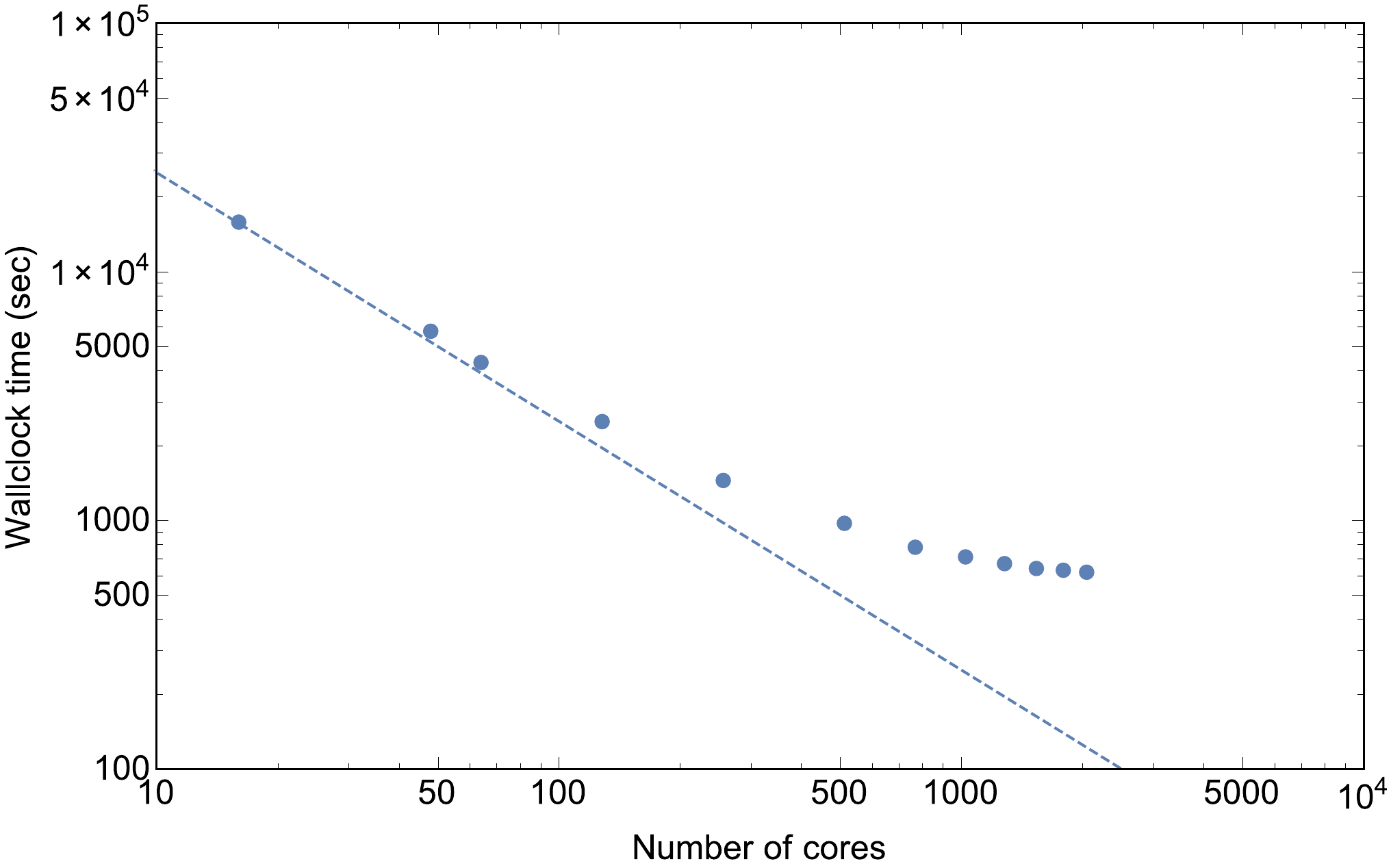}
\caption[Strong Scaling test]{Strong scaling behaviour of GRChombo on the SuperMike-II cluster at the Louisiana State University. The code achieves excellent strong scaling up to 200 cores, and a useful scaling up to around 1000 cores.
\label{fig-StrongScaling}}
\end{center}
\end{figure}

\begin{figure}[htp]
\begin{center}
\includegraphics[height=8cm]{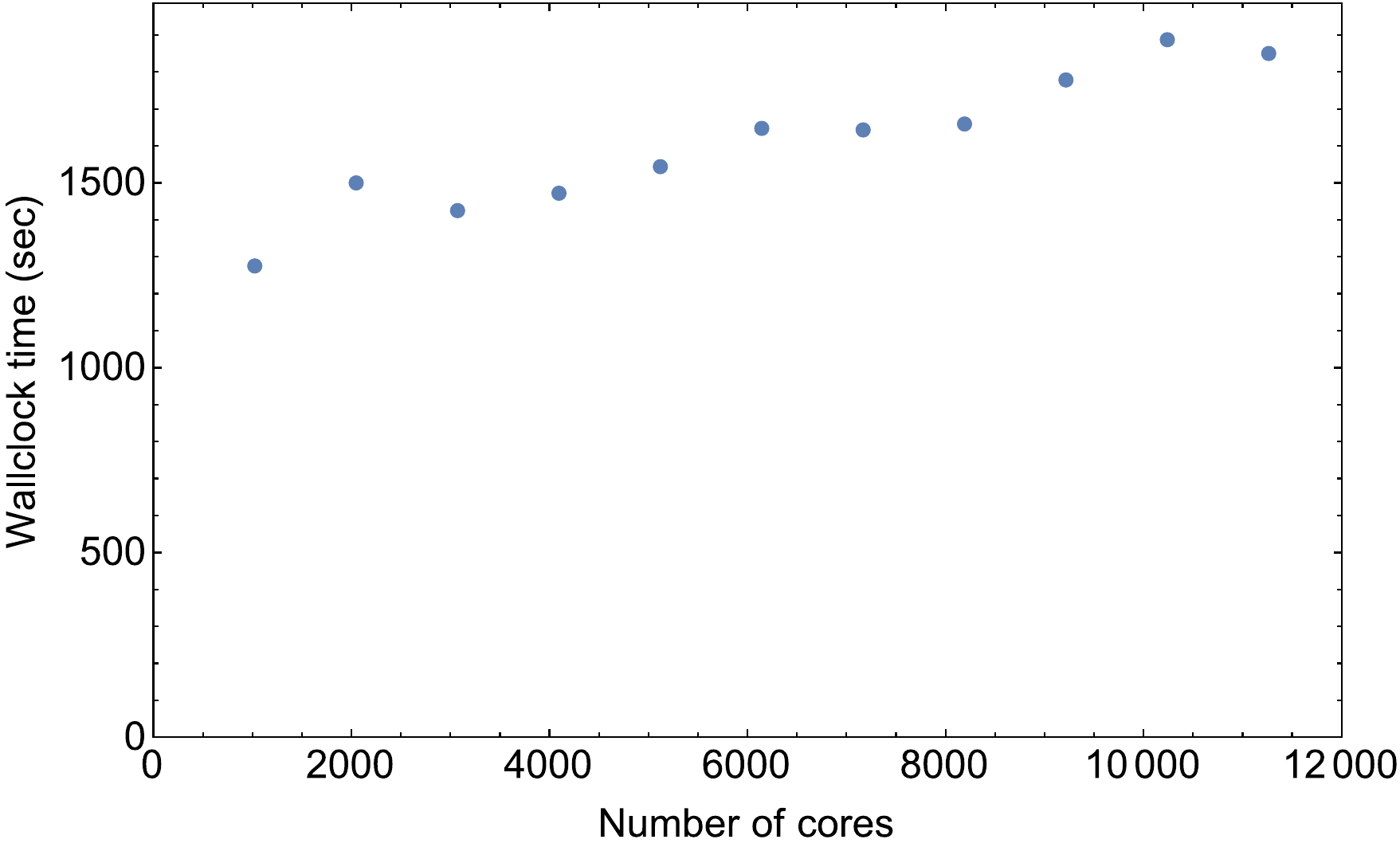}
\caption[Weak scaling test]{Weak scaling behaviour of GRChombo on the Mira Blue Gene/Q cluster at the Argonne National Laboratory over a 10x increase in core count.
\label{fig-WeakScaling}}
\end{center}
\end{figure}

Our strong scaling test is performed using a head on binary black hole system. We set up Brill-Lindquist initial data for two static black holes of mass $0.5M$, with a separation of $6M$. Our overall computational domain is a box of size $160M$, and at the coarsest level, we fix the total number of grid points to $320$ in each direction, giving a grid spacing of $0.5M$. The centre of mass of the system is at the centre of the domain. For the mesh refinement, we fix the total number of levels to six. The simulation is allowed to run up to the time of $2M$. The bulk of this test was performed on the SuperMike-II cluster at the Louisiana State University. Each compute node consists of two 2.6GHz 8-core Sandy Bridge Xeon processors, connected via a InfiniBand QDR fabric. We fix the computational load across all jobs and vary the core count from 16 to 2048. Our data in Figure \ref{fig-StrongScaling} shows excellent strong scaling up to 200 cores on this cluster. We continue to see a reasonable speedup up to around 1000 cores for this particular problem.
Of course, in a production environment, it is often desirable to use additional cores to be able to run a larger simulation, rather than to speed up a problem of fixed size. In this scenario, weak scaling behaviour is of interest. We begin at 1024 cores with an identical setup to that in the strong scaling test. We then scale up the number of grid points at the coarsest level proportional to the increase in core count up to 10240, whilst adjusting the tagging threshold in order to maintain the shape and size of the refined regions. We also adjusted the time step size (i.e. the Courant factor) so that each simulation would reach the target stop time in the same number of steps. We use the Mira Blue Gene/Q cluster at the Argonne National Laboratory for this due to the larger number of cores available. Figure \ref{fig-WeakScaling} shows a less-than-perfect scaling behaviour in this setup, with the main bottleneck appearing in the regridding and box generation stages. We are working together with the developers of {\tt Chombo} to improve this aspect of the code performance. It is worth noting, however, that even in its current state the code still shows a useful level of scalability: the wallclock time increases by less than 2x over the 10x increase in core count.
\subsection{Performance comparison}
\label{sec-Tests-performance}
Lastly, we demonstrate that $\grchombo$'s performance on standard $3+1$ black hole problems is comparable to that of an existing numerical relativity code.

Our comparison target is the $\Lean$ code~\citep{Sperhake:2006cy,Zilhao:2010sr}, a $3+1$ numerical relativity code designed to evolve four and higher dimensional vacuum spacetimes. $\Lean$ is based on the {\texttt{Cactus}} computational toolkit~\citep{Cactuscode:web} and realises moving-box mesh refinement via the \texttt{Carpet} package~\citep{Schnetter:2003rb,CarpetCode:web}, both of which are part of the open-source {\texttt{Einstein Toolkit}}~\citep{Loffler:2011ay,EinsteinToolkit:web}. Initial data is constructed either analytically or numerically by employing the {\texttt{TwoPunctures}} spectral solver~\citep{Ansorg:2004ds}. \\
In order to track apparent horizons, $\Lean$ makes use of~\texttt{AHFinderDirect}~\citep{Thornburg:2003sf,Thornburg:1995cp}.

The $\grchombo$ setup is identical to that in the strong scaling test as detailed in Sec. \ref{sec-Tests-scaling}. The $\Lean$ code is subject to the limitation of Carpet, where successive levels may only occur in a collection of nested-box hierarchies, whose sizes are typically related by a power of two. In this case, we first fix boxes of side lengths $160$, $80$, $40$ and $20M$ at the centre of the domain, encompassing both black holes, then fix further boxes of side lengths 5 and $2.5M$ centred at each of the black holes. During the evolution, $\Lean$ has the capability to track the black holes and move or merge the finer boxes as appropriate, however the shape and size of the boxes remain unchanged. The $\grchombo$ code is not subject to this box structure limitation, and therefore we simply tune the regridding threshold so that the size of the finest level matches that of the $\Lean$ setup. We make no attempt to match the sizes of the intermediate levels as this would defeat the spirit of fully-flexible AMR.
\begin{figure}[htp]
\begin{center}
\includegraphics[height=8cm]{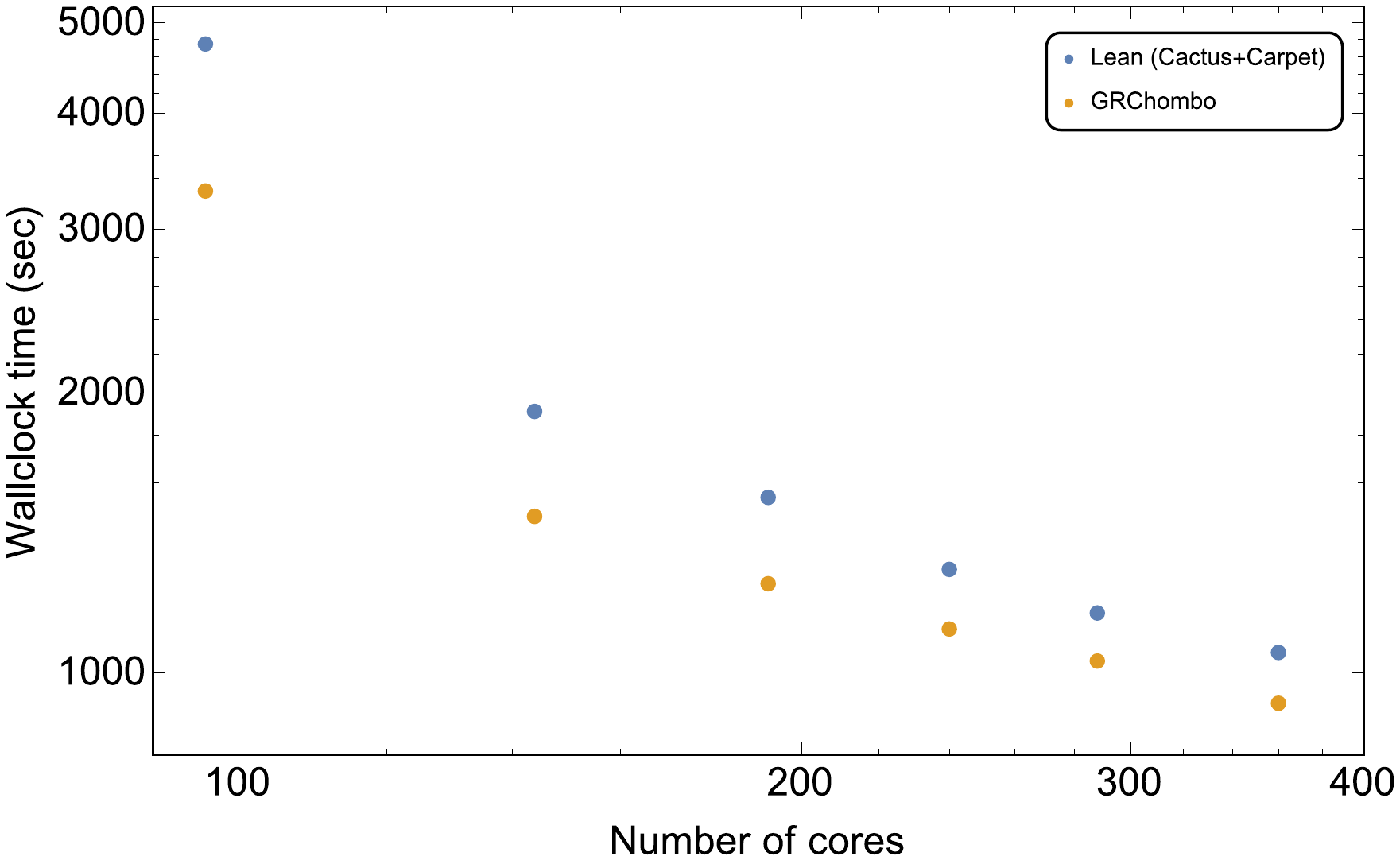}
\caption[Comparison of GRChombo and Lean]{Runtime and scaling comparison between $\grchombo$ (orange) and $\Lean$ (blue). The leftmost data points show disproportionately large wallclock times as the machine becomes memory-limited at this core count.
\label{fig-LeanComparison}}
\end{center}
\end{figure}
Our comparison tests were performed on the COSMOS VIII shared memory facility. Both codes were executed on the same SGI UV1000 machine, utilising up to 60 Nehalem EX 2.67GHz CPUs with 6 cores per CPU, giving up to 360 cores in total. In all of these runs, we pin one MPI rank to each core and disabled all checkpointing activity since we wish to exclude I/O bottlenecks. We allowed the simulation to run up to coordinate time $t = 2$, and measured the wall-clock time taken to execute the time evolution portion of the code (i.e. we excluded the time spent during initial setup).

Within the range of 150-360 cores, both $\grchombo$ and $\Lean$ exhibit similar performance and strong scaling characteristics (figure \ref{fig-LeanComparison}). Below 150 cores, we cannot meaningfully test the strong scaling behaviour as the machine becomes memory-limited. We have not performed this comparison on a larger cluster due to the lack of resource availability, but we have no reason to expect any significant difference provided that the problem size is also scaled up appropriately. Having said this, we believe that a framework like Cactus probably remains the better choice when it comes to these standard problems, owing to the wealth of existing tools and resources and a more mature community of users. Instead, we intend for $\grchombo$ to be complementary to existing numerical relativity codes in order to open up new avenues of research by enabling a wider range of problems to be tackled at a feasible level of resources.

%% file: Chapter4/chapter4.tex
\chapter{Inhomogeneous Inflation}
\label{ch-Inflation}

\ifpdf
    \graphicspath{{Chapter4/Figs/Raster/}{Chapter4/Figs/PDF/}{Chapter4/Figs/}}
\else
    \graphicspath{{Chapter4/Figs/Vector/}{Chapter4/Figs/}}
\fi

\section{Introduction}

Cosmic Inflation \citep{Guth:1980zm,Linde:1981mu,Albrecht:1982wi,Starobinsky:1980te} is thought to provide a solution to several problems in standard Big Bang theory by dynamically driving a ``generic'' initial state to a flat, homogeneous and isotropic Universe, while generating a nearly scale-invariant power spectrum of primordial perturbations which is consistent with observations. 
The question of what constitutes a ``generic'' initial state is a difficult one, and can only be understood in the context of a quantum theory of gravity. However, regardless of the nature of quantum gravity, a random realisation from the set of all possible initial conditions will not look like an inflationary spacetime, at least initially~\citep{Hollands:2002yb}, and one should expect the initial conditions from which inflation begins to contain some measure of inhomogeneity.

The issues concerning initial conditions and the stability of de Sitter and inflationary spacetimes have been under investigation for as long as inflation itself, and there are many analytic and semi-analytic~\citep{Gibbons:1977mu,Hawking:1981fz,Wald:1983ky,Starobinsky:1982mr,Barrow:1984zz,Albrecht:1984qt,Barrow:1985,Gibbons:1986xk,Jensen:1986nf,Hawking:1987bi,Penrose:1988mg,Muller:1989rp,Kitada:1991ih,Kitada:1992uh,Bruni:1994cv,Maleknejad:2012as,Gibbons:2006pa,Boucher:2011zj,Bruni:2001pc,Muller:1987hp,Barrow:1989wp,Bicak:1997ne,Capozziello:1998dq,Vachaspati:1998dy,Barrow:1987ia,Barrow:1986yf,Polyakov:2009nq,Marolf:2010nz,Tsamis:1992sx,Brandenberger:2002sk,Geshnizjani:2003cn,Marozzi:2012tp,Brandenberger:1990wu,Carroll:2010aj,Corichi:2010zp,Schiffrin:2012zf,Remmen:2013eja,Corichi:2013kua,Mukhanov:2014uwa,Remmen:2014mia,Berezhiani:2015ola,Kleban:2016sqm,Alho:2011zz} as well as numerical studies~\citep{Albrecht:1985yf,Albrecht:1986pi,KurkiSuonio:1987pq,Feldman:1989hh,Brandenberger:1988ir,Goldwirth:1989pr,Goldwirth:1989vz,Brandenberger:1990xu,Laguna:1991zs,Goldwirth:1991rj,KurkiSuonio:1993fg,Easther:2014zga,East:2015ggf,Braden:2016tjn} (see~\citep{Brandenberger:2016uzh} for a short review). Goldwirth and Piran~\citep{Goldwirth:1989pr,Goldwirth:1989vz,Goldwirth:1991rj} were the first to study the robustness of inflation to spherically symmetric perturbations using general relativistic 1+1D simulations.\footnote{An earlier pioneering work \citep{Albrecht:1985yf} showed that inhomogeneous scalar fields will homogenise in a fixed FRW background. See also \citep{Easther:2014zga} for a recent follow up work in this direction.} In modern terminology, their conclusion was that large field models, in which the inflaton traverses more than a Planck mass during the inflationary period, $\delta \phi \gtrsim \mpl$, are more robust than small field models, $\delta \phi \ll \mpl$. Their results are often taken to imply that inflation requires a homogenous patch of size roughly~$1/H$ to begin. This work was later followed by 3+1D numerical simulations in Refs.~\citep{KurkiSuonio:1993fg,Laguna:1991zs} showing large field inflation to be robust to simple inhomogeneous (and anisotropic) initial conditions with large initial gradient energies in situations in which the field is initially confined to the part of the potential that supports inflation. This was confirmed recently in Ref.~\citep{East:2015ggf}, which demonstrated that large field inflation is robust even if the average energy density due to spatial gradients in the field $\rho_{\mathrm{grad}} \approx 1000\,\rho_{V}$, where $\rho_{V}$ is the vacuum energy density, at least if the universe initially expands at the same rate everywhere.

In this work, we continue this line of research and test the robustness of inflation to a slightly more general but still very simple class of inhomogeneous initial conditions both in the scalar field profile and the extrinsic curvature. We use $\grchombo$, setting up the machinery that will allow us to study more general classes of initial conditions in the future.  Since the degree of robustness to inhomogeneities depends on the exact model of inflation, this provides us with an approach to checking model viability. According to the Lyth bound~\citep{Lyth:1996im,Turner:1996ck}, inflation occurs at high energies and involves large field excursions in models that produce observable amounts of primordial gravitational waves, whereas the energy scale and field excursion are small in models that do not. 
Our results are summarised as follows:
\begin{itemize}
\item{For the initial conditions we consider, we find that large field inflation is robust to large gradient energies of $\rho_{\mathrm{grad}}/\rho_V \gg 1$, in agreement with~\citep{Laguna:1991zs,KurkiSuonio:1993fg,East:2015ggf}.}
\item{{Small field inflation is less robust than large field inflation.} It can fail even when the energy density in gradients is subdominant $\rho_{\mathrm{grad}}/\rho_V \ll 1$. We show that small field inflation fails when a large enough local fluctuation ends inflation early in that particular region, with the gradients quickly dragging the rest of the spacetime from the inflating part of the potential. However, the size of local fluctuation required to end inflation must be large enough to explore the boundary of the inflationary regime of the potential, making small field inflation somewhat more robust than might be expected.}
\item{{Large inhomogeneities do not form dominant black hole spacetimes.} In the large field case, the potential is sufficiently wide to support inhomogeneities which result in collapse to form black holes. However, in the case where the initial spacetime is flat on average, increasing gradient energy implies an increase in average initial expansion. This expansion prevents the formation of inflation-ending black hole spacetimes. We found that there exists a maximum black hole mass which is subdominant to the inflationary spacetime, which we derived both analytically and numerically.  }
\item{We show that for initial spacetimes containing both expanding and collapsing regions local regions may collapse into black holes. However, inflation will occur as long as the spacetime is on average initially expanding. This is consistent with the theoretical expectations of \citep{Barrow:1985} and \citep{Kleban:2016sqm}.}
\end{itemize}

This chapter is organised as follows. In Section \ref{sect:method} we present the theory and methodology of our approach. In Sections \ref{sect:SF} and \ref{sect:LF}, we present the numerical results, and discuss their implications for the small field and large field cases respectively. We conclude in Section \ref{sect:conclusions}. The work presented in this chapter is derived from the paper ``Robustness of Inflation to Inhomogeneous Initial Conditions'' \citep{Clough:2016ymm}. Note that in this chapter, as in section \ref{sec-Cosmology}, we do not set $G=1$ but follow the convention in \citep{WeinbergBook} and replace it with (non-reduced) Planck units $\mpl = \sqrt{\hbar c/G} = 2.17 \times 10^{-8} ~ \rm{kg}$, with $\hbar=c=1$.

\section{Theory and methodology} \label{sect:method}

We consider single-field inflation with a canonical kinetic term in the Lagrangian
\begin{equation}
L_{\phi} = -\frac{1}{2}g^{\mu\nu}\partial_\mu \phi\partial_\nu\phi - V(\phi) \label{eqn:singlefield},
\end{equation}
For a spatially homogeneous configuration, inflation occurs if $V>0$ and the potential slow-roll parameters satisfy 
\begin{equation}
\epsilon_V  = \frac{\mpl^2}{16\pi}\left(\frac{V'}{V}\right)^2 \ll 1~, \quad \eta_V = \frac{\mpl^2}{8\pi}\left|\frac{V''}{V}\right|^2 \ll 1\, ,
\end{equation}
with  $\mpl^2 = \hbar c/G$.
In this case the field is slowly rolling and $V\approx \mathrm{constant}$ acts as a cosmological constant resulting in an inflating spacetime. The second condition $\eta_V \ll 1$ is required to ensure that inflation occurs for the sufficient amount of $e$-foldings. 

\nomenclature[a-pi]{$L$}{length of the numerical grid, in cosmology equal to $1/H_V$}
\nomenclature[g-pi]{$\delta \phi$}{in cosmology, the distance the inflaton rolls during inflation}

\begin{figure}
\begin{center}
\includegraphics[width=0.6\textwidth]{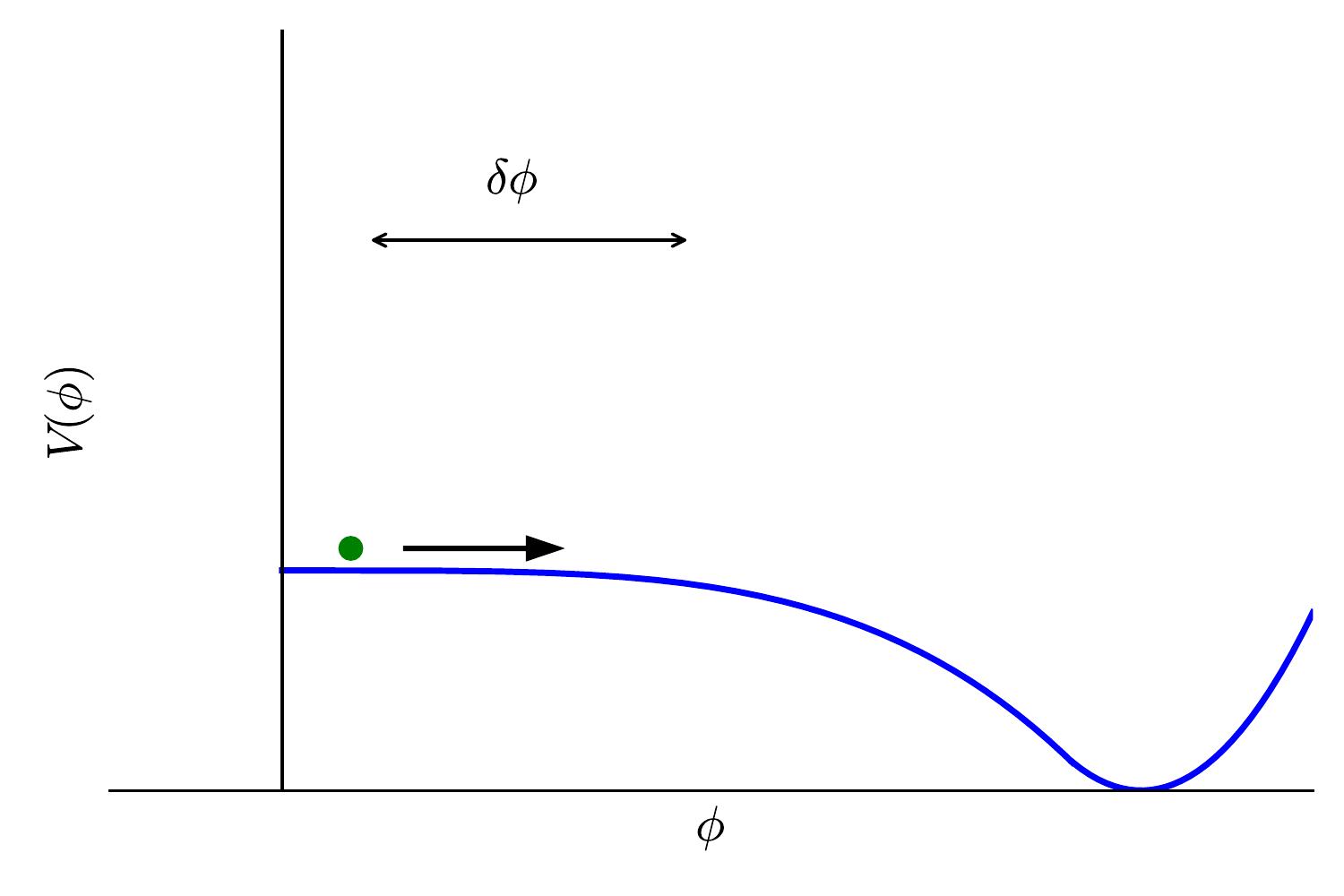}
\caption[Small field inflation, potential]{In small field inflation the width of the inflationary ``slow-roll'' part of the potential $\delta \phi \ll \mpl$, whereas in large field $\delta \phi \gg \mpl$.
\label{fig:SFvsLF}}
\end{center}
\end{figure}

In the large-field models, the region of the potential where this occurs is super-Planckian, i.e. the field needs to roll $\delta \phi \gtrsim \mpl$ for sufficient inflation, while in the small field model the field traverses a sub-Planckian distance in field space $\delta \phi  \ll \mpl$. This is illustrated in Figure \ref{fig:SFvsLF}. In the context of single field inflation, the Lyth bound~\citep{Lyth:1996im} implies that high/low-scale inflation is associated with large/small field inflation. 

\subsection{Initial conditions}

\begin{figure}
\begin{center}
\includegraphics[width=0.8\textwidth]{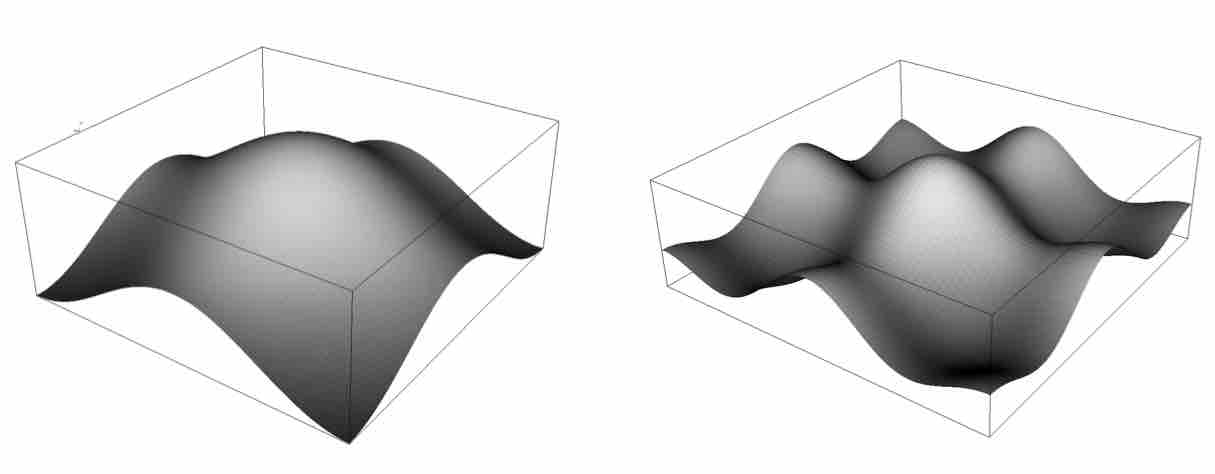}
\caption[Scalar field profiles, one and two modes]{Illustration of the cases $N=1$ and $N=2$ showing the values of $\phi$ on a 2D slice through the y axis. On this slice the maximum value of $\phi$ in each case is $4 \times 10^{-4} \mpl$ and the minima are $-9.5 \times 10^{-5} \mpl$ and $-6.0 \times 10^{-5} \mpl$ respectively.
\label{fig:Nis12}}
\end{center}
\end{figure}

We impose very simple inhomogeneous initial conditions similar to those in~\citep{East:2015ggf} by specifying the initial condition for the scalar field as follows
\begin{equation}
\phi(t = 0,x^i) = \phi_0 + \frac{\Delta \phi}{N} \sum_{n=1}^{N} \left( \cos{\frac{2 \pi nx}{L}} + \cos{\frac{2 \pi ny}{L}} + \cos{\frac{2\pi nz}{L}} \right) \label{eqn:phi}\,,
\end{equation}
and
\begin{equation} \label{eqn:phidot}
\frac{\partial \phi(t = 0,x^i)}{\partial t} = 0 \,,
\end{equation}
where $x^i$ is the spatial coordinate of a foliation labeled by the time coordinate $t$, and $\Delta \phi$ is the amplitude of the initial inhomogeneities. The value $\phi_0$ is chosen such that we have 100 $e$-folds of inflation in the absence of any inhomogeneities. Since there are three modes each with amplitude $\Delta \phi$, the maximal total amplitude of the fluctuations about $\phi_0$ is $3\Delta \phi$. We chose not to include random phases in this work as we have found that random phases do not materially change the overall results. Note that we have normalised the total $\Delta \phi$ by the number of modes $N$ -- this means that the average gradient energy is slightly higher for larger $N$, but that the maximum traverse from $\phi_0$ towards the inflationary minimum is the same. See Figure \ref{fig:Nis12} for an illustration of the cases $N=1$ and $N=2$. 

\nomenclature[g-pi]{$\Delta \phi$}{in inflation, the amplitude of the spatial fluctuations in the scalar field}
\nomenclature[a-pi]{$N$}{in inflation, the total number of modes of spatial fluctuations in the scalar field}
\nomenclature[g-pi]{$\phi_0$}{in inflation, the average initial position of the scalar field in field space}

We set $L$ to be the length of the simulation domain, and use periodic boundary conditions to simulate a space composed of periodic fluctuations of this length and amplitude. $L$ is chosen to be the Hubble length in the absence of inhomogeneities ($\Delta \phi=0$), that is
\begin{equation}
L = \frac{3 \mpl}{\sqrt{24 \pi V(\phi_0)}} \label{eqn:Lformula}\, .
\end{equation}
Hence, our model of initial inhomogeneities depends on the integer $N$, the amplitude of inhomogeneities $\Delta \phi$, and the potential $V(\phi_0)$. The potential $V(\phi_0)$ sets the inflationary Hubble scale, the integer $N$ sets the wavelength of the shortest perturbations relative to this scale, and $\Delta\phi$ sets the amplitude of the inhomogeneities. In this work we focus on $N$ of order unity and leave a more systematic study of the space of initial conditions for future work. In the limit in which the gradient energy dominates, i.e. $\rho_{\mathrm{grad}} \gg V(\phi_0)$, changing $\Delta \phi$ is equivalent to changing the wavelength of the mode relative to the actual Hubble length $H^{-1}$ (including the energy density from fluctuations). For $N=1$ and a Euclidean metric on the initial slice the wave number $k=2\pi/L$ satisfies 
\begin{equation}\label{eqn:equiv}
\frac{k}{H} = \frac{\mpl}{\sqrt{2\pi}\Delta \phi}\,.
\end{equation}
In large field inflation we use a single mode, i.e. $N=1$, and vary $\Delta \phi$. In small field inflation, we consider the two cases $N=1$, and $N=2$, a superposition of two modes, in addition to the variation of $\Delta \phi$. 

\nomenclature[a-pi]{$k$}{in inflation, the wave number of the fluctuations $2\pi/L$}

We use the BSSN formalism, with $\alpha=1$ and $\beta^i=0$ on the initial hypersurface, and hence the initial gradient energy on this hypersurface is
\begin{equation}
\rho_{\mathrm{grad}} \equiv \frac{1}{2}\gamma^{ij}\partial_i \phi \partial_j \phi \label{eqn:grad} \, .
\end{equation}
We evolve the lapse and shift in the moving puncture gauge \citep{vanMeter:2006vi,Campanelli:2005dd}, which allows us to stably form and evolve black holes in the spacetime, as 
\begin{equation}
\partial_t \alpha = -\mu_{\alpha 1}\alpha K + \beta^i\partial_i \alpha \, , \label{eqn:alphadriver3}
\end{equation}
\begin{equation}
\partial_t \beta^i = B^i \, , \label{eqn:betadriver3}
\end{equation}
\begin{equation}
\partial_t B^i = \frac{3}{4} \partial_t \tilde\Gamma^i-\eta_{\beta 2} B^i \, . \label{eqn:gammadriver3}
\end{equation}
The exact values of $\mu_{\alpha 1}$ and $\eta_{\beta 2}$ are chosen to improve stability in any particular numerical simulation.

Next, we have to specify the initial conditions for the metric $\gamma_{ij}$ and extrinsic curvature $K_{ij}$, so as to satisfy both the Hamiltonian and momentum constraints on the initial hypersurface. Introducing the notation 
\begin{equation}
\xi \equiv \frac{1}{\alpha}\left(\partial_t{\phi} - \beta^k\partial_k \phi\right),
\end{equation}
so that the energy density at any point in the hypersurface is 
\begin{equation}
\rho = \frac{1}{2}\xi^2 + \frac{1}{2}\gamma^{ij}\partial_i \phi \partial_j \phi + V\,,
\end{equation}
the constraint equations become
\begin{equation}
\chi \tilde{D}^2\chi -\frac{3}{2}\tilde{\gamma}^{ij}\tilde{D}_i\chi \tilde{D}_j\chi + \frac{\chi^{2}\tilde{R}}{4} + \frac{K^2}{6} - \frac{1}{4\chi} \tilde{A}_{ij}\tilde{A}^{ij} = 4\pi G \rho \label{eqn:HamCon}\,,
\end{equation}
and
\begin{equation}
\tilde{D}_j \tilde{A}^{ij} -\frac{3}{\chi} \tilde{A}^{ij}\tilde{D}_j \chi -\frac{2}{3}\tilde{\gamma}^{ij}\tilde{D}_j K = 8\pi G \xi \tilde{\gamma}^{ij} \partial_j \phi\,. \label{eqn:MomCon}
\end{equation}
$K$ is the local expansion rate of spacetime, and, as noted previously, in the special case of the Friedmann-Robertson-Walker metric, $K = -3H$ where $H$ is the Hubble constant.

\nomenclature[g-pi]{$\xi$}{in inflation, the kinetic part of the energy density for the field}

This is a set of coupled elliptic equations and is non-trivial to solve in general. Throughout this work, we will make the simplifying assumption that the metric is conformally flat and the traceless part of the extrinsic curvature $K_{ij}$ is zero everywhere on the initial slice
\begin{equation}
\tilde{\gamma}_{ij} = \delta_{ij}\,,  \label{eqn:conformalcond}
\end{equation}
and
\begin{equation}
\tilde{A}_{ij} = 0\,. \label{eqn:tracelessK}
\end{equation}
In this special class of initial conditions, we consider two possible solutions, that of uniform initial expansion $K$, and one with spatially varying $K$.

\subsubsection{Uniform initial expansion $K=\mathrm{constant}$}

\noindent For spatially varying $\phi$,  the momentum constraint \eqn{eqn:MomCon} is trivially satisfied for $\xi=0$ and $K=$const. $K$ is in principle a free parameter, corresponding to a uniform local expansion rate across the initial hypersurface. However, in order to satisfy periodic boundary conditions for $\chi$ and the Hamiltonian constraint, $K^2/24 \pi$ needs to lie close to the average initial energy density for the hypersurface. For simplicity, we choose it to be equal to the average initial energy density, approximating the metric to be Euclidean 
\begin{equation}
K = -\sqrt{24 \pi G \langle \rho \rangle}\,, \label{eqn:Kavg}
\end{equation}
with
\begin{equation}
\rho = \frac{1}{2}(\partial_i \phi)^2 + V(\phi)\,,
\end{equation}
where $\langle X \rangle = {\cal V}^{-1} \int X~d{\cal V}$ indicates the average over the spatial volume ${\cal V}$ of the quantity $X$.  Once $K$ is chosen, the initial field profile and the Hamiltonian constraint then fully determine the conformal factor $\chi$ (which we solve for using numerical relaxation).

In cases where the gradient energy dominates i.e. $\rho \approx \rho_{\mathrm{grad}} \gg V(\phi)$, the initial expansion rate is large compared to the Hubble rate associated with inflation. This large initial uniform expansion means that we are stacking the deck against ending inflation. In general, we should expect the local expansion rate to be a function of spatial position that can be both initially expanding or collapsing. To study the general case will require relaxing some combination of the conformal condition \eqn{eqn:conformalcond}, the condition on $\tilde{A}_{ij}$, \eqn{eqn:tracelessK}, and the condition of zero initial scalar field velocity, $\xi = 0$. We reserve the general case for future work, but there exists a second solution consistent with equations~\eqref{eqn:conformalcond} and~\eqref{eqn:tracelessK}, given our assumptions. By imposing $\xi \neq 0$, we can obtain a non-uniform initial expansion. We turn to this solution next.

\subsubsection{Expanding/contracting initial condition $K\neq \mathrm{constant}$}

\noindent For constant initial scalar velocity $\xi$
\begin{equation}
\xi = -\frac{C}{12\pi G}, \label{eqn:xiansatz}
\end{equation}
with $C$ some constant, the momentum constraint~\eqn{eqn:MomCon} relates the extrinsic curvature $K$ to the initial scalar field profile $\phi$
\begin{equation}
K = -C\phi + K_0, \label{eqn:Kansatz}
\end{equation}
where $K_0$ is an integration constant.  This initial condition means that a \emph{constant} initial scalar velocity $\xi$ and a varying field $\phi$ will lead to a spatially varying $K$. If $K_0$ is chosen to be approximately the average value of $C \phi$, the spacetime will be locally initially expanding or contracting depending on its position. (Note that as above, the choice of $K_0$ is not completely free because of the need to satisfy periodic boundary conditions for $\chi$.) We can then again solve the Hamiltonian constraint for the conformal factor $\chi$ in order to complete the specification of the initial conditions. 

\subsection{Numerical set-up}

We rescale our simulations (by choosing the geometrised mass unit $M$ to represent some convenient multiple of $\mpl$) such that the size of our physical domain is covered by $(32M)^3$. We turn on $\grchombo$'s adaptive mesh refinement, using the gradients of $K$ and $\phi$ as refinement threshold conditions, with a coarsest level grid size of $64^3$, allowing up to 6 levels of refinement with a refinement ratio of $2$ per level. We check convergence approximately in this case by checking that the same results are obtained when starting from a coarsest grid of $128^3$, increasing the number of grids by one and using a more aggressive regridding condition (approximately halving the thresholds). It was found that the difference in the results was small -- for example, the number of $e$-folds at failure in the small field cases were different by $\pm 0.1\%$.

We can track inflationary simulations for around 23 \mbox{$e$-folds}. After this point numerical error begins to dominate as the conformal factor $\chi$ (equal to the inverse of the scale factor) falls below working precision.

\section{Small field inflation} \label{sect:SF}

\begin{figure}
\begin{center}
\includegraphics[width=0.6\textwidth]{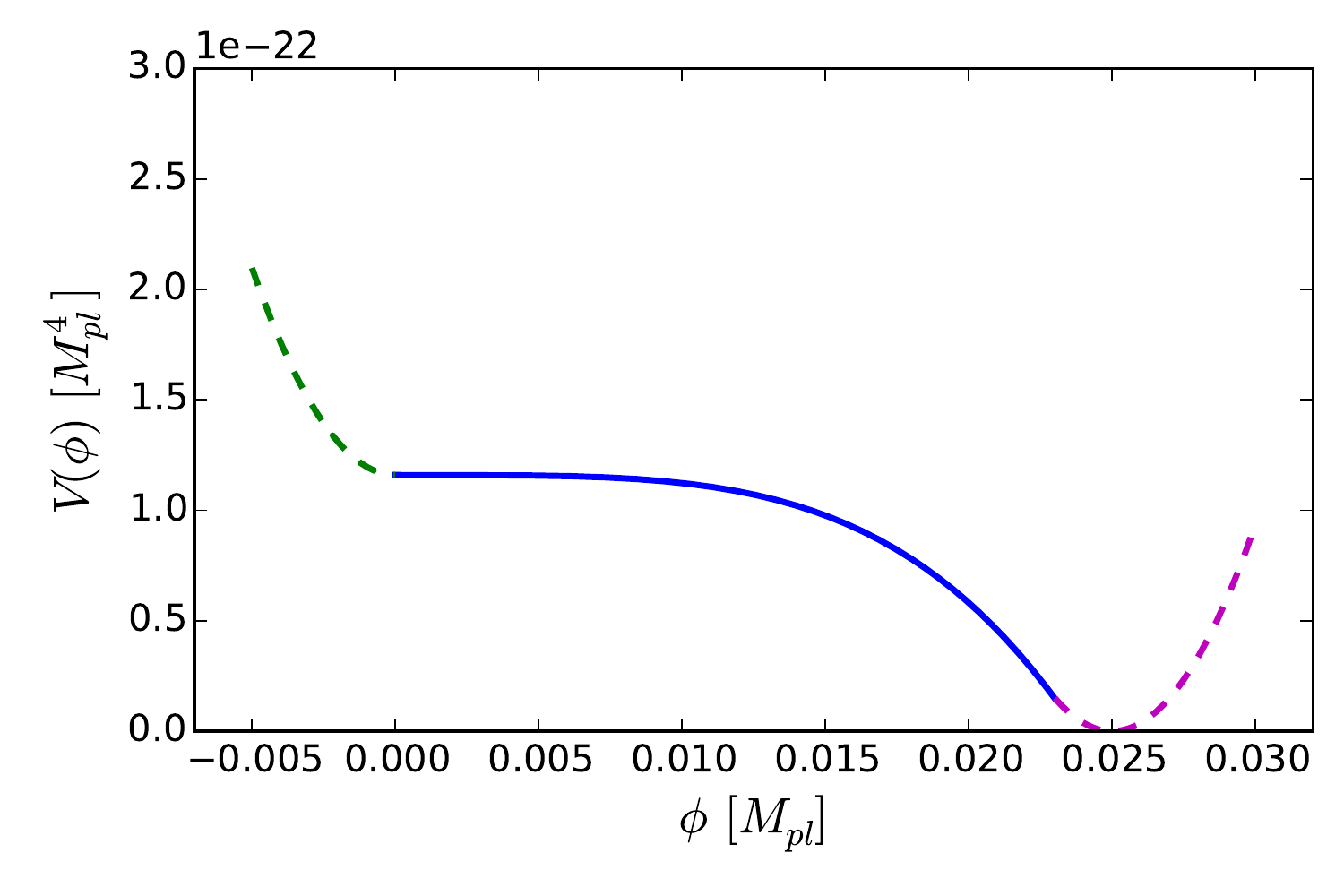}
\caption[Small field potential function - short plateau]{Small field potential function $V(\phi)$ without an extended flat region, showing the three regions, the central solid line (blue) region gives rise to the slow-roll inflationary period
\label{fig:PotSFsteep}}
\end{center}
\end{figure}

As discussed in the Introduction, the inflating plateau of the small field potential can be relatively narrow, with $\delta \phi \ll \mpl$. The reason is as follows. The scalar power spectrum for single field inflation is given by 
\begin{equation}
\Delta_R^2=\frac{H^2}{\pi \mpl^2 \epsilon} \approx 2\times 10^{-9} \label{eqn:Ps}\,.
\end{equation}
For low scale inflation $(H/\mpl)^2$ is small, which means that $\epsilon$ must be small to achieve sufficient amplification of the observed scalar power. This means that the inflaton has to roll very slowly (when compared to the large field case). The number of $e$-folds $\mathcal{N}$, is given by 
\begin{equation}
\mathcal{N} \approx  \int \sqrt{\frac{4\pi}{\epsilon}} \frac{|d\phi|}{\mpl}, \label{eqn:Nfolds}\,.
\end{equation}
Assuming $\epsilon$ and $H$ constant, we estimate the field range with the help of equation~\eqref{eqn:Ps} 
\begin{equation}
\delta \phi \approx \frac{\mathcal{N}}{2}\frac{H}{\mpl}10^{5} \mpl. \label{eqn:totaltraverse}
\end{equation}
For a typical low scale inflation $H/\mpl \sim 10^{-10}$, we see that $\delta \phi/\mpl\sim 3 \times 10^{-4}$ so $\delta \phi$ is sub-Planckian as we argued. A typical small field inflation model is shown in Fig. \ref{fig:PotSFsteep}, where the inflating domain is around the inflection point of the potential.

\nomenclature[g-pi]{$\Delta_R$}{in inflation, the scalar power spectrum}

\begin{figure}
\begin{center}
\includegraphics[width=0.6\textwidth]{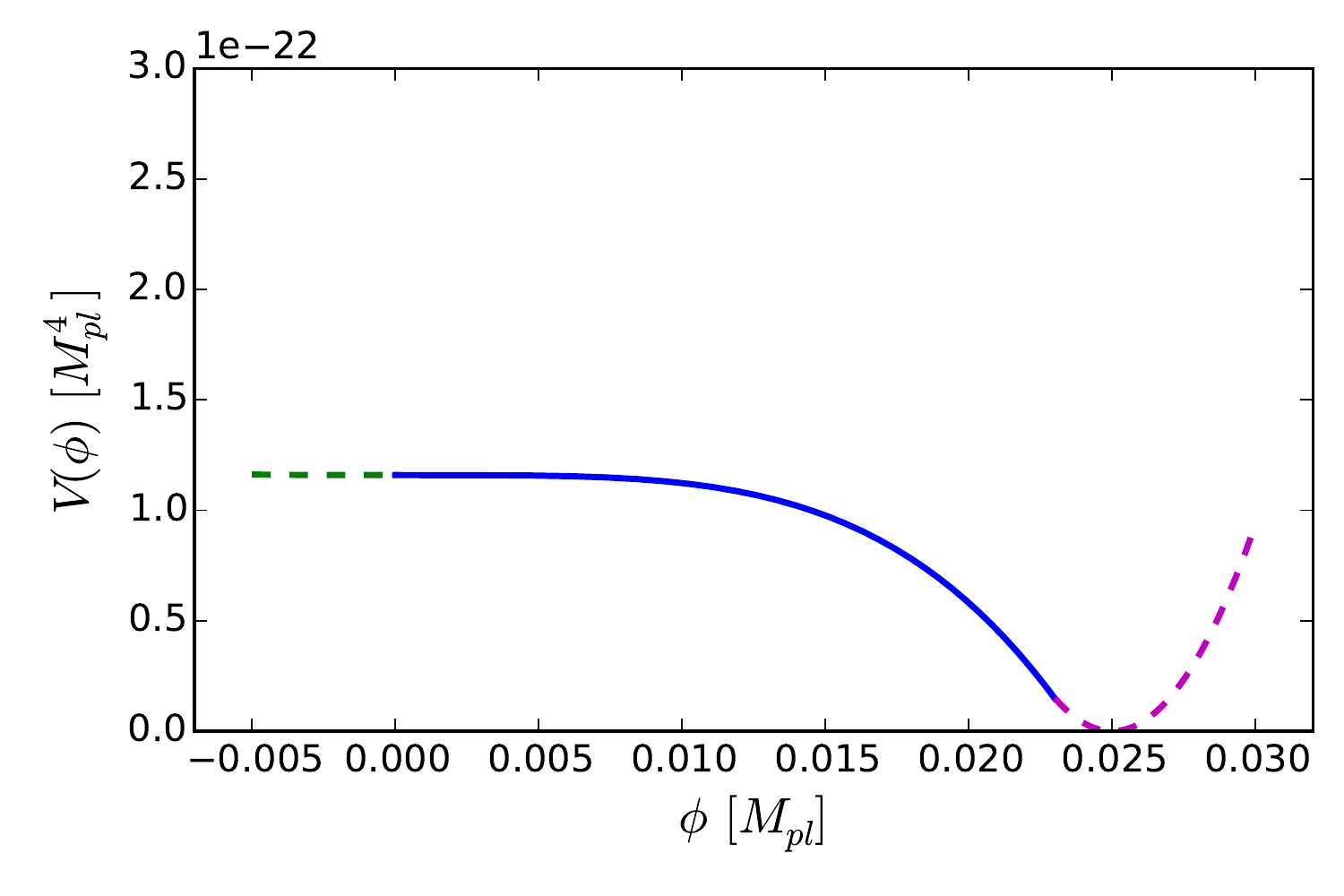}
\caption[Small field potential function - long plateau]{Small field potential function $V(\phi)$ with an extended flat region, showing the three regions, the central solid line (blue) region gives rise to the slow-roll inflationary period
\label{fig:PotSFflat}}
\end{center}
\end{figure}

Inflation could occur for much longer than 60 \mbox{$e$-folds}. Therefore the inflection point could be quite broad, while still accommodating the small field requirement, so the potential could instead look like Figure~\ref{fig:PotSFflat}, with a wider plateau.  In the context of inhomogeneous inflation, this distinction is important -- the presence of large gradients means that the scalar field can now sample a large domain of the potential, which could include the non-inflating ``cliff'' on the left side of the potential in Figure~\ref{fig:PotSFsteep}. The additional potential energy in such regions is converted to scalar kinetic energy as the field rolls down the hill towards the inflection point, which can disrupt slow roll sufficiently to end inflation.

In this section, we will explore and compare the two cases, a potential with an extended flat direction, and one with a steeper rise. Note that we do not consider the effect of varying $K$ on small field inflation. This is because a profile for $K$ which covers both negative and positive values (i.e. with both expanding and contracting regions), requires the addition of a relatively large kinetic energy term $\xi$, which immediately pushes the field into the minimum, ending inflation. Thus the case where $K$ is constant represents a best case scenario -- adding variation in $K$ will only end inflation sooner. This is consistent with initial kinetic energy being the most important failure mode, as shown by Goldwirth and Pirin \citep{Goldwirth:1989pr,Goldwirth:1989vz,Goldwirth:1991rj}.

\subsection{Small field model with extended flat direction} \label{sec-smallflat}

In this section, we will investigate the robustness of small field inflation for the case depicted in Figure \ref{fig:PotSFflat}.

We model the inflationary potential as
\begin{equation}
V(\phi)=
\begin{cases}
V_0& \phi<0\\
V_0 \left(1 - \left(\frac{\phi}{\mu}\right)^4\right) & 0<\phi<0.023 \mpl\\
m^2 (\phi - \phi_*)^2 & \phi > 0.023 \mpl \, ,
\end{cases}
\end{equation}
with $\mu = 0.0238\mpl$, $V_0 = 1.15949 \times 10^{-22} \mpl^{4}$, \mbox{$m^2 = 3.75 \times 10^{-18} \mpl^2$} and $\phi_* = 0.025\mpl$. The Hubble rate during inflation for this choice of parameters is $H_{\rm inf} = 3.125 \times 10^{-11} \mpl$.

These rather specific looking values are chosen such that for a homogeneous initial value of the field of $\phi_0 = 10^{-4} \mpl$, they would result in 100 $e$-folds of inflation, $\Delta_R = 10^{-5}$ and $n_s = 0.95$ for modes that exit the horizon 60 $e$-folds before the end of inflation. We find the end of the inflationary plateau, the point at which the potential is no longer ``slow roll'', to be at approximately $\phi=0.008 \mpl$, with all but the last $e$-fold taking place for $\phi<0.001\mpl$.  

\nomenclature[g-pi]{$n_s$}{in inflation, the spectral index}

The length scale for the fluctuations $L$, set to the Hubble length in the absence of fluctuations, is then \mbox{$L=3.2 \times 10^{10} \mpl^{-1}$}, and the value of $K$ is constant across the grid as described in section \ref{sect:method}. This satisfies the Hamiltonian constraint, assuming that the initial value of the conformal factor of the metric, $\chi$, is approximately of order 1. In our simulations, we set this constant value of $K$ across the grid and then relax the value of $\chi$ from a value of 1 everywhere to satisfy the Hamiltonian constraint exactly\footnote{Although in the small field case $\chi$ remains very close to 1 as the fluctuations are small, and the space is approximately flat.}.

\begin{figure}
\begin{center}
\includegraphics[width=0.8\textwidth]{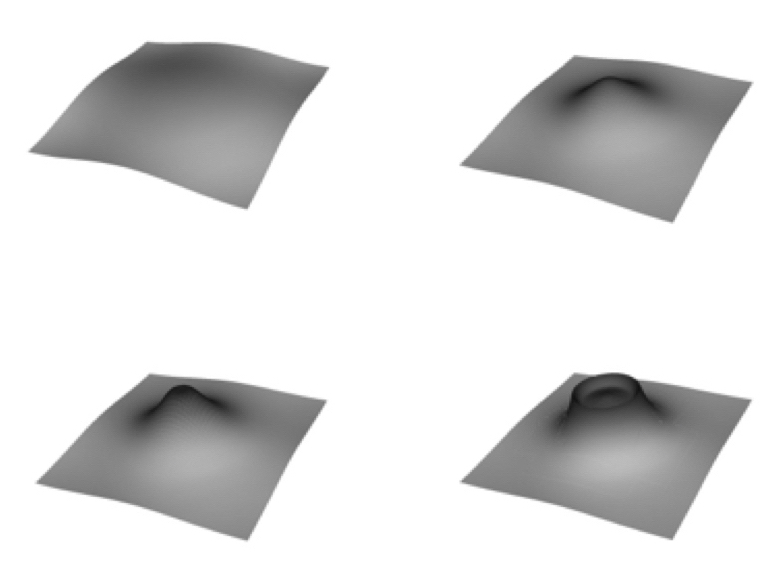}
\caption[Small field failure, field evolution]{A time series of the value of the field $\phi$ is shown on a 2 dimensional spatial slice. The maximum initial value of the field is at the centre, and it can be seen how this point ``falls off'' the inflationary potential, and subsequently drags the remaining space down with it. This means that the failure of a single point quickly ends inflation throughout the spacetime.
\label{fig:FallSpread}}
\end{center}
\end{figure}

\begin{figure}
\begin{center}
\includegraphics[width=0.75\textwidth]{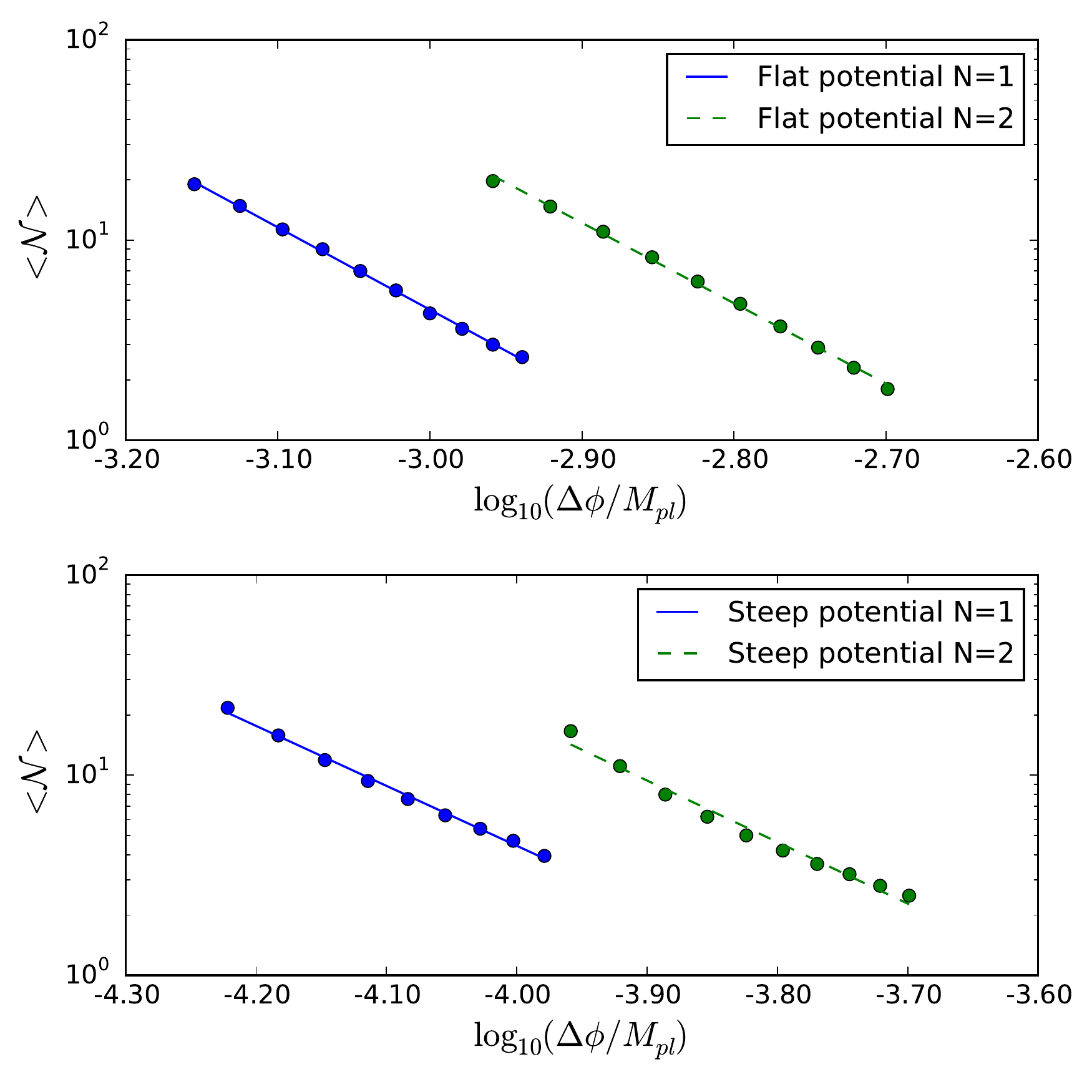}
\caption[Reduction in $e$-folds with increasing amplitude of fluctuations]{Plot showing the failure of inflation for small field inflation in the steep and flat cases for $N=1$ and $N=2$. The average number of $e$-folds $\langle \mathcal{N} \rangle$ decreases as $\Delta \phi$ is increased, with an approximate relationship of $\langle \mathcal{N} \rangle \propto \Delta \phi^{-4}$ for the flat case and of $\langle \mathcal{N} \rangle \propto \Delta \phi^{-3}$ for the steep case.
\label{fig:NefoldsSF}}
\end{center}
\end{figure}

We then evolve the initial conditions forward in time until inflation ends, or we reach the maximum number of $e$-folds we can simulate. We define the end of inflation as being the point at which a single point in the space falls to the minimum of the potential, that is, when the value of $\phi = \phi_*$ somewhere on the grid. The rest of the space will subsequently be pulled in by gradients, as illustrated in Figure \ref{fig:FallSpread}, and as we will discuss in more detail in the next section. The average number of $e$-folds $\langle \mathcal{N} \rangle$ is measured on this time slice\footnote{While the remaining spacetime can achieve several more $e$-folds before falling to the minimum, we treat this point as having ended inflation for measurement purposes. Allowing the simulations to run until the whole spacetime has fallen to the minimum and fully ceased inflating would displace the lines in Figure \ref{fig:NefoldsSF} vertically, but the trends would be the same. The actual values of $\langle \mathcal{N} \rangle$ are, in any case, model specific.}. We do this for a range of $\Delta \phi$. The results are shown in Figure \ref{fig:NefoldsSF} for the cases $N=1$ and $N=2$. 

\nomenclature[g-pi]{$\phi^*$}{in inflation, the minimum of the potential, corresponding to the end of inflation, and start of reheating}

For $N=1$ we find that inflation ends with less than 20 $e$-folds (which we call ``failure'' for our purposes) for initial amplitudes of around $\Delta \phi> 0.0007\mpl$. These values of $\Delta \phi$ correspond to $\rho_{\mathrm{grad}}/\rho_{V_0} \geq 1\times 10^{-4} $. Hence the gradient energy is still sub-dominant to the length scale $L \approx 1/H$. This is highly ``homogeneous''. Note that the density contrast in inflationary primordial perturbations are expected to be of order $10^{-5}$ which is only an order of magnitude smaller than this. However, the perturbations here are concentrated in one or two modes and it is the field excursions that should be compared. The typical displacement due to quantum fluctuations is $\Delta\phi_{\rm QM}\sim H_{\rm inf}/2\pi\ll0.0007\mpl$. We will explore the robustness of inflation to initial perturbations with more general power spectra in future work.

For $N=2$ we find failure when $\Delta \phi > 0.0011\mpl$. It can be seen that adding the additional mode makes inflation more robust. Recall from the definition \eqn{eqn:phi} that we have normalised $\Delta \phi$ to the total number of modes so adding modes adds to the gradient energy but not to the maximum field value -- \emph{this suggests that inflationary failure scenarios are more dependent on single long wavelength inhomogeneities rather than multiple short wavelength ones}. The number of $e$-folds decreases with an approximate relationship of $\langle \mathcal{N} \rangle \propto \Delta \phi^{-4}$ in both the $N=1$ and $N=2$ cases. 

\subsection{Pull back effects in small field inflation}

\begin{figure}
\begin{center}
\subfigure[The field value in physical space]{
\includegraphics[width=0.45\textwidth]{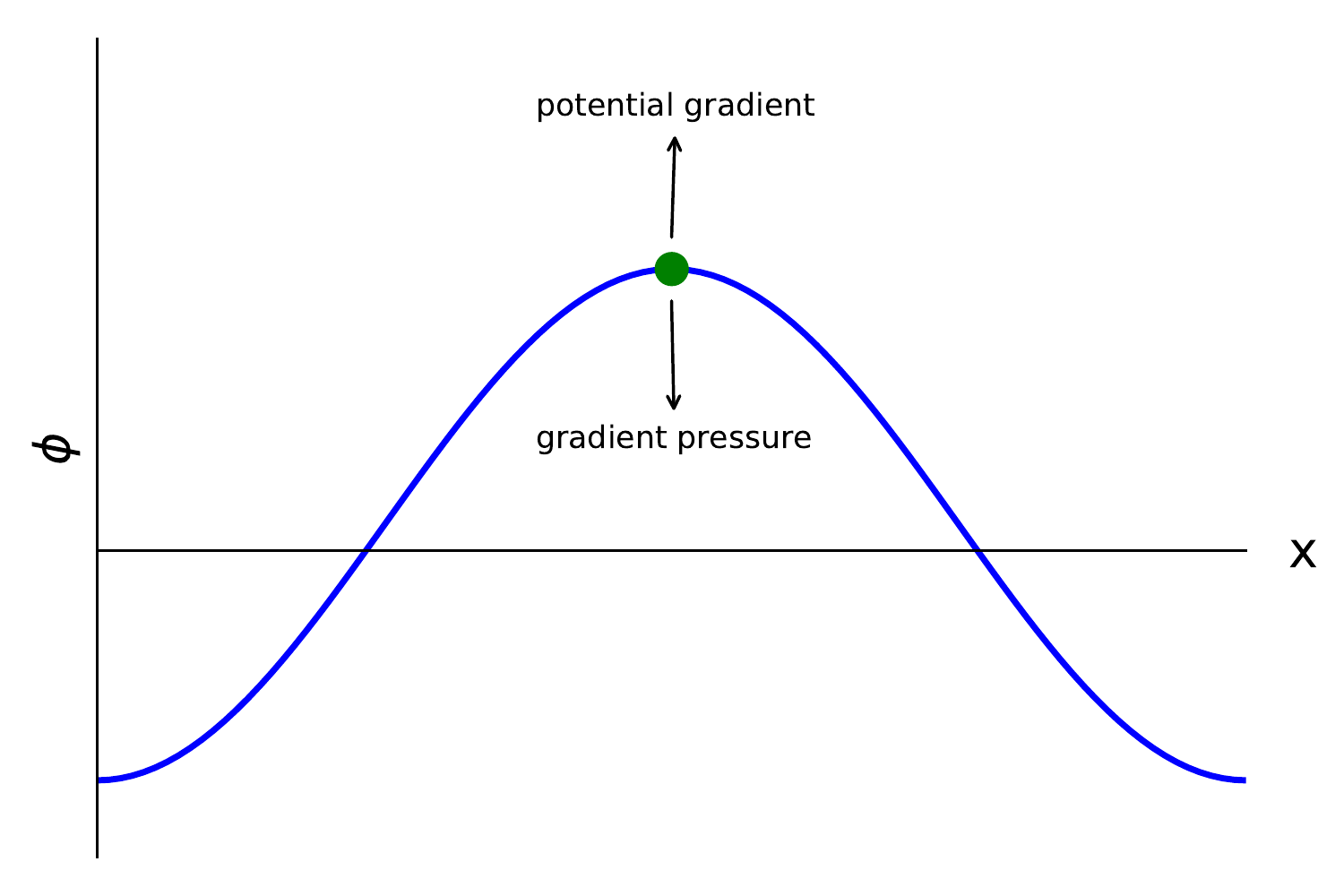}
}
\hspace{0.5cm}
\subfigure[The field value in field space]{
\includegraphics[width=0.45\textwidth]{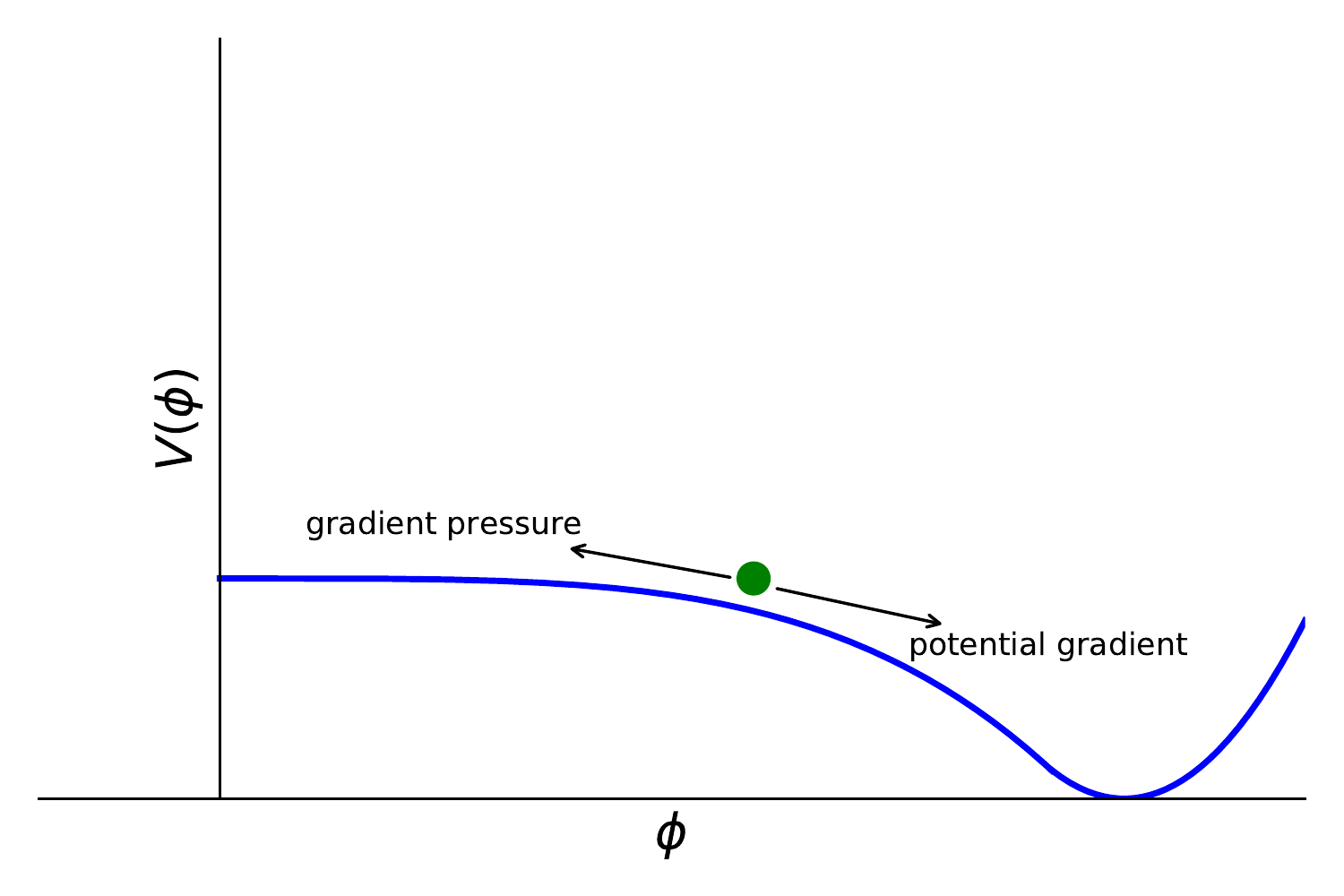}
}
\caption[The pull back effect]{Illustration of competition between gradient pressure and potential gradient for a concave model - the point in space marked with a green dot corresponds to the point on the potential which is closest to falling into the minimum. The two competing effects are the potential gradient, which tends to increase the field value, and the gradient pressure, which tends to pull it back up the potential hill.}
\label{fig:Compete}
\end{center}
\end{figure}

As was mentioned above, once one part of the field falls into the minimum, it quickly ``drags down'' the remaining spacetime, as shown in Figure \ref{fig:FallSpread}. It is instructive to consider the scalar field dynamics which leads to the failure as the naive expectation that the part of the field which has the maximum initial value (and hence is closer to the point where inflation ends) is that which falls to the minimum first is not always correct. There is some initial resistance from gradient pressure which, for a range of $\Delta \phi$, pulls the field back up the hill away from the minimum, ``saving'' inflation and making it more robust than one might expect. See Figure \ref{fig:Compete}.

Using the Klein-Gordon equation, we can show that local gradient pressure should temporarily ``save'' inflation up to some critical value, above which the maximum field value will fall directly to the minimum. The critical value for $\Delta \phi$ where this happens can be approximated quite accurately as follows. Consider the Klein-Gordon equation,
\begin{equation}
\partial_t^2{\phi} - \gamma^{ij}\partial_i \partial_j\phi + \frac{d V}{d \phi}=0,
\end{equation}
where we have ignored the friction term due to the expansion, and let 
\begin{equation}
\phi_{\max}(t) = \mathrm{max}(0,\phi({\bf x},t)).
\end{equation}
Initially, $\phi_{\rm max} = \phi_0 + 3 \Delta \phi$ and $\gamma^{ij}\sim {\cal O}(1)$. For this point, and assuming that we are still in the concave part of the potential (ie, the ``hilltop'' part), the field value should fall initially towards the minimum if 
\begin{equation}
-\nabla^2 \phi  \lsim \left| \frac{dV}{d\phi} \right|.
\end{equation}
For the initial conditions and the potential we consider, this is the case (assuming $N=1$) for
\begin{equation}
3 k^2 \Delta \phi \lsim -\frac{dV}{d\phi} \bigg|_{\phi=\phi_{\rm max}}. \label{eqn:pullback}
\end{equation}
Using the relation \eqn{eqn:Lformula}, $\phi_0 \ll 3 \Delta \phi$, and $V(\phi_0) \approx V_0$ for our small field case this becomes
\begin{equation}
\frac{32 \pi^3 n^2 V_0 \Delta \phi}{\mpl^2} \lesssim \frac{4 V_0 (3 \Delta \phi)^3}{\mu^4},
\end{equation}
which simplifies to
\begin{equation}
\Delta \phi \gtrsim \sqrt{\frac{8 \pi^3}{27}} \frac{n \mu^2}{\mpl}.
\end{equation}
The critical value for our chosen values and $n=1$ is $\Delta \phi \approx 0.0017\mpl$, which corresponds to $\phi_{\rm max, crit} =0.00525\mpl$, beyond the part of the potential that supports an extended period of inflation. Small field inflation is thus more robust than one might naively expect because local excursions towards the edge of the inflationary plateau are pulled back onto it. 

\begin{figure}
\begin{center}
\includegraphics[width=0.8\textwidth]{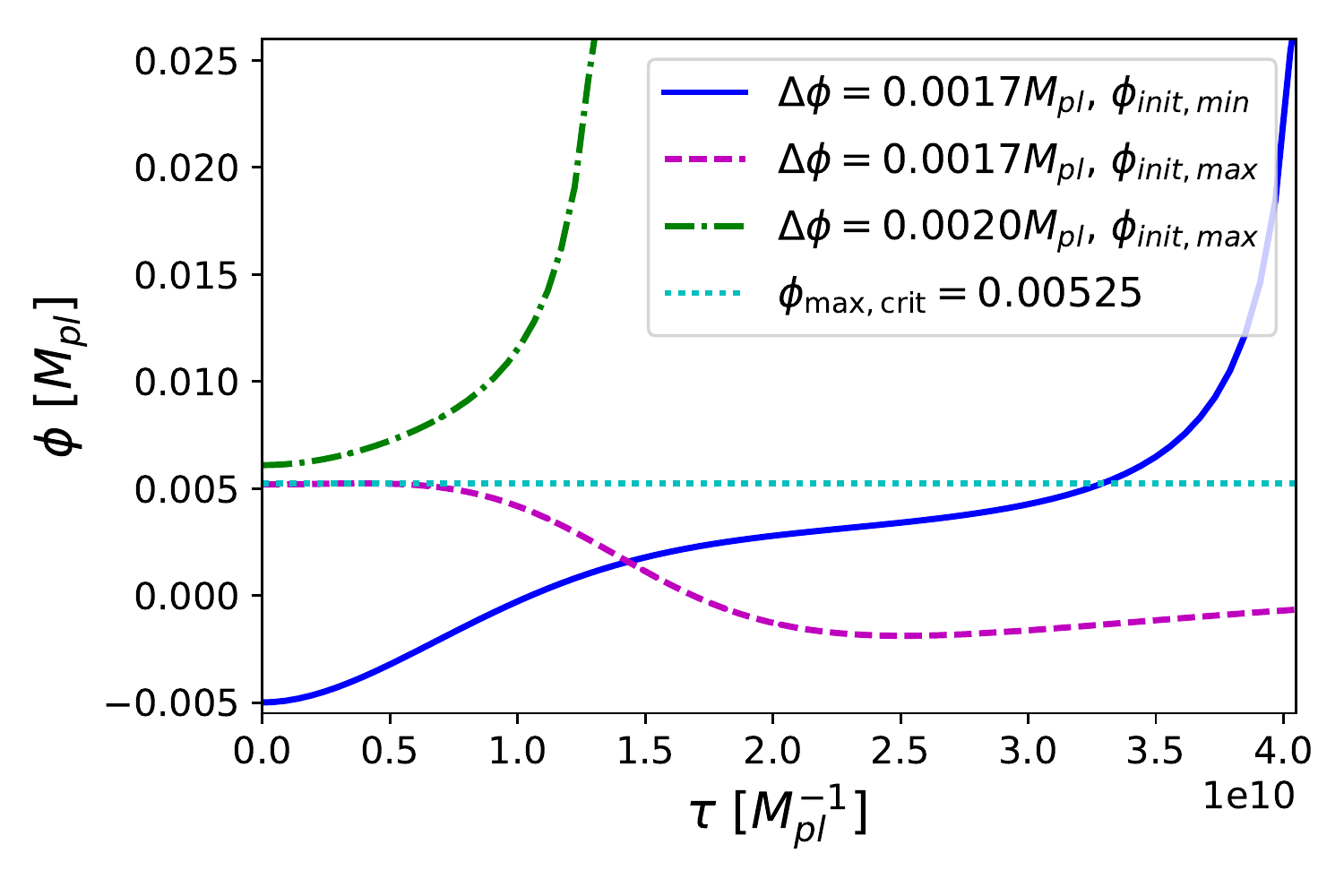}
\caption[Small field with extended plateau, critical maximum value of the field]{The evolution of \emph{initial} $\phi_{\max}$ and $\phi_{\min} = \min(\phi({\bf x},t_0)$  points versus proper time $\tau$ for small field inflation (extended flat case). At the derived ``critical value'' of $\Delta \phi =0.0017\mpl$ and $\phi_{\rm max, crit} = 0.00525\mpl$, shows a very small initial increase at the maximum value (in dashed/pink) but then is pulled back, so that it is in fact the initial minimum point (shown in blue/solid) which fails first. A slightly larger value of $\Delta \phi =0.002\mpl$ fails immediately at the maximum point (in green/dotted), as expected. Note that $\tau$ is the proper time experienced by an observer at the specified coordinate location. 
\label{fig:SFphi_vs_t}}
\end{center}
\end{figure}

The results of several simulations are illustrated in \mbox{Figure~\ref{fig:SFphi_vs_t}}. We see that the predicted critical value for $\Delta \phi$ at which immediate failure occurs is approximately correct. In fact the field can even resist some initial movement towards the minimum at the maximum point, before being pulled back up the potential hill. 

\subsection{Small field model without extended flat direction} \label{sec-smallsteep}

In this section, we will investigate the robustness of small field inflation in the case in which the negative $\phi$ direction is a ``cliff''. 
We model the potential as
\begin{equation}
V(\phi)=
\begin{cases}
V_0+m^2\phi^2& \phi<0\\
V_0 \left(1 - \left(\frac{\phi}{\mu}\right)^4\right) & 0<\phi<0.023 \mpl\\
m^2 (\phi - \phi_*)^2 & \phi > 0.023 \mpl \, ,
\end{cases}
\end{equation}
with the same parameters and initial conditions as in Section \ref{sec-smallflat}. 
The is illustrated in Figure \ref{fig:PotSFsteep}.

For the $N=1$ case, we find failure for initial amplitudes of $\Delta \phi \gtrsim 5 \times 10^{-5}\mpl$, which corresponds to $\rho_{\mathrm{grad}}/\rho_{V_0} \approx 6 \times 10^{-7} $. Note that inflation now fails for amplitudes roughly an order of magnitude smaller than in the case of the extended flat region, showing that small field inflation is highly sensitive to changes in the potential around the flat region. Although the energy density in these fluctuations is now smaller than those expected from inflationary primordial perturbations, recall that it is concentrated in one mode, rather than being a scale invariant spectrum. In particular, the typical displacement due to quantum fluctuations for a given mode is still significantly smaller than the fluctuations considered here $\Delta\phi_{\rm QM}\sim H_{\rm inf}/2\pi\ll 5\times 10^{-5}\mpl$. 

The results are shown in Figure \ref{fig:NefoldsSF} for the cases $N=1$ and $N=2$, below those for the case with the flatter potential. Again it can be seen that adding the additional mode makes inflation slightly more robust. The number of $e$-folds decreases as a power law with an approximate relationship of $\langle \mathcal{N} \rangle \propto \Delta \phi^{-3}$.

\begin{figure}
\begin{center}
\includegraphics[width=0.8\textwidth]{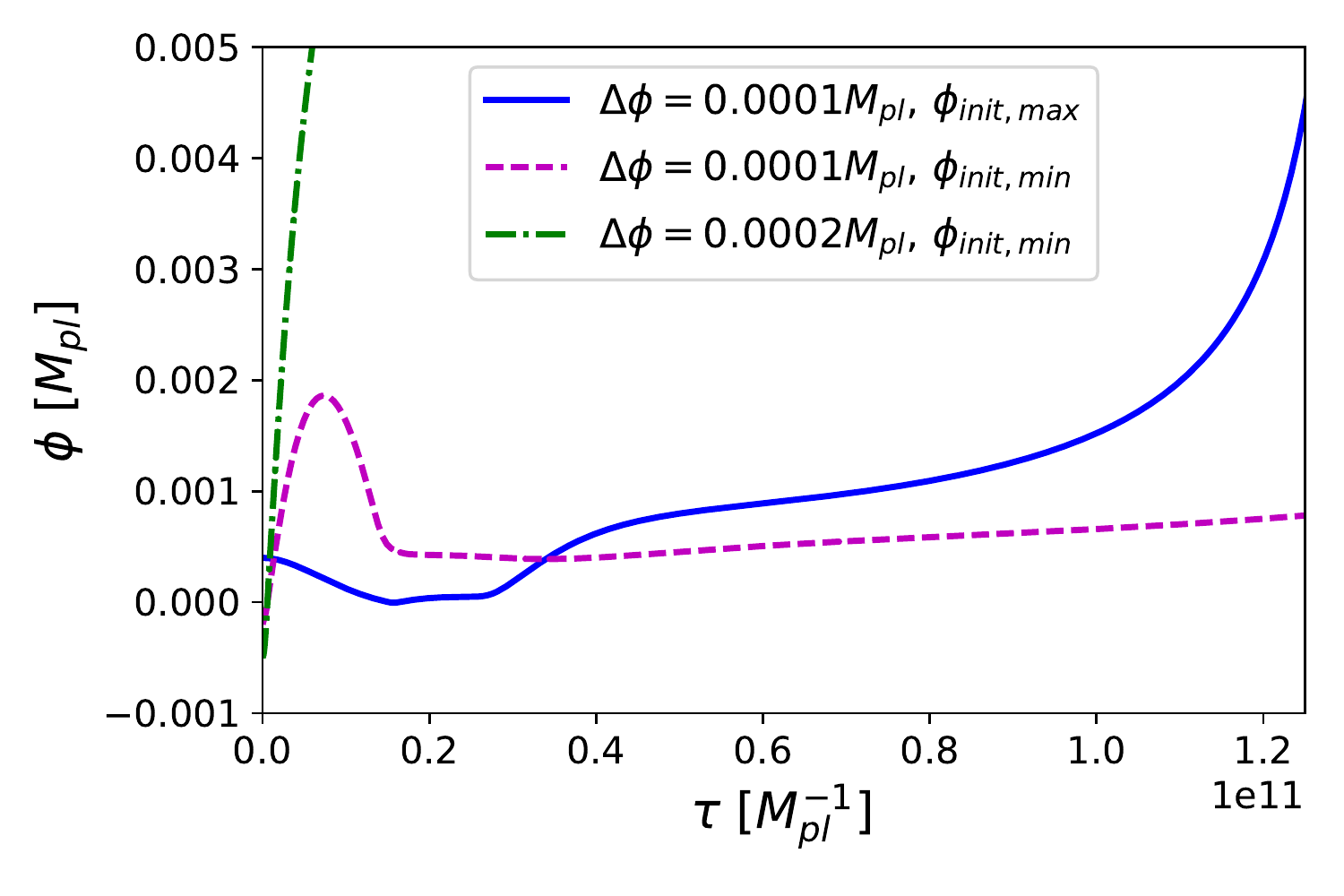}
\caption[Small field with short plateau, critical maximum value of the field]{The evolution of \emph{initial} $\phi_{\max}$ and $\phi_{\min} = \min(\phi({\bf x},t_0)$  points versus proper time $\tau$ for small field inflation (steep potential case). The spatial point which is initially at the maximum (in blue/solid) fails first, after hitting the ``wall'' to the left of the inflationary potential, and rebounding into the minimum of the potential. While the minimum point does rapidly overshoot due to its kinetic energy (in pink/dashed), it is pulled back by gradients and ultimately ``saved'' from failure. However, increasing $\Delta \phi$ further does eventually lead to the spatial point with the minimum initial field value dominating the collapse due to its kinetic energy (in green/dot-dash). Note that $\tau$ is the proper time experienced by an observer at the specified coordinate location. 
\label{fig:SFphi_vs_t2}}
\end{center}
\end{figure}

In this case the dynamics of the failure is driven by the most positive point. One might expect the most negative point to rapidly gain kinetic energy and overshoot the inflationary plateau, but this is only observed for higher values of $\Delta \phi$; for small $\Delta\phi$ the field is ``pulled back'' by the gradients in the field. The most positive point in this case gets pulled back as before, but then hits the steep ``wall'' and proceeds to roll off the plateau. This is illustrated in Figure \ref{fig:SFphi_vs_t2}.

Again thinking solely about the scalar field dynamics of the extremal points, one can estimate at which point the most negative point will fail directly.  Consider the most negative value of $\phi$ initially, $\phi_{\rm min} = \phi_0 - 3 \Delta \phi$. We can see that it will fail if, having oscillated through $\phi_0$, the point at which it would be brought to rest by gradient pressure exceeds the critical point derived in the previous section of $\phi_{\rm max, crit} =0.00525\mpl$. If the potential were flat in this region, this value would be the same as $\phi_{\rm max} = \phi_0 + 3 \Delta \phi$ (since it is effectively in simple harmonic motion). However, the initial slope in $V(\phi)$ gives it an extra ``push'', which we can equate to having started with a larger value of $\Delta \phi$. Considering the initial energy density $\rho_0$ at the minimum point, relative to the point $\phi_0$,
\begin{equation}
\rho_0 = (V(\phi_{\rm min}) -V(\phi_0)) + \frac{1}{2} (\nabla\phi)^2 \, ,
\end{equation}
then by making the approximations $\phi_0 \ll 3 \Delta \phi$ and $V(\phi_0) \approx V_0$, this becomes
\begin{equation}
\rho_0 = m^2 (-3 \Delta \phi)^2 + \frac{6 \pi^2 n^2 \Delta \phi^2 }{L^2} \, ,
\end{equation}
and we can find an ``effective'' initial value of $\Delta \phi$, which would have the same initial energy density
\begin{equation}
\Delta \phi_{\rm eff} = \Delta \phi \sqrt{1+ \frac{3 m^2 L^2}{2 \pi^2 n^2}} \, .
\end{equation}
Setting this equal to $\phi_{{\rm max, crit}}$ gives a rough estimate for the initial value of $\Delta \phi$ leading to immediate failure at the minimum point, which using our specific values gives $\Delta \phi \approx 0.0002$. As shown in Figure \ref{fig:SFphi_vs_t2}, this value is consistent with our findings, although failure will occur slightly below this value, due to the various assumptions made, in particular that the potential is flat after $\phi_0$, when it is in fact sloped downwards. 

\section{Large field inflation} \label{sect:LF}

In large field inflation the inflationary part of the potential is $\delta \phi \gg \mpl$. It thus supports larger fluctuations in the field while still keeping the entire space within the inflationary regime.

The robustness of inflation in large field inflation was tested in ~\citep{KurkiSuonio:1993fg}, and more recently in \citep{East:2015ggf} who found that large field inflation with uniformly expanding initial conditions is very robust to large inhomogeneities of up to $\rho_{\rm grad} = 1000\rho_{V_0}$. In this section we broadly reproduce their results, before extending the work to consider the limit of very large fluctuations and non-uniform initial expansion rates which include initially contracting regions. 

\begin{figure}
\begin{center}
\includegraphics[width=0.6\textwidth]{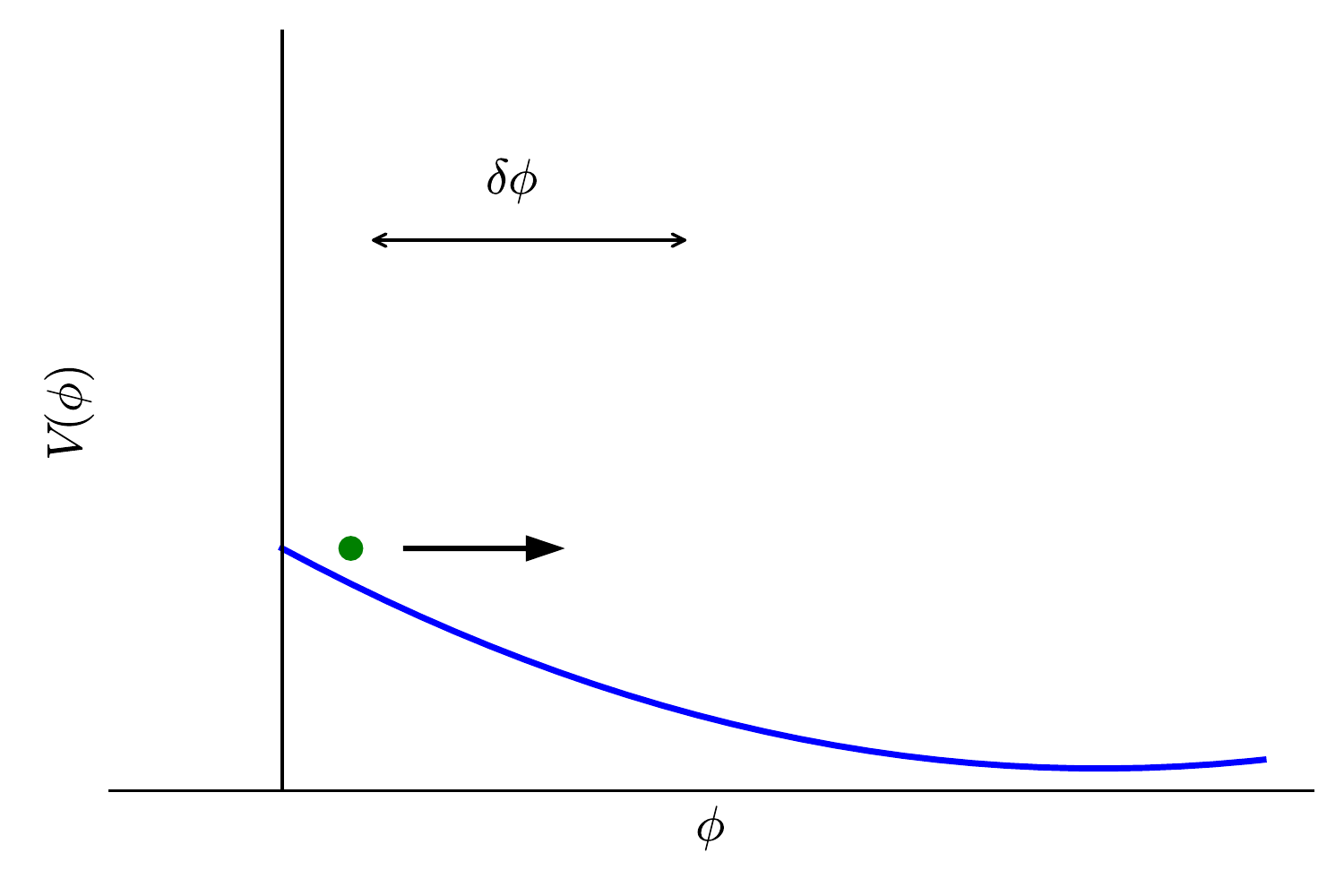}
\caption[Large field inflation, potential]{In large field inflation the width of the inflationary ``slow-roll'' part of the potential $\delta \phi \gg \mpl$. We use $m^2 \phi^2$ as illustrated, which as can be seen requires a longer stretch, but permits less flatness than small field inflation.
\label{fig:SFvsLF2}}
\end{center}
\end{figure}

We use $m^2 \phi^2$ inflation as a generic model for large scale inflation\footnote{While this model is marginally ruled out by the latest {\em Planck} data \citep{Ade:2015lrj}, we chose it for its simplicity of implementation. More complicated models will not lead to any drastically different results since the key feature is the flatness of the potential and the long traverse to the minimum.}, see figure \ref{fig:SFvsLF2}, i.e.
\begin{equation}
V(\phi) = m^2 \phi^2 \, ,
\end{equation}
with $m=1.07967 \times 10^{-7} \mpl$, leading to an inflationary scale of $H_{\rm inf} = 1.25 \times 10^{-6} \mpl$. For an initial value of the field of $\phi_0 = 4\mpl$, these values would result in 100 $e$-folds of inflation, with the scalar perturbation amplitude $\Delta_R = 10^{-5}$ and the spectral index $n_s \approx 0.97$ for modes that exit the horizon 60 $e$-folds before the end of inflation. 

The length scale for the fluctuations $L$ is set to the Hubble length in the absence of fluctuations, for our choice of parameters $L=8.0 \times 10^5 \mpl^{-1}$.

\subsection{Large field inflation with constant K}

In this section the value of $K$, the extrinsic curvature, is set as a constant across the grid using \eqn{eqn:Kavg} as above. We first considered a range of initial amplitudes of the perturbations $\Delta \phi$, and found that values below $0.2 \mpl$ resulted in inflation everywhere.  

We find that at larger values of $\Delta \phi$, we form black holes. As in \citep{East:2015ggf}, we can argue for their formation at this scale using the hoop conjecture. The black hole mass $M$ must be enclosed within a hoop of radius $R$. Assuming spherical symmetry, this is
\begin{equation}
R = 2GM \, .
\end{equation}
For a perturbation of wavelength $L$, given that a single mode in each spatial direction necessarily creates two  black holes by symmetry, the greatest radius from which each black hole can accrete is approximately
\begin{equation}
R = \frac{L}{4} \label{eqn:BHmass}\, .
\end{equation}
Note that $L$ is an arbitrary length (the wavelength of the perturbation) and not necessarily the Hubble length. The mass enclosed, $M$, is
\begin{equation}
M \approx \frac{4}{3} \pi R^3 \langle \rho_{\mathrm{grad}} \rangle \, ,
\end{equation}
where
\begin{equation}
\langle \rho_{\mathrm{grad}}\rangle  \approx  3 \pi^2 \frac{\Delta \phi^2}{L^2}  \label{eqn:rhograd} \, ,
\end{equation}
which is obtained from the volume average of $(\nabla \phi)^2$ from \eqn{eqn:phi} over a volume $L^3$. The \emph{maximum} mass is then linearly proportional to $L$, i.e.
\begin{equation}
M = \frac{\pi^3}{16} L \left(\frac{ \Delta \phi}{\mpl}\right)^2 \, . \label{eqn:MassBH}
\end{equation}
Combining these gives the condition
\begin{equation}
\frac{\Delta \phi}{\mpl} \geq \sqrt{\frac{2}{\pi^3}}
\end{equation}
as the critical case for black hole formation, independent of the length $L$. This value of approximately $0.25\mpl$ is consistent with our findings above that the critical $\Delta \phi \approx 0.2$ (given the approximate nature of the calculation).

Using equation~\eqn{eqn:equiv}, this value of $\Delta \phi \approx 0.25\mpl$ gives $k/H \approx 1.6$.  This result is also consistent with the findings of \citep{East:2015ggf}.  In other words, black holes will form when the wavelength of the perturbation is four times the Hubble length $H^{-1} \approx 3/\sqrt{24\pi \rho_{\mathrm{grad}}} \mpl$ or larger. 

\begin{figure}
\begin{center}
\includegraphics[width=0.8\textwidth]{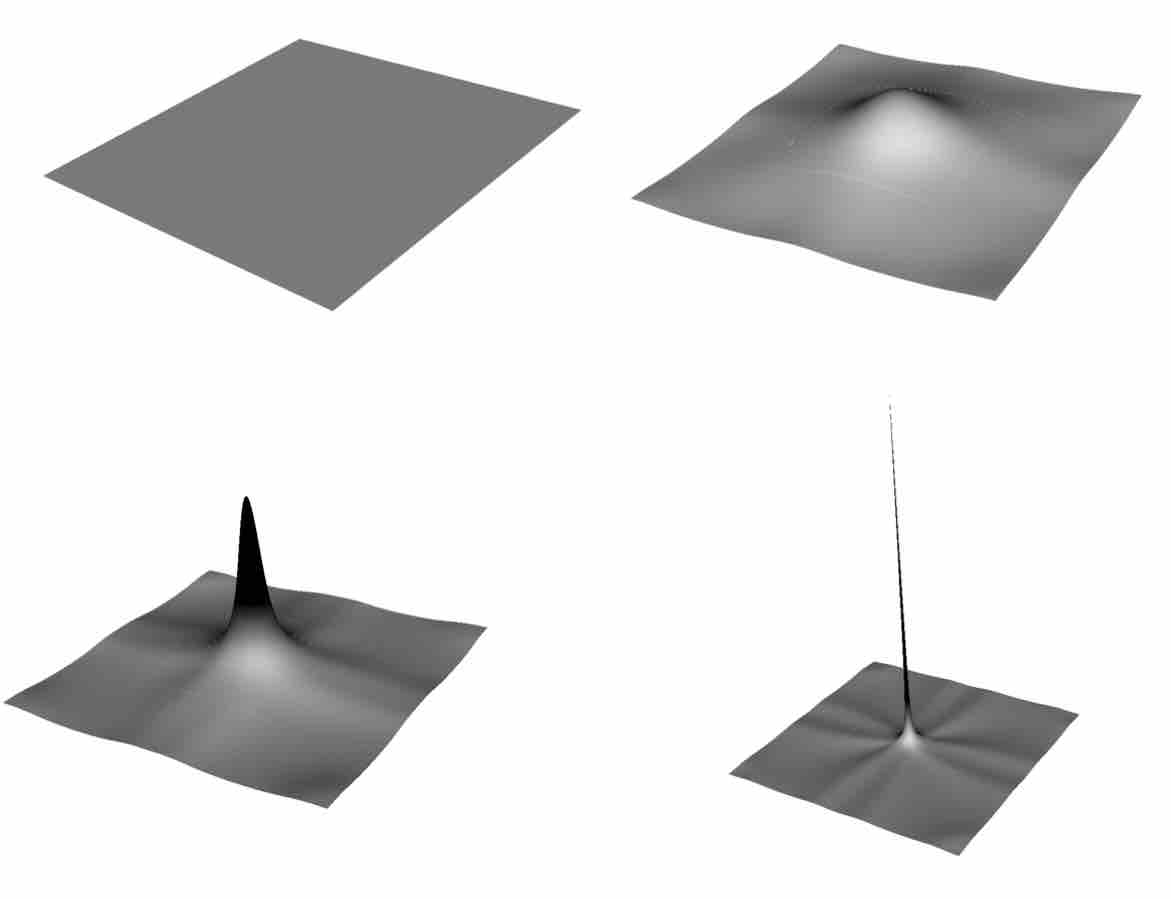}
\caption[Large field black hole formation, extrinsic curvature]{A time series of the value of the extrinsic curvature $K$ is shown on a 2 dimensional spatial slice in the case $\Delta \phi = 0.5 \mpl$. Black areas are collapsing (within the black hole) and the remaining areas are inflating, with the colour scale from black to white varying between $K = \pm 5 \times 10^{-5} \mpl$. We confirm the formation of a black hole by using an apparent horizon finder, but are able to continue evolving the black hole until it ``inflates out'' of the space. Although this is in part due to the gauge conditions (the moving puncture method tends to move the coordinate points away from the black hole singularity), the dominant effect is the inflation of the surrounding space. Eventually the spike will disappear once it falls below the coordinate grid resolution. 
\label{fig:LF_BH}}
\end{center}
\end{figure}

Above the critical value, black holes were formed, but these only created locally collapsing regions, and did not dominate the overall inflationary behaviour. As such they were quickly ``inflated out'' of the spacetime, see Figure \ref{fig:LF_BH}.

The robustness was, as in the small field case, due in part to the fact that the most extreme value is quickly ``pulled back'' up the hill by gradient energy, resulting in initial inhomogeneities being smoothed out.  For the $m^2\phi^2$ potential we are using,  $\phi_0$ and $\Delta \phi$ are both large and so the field will move towards the minimum when
\begin{equation}
\frac{32 \pi^3 n^2 m^2 \phi_0^2 \Delta \phi}{\mpl^2} < 2 m^2 (\phi_0 - 3 \Delta \phi) \, ,
\end{equation}
which reduces to
\begin{equation}
\Delta \phi < \frac{\phi_0\mpl^2}{16 \pi^3 n^2 \phi_0^2+3\mpl^2}\approx\frac{\mpl^2}{16\pi^3 n^2\phi_0} \, .
\end{equation}
Thus there is a \emph{minimum} value beyond which pullback always occurs. This is because $dV/d\phi$ approaches zero at the minimum for $m^2\phi^2$ type potentials -- i.e. it is a convex potential\footnote{Note that in our convention, hilltop type models are concave, whereas $m^2 \phi^2$ type potentials we call convex.}. For concave type potentials (e.g. hill-top models \citep{Boubekeur:2005zm}), there will be a maximum $\Delta \phi$ instead, as we have discussed in the small field case. The value of this bound is small, in our model $\Delta \phi \approx 0.0005$, and in the limit of a very flat extended potential it is zero. Thus almost any perturbations will tend to be pulled back to a (potentially) more homogeneous configuration in this potential. This more homogeneous configuration then continues to inflate, with the number of $e$-folds approximately equal to that given by $\phi_0$. 

Thus, as expected, inflation eventually wins out, even with fluctuations which reach almost to the minimum of the $m^2 \phi^2$ potential -- \emph{large field inflation is very robust to scalar field inhomogeneities}. 

\subsection{Can black holes stop inflation?}

Naively, one might imagine that one can continuously increase the size of the fluctuations to generate black holes of increasing mass, to the point that the Schwarzschild radius dominates over the scale of inflation $R_{S} \gg H_V^{-1}$, ending inflation. The critical limit for this to occur is the so-called Nariai limit, which occurs when the black hole horizon and the de Sitter horizon coincide in a Schwarzschild-de Sitter spacetime. In our units, the critical mass of the black hole is
\begin{equation}
M_{\rm N} = \frac{1}{3}\frac{1}{\sqrt{8\pi}}\frac{\mpl^2}{\sqrt{V_0}}\mpl \label{eqn:nariai}\, .
\end{equation}
We will show in this section that this is not possible. 

\nomenclature[a-pi]{$M_{\rm N}$}{in inflation, the Nariai mass of a black hole}

\begin{figure}
\begin{center}
\includegraphics[width=0.8\textwidth]{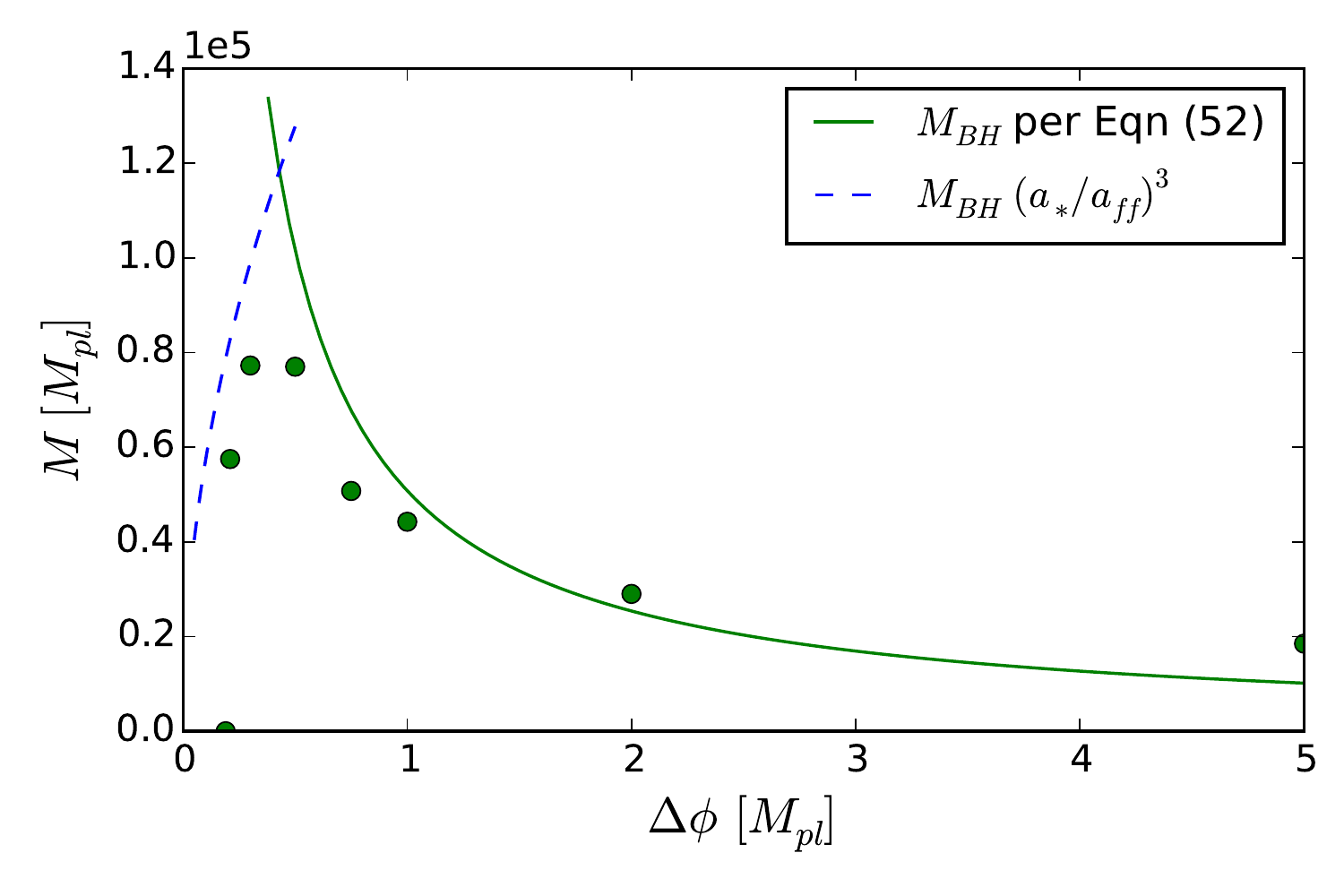}
\caption[Large field black hole mass versus initial perturbation size]{Plot showing the mass of the black holes formed versus the size of the initial perturbations. Although the mass initially increases, it reaches a maximum (at a size of about $R_{\rm S}=0.2L$) after which it falls off as $1/\Delta\phi$, meaning the maximum mass of the black hole which can be formed is bounded. The green dots are results from numerical simulations, and the lines illustrate the approximate agreement to our analytic calculations. The green solid line shows the maximum mass $M_{\rm BH}$ predicted per \eqn{eqn:LFBHmass}. The blue dotted line shows $M_{\rm BH} (a_*/a_{\rm ff})^3$, reflecting the effect of the transition to de Sitter on the radius from which energy can fall in. The two lines meet at the critical point $\Delta \phi \approx 0.43 \mpl$.
\label{fig:LFBHMass}}
\end{center}
\end{figure}

In Figure \ref{fig:LFBHMass}, we show the mass of the black holes formed as we gradually increase the amplitude $\Delta \phi$  obtained from numerical simulations. We see that while the mass increases with $\Delta \phi$ initially, at some point, the black hole mass begins to \emph{decrease} as $M \propto \Delta \phi^{-1}$. Thus there is a maximum black hole mass which can be formed (which in our case had a Schwarzschild radius of about 20 per cent of the Hubble radius related to the initial $V(\phi)$, that is, $L$).  This can be understood as follows.

Since $\rho_{\mathrm{grad}} \gg V_0$ initially, the early expansion of the spacetime is roughly that of a radiation dominated universe, i.e. $\rho_{\mathrm{grad}} \propto a^{-4}$ and $H^2 \propto \rho_{\mathrm{grad}}$.  At late times, the expansion rate is that of de Sitter, i.e. $a\propto e^{H_V t}$ where $H_V^2 = (8\pi G/3)V_0$. 

Meanwhile, the free fall time-scale for some matter distribution of average density $\rho$ is given by
\begin{equation}
\Delta t_{\rm ff} \approx \sqrt{\frac{1}{G \rho}} \label{eqn:tff}\, .
\end{equation}
If there is no expansion, then this is roughly the timescale for some cloud of density $\rho$ to collapse to form a black hole as long as the initial distribution is supercritical. However, due to the presence of the large gradient energy density, the spacetime is roughly expanding as a radiation dominated universe, dissipating some of the energy away from forming a black hole. Once the spacetime is dominated by vacuum energy, it is safe to assume that any remaining energy that has not collapsed into a black hole will be rapidly dissipated. The time scale for this to happen, $a_*$, occurs at vacuum-gradient energy equality $\rho_{\mathrm{grad}}a_*^{-4} = V_0$, i.e.\footnote{Recall that in our conventions $a\approx 1$ on the initial slice.}
\begin{equation}
a^2_* = \sqrt{8\pi^3} \frac{\Delta \phi}{\mpl} \label{eqn:astar}\,,
\end{equation}
where we have used \eqn{eqn:Lformula} and \eqn{eqn:rhograd}. 

\nomenclature[a-pi]{$a_*$}{in inflation, the scale factor when the gradient energy becomes subdominant}
\nomenclature[a-pi]{$a_{\rm tt}$}{in inflation, the scale factor at the free fall timescale}

Converting $\Delta t_{\rm ff}$ into the scale factor by solving the Friedmann equation, we get
\begin{equation}
a_{\rm ff}^2 = 2\sqrt{\frac{8\pi}{3}} +1 \label{eqn:aff}\,,
\end{equation}
which is independent of $\rho_{\mathrm{grad}}$ as expected. This predicted value of $a_{\rm ff}=2.6$ provides a good approximation to the value of $\langle a \rangle \approx 3$ measured at black hole formation in the simulations over the range of $\Delta \phi$ tested. If $a_* < a_{\rm ff}$, then de Sitter space will take over before the collapse has finished, leading to a lower mass black hole, and this is the case for \emph{smaller} values of $\Delta \phi$.  By equating these two times from \eqn{eqn:astar} and \eqn{eqn:aff}, we obtain 
\begin{equation}
\Delta \phi \approx 0.43 \mpl\,,
\end{equation}
as the point when the free fall is no longer stopped by de Sitter expansion, resulting in a maximum mass of the black hole. This is in good agreement with the numerical value which gives the maximum mass at $\Delta \phi \approx 0.4 \mpl$, as seen in Figure \ref{fig:LFBHMass}.

Above this point, the free fall timescale falls fully within the radiation dominated era. Consider the mass enclosed in a spherical distribution of matter of size $r$
\begin{equation}
M(r) = \frac{4}{3}\pi r^3 \rho\,.
\end{equation}
Since the collapse occurs well within the radiation domination era, the largest radius from which matter can still collapse into a black hole is the Hubble radius, $r \sim H^{-1}$, with the largest mass occurring when $\rho = \rho_{\mathrm{grad}}$, giving us 
\begin{equation}
M_{\rm BH} \lsim \frac{4 \sqrt{3}}{(8\pi)^2} \left(\frac{\mpl}{\Delta \phi}\right) \left(\frac{\mpl^2}{\sqrt{V_0}}\right) \mpl, \label{eqn:LFBHmass}
\end{equation}
which scales like $M \propto \Delta \phi^{-1}$ as our numerical results indicate, and has a Schwarzschild radius of
\begin{equation}
R_{\rm S} \lsim \frac{L}{\sqrt{8\pi^3}} \frac{\mpl}{\Delta \phi} \, ,
\end{equation}
which agrees to the maximum observed size of about $R_{\rm S}=0.2L$ for $\Delta\phi=0.4$. This means that one cannot make a ``Giant Death Black Hole'' using the methods we outlined in this work -- there is a maximum mass, roughly $1/3$ of the mass of the Nariai black hole \eqn{eqn:nariai}, after which increasing $\Delta \phi$ leads to a reduction in BH size. While our analysis and numerical simulations have focused on the specific case where the initial expansion is uniform and scaled to the gradient energy, we expect similar no-go results to hold as long as the initial hypersurface is approximately flat  i.e. $\chi \sim 1$ and $A_{ij} \approx 0$ since the Hamiltonian constraint \eqn{eqn:HamCon} implies that the initial expansion will be, on average, uniform and large.

\nomenclature[a-pi]{$R_{\rm S}$}{the Schwarzschild radius}

\subsection{Large field inflation with spatially varying K}

We now consider spatially varying $K$ in the case where \mbox{$\Delta \phi=0.1\mpl$} and study the effect on inflation. The potential is now set to be simply a cosmological constant with the value $V(\phi_0)$ from the previous large field case, to allow inflation to continue indefinitely.

\begin{figure}
\begin{center}
\includegraphics[width=0.6\textwidth]{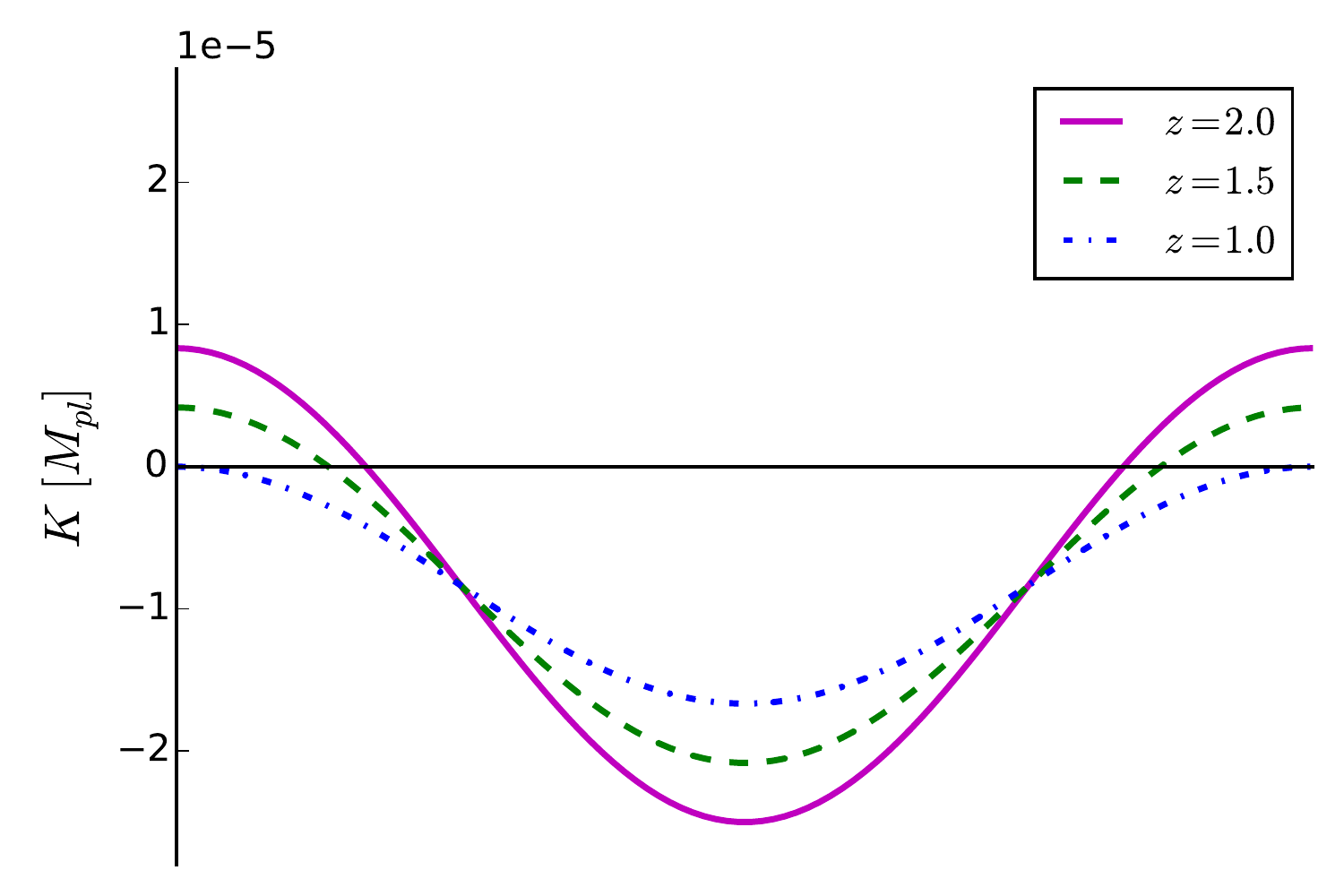}
\caption[Spatial variation of extrinsic curvature]{Illustration of the change in the spatial variation in $K$ when the parameter $z$ is varied, showing a slice through the maximum and minimum values of the profile. 
\label{fig:Kprofiles}}
\end{center}
\end{figure}

For our purposes it is useful to recast \eqn{eqn:Kansatz} for $K$ in the form
\begin{equation}
K = - z \bar{C}(\phi-\phi_0) + \langle K \rangle \label{eqn:Kansatz2}\, .
\end{equation}
We set
\begin{equation}
\langle K \rangle  = -\sqrt{24 \pi G \langle \rho \rangle}\,, \label{eqn:Kavg2}
\end{equation}
where the value of $\rho$ now includes the contribution from $\xi$.
Without loss of generality, we set $\bar{C}=2.78 \times 10^{-5}$ so that the maximum value of $K$ is zero for $z=1$. Increasing $z$ increases the amplitude of the fluctuations in $K$ and allows us to consider larger regions of spacetime that are initially collapsing, $K>0$. The profile for $K$ and the dependence on $z$ is illustrated in Figure \ref{fig:Kprofiles}. 

We test a range of values of $z$ between $1.0$ and $2.0$, and find that in cases of smaller $z$ (where most of the spacetime is expanding initially) the collapsing part of the spacetime ``bounces back'', such that $K$ quickly becomes approximately constant with a negative value everywhere. Inflation then continues, and over 20 $e$-folds are reached. 

\begin{figure}
\begin{center}
\includegraphics[width=0.6\textwidth]{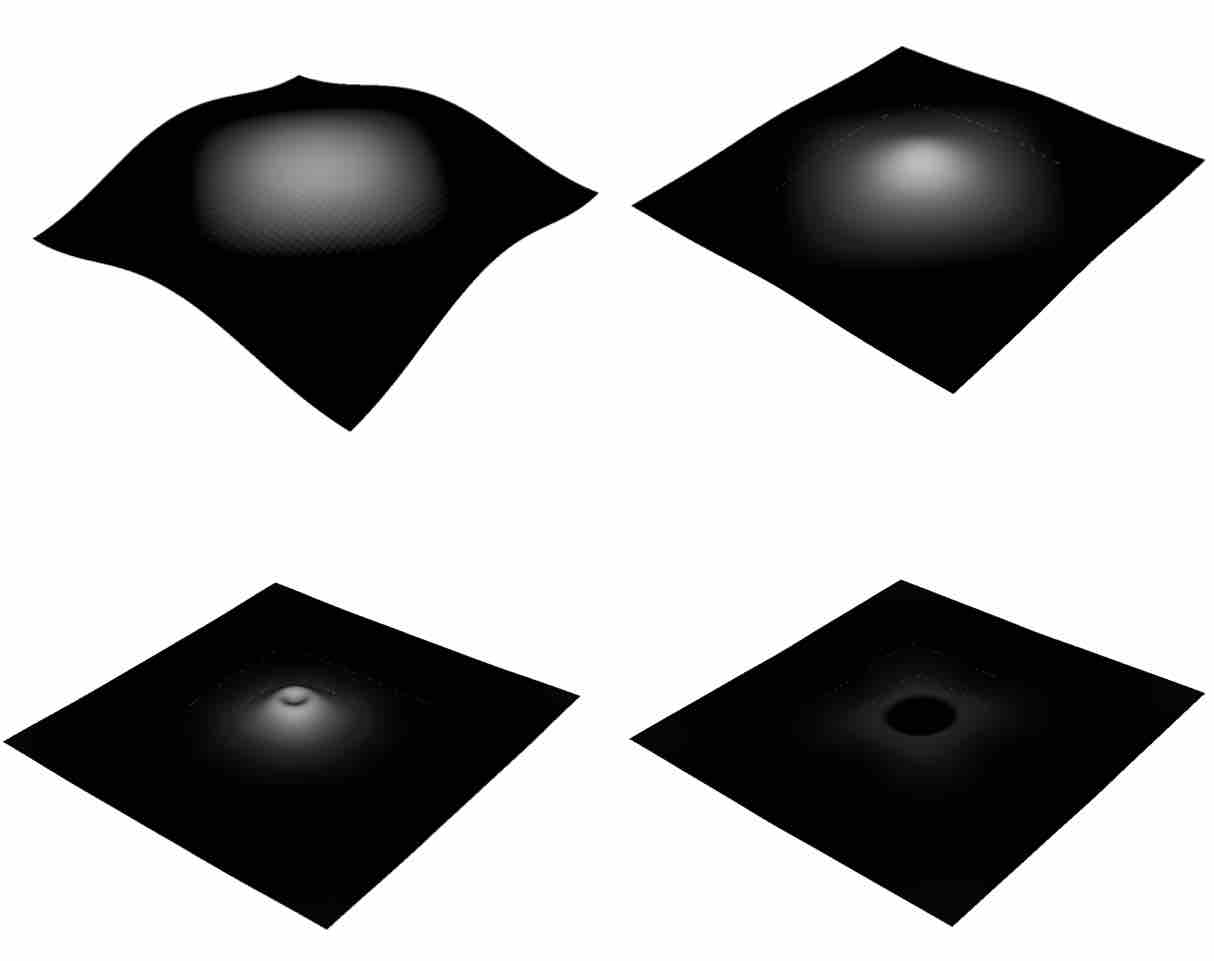}
\caption[Spatially varying extrinsic curvature, evolution, dispersal]{Illustration of the evolution of $K$ in the large field varying $K$ case for $z=1.5$. The initial inhomogeneities in $K$ quickly disperse and it settles into an inflating spacetime everywhere. White areas are collapsing and black areas are inflating, with the colour scale from black to white varying between $K = \pm 5 \times 10^{-5} \mpl$. 
\label{fig:LFK_vs_xt}}
\end{center}
\end{figure}

\begin{figure}
\begin{center}
\includegraphics[width=0.6\textwidth]{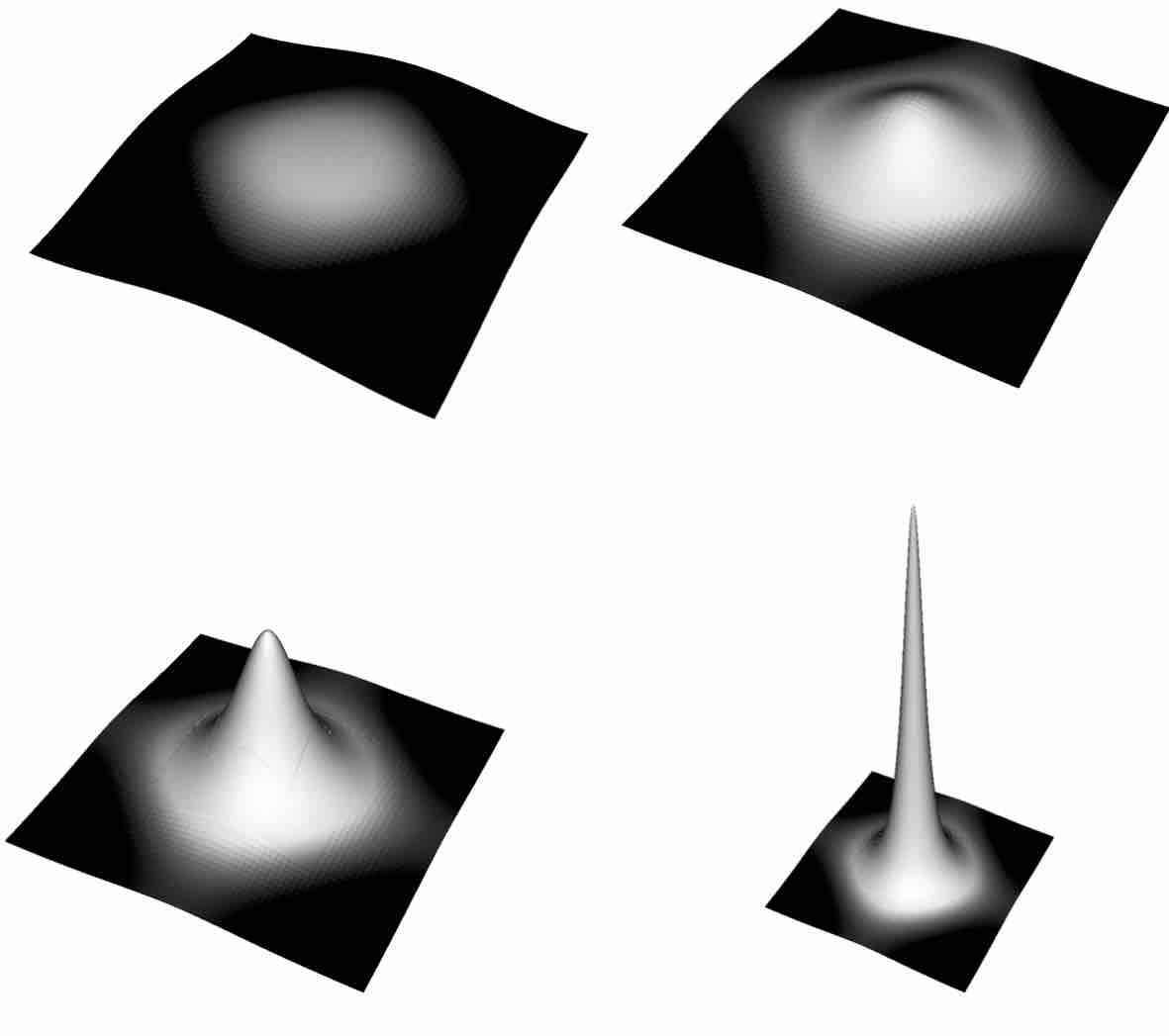}
\caption[Spatially varying extrinsic curvature, evolution, black hole]{Illustration of the evolution of $K$ in the large field varying $K$ case for $z=2.0$, showing the black hole formation at the centre. White areas are collapsing and black areas are inflating, with the colour scale from black to white varying between $K = \pm 5 \times 10^{-5} \mpl$. The peak value in the last frame is $K = 0.005 \mpl$, and the average number of $e$-folds across the grid at this point is roughly $0.4$. 
\label{fig:LFK_vs_xt2}}
\end{center}
\end{figure}

In cases of higher $z$, with $z > 2.0$, where more of the spacetime (but still less than half) is collapsing initially, we find that black holes form at the initially contracting point. We therefore find that for a spacetime that would have resulted in inflation everywhere for constant $K$ ($\Delta \phi=0.1$ is subcritical for black hole formation in the constant $K$ case), we are now able to generate regions of collapse by introducing variations of $K$. This is illustrated in Figures \ref{fig:LFK_vs_xt} and \ref{fig:LFK_vs_xt2}.

However, even in these cases, the remaining spacetime continues to expand and inflate. Since $\langle K \rangle <0$ in all cases here, this result is consistent with what would be expected from \citep{Barrow:1985} and \citep{Kleban:2016sqm} (note that our sign convention means that $K<0$ denotes locally expanding spacetime). 

\section{Conclusions} \label{sect:conclusions}

We investigated the robustness of small and large field models of inflation, subjecting it to several simple inhomogeneous initial conditions both in the scalar field profile and in the extrinsic curvature. In doing so we have set up a framework that will allow us to study more general initial conditions in the future. As expected, we found that large field inflation was far more robust than small field inflation. In particular, small field inflation can fail even for small subdominant gradient energies $\rho_{\mathrm{grad}}/\rho_{V} \approx 10^{-4}$ while large field inflation is robust even to dominant gradient energies of $\rho_{\mathrm{grad}}/\rho_{V}  \gg 1$. This implies that small field inflation requires at least some level of tuning to begin or a dynamical mechanism that sets up appropriate initial conditions.

\subsection{Robustness of small field inflation}

The primary failure mode for small field inflation is the disruption of coherent slow roll dynamics, causing some parts of the scalar field to irrevocably fall into the non-inflating minimum. Once a region of the scalar field falls into the minimum, this region will expand and dominate the rest of spacetime ending inflation for the entire hypersurface. This failure mode can be induced in the following ways:

\begin{itemize}
\item \emph{Adding large amplitude scalar fluctuations.} Large amplitude scalar fluctuations can create excursions outside the inflationary part of the potential which lead to one region falling to the minimum and dragging the rest of the field down with it. However, local excursions of the field towards the edge of the inflationary part of the potential get pulled back by gradient pressure, making small field inflation more robust than one might expect (see below).
\item \emph{Converting additional potential energy into kinetic energy.} If the inflationary region of the potential is small -- in the case of typical small field models it is often just an inflection point -- then a large initial fluctuation may have support on the steep part of the potential (i.e. the green dotted line in Figure \ref{fig:PotSFsteep}). This additional potential energy will be converted to scalar kinetic energy, generating a large fluctuation and pushing the scalar closer to the minimum, thus ending inflation.
\end{itemize}

Nevertheless, we found that small field inflation is more robust than one might naively expect. In particular, we find the following:

\begin{itemize}
\item \emph{Pullback effect of gradients.} We show that perturbations tend to ``homogenise'', i.e. gradients tend to flatten out. This means that some initial conditions which have regions in the non-inflating regime can still inflate as the scalar field gets pulled back into the inflating regime. We provide a formula for this critical point for any given potential, and demonstrate its effect numerically.
\item \emph{Adding additional shorter wavelength modes makes inflation more robust for a given maximum initial value of $\phi$.} Adding a second mode with half the wavelength, but normalised to keep the same value of $\phi_{\rm max}$, resulted in a higher threshold for inflation. This is somewhat unexpected since adding an $N=2$ mode \emph{increases} the gradient energy. We propose that this could be related to the pullback effect described above, which is stronger for higher wavenumber modes.
\end{itemize}

\subsection{Robustness of large field inflation}
In the large field case, except for the trivial case of an initial hypersurface which is contracting everywhere, i.e. $K > 0$, we did not find a viable failure mode for the initial conditions we considered -- large field inflation is robust to very large gradient energies $\rho_{\mathrm{grad}}/\rho_V > 10^2$. The primary reason for its robustness is the potential's large support for slow roll i.e. $\delta \phi \gg \mpl$, which combined with the rapid dissipation of gradient energy due to expansion, makes it difficult for the scalar to reach a non-inflating region. Furthermore, we find the following:

\begin{itemize}
\item \emph{No ``Giant Death Black Holes''}. Given a uniformly expanding initial condition scaled to the total initial gradient energy, there is a maximum mass black hole that is formed for which the radius is of order $0.2$ times the size of the vacuum energy Hubble radius. Increasing initial gradients beyond this point \emph{decreases} the final black hole mass -- this is caused by the dissipation of gradient energies due to the large initial expansion. We calculate the maximum mass, which occurs when the transition to de Sitter expansion no longer limits the black hole mass, and confirm it with numerical simulations. We find that it is roughly $1/3$ of the mass of the Nariai black hole.
\item \emph{Pullback effect of gradients}. Similar to the small field model, large gradients tend to homogenise. We show that for convex potentials, even with initial fluctuations which reach to the minimum of the potential, inflation can eventually succeed.
\item \emph{If $\langle K \rangle <0$, then inflation wins}.  If the initial hypersurface (a Cauchy surface) has a net negative (expanding) value of $K$ there will always be an expanding region, as predicted in analytic studies \citep{Barrow:1985} and \citep{Kleban:2016sqm}.
\end{itemize}

%% file: Chapter5/chapter5.tex
\chapter{Critical Bubble Collapse}
\label{ch-CriticalCollapse}

\ifpdf
    \graphicspath{{Chapter5/Figs/Raster/}{Chapter5/Figs/PDF/}{Chapter5/Figs/}}
\else
    \graphicspath{{Chapter5/Figs/Vector/}{Chapter5/Figs/}}
\fi

\section{Introduction}

\nomenclature[z-pi]{QCD}{Quantum Chromo Dynamics}

Since their discovery by Choptuik \citep{Choptuik:1992jv}, critical phenomena have been studied in many different contexts \citep{Abrahams:1993wa, Healy:2013xia, Choptuik:2003ac, Hilditch:2013cba, Evans:1994pj, Brady:1997fj, Honda:2001xg, Akbarian:2015oaa} -- for a review see \citep{Gundlach:2007gc}. 

Restating briefly the key points from section \ref{sec-CriticalCollapse}, any one parameter ($p$) family of initial configurations of the scalar field will evolve to one of the two final end states -- a black hole or the dispersal of the field to infinity. The transition between these two end states occurs at a value of the parameter $p^*$, at which the critical solution exists. 

In a spherically symmetric collapse, the mass of any black hole that is formed from such a collapse follows the critical relation 
\begin{equation}
M \propto (p - p^*)^{\gamma_S},
\end{equation}
where the scaling constant $\gamma_S$ is universal in the sense that it does not depend on the choice of family of initial data, although it does depend on the type of matter. 

The other key phenomenon which is observed is that of self-similarity in the solutions, or ``scale-echoing''. Close to the critical point, and in the strong field region, the value of any gauge independent field $\phi$ at a position $x$ and time $T$ exhibits the following scaling relation,
\begin{equation}
\phi (x,T) = \phi(e^{\Delta_S} x, e^{\Delta_S} T) , \label{eqn:echo}
\end{equation}
where $\Delta_S$ is a dimensionless constant with another numerically determined value of 3.44 for a massless scalar field in the spherical case. 

In spherical symmetry, the system of the Einstein equations coupled to matter can be reduced to a 1+1D system, and hence it is widely studied. Beyond spherical symmetry, there has been some recent progress in studying the phenomenon, for example, \citep{Abrahams:1993wa, Healy:2013xia, Choptuik:2003ac, Hilditch:2013cba}, but progress in making firm conclusions has been slower than expected due to the extremely high refinements required to study the stages of the collapse, which are magnified three-fold in full 3+1 codes. 

This universality in the critical behaviour can be derived by assuming that the critical case is an intermediate attractor of co-dimension one, which, when perturbed, has exactly one unstable mode \citep{Koike:1995jm}. It is not clear whether this will be true in more complex cases beyond spherical symmetry. Linear perturbations of the spherically symmetric case \citep{PhysRevD.59.064031} do not show additional unstable modes, but numerical studies such as that of Choptuik et al. \citep{Choptuik:2003ac} gave hints of further unstable modes that may be present in the full non-linear regime. Beyond spherical symmetry, work has generally focussed on axisymmetric vacuum collapses of gravitational waves, massless scalar fields and radiation fluids, as in \citep{Abrahams:1993wa, Choptuik:2003ac, Hilditch:2013cba, Gundlach:1999cw,Baumgarte:2015aza}. More recently, Healy and Laguna \citep{Healy:2013xia} considered a fully asymmetric case in full three dimensional simulations for a massless scalar field with $Y_{2,1}$ spherical harmonic perturbations on a spherical bubble shaped potential.

Simply adding a $\phi^4$ self-interaction potential to a massless scalar field is not expected to change the universality of the solution or the critical solution which is approached, as in principle no new mass scale is introduced, therefore one expects that the constants $\gamma_S$ and $\Delta_S$ will be unchanged \citep{Chop_conf}. Due to the assumed self similarity of the critical solution, the field values, and hence the potential, should remain bounded during the evolution \citep{Gundlach:1996eg}. The behaviour is then dominated by the gradient terms in the Lagrangian as the critical solution is approached, rather than the potential. The main cases which have been explored beyond the massless case have been massive fields with an $m^2$ mass term such as \citep{Brady:1997fj}, in which Type I critical collapse is observed when the mass of the field becomes comparable to the mass scale of the initial perturbation. Honda and Choptuik \citep{Honda:2001xg} also studied sub-critical oscillons in a $\phi^4$ potential. Few papers have considered a more general potential, as it is expected that the results will be the same as those for the massless or massive case depending on the relative size of the mass scales present. However, the particular case of bubbles set up in multi minima potentials is of interest to early universe cosmologists, as they are natural consequences of phase transitions, as will be discussed below. 

It has not been clear that the standard BSSN formulation of the Einstein equations, combined with the moving puncture gauge commonly used for black hole evolutions, \citep{Campanelli:2005dd, Baker:2005vv}, would be well adapted to the study of this problem. It can be challenging to find a stable choice of gauge parameters near the critical point, and moreover, it is not clear that any chosen gauge will be ``symmetry seeking'' \citep{Garfinkle:1999cm}, that is, that it will be adapted to observing the scale-echoing phenomena. Recent work by Healy et al \citep{Healy:2013xia} and Akbarian et al. \citep{Akbarian:2015oaa} have indicated that these standard choices may indeed be well adapted for the study of critical phenomena, although the latter paper used a special choice of coordinate grid and made some amendments to the standard puncture gauge. 

\begin{figure}
\begin{center}
\includegraphics[width=15cm]{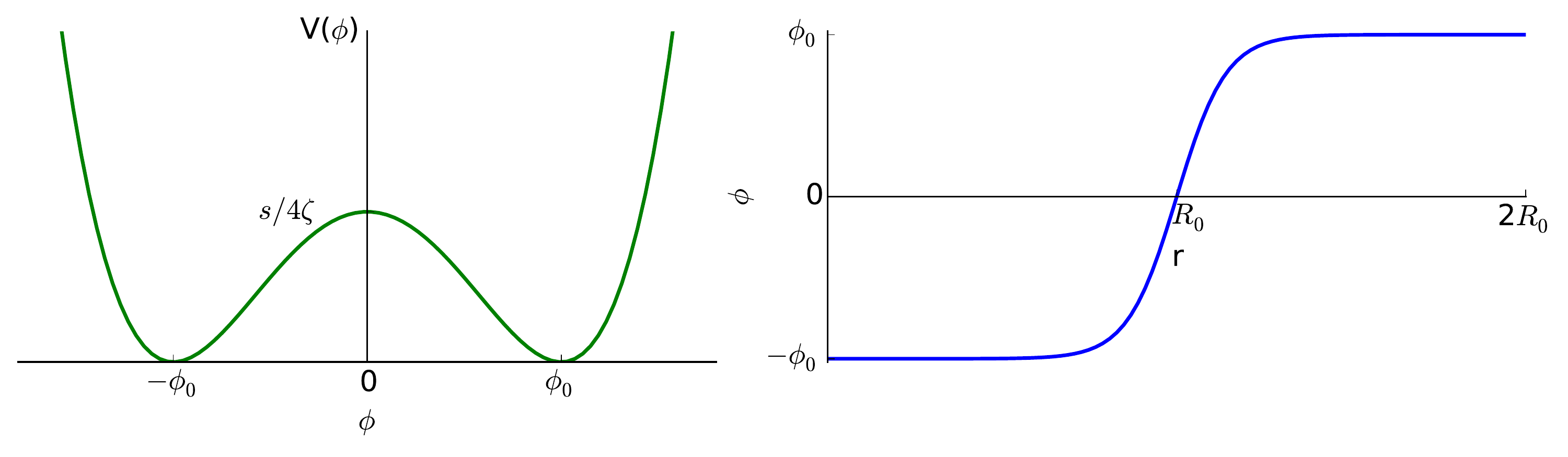}
\caption[Double well potential and bubble wall profile]{The double well potential considered in this work, and the initial bubble wall profile. 
\label{fig-potential}}
\end{center}
\end{figure}

In this work, we study the critical collapse and formation of black holes of ``Bubbles'', in both spherically symmetric and perturbed asymmetric cases, in 3+1D using cartesian coordinates and a slightly modified version of the puncture gauge. Bubbles are regions of space bounded by a scalar field domain wall. The domain wall interpolates between the two minima.  We consider the case of a minimally coupled scalar field, subject to the following potential, as shown in Figure \ref{fig-potential}, with two degenerate minima
\begin{equation}
V(\phi) = \frac{s}{4\zeta \phi_m^4} (\phi^2-\phi_m^2)^2 \label{eqn:potential} \, .
\end{equation}

The potential barrier between the two minima generates a tension $\sigma_{wall}$ on the bubble wall 
\begin{equation}
\sigma_{wall} \approx \int_{-\phi_m}^{\phi_m}d\phi~ \sqrt{2V}.
\end{equation}
The presence of this tension is an important difference between that of a simple ``bubble-like configuration'', say a top-hat profile, of a scalar-field (either massive or massless) as the field configuration seeks to maintain this tension even as the bubble collapses. Indeed, the tension increases as the bubble becomes smaller, causing the wall's gradient to rapidly increase. Even though the vacuum is degenerate (both minima have the same $V(\phi)$), the bubble collapses due to the pull of both gravity \emph{and} this tension. In other words, without gravity a massless ``bubble'' will disperse while bubbles with tension will coherently collapse.

\nomenclature[g-pi]{$\sigma_{wall}$}{In bubble collapse, the tension in the bubble wall}
\nomenclature[g-pi]{$\zeta$}{In bubble collapse, parameter defining the shape of the potential $V(\phi$)}
\nomenclature[g-pi]{$s$}{In bubble collapse, parameter defining the shape of the potential $V(\phi$)}
\nomenclature[g-pi]{$\phi_m$}{In bubble collapse, the location of the minima in field space}

Our work is motivated by the desire to understand transitions in scalar fields that are subject to multi minima potentials, which are found on cosmological scales in both the early and late universe. 

In the late universe,  cosmological axion fields \citep{Marsh:2015xka} are candidates for Dark Matter. Cosmological axions (as opposed to the QCD axion) are pseudo-bosons which can be described by a real scalar field. If the axion decay constant $f<H_i$, where $H_i$ is the scale of inflation, then its global symmetry will be broken after inflation, and the subsequent phase transition will populate the universe with bubbles of different vacuum expectation values, forming a network of domain walls. These bubbles are expected to collapse to form structures called ``mini-clusters'' which can be the source of cosmological structure formation and/or black holes which may be the origin of the super massive black holes in the centres of galaxies \citep{Hogan:1988mp}.

In the early universe, potentials with many minima are often considered to be candidates for inflationary models (see \citep{Martin:2013tda} for an exhaustive review). In the context of Type IIB string theories, the low energy effective theory can often be described by a potential landscape of many different minima, arising from the different choice of fluxes used for its compactification to 4 dimensions \citep{Feng:2000if}. In \citep{Wainwright:2014pta}, collisions between pairs of bubbles are studied in 1+1D, and the observables resulting from such a ``multiverse'' scenario are quantified, such that possible models may be constrained by observables. However, in the early universe, many such bubbles may have formed and collided simultaneously, leading to more random and asymmetric configurations. This work represents a first step towards understanding these more complex interactions. In this context, a single asymmetric bubble collapse can act as a simple model for the collapsing shapes formed when multiple spherical bubbles collide randomly and simultaneously. As will be discussed in our Conclusions in chapter 6, we will study such multiple bubble collisions in a future work. 

The chapter is organised as follows:
\begin{itemize}
\item In section \ref{sect:formcode}, we describe the methodology and formalism used in the 3+1 simulations. 
\item In section \ref{sect:1Dout}, we describe the spherically symmetric case, using both a 1+1D code and the full 3+1D simulations with GRChombo.
\item In section \ref{sect:3Dout}, we describe collapse in two different asymmetric cases. 
\item In section \ref{sect:discuss}, we discuss the results. Areas for further work will be suggested in Chapter 6.
\end{itemize}

The work presented in this Chapter is derived from the paper ``Critical Phenomena in Non-spherically Symmetric Scalar Bubble Collapse'' \citep{Clough:2016jmh}. We work in geometric units with $G=c=1$.

\section{Methodology}
\label{sect:formcode}

In this section we briefly describe the methodology and formalism used in the 3+1 simulations. 

\subsection{Gauge conditions}

\begin{figure}
\begin{center}
\includegraphics[width=0.7\textwidth]{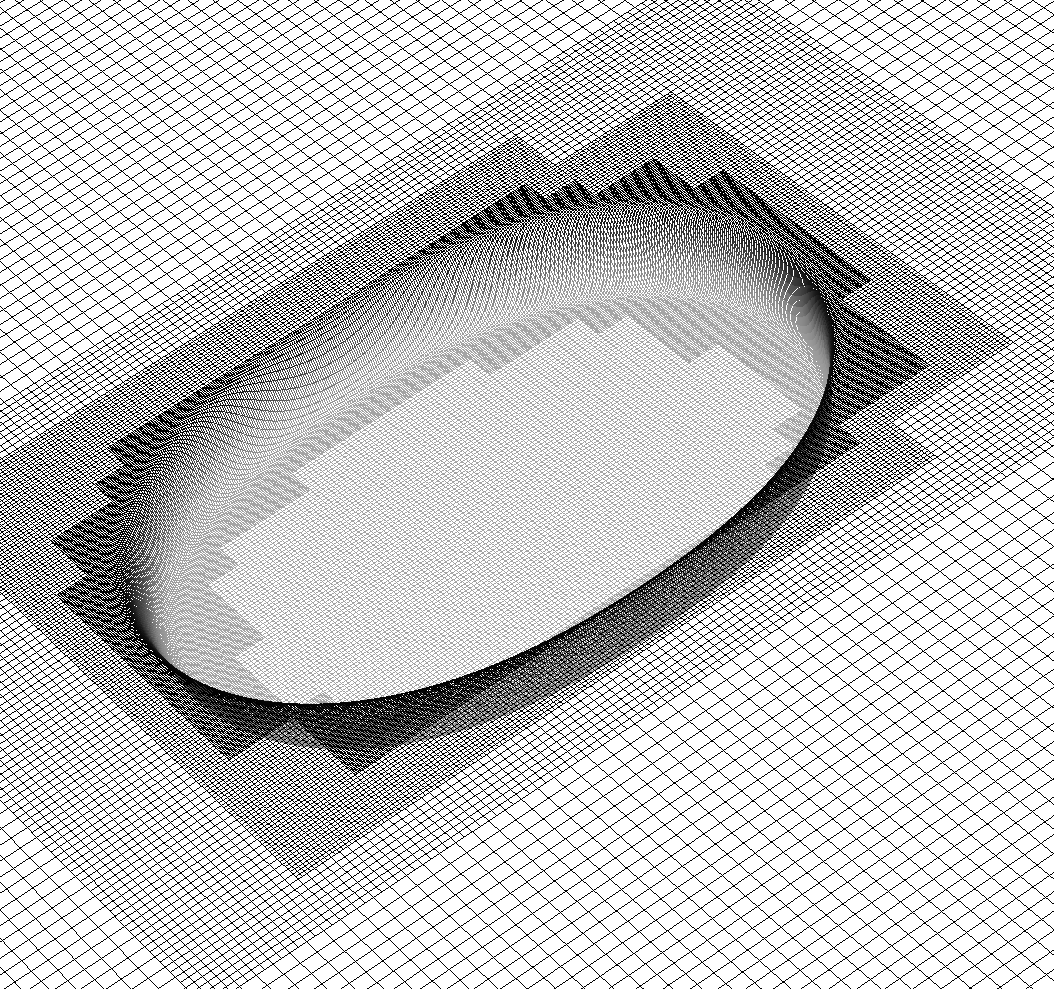}
\caption[Scalar field bubble, mesh]{The mesh for a scalar field bubble in GRChombo, which is adapted to give greatest curvature at the steep walls of the bubble. The image shows a slice through the x-z plane with the elevation corresponding to the value of the field $\phi$ at that point.
\label{fig-mesh}}
\end{center}
\end{figure}

A full description of the GRChombo code can be found in Chapter 3. As discussed, the code implements the BSSN formulation of the 3+1D decomposition of the Einstein equation, with full AMR. An illustration of the adaptive mesh responding to the curvature in an axisymmetric bubble is shown in Figure \ref{fig-mesh}.

In this work, a modification of the moving puncture gauge condition, see \citep{Campanelli:2005dd, Baker:2005vv}, was used in all the evolutions. It was found that having steep walled bubbles, and adding in a potential term, made the system of equations significantly more challenging to evolve stably near the critical point. After some experimentation with different gauges, the most suitable gauge was found to be given by the gamma driver condition for the shift,
\begin{eqnarray}
\partial_t \beta^i &=& B^i \, , \label{eqn:betadriver2} \\
\partial_t B^i &=& \partial_t \hat\Gamma^i-\eta_{\beta 2} B^i \, , \label{eqn:gammadriver2}
\end{eqnarray}
where $\eta_{\beta 2}$ is of order $1/2M$, where $M$ is the ADM mass of the initial spacetime, and the alpha driver condition for the lapse,
\begin{equation}
\partial_t \alpha = -\mu_{\alpha_1}\alpha^2 K + \beta^i\partial_i \alpha, \label{eqn:alphadriver2}\,
\end{equation}
with $\mu_{\alpha_1}$ of order 1, but set as described below. 

Note that this lapse condition would be harmonic slicing if coupled with a zero shift vector. It was found that using $\alpha^2$ as opposed to simply $\alpha$ (as in the standard puncture gauge) kept the lapse around ${\cal O}(1)$ during the final stages of the collapse and thereby increased stability.

As noted, it was non-trivial to find a stable gauge for evolutions near the critical point, possibly due to the additional non-linearity introduced by the self interaction potential, and a certain amount of trial and error was required. The key requirements for stability are listed below.

\begin{itemize}
\item The coefficients in the lapse condition were chosen so that the lapse remained between values of 0.1 and 1.0 until the field settled into a black hole or dispersed, as freezing of the coordinates or large forward steps resulted in instabilities developing. In the late stages of collapse, high scalar field gradients resulted in the lapse being driven near zero at the centre of the grid, whereas it needed to be free to oscillate further in order to prevent slice-stretching. The $\mu_{\alpha_1}$ parameter for the lapse condition was therefore chosen to keep the lapse of order one whilst the solution was still evolving (i.e. before an apparent horizon formed). This was found to be important for stability close to the critical point.
\item A non-zero, evolving shift was required in supercritical cases for the evolution to remain stable. The $\eta_{\beta 2}$ coefficient of the shift was of order $1/2M$ where $M$ is the ADM mass of the initial spacetime, but the stability of simulations was not particularly sensitive to its exact value. 
\item Some level of damping of high frequency numerical noise, such as by Kreiss-Oliger dissipation \citep{TUS:TUS1547}, was needed, but again the exact coefficient was not crucial to stability.
\end{itemize}

\subsection{Initial data}

In the spherically symmetric case the initial conditions were derived from a numerical Mathematica solution in the areal polar gauge. 

In asymmetric cases, we chose for the initial conditions a moment of time symmetry, such that $K_{ij} = 0$, and a conformally flat metric. The remaining degree of freedom, the conformal factor $\chi$, was solved for using a relaxation of the Hamiltonian constraint $\mathcal{H}$ over some chosen relaxation time, i.e.
\begin{eqnarray}
\partial_t \chi = C_R \mathcal{H} ,
\end{eqnarray}
with $C_R$ a user defined constant.

\subsection{Resolution and convergence}

The coarsest level of refinement was $dx=M$, with $M$ an arbitrary mass scale in the simulation (we worked in geometric units in which $G=c=1$). The physical domain was $(128M)^3$ and regridding was triggered by the change in $\phi$ or $\chi$ across the cell exceeding a certain empirically determined threshold (see section \ref{sec-BRAMR})). The maximum number of regriddings was 8, 9 or 10 depending on how close the simulation was to the critical point. In supercritical evolutions, the number of grid points across the event horizon needed to be, at minimum, around 20 in order for the horizon to be well resolved. A Courant factor of 0.25 was used in the fourth order Runge-Kutta update.

It was found that increasing the number of levels of refinement, and reducing the threshold at which regridding occurred, did not significantly change the mass of the black hole formed, so that the results had converged at the levels used. The black hole masses were measured from the area of the apparent horizon, once the black hole had settled into a roughly spherically symmetric configuration (for asymmetric initial conditions). At this point the majority of the scalar field radiation had either dispersed or fallen into the black hole, but the simulation had not yet run for long enough for boundary effects to contaminate the results.

The typical run time for a full collapse into a black hole was of the order of 24-36 hours when run on 256 cores. Closer to the critical point the simulations were more challenging and above 11 levels (10 regriddings above the coarsest level) the simulations became impractically slow and memory intensive with the resources we had available. However, there was in principle no barrier to increasing the levels of refinement further to study smaller mass black holes.

\section{Spherical symmetry}
\label{sect:1Dout}

\subsection{Spherical symmetry in 1+1D}

In preparation for the full 3+1D GR simulations, we simulated bubble collapses in spherical symmetry with a separate 1+1D numerical code, with an initial configuration chosen to be a moment of time symmetry with
\begin{equation}
\phi = \phi_m \tanh\left(k_{wall}(r-R_0)\right) \, , \quad \Pi=0 \, ,  \label{eqn:phiinit}
\end{equation}
where $k_{wall}$ defines the steepness of the wall between the two vacua and is set by the form of the potential in \ref{eqn:potential} as
\begin{equation}
k_{wall} = \sqrt{\frac{s}{2 \phi_m^2 \zeta}} \label{eqn:littlek} \, ,
\end{equation}
with $s=1$, $\phi_m=0.01$, and $\zeta = 10,000$. 

\nomenclature[a-pi]{$k_{wall}$}{in bubble collapse, the parameter defining the steepness of the bubble wall}
\nomenclature[a-pi]{$R_0$}{in bubble collapse, the initial radius of the bubble wall}

\begin{figure}
\begin{center}
\includegraphics[width=10cm]{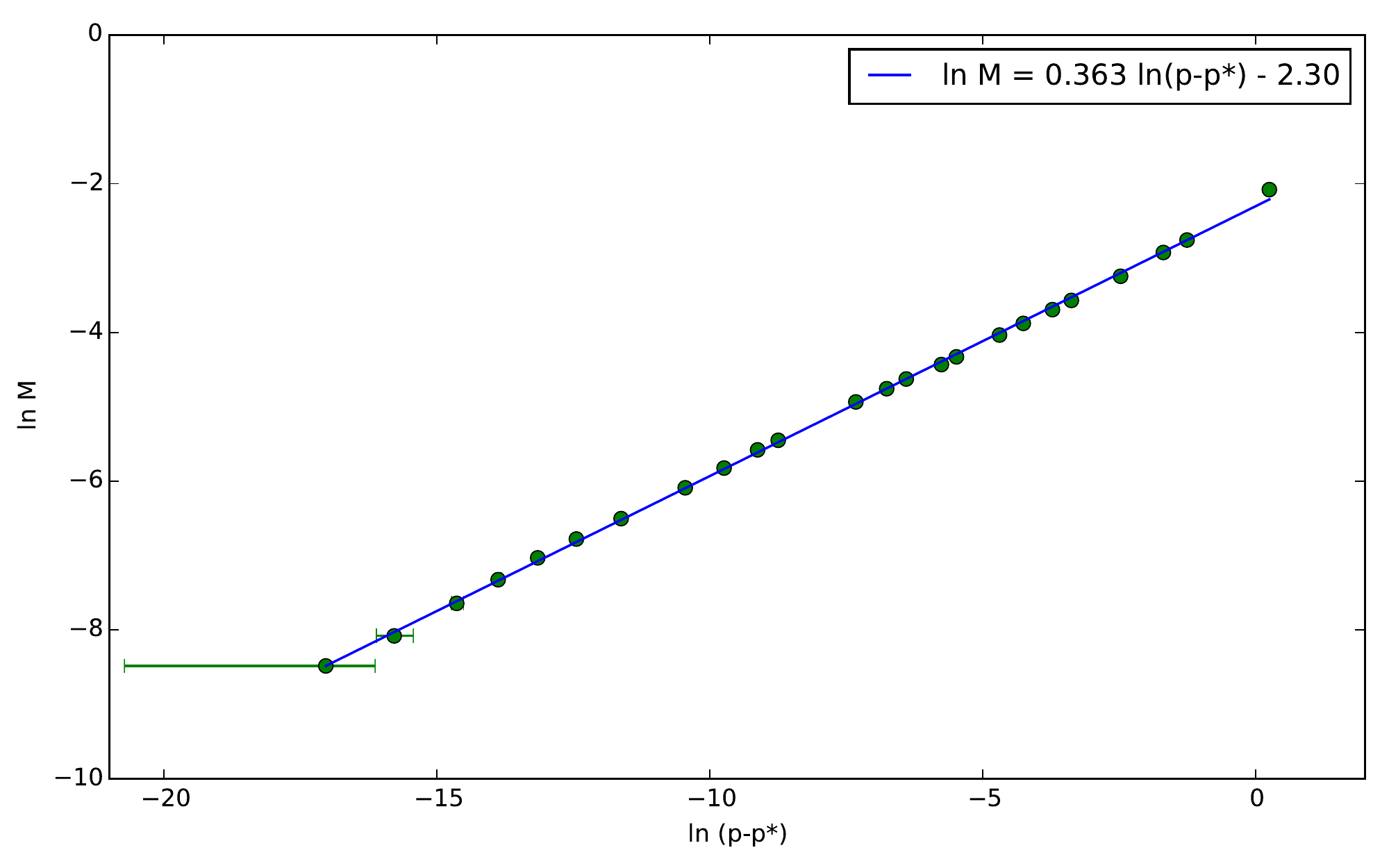}
\caption[1+1D critical collapse, scaling results]{Results from the simulations of a 1+1D spherically symmetric code. The critical index is $\gamma_S=0.363$, which is consistent with the value obtained in the full 3+1D simulations presented elsewhere.}
\label{fig-1Dcode}
\end{center}
\end{figure}

The value of the initial bubble radius $R_0$ was chosen as the critical parameter $p$, and was gradually increased from subcritical to supercritical. The masses of the resulting black holes were recorded.  The results are given in figure \ref{fig-1Dcode}. The critical index from these simulations is $\gamma_S=0.363$, consistent over a range of $\ln (R_0-R_0^*)$ from $0$ to $-15$. There is an uncertainty of between $+0.008$ and $-0.028$, which arises from considering the uncertainty in the critical point, as bounded by the smallest radius simulated for which a black hole formed, and the largest for which it did not. The quoted critical index is based on the critical value for which the residuals in the best fit line were minimised. There are hints of a periodic oscillation of the black hole masses as predicted in \citep{Gundlach:1996eg} but we did not pursue this further.
A technical difficulty encountered during the simulations was the rapid Lorentz contraction of the bubble wall during the collapse, which necessitated very high resolution meshes. Since most of the energy of the bubble wall is concentrated along a small region of the simulation domain, this posed a challenge even in 1+1D. We used a time-independent variable mesh 1+1D code (as opposed to general adaptive mesh) to push as close to the critical point as we could within reasonable expenditure of computational resources. This portended the difficulties we would face when we moved to full 3+1D simulations. 

\subsection{Spherical symmetry in 3+1D}

Initially, we repeated the $\zeta = 10,000$ tests undertaken with the 1+1D code in order to confirm that the 3+1D code gave consistent results.

We looked for a consistent scaling relation for black hole masses in supercritical collapses, and for evidence of scale echoing. For the latter, we observed the values of scale invariant quantities like $\chi$, $K$ and $\rho$ at the centre of the bubble in a slightly subcritical evolution, and how they evolved in proper time before the critical accumulation point was reached (the point at which the field began to disperse). We also looked at radial profiles of $dm/dr = 4\pi r^2 \rho$ at and around the critical time.

\begin{figure}
\begin{center}
\includegraphics[width=10cm]{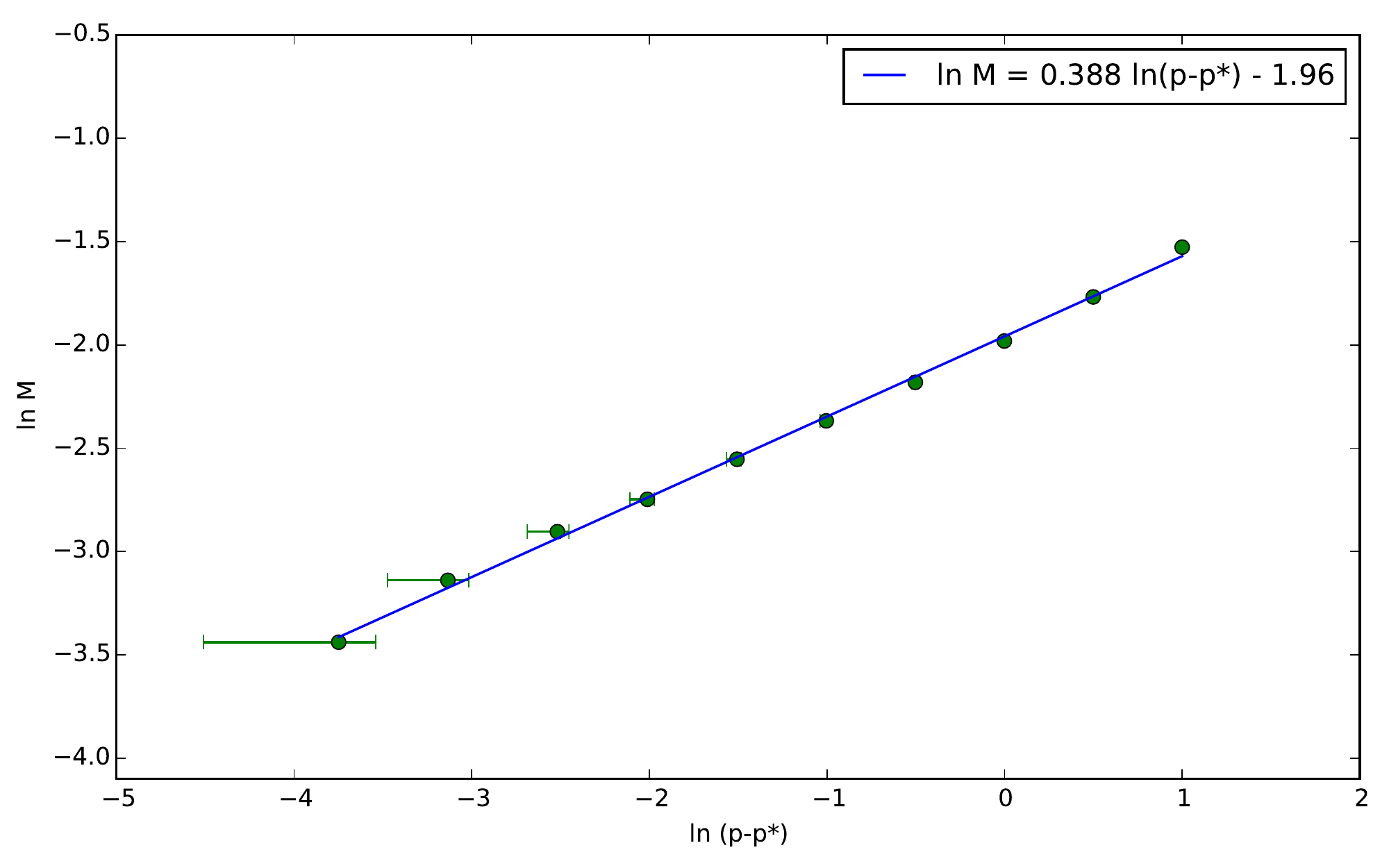}
\caption[3+1D spherically symmetric critical collapse, scaling results]{Plot of ln$(M_{BH})$ vs ln$(R-R^*)$ for the symmetric 3+1D bubble in a potential with $\zeta =$ 10,000. The horizontal error bars represent the possible error in the plotted values due to the uncertainty in the critical value of $p^*$. The best fit line is based on the critical value for which the residuals of the fit were minimised.
\label{fig-scale1}}
\end{center}
\end{figure}

We found that, as was expected, spherical bubbles subject to a double well potential showed the same critical phenomena as in the massless case and were consistent with our 1+1D simulations. Figure \ref{fig-scale1} shows a plot of the scaling relation between ln$(M_{BH})$ and ln$(R-R^*)$ which was obtained, with the best fit line giving a value for the critical exponent of $\gamma_S = 0.39$. There is an uncertainty of $+0.01$ and $-0.04$ which arises from considering the uncertainty in the critical point, as bounded by the smallest radius simulated for which a black hole formed, and the largest for which it did not. As in the 1D case, the quoted critical index is based on the critical value for which the residuals in the best fit line were minimised. 

\begin{figure}
\begin{center}
\includegraphics[width=15cm]{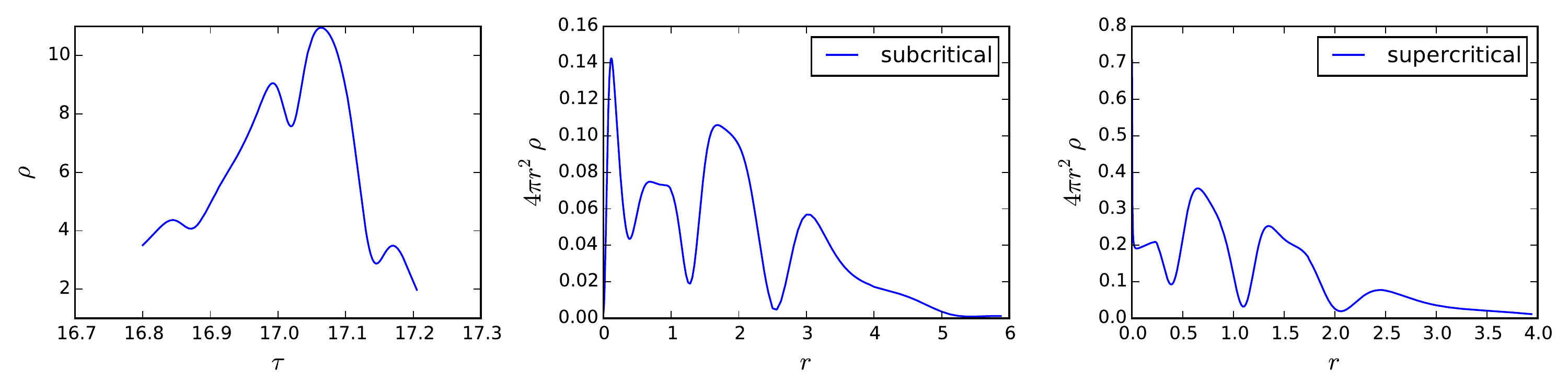}
\caption[3+1D spherically symmetric critical collapse, echoing results]{Plots of echoing for the symmetric bubble in a potential with $\zeta =$ 10,000. The first shows the evolution of the energy density $\rho$ against proper time $\tau$. If one applies a fit to $\tau^*$ and $\Delta_S$ using the relation in the echoing equation with respect to the extremal values in this plot one finds a value of $\Delta_S$ of approximately 0.7, which is inconsistent with the massless case value of 3.4 which is expected. The second and third plots show radial profiles of $dm/dr = 4\pi r^2 \rho$ in the marginally subcritical and supercritical cases simulated, and whilst there are hints of echoing, they are not conclusive.
\label{fig-echo1}}
\end{center}
\end{figure}

Figure \ref{fig-echo1} shows the echoing plots that were produced in near critical evolutions. As can be seen, there is some evidence for echoing, but insufficient to be conclusive. If one applies a fit to $\tau^*$ and $\Delta_S$ using the relation in Equation \ref{eqn:echo}, one finds a value of $\Delta_S$ of approximately 0.7, which is inconsistent with the massless case value of 3.4 which is expected. It is likely that we are still too far from the critical point to observe true echoing. It is also possible that, in the case of the radial profiles, the chosen coordinates are not well adapted to the echoing symmetries. 

\section{Beyond spherical symmetry - 3+1D simulations}
\label{sect:3Dout}
\subsection{Radial perturbations of a spherical bubble - axisymmetric bubbles}

\begin{figure}
\begin{center}
\includegraphics[width=10cm]{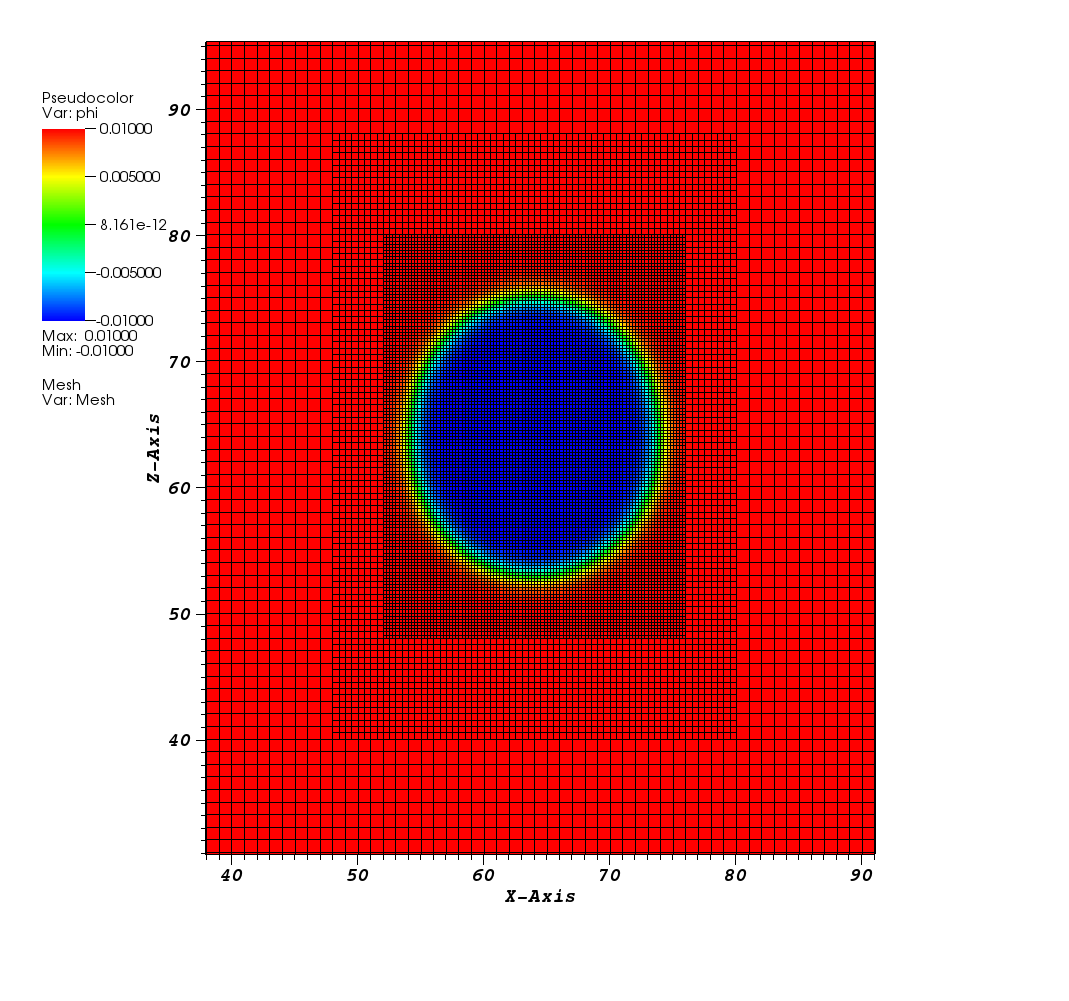}
\caption[3+1D axisymmetric critical collapse]{Initial configurations for the axisymmetric radially perturbed bubble case, showing the adaptive mesh. The image shows a slice through the $x-z$ plane with the colour corresponding to the value of $\phi$ at that point per the legend.
\label{fig-bubaxi}}
\end{center}
\end{figure}

\begin{figure}
\begin{center}
\includegraphics[width=10cm]{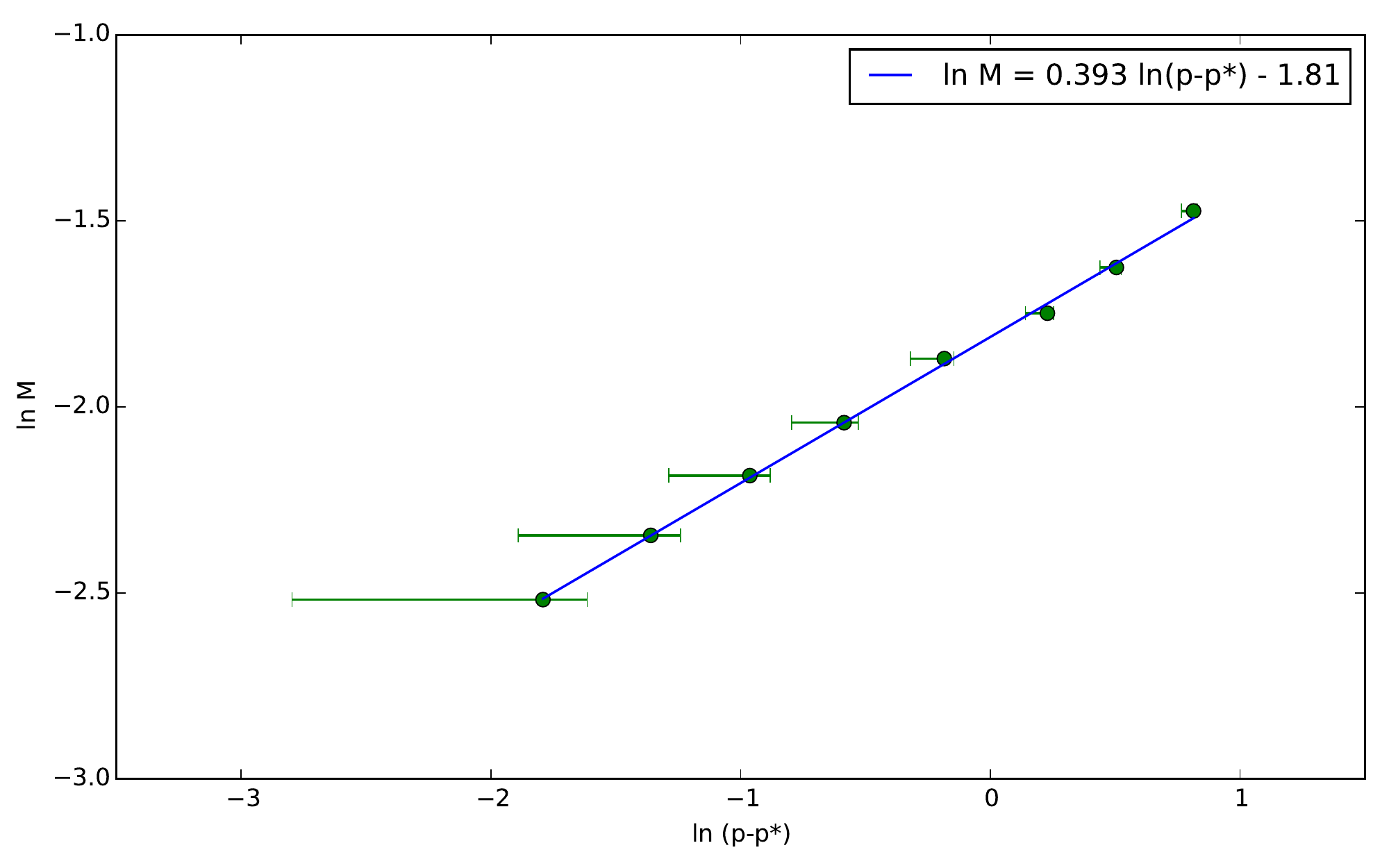}
\caption[3+1D axisymmetric critical collapse, scaling results]{Plot of ln$(M_{BH})$ vs ln$(R_0-R_0^*)$ for the axisymmetric, radially perturbed bubble in a potential with $\zeta =$ 5,000. The horizontal error bars represent the possible error in the plotted values due to the uncertainty in the critical value of $p^*$. The best fit line is based on the critical value for which the residuals of the fit were minimised.
\label{fig-scale2}}
\end{center}
\end{figure}

We set up bubbles in a $s=1$, $\phi_m=0.01$, and $\zeta = 5,000$ potential for which the radius of the bubble wall was perturbed by the $Y_{1,0}$ spherical harmonic, such that the initial configuration for the field is (see Figure \ref{fig-bubaxi}))
\begin{equation}
R_{asym} = \left( 1+\epsilon_{asym} | Re(Y_{1,0}) |^2 \right) R_0 \, ,
\end{equation}
\begin{equation}
\phi = \phi_m \tanh \left[k_{wall}(r-R_{asym})\right] \, , \label{eqn:phiinit2}
\end{equation}
with $k_{wall}$ as in equation \ref{eqn:littlek}, and initially static, i.e. 
\begin{equation}
\Pi = 0 \, .
\end{equation}
We considered the case in which $\epsilon_{asym} = 0.5$, and again, we varied the initial radius of the bubbles $R_0$ until we had bounded the critical point. We investigated the scaling relations in the black hole masses which resulted in supercritical evolutions, and scale echoing in evolutions above and below the critical point. 
We found evidence for critical phenomena consistent with that in the massless and spherically symmetric cases. Figure \ref{fig-scale2} shows a plot of the scaling relation between ln$(M_{BH})$ and ln$(R_0-R_0^*)$ which was obtained, with the best fit line giving a value for the critical exponent of $\gamma_S = 0.39$, using the method described in spherical symmetry above. This value is consistent with the spherically symmetric case. The uncertainty in this case was between $+0.02$ and $-0.10$. 

\begin{figure}
\begin{center}
\includegraphics[width=15cm]{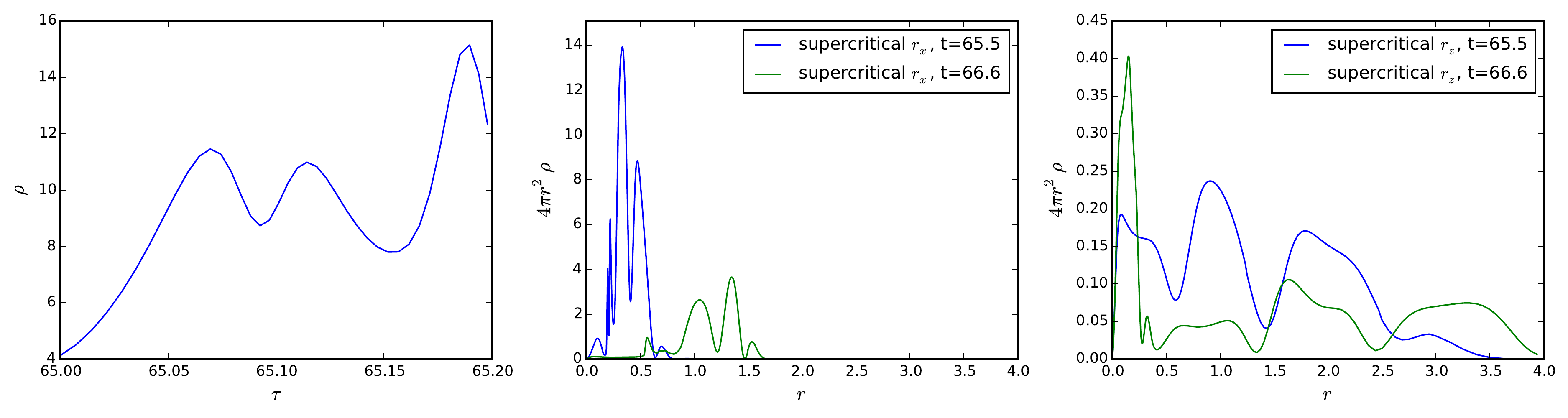}
\caption[3+1D axisymmetric critical collapse, echoing results]{Plots of echoing for the axisymmetric, radially perturbed bubble in a potential with $\zeta =$ 5,000. The first graph shows the evolution of the energy density $\rho$ against proper time $\tau$. The peaks are not consistent with the relation in the echoing equation since the proper time between the peaks appears to increase as the critical time is approached. The second plot shows radial profiles of $4\pi r^2 \rho$ along the z axis in a subcritical case, at two times. The third shows the same profiles in the x axis, perpendicular to the axis of symmetry.
\label{fig-echo2}}
\end{center}
\end{figure}

Figure \ref{fig-echo2} shows the echoing plots that were produced. Again, there is some evidence for echoing, but insufficient to be conclusive, for the reasons discussed above.

\begin{figure}
\begin{center}
\includegraphics[width=10cm]{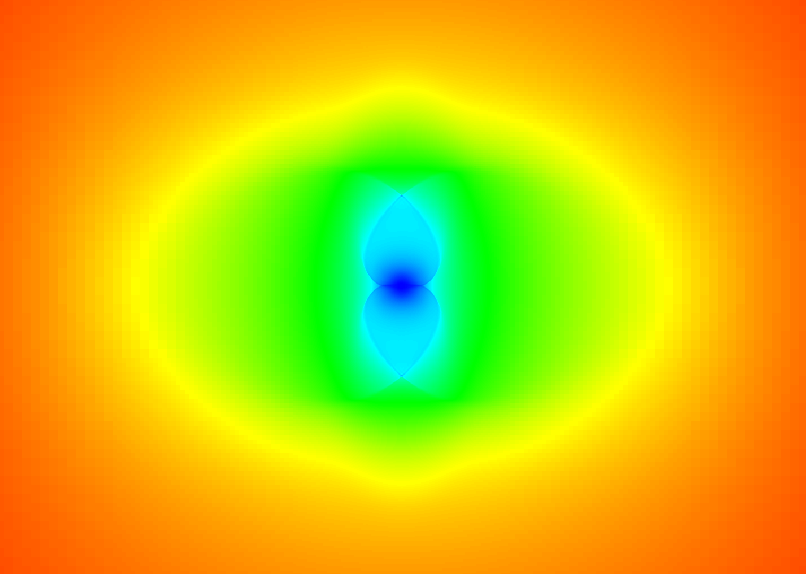}
\caption[3+1D axisymmetric critical collapse, shock fronts]{We find that ``shock waves'' (blue regions transitioning into green) develop in some parameters in the axisymmetric case close to the critical point, particularly in $\alpha$, the lapse parameter, which is shown here, with very steep gradients and asymmetric configurations. The blue parts correspond to small values of the lapse (the lapse has ``collapsed''), whilst the red parts to $\alpha=1$.
\label{fig-shock}}
\end{center}
\end{figure}

Although we saw no definitive evidence in the simulations for other growing asymmetric modes, it seemed clear that the behaviour was becoming more strongly asymmetric as the critical radius was approached. The simulations became extremely challenging and difficult to evolve as the critical point was neared, and we saw that asymmetric ``shock waves'' developed in some parameters, particularly in the lapse, with extremely steep gradients and clearly non-spherical forms, as shown in figure \ref{fig-shock}. To enable us to probe this asymmetric behaviour further, we are considering amending the gauge conditions further to ``smooth'' the lapse, and perhaps to implement some of the shock avoidance techniques applied in \citep{Figueras:2015hkb}.

\subsection{Amplitude perturbations of a spherical bubble - asymmetric bubbles}

\begin{figure}
\begin{center}
\includegraphics[width=10cm]{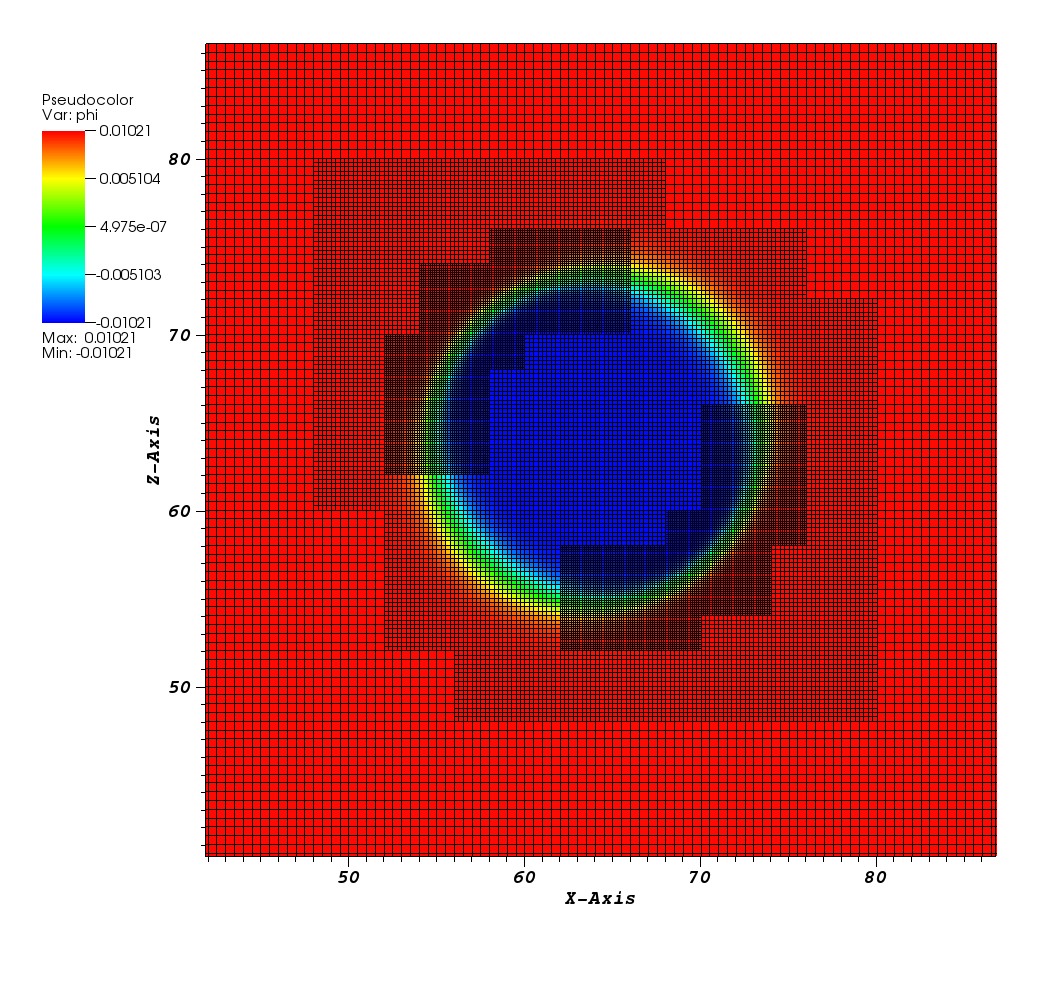}
\caption[3+1D asymmetric critical collapse]{Initial configurations for the asymmetric amplitude perturbed bubble case, showing the adaptive mesh. The image shows a slice through the x-z plane with the colour corresponding to the value of $\phi$ at that point per the legend.
\label{fig-bubasym}}
\end{center}
\end{figure}

We set up bubbles in a $s=1$, $\phi_m=0.01$, and $\zeta = 5,000$  potential for which the amplitude of the bubble at the wall was perturbed by the $Y_{2,1}$ spherical harmonic, which was similar to the configuration studied in the massless case by Healy et. al. \citep{Healy:2013xia}. The initial configuration for the field is (see Figure \ref{fig-bubasym})
\begin{equation}
\phi = \left(1+\epsilon_{asym} Re(Y_{2,1})  e^{\frac{r-R_0}{\sigma}}\right)\phi_m \tanh\left[k_{wall}(R-R_0)\right], \label{eqn:phiinit3}
\end{equation}
with $k_{wall}$ as in equation \ref{eqn:littlek}, $\sigma=0.5$, and
\begin{equation}
\Pi = 0.
\end{equation}
We considered the case $\epsilon_{asym} = 1$, and again, we varied the initial radius of the bubbles $R_0$ until we had bounded the critical point above and below. We investigated the scaling relations in the black hole masses which resulted, and evidence for scale echoing. 

\begin{figure}
\begin{center}
\includegraphics[width=10cm]{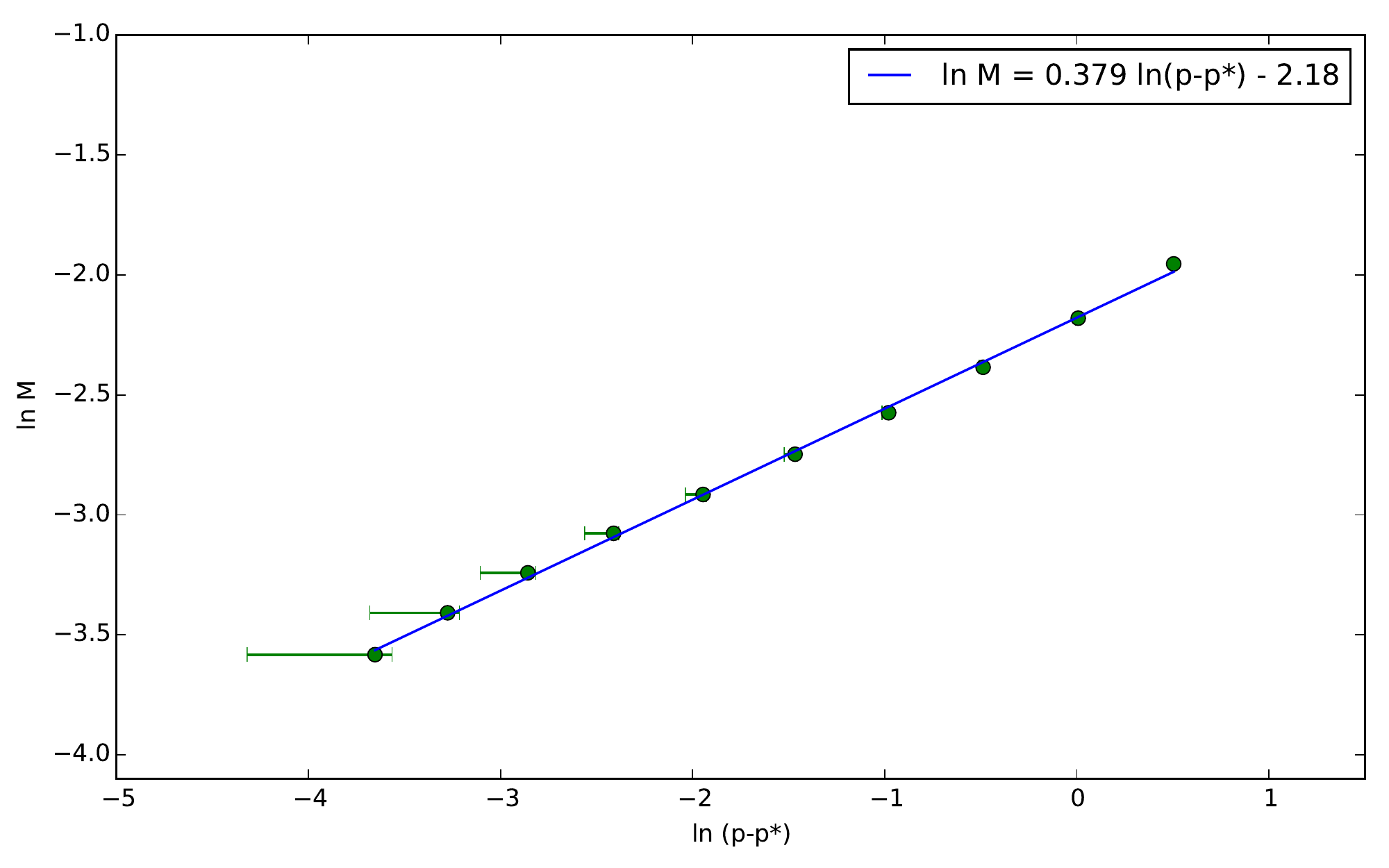}
\caption[3+1D asymmetric critical collapse, scaling results]{Plot of ln$(M_{BH})$ vs ln$(R -R_0^*)$ for the asymmetric amplitude perturbed bubble in a potential with $\zeta =$ 5,000. The horizontal error bars represent the possible error in the plotted values due to the uncertainty in the critical value of $p^*$. The best fit line is based on the critical value for which the residuals of the fit were minimised.
\label{fig-scale3}}
\end{center}
\end{figure}

\begin{figure}
\begin{center}
\includegraphics[width=15cm]{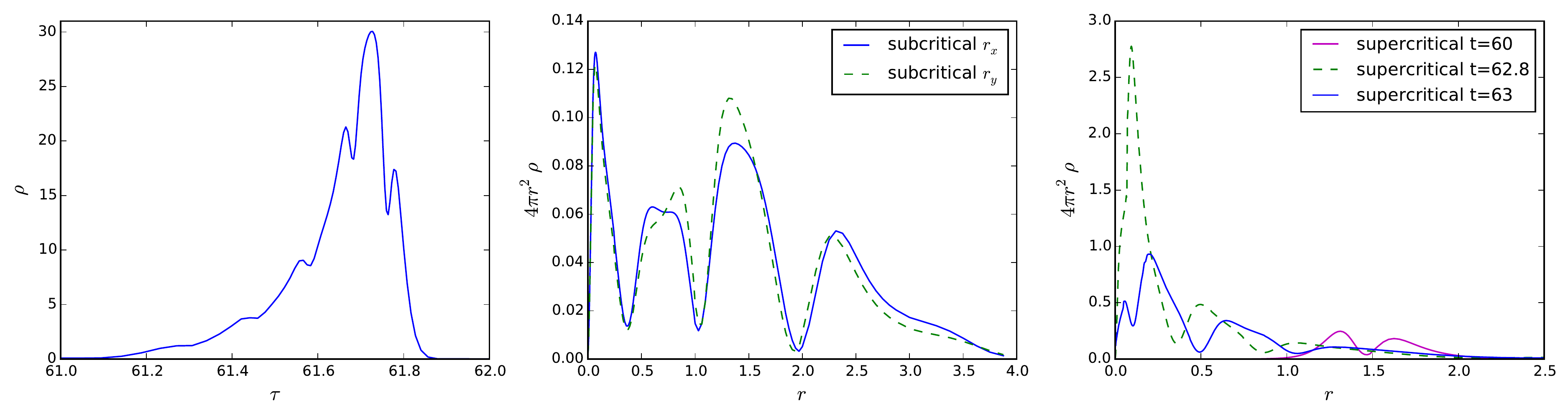}
\caption[3+1D asymmetric critical collapse, echoing results]{Plots of echoing for the asymmetric amplitude perturbed bubble in a potential with $\zeta =$ 5,000. The first graph shows the evolution of the energy density $\rho$ against proper time $\tau$. If one applies a fit to $\tau^*$ and $\Delta_S$ using the relation in the echoing equation with respect to the extremal values in this plot one finds a value of $\Delta_S$ of approximately 0.7, which is inconsistent with the massless case value of 3.4 which is expected. The second plot shows radial profiles of $4\pi r^2 \rho$ along and perpendicular to the z axis in a subcritical case. The third shows the z axis profile in a supercritical case at several times close to the critical time. Whilst there are hints of echoing in these plots, they are not conclusive.
\label{fig-echo3}}
\end{center}
\end{figure}

We found the same critical phenomena as in the symmetric and axisymmetric cases. Figure \ref{fig-scale3} shows a plot of the scaling relation between ln$(M_{BH})$ and ln$(R_0-R_0^*)$. This was obtained, with the best fit line which minimises the residuals giving a value for the critical exponent of $\gamma_S = 0.38$, with an uncertainty of $+0.01$ and $-0.05$.
Figure \ref{fig-echo3} shows the echoing plots that were produced in near critical evolutions. As above, there is some evidence for echoing, but insufficient to be conclusive. 

Interestingly, although the initial asymmetry was larger in this case than in the axisymmetric case, there was much less evidence of any growing asymmetry near the critical point, and in fact the asymmetry appeared to decay during collapse even as the critical point was neared. This is consistent to the findings of Choptuik et. al. in \citep{Choptuik:2003ac} in the massless axisymmetric case, where they also found that radial perturbations produced more asymmetry than amplitude perturbations. Thus there is not a universality in the parameter to which the asymmetry is applied - some parameters produce more asymmetry than others, and future efforts to find asymmetric growing modes are best focussed in these areas.

\section{Discussion}
\label{sect:discuss}

We have shown that in spherical symmetry, bubble collapse behaviour with a non-trivial self interaction shows the same critical scaling behaviour as in the massless case.

In the axisymmetric and asymmetric cases studied, in which the radius of the initial bubble and its amplitude, respectively, were perturbed, we see a scaling relation which is again consistent with the massless spherically symmetric case. This seems to imply that, at least for the cases we considered, even in the presence of fairly significant initial asymmetry in the configuration, the final state of the collapse is dependent on a single dominant mode. 

There are some hints of echoing, but insufficient to be conclusive. It is likely that we are still too far from the critical point to observe echoing. It is also possible that the chosen coordinates are not well adapted to the symmetries of the echoing.

The case in which the asymmetry was introduced via the radial perturbations appeared to produce greater asymmetry near the critical point than the case in which the amplitude of the bubbles was varied. This is then a better candidate for future investigations to look for growing but subdominant asymmetric modes beyond spherical symmetry. 

%% file: Chapter6/chapter6.tex
\chapter{Conclusions and further work}
\label{ch-Conclusions}

\ifpdf
    \graphicspath{{Chapter6/Figs/Raster/}{Chapter6/Figs/PDF/}{Chapter6/Figs/}}
\else
    \graphicspath{{Chapter6/Figs/Vector/}{Chapter6/Figs/}}
\fi

In this thesis, we introduced and described $\grchombo$, a new multi-purpose numerical relativity code built using the {\tt Chombo} framework. It is a $3+1$D finite difference code based on the BSSN/CCZ4 evolution scheme. It supports Berger-Collela type AMR evolution with Berger-Rigoutsos block structured grid generation and is fully parallelised via the Message Passing Interface. Time evolution is via standard 4th order Runge-Kutta time-stepping.

We showed that $\grchombo$ successfully passes the standard ``Apples with Apples'' tests, evolved standard single black hole spacetimes (Schwarzschild and Kerr) and showed that they are stable to more than $T =10,000M$. Using the moving puncture gauge, we also show that $\grchombo$ stably evolves the merger of two and three black holes in inspiral and head-on collisions. We simulated the supercritical collapse of a scalar field configuration and found that it forms a black hole, as expected, to show that the code supports non vacuum spacetimes. Finally we tested the MPI scaling properties of the code, both strongly and weakly, and compared this with an alternative numerical relativity code based on the popular Cactus framework.

Many areas of physics can potentially benefit from this code, such as multiple black hole mergers and scalar field collapse. Such fields require a code which adapts to changes in the range and location of scales at different points in space and time in the simulation. We emphasise that setting the initial conditions for these mergers are trivial -- $\grchombo$ automatically remeshes the grid given a set of analytic initial conditions without further user intervention.

In this thesis two applications were considered: Inhomogeneous inflation and critical bubble collapse.

In the first, we investigated the robustness of small and large field models of inflation, subjecting it to several simple inhomogeneous initial conditions both in the scalar field profile and in the extrinsic curvature. In doing so we have set up a framework that will allow us to study more general initial conditions in the future. As expected, we found that large field inflation was far more robust than small field inflation. In particular, small field inflation can fail even for small subdominant gradient energies $\rho_{\mathrm{grad}}/\rho_{V} \approx 10^{-4}$ while large field inflation is robust even to dominant gradient energies of $\rho_{\mathrm{grad}}/\rho_{V}  \gg 1$. This implies that small field inflation requires at least some level of tuning to begin or a dynamical mechanism that sets up appropriate initial conditions. A full summary of findings is given in section \ref{sect:conclusions}.

In the second, we showed that in spherical symmetry, bubble collapse behaviour with a non-trivial self interaction shows the same critical behaviour as in the massless case. In the axisymmetric and asymmetric cases studied, in which the radius of the initial bubble and its amplitude respectively were perturbed, we see a scaling relation that is again consistent with the massless spherically symmetric case. This implies that, at least for the cases we considered, even in the presence of fairly significant initial asymmetry in the configuration, the final state of the collapse is dependent on a single dominant mode. There were some hints of echoing, but insufficient to be conclusive. It is likely that we are still too far from the critical point to observe echoing. It is also possible that the chosen coordinates are not well adapted to the echoing. 

In the following sections, directions for further research will be proposed. The first three relate directly to the work presented in this thesis. The final section considers a separate topic which is also under investigation.

\section{Development of GRChombo}

Despite its power, the AMR capability of $\grchombo$ has to be treated with care. As we mentioned earlier, coarse-fine boundaries can be a significant source of inaccuracy, even though the Hamiltonian constraint may still be kept under control. We wish to investigate how well angular momentum is conserved during an evolution, which is known to be a problem in cartesian codes. This will be particularly important for the investigation of rotating systems in future.

There are also several code development projects which are work in progress by the $\grchombo$ collaboration.

We are currently in the process of rewriting $\grchombo$ in a more modular way, so as to hide more of the $\mathtt{Chombo}$ functionality from the user, and make the NR sections of the code more reusable in different applications. In addition, sections have been rewritten entirely in C++, with no Fortran calls, which makes the code more readable, and enables support for vectorisation of the data updates (so that several gridpoints can be processed simultaneously).

In addition to rewriting the existing matter code in the new format, it would be useful to add other types of matter, such as fluid matter, and ultimately an MHD module. We would like to introduce more general boundary conditions, in particular ones which better damp outgoing radiation and reduce reflections, allowing us to undertake longer runs without reflections contaminating the results. We plan to write a general initial condition solver, using the $\mathtt{chombo}$ multigrid Poisson solver, to allow us to consider more general, non symmetric initial data. A non spherically symmetric apparent horizon finder would allow us to measure masses of black holes without waiting for them to settle into spherical symmetry. 

The ultimate aim is to release the code to the public and, with this in mind, documentation (a wiki, and doxygen manual) is under development.

\section{Inhomogeneous inflation}

Our work on inflation was only a starting point for a wider investigation of more general initial conditions. We investigated a very restrictive class of initial conditions; only one or two modes of horizon scale, with significant symmetries still assumed in the metric components. 

The work can easily be extended by relaxing some of these assumptions. First, we wish to consider the effect of smaller wavelengths of the scalar field fluctuations, building up to a scale invariant spectrum of perturbations. In our results adding modes seemed to increase robustness, and it would be interesting to confirm that this trend continues for more than two modes and over a range of scales.

The initial metric can be made more general by breaking isotropy - i.e. by having a non conformally flat metric. In addition, we can introduce a non zero traceless part of the extrinsic curvature by decomposing it, as in the CTT and CTS decompositions, and specifying particular components.

Ideally, we might like to generate a range of simulations with randomly generated initial data (which still satisfies the constraints). For this we require the general initial condition solver for $\grchombo$; which is work in progress, as above.

Furthermore, we focused on the case of single field inflation. It will be interesting to study whether the presence of additional degrees of freedom renders inflation more or less robust to inhomogeneities.  We will pursue these and other questions in future work.

\section{Critical collapse in asymmetry}

The study of the axisymmetric radially perturbed case merits further study, as there seemed to be evidence of asymmetric modes near the critical point. As was mentioned previously, linear perturbations of the spherically symmetric case \citep{PhysRevD.59.064031} do not show additional unstable modes, but numerical studies such as that of Choptuik et al. \citep{Choptuik:2003ac} gave hints of further unstable modes in the full non-linear regime. Given the additional challenges introduced by the bubble case (the steepness of the wall at collapse), it would be useful initially to take a step back and first recreate the results of Choptuik in \citep{Choptuik:2003ac} in the massless case. Choptuik's results are still unconfirmed by any other simulations and it is not certain that they do not result from numerical sources, therefore there is strong interest in the community in confirming them with an independent code.

It seems reasonable to assume that in strongly asymmetric cases the spherical symmetry will no longer dominate, and the scaling relation will eventually break down. For example, if we set up an ellipsoidal shaped bubble,  then one can imagine that the bubble can collapse into two black holes if one of the semi-principal axes is much longer than the other two (i.e. a ``long sausage''). Nevertheless, these two black holes will eventually collide and merge to form a single black hole as the final state. It is an interesting question to ask whether this final state black hole will still follow the scaling relation. 

It is hoped that with the new version of the $\grchombo$ code, combined with a greater appreciation of efficiencies in regridding\footnote{Use of $K$ as the regridding trigger is more efficient than $\chi$, which was previously used, and we have found that we can reduce the regridding frequency significantly and still maintain good resolution.}, we can get sufficiently close to the critical point to properly observe echoing. As mentioned, we intend to test this first in the less challenging massless case, before returning to bubbles in a multi minima potential. This is ongoing work.

A further related area is the study of bubble collisions, which was the motivation for the original work. We would like to investigate whether collision of bubble walls can create black holes via non trivial self interactions, and the gravitational wave signals which are emitted in multiple collisions.

\section{Axion stars}

Axions are pseudo-Goldstone bosons of spontaneously broken global $U(1)$ ``Peccei-Quinn'' (PQ) symmetries~\citep{pecceiquinn1977}. The complex PQ-field, $\varphi$, has the potential
\begin{equation}
V(\varphi) = \lambda_\varphi \left(|\varphi|^2-\frac{f_a^2}{2}\right)^2 \, .  \label{eqn:U1pot}
\end{equation}
The $U(1)_{\rm PQ}$ symmetry is broken at some constant scale $f_a$. After symmetry breaking, writing the PQ field as $\varphi=(\varrho/\sqrt{2}) e^{i\phi/f_a}$, the radial field $\varrho$ acquires a vacuum expectation value such that: $\langle\varphi\rangle=(f_a/\sqrt{2})e^{i\phi/f_a}$. The angular degree of freedom, the axion $\phi$, is the Goldstone boson of the broken symmetry. 

\nomenclature[g-pi]{$\varphi$}{the complex PQ-field (for axions)}
\nomenclature[a-pi]{$f_a$}{the symmetry breaking scale for the PQ-field (for axions)}
\nomenclature[a-pi]{$m_a$}{the axion mass defined from the potential as $\Lambda_a^2/f_a$}
\nomenclature[g-pi]{$\Lambda_a$}{parameter used in defining the shape of the axion potential}

As a Goldstone boson, the axion enjoys a shift symmetry, i.e. the action contains only terms in $\partial_\mu\phi$ and there is a symmetry under $\phi\rightarrow\phi+c$ for any real number $c$. In general, this shift symmetry is anomalous and is broken to a discrete symmetry, $\phi\rightarrow\phi+2\pi n$ for some integer $n$. The breaking of the axion shift symmetry selects a particular direction in the field space $\varphi=\varphi_1+i\varphi_2$. In the potential we can write this as
\begin{equation}
V(\varphi) = \lambda_\varphi \left(|\varphi|^2-\frac{f_a^2}{2}\right)^2+\epsilon_{PQ}\varphi_1 \, ,
\end{equation}
for some parameter $\epsilon_{PQ}$ of mass dimension three, which is ``small'' in the sense that $\epsilon_{PQ}/f_a^3\ll 1$. In some limits, we can ignore the radial mode and consider simply a periodic potential for the axion
\begin{equation}
V(\phi)=\Lambda_a^4\left[1-\cos\left(\frac{\phi}{f_a}\right)\right] \equiv m_a^2f_a^2\left[1-\cos\left(\frac{\phi}{f_a}\right)\right] \, ,  \label{eqn:U2pot}
\end{equation}
from which we find that $\epsilon_{PQ} = \sqrt{2} m_a^2 f_a$. The minimum of the potential at $\phi=0$ is used to define the ``axion mass'', $m_a=\Lambda_a^2/f_a$. Non-perturbative effects generally switch on at scales far below the fundamental scale, while we expect $f_a$ to be of the order of the fundamental scale. Thus axions are naturally extremely light via the seesaw mechanism as long as the shift symmetry breaking is small, $\epsilon/f_a^3 =\sqrt{2} (m_a/f_a)^2 \ll 1$. The axion is also hierarchically lighter than the radial field, $\varrho$. Due to the hierarchy of scales between the axion mass and the radial mode, it is possible to simulate the axion field as a real valued scalar field in a cosine potential. 

\begin{figure}
\begin{center}
\includegraphics[width=.8\textwidth]{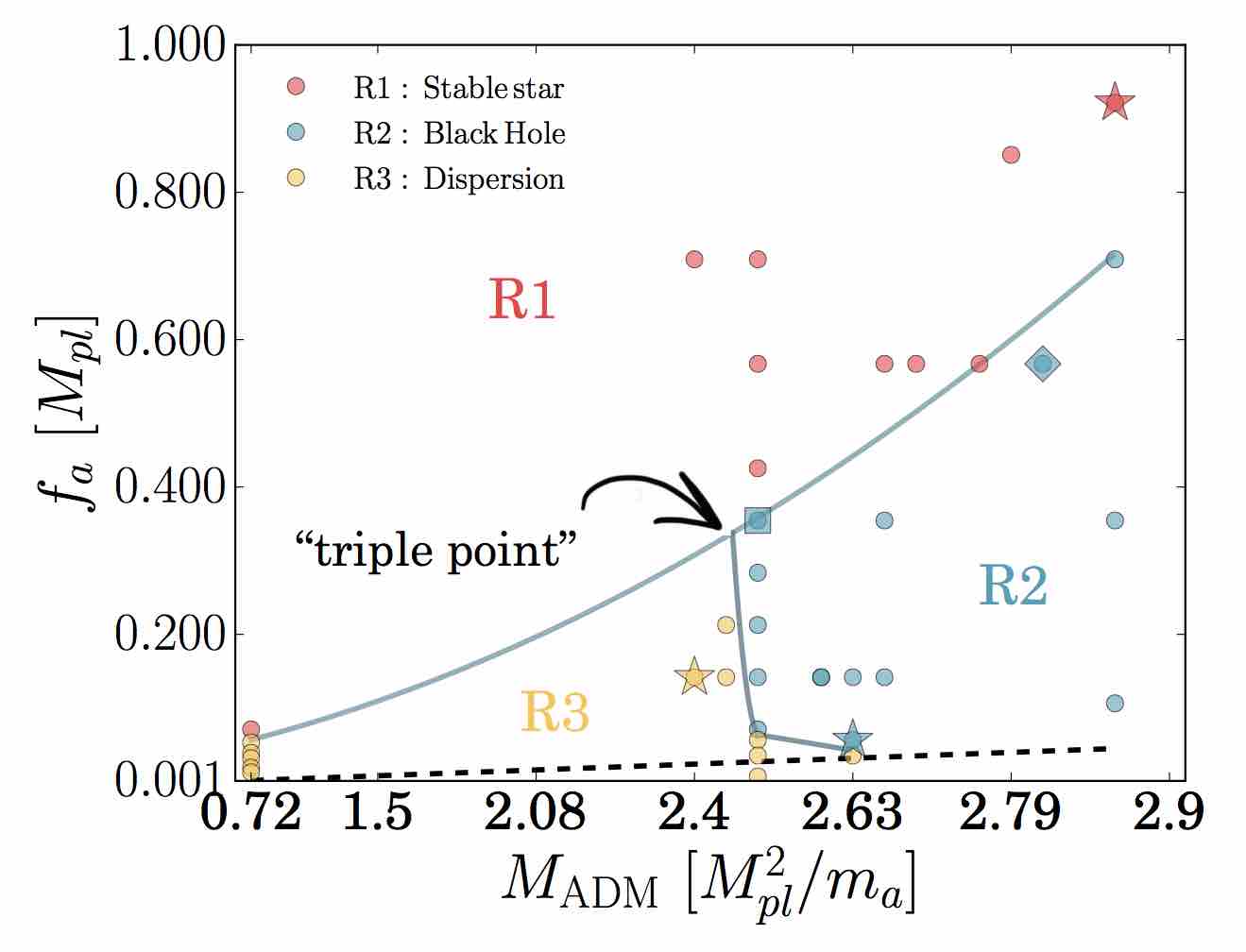}
\caption[The axion star stability diagram]{The stability diagram is parameterised by the axion decay constant, $f_a$, and the initial condition $M_{\rm ADM}$, the initial mass of the axion star (which we set using the initial field velocity, $\Pi$, at the centre). Solid lines mark the approximate boundaries between three regions of the axion star parameter space: quasi-stability (R1), collapse to a BH (R2), and dispersal (R3). We postulate the existence of a ``triple point'' between these regions. The dashed line marks the region below which axion mass is effectively negligible. Simulated axion stars are marked as circles; other symbols mark points explored in more detail. Below the triple point, for $f_a\ll M_{pl}$, under an increase in mass, dispersal of the star via winding of the axion field occurs before collapse to a BH. Above the triple point, stable axion stars can collapse to BHs by acquiring mass e.g. by accretion.}
\label{fig-money_plot}
\end{center}
\end{figure}

The classical equations of motion for an axion with potential given by \eqn{eqn:U2pot} possess quasi-stable, localised, oscillating solutions, which are sometimes referred to as ``axion stars''. In the paper \citep{Helfer:2016ljl}, the author and collaborators studied, for the first time, collapse of axion stars numerically using the full non-linear Einstein equations of general relativity and the full non-perturbative cosine potential. Regions were mapped on an ``axion star stability diagram'', parameterised by the initial ADM mass, $M_{\rm ADM}$, and axion decay constant, $f_a$. Three regions of the parameter space were identified:
\begin{enumerate}
\item{Long-lived oscillating axion star solutions, with a base frequency, $m_a$, modulated by self-interactions}
\item{Collapse to a black hole}
\item{Complete dispersal due to gravitational cooling and interactions}
\end{enumerate}
We located the boundaries of these three regions and an approximate ``triple point'' $(M_{\rm TP},f_{\rm TP})\sim (2.4 M_{pl}^2/m_a,0.3 M_{pl})$. See figure \ref{fig-money_plot}.

\begin{figure}
\begin{center}
\includegraphics[width=.8\textwidth]{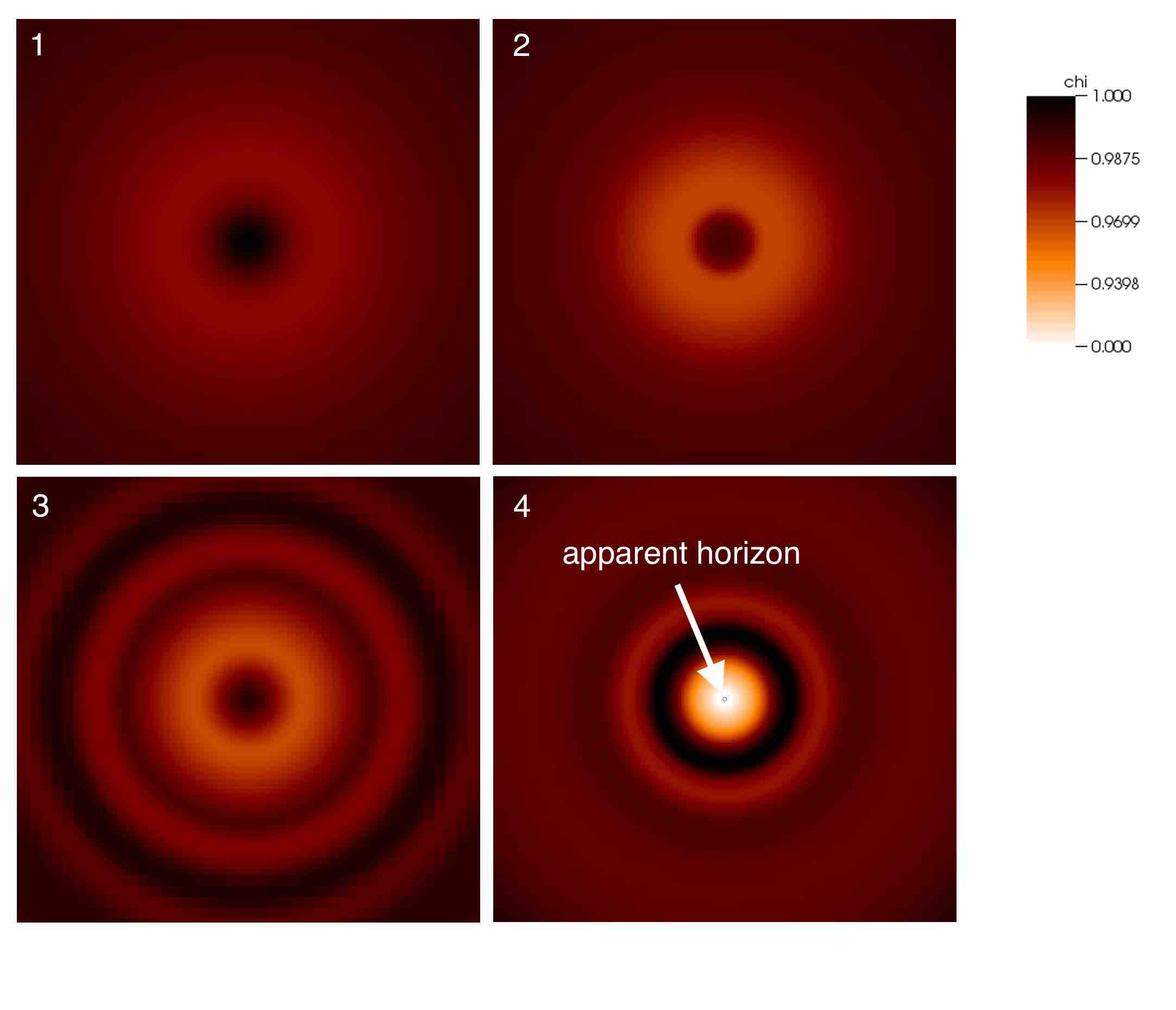}
\caption[Black hole formation near the Axion star dispersal region]{$(M_{\rm ADM},f_a)=(2.63,0.055)$, R2 star in Fig.~\ref{fig-money_plot}. The parameter shown is the conformal factor of the metric, $\chi$. The first panel shows the initial data, and subsequent panels show the evolution. The initial state is spatially extended, with low curvature. As collapse proceeds, the axion star becomes smaller, and some scalar radiation is emitted in waves. The final BH has an apparent horizon that is very small compared to the initial axion star size. Some of the axion field remains outside the BH, and gravitationally bound to it, in ``scalar wigs''.} 
\label{fig-BlackHoleFormation}
\end{center}
 \end{figure}

For $f_a$ below the triple point BH formation proceeds during winding (in the complex $U(1)$ picture) of the axion field near the dispersal phase. This could prevent astrophysical BH formation from axion stars with $f_a\ll M_{pl}$. For larger $f_a\gtrsim f_{\rm TP}$, BH formation occurs through the stable branch and we estimate the mass ratio of the BH to the stable state at the phase boundary to be $\mathcal{O}(1)$ within numerical uncertainty, see figure \ref{fig-BlackHoleFormation}. Our findings have observational relevance for axion stars as BH seeds, which are supermassive in the case of ultralight axions. For the QCD axion, the typical BH mass formed from axion star collapse is $M_{\rm BH}\sim 3.4 (f_a/0.6 M_{pl})^{1.2} M_\odot$.

Much work remains to be done in this area. Firstly, one can study the collisions of axion stars or axion stars with black holes, to obtain gravitational wave signals. One can also consider gradual accretion of scalar matter onto a single axion star, and the way in which the axion star then moves between the different regions of the parameter space that we have identified above. Extension to the rotating case may also lead to interesting phenomena and potentially a more complex solution space.

%% file: Appendix1/appendix1.tex
\chapter{Technical points and summary of equations} 
\label{sec-appendix1}

\ifpdf
    \graphicspath{{Appendix1/Figs/Raster/}{Appendix1/Figs/PDF/}{Appendix1/Figs/}}
\else
    \graphicspath{{Appendix1/Figs/Vector/}{Appendix1/Figs/}}
\fi

\section{Geometric and Planck Units}
\label{sec-appendix1A}

In the code $\grchombo$, and NR in general, one uses geometrised units, whereas in most other areas of physics, including cosmology, one uses Planck units. This section aims to briefly summarise the two systems and their conversions. The reader is referred to the Appendix of Wald \citep{wald1984general}, which has a good introduction to geometrised units. Most physics textbooks contain an introduction to Planck units. 

Note that where we refer to the Planck mass throughout this thesis, we mean $\mpl = 2.17651 \times 10^{-8}$ kg, i.e. $\mpl^2 = \hbar c/G$, rather than the reduced planck mass $\mpl^2 = \hbar c/8\pi G$. The reduced Planck mass, which is heavily used in cosmology, has an additional factor of $8\pi$. In $\grchombo$, factors of $8\pi$ are included explicitly and not set to one, so we prefer to work with the non-reduced Planck mass. Future versions of the code will enable the user to set $G$ so as to eliminate the factors of $8\pi$ if desired.

\subsection{Geometric Units}
\label{Sec-Geo}

These are units in which $G=c=1$, which means that lengths, masses and times all have the same unit (recall that the SI units of $G$ are $\rm{m^3 kg^{-1}s^{-2}}$ so setting length equal to time from $c=1$ leads to mass and length having the same dimension). For our purposes in $\grchombo$, we usually work with a Mass $M$ as the standard geometrised unit (this is equivalent to using a length $L$, which is often preferred). Then all lengths and all time intervals are expressed in these units of mass $M$. 

To convert to real SI units one must multiply by the appropriate factors of $G$ and $c$, and then by kg to the appropriate geometrised mass dimension, to give the correct units. For example, the length in meters is obtained by multiplying the length in geometrised (mass) units by $G/c^2$ (which has units m/kg), and then by the value of $M$ in kg. Notice that the geometrised mass unit $M$ can be chosen freely. Thus if we simulate a black hole of geometrised mass one, $M=1$, we can describe the spacetime for any mass of black hole simply by scaling the results according to what we want this unit mass to physically represent in kg. However, once one physical value has been fixed - for example, in inflation, the energy density scale associated with the field - everything else must be consistent with this scale.

The geometrised unit of any other dimensionful quantity in the simulation can be expressed by considering its real physical dimension and replacing any $L$ and/or $T$ with $M$. The value in real SI units must again be calculated using the appropriate conversion factors of $G$ and $c$, and the geometrised mass dimension. Sometimes one wishes to find the geometrised quantity from a real physical value, which is the reverse process. 

Two useful examples in this thesis are:
\begin{enumerate}
\item{\emph{Energy Density} The energy density has real dimensions $M L^{-1} T^{-2}$ so in geometrised units it has dimension $M^{-2}$. The geometric quantity must be multiplied by a factor of $c^8/G^3$ and then by the value of $1/M^2$ in $\rm{kg}^{-2}$ to obtain the SI value.}
\item{\emph{Scalar Field} Using the fact that the action has dimensions of $\hbar$ we can show that the scalar field has units of $L^{1/2} M^{1/2} T^{-1}$ (alternatively $(\partial_x \phi)^2$ has the same units as energy density), thus in geometrised units it is dimensionless. The geometric quantity must be multiplied by a factor of $c^2/G^\onehalf$ to obtain its SI value (note that the value of $M$ does not affect this).}
\end{enumerate}

For convenience we also usually choose $M$ to be $\mpl$, which makes the conversion to Planck units more obvious, as we will discuss below. However, in some cases we prefer to choose that $M = q \mpl$ where $q$ gives the fraction or multiple of $\mpl$ that is represented by one geometrised unit in our code. This can lead to more manageable numbers in our simulations, for example, we can make the length of our grid of order 10.

\subsection{Planck Units}
\label{Sec-planck}

The aim of Planck units is to express all quantities as multiples of a set of base units defined by appropriate dimensionful combinations of the fundamental constants $\hbar$, $c$ and $G$, which are the Planck length $l_p$, the Planck mass $\mpl$ and the Planck time $t_p$.

Any other dimensionful quantities are expressed as multiples of the appropriate combination of $l_p$, $\mpl$ and $t_p$, given their real physical dimension, for example, one ``Planck force" $F_p$ is equal to $\mpl l_p / t_p^{-2}$.

Quantities are then expressed in dimensionless form in equations, e.g. one can write Newton's law of gravitation
\begin{equation}
F = \frac{G m_1 m_2}{r^2} \, ,
\end{equation}
as
\begin{equation}
F = \frac{m_1 m_2}{r^2} \, ,
\end{equation}
where this is understood to mean
\begin{equation}
\frac{F}{F_p} = \frac{(m_1/\mpl) (m_2/\mpl)}{(r/l_p)^2} \, .
\end{equation}
However, what conventionally happens when using Planck units is that $c$ and $\hbar$ are set to one, and then, since all powers of $l_p$ and $t_p$ can be expressed in terms of $\mpl$, all figures in Planck units are expressed as a multiple of $\mpl$. This is to contain the uncertainty in $G$, which is not as well measured as the other constants. What this means in effect is that $\mpl$ replaces $G$ (as $G = \mpl^{-2}$) in all the equations, thus the above Newton's equation would be written
\begin{equation}
F = \mpl^{-2} \frac{m_1 m_2}{r^2} \, .
\end{equation}
In this formulation of Planck units, lengths and times have units of inverse mass $\mpl^{-1}$. 

To convert back to SI units one must multiply the value in Planck units by the appropriate factors of $\hbar$ and $c$ to get a result in the SI units, and then by $\mpl$ in kg to the appropriate mass dimension. For example, an energy in Planck units has mass dimension 1 (it is expressed as a multiple of $\mpl$), so its value in joules is found by multiplying the value by $c^2$ to obtain the correct SI unit, and then by $\mpl$ in kg.

\subsection{Conversion between units}
\label{Sec-convert}

\subsubsection{Using Planck Units directly}

If one uses the geometrised unit $M = \mpl$ in simulations, one can extract directly the values in Planck units of other quantities - the numerical value in geometrised units is the same as that in Planck units. 

This can be shown by converting the value first into SI units using appropriate factors of $G$ and $c$, and then into Planck units by adding factors of $\hbar$ and $c$. The factors of $\hbar$, $c$ and $G$ combine into some multiple of the Planck mass such that the correct units are given but the numerical value is the same.

That is, if one models a black hole of (geometrised) mass $2M$, where $M=\mpl$, then the lengths are Planck lengths and the times are Planck times.  To be totally explicit, this means that if the radius is $4M$, one finds that the radius in SI units is $4 l_p$ (with $l_p$ expressed in meters) and thus equal to $4\mpl^{-1}$ in Planck units where $\hbar=c=1$. Initially this may seem a bit counterintuitive as one unit contains an inverse $\mpl$ whilst the other contains $\mpl$, but one can show that this is correct by first converting to SI units from geometrised ones and then on to Planck units as above. In addition, if $\mpl$ is set to one, which is equivalent to setting $G=1$, both units agree on the numerical value, as we would expect. 

Measures of time behave in exactly the same way, with $1M$ equal to $t_p$ and thus $1\mpl^-1$ in Planck units, where $\hbar=c=1$. Once we know how to treat mass, length and time, other quantities that are combinations of these follow in the natural way.

A useful example to consider is the scalar field $\phi$, which is dimensionless in geometrised units as above. This means that a change in $\phi$ of 1 in our simulation is equivalent to a change of $1 \mpl$ in Planck units, where $\hbar=c=1$. This is invariant whatever we choose our geometrised mass unit to represent, which initially seems surprising, but can be explained by the fact that the physical meaning of the scalar field is effectively absorbed into the energy density $V(\phi)$, and its spatial and temporal gradients.

Similarly, an energy density of $1M^{-2}$, where $M=\mpl$, corresponds to an energy density of $1\mpl^4$ in Planck units. However, care needs to be taken (especially in this case) when $M \neq \mpl$, as explained below.

\subsubsection{Using scaled Planck Units}

As noted above, in some cases we prefer to choose that $M = q \mpl$ where $q$ gives the fraction or multiple of $\mpl$ that is represented by one geometrised unit in our code. This can lead to more manageable numbers in our simulations, for example, we can make the length of our grid of order 10, which tends to be easier to work with.

The strategy in this case is to first recover the case above, where the geometrised unit is $\mpl$. Then, whatever number you have is again the correct number in Planck units. This is best seen by an example:

One chooses a geometrised unit $M$ to represent a physical mass of $10 \mpl$.

Now for a length of, say, $2M$, this corresponds to $2 \times 10 \mpl= 20 \mpl$, and therefore it represents a length of $20 l_p$ or $20 \mpl^{-1}$ in Planck units. The same thing happens with time, and other dimensionful quantities follow in the same way. 

For another example, consider an energy density of $\rho = 2$ in geometric units. This is actually $\rho = 2M^{-2}$ because energy density has (geometrised) mass dimension of $-2$ as discussed above. Therefore the correct conversion when one has chosen $M=10 \mpl$ is to first say that (still in geometrised units)
\begin{equation}
\rho = 2 (10 \mpl)^{-2} = 0.02 \mpl^{-2} \, .
\end{equation}
Therefore the value in Planck units is 0.02 $\mpl^4$. 

Note that this conversion can also be thought of in terms of an effective mass $m_{eff}$, which is used to ``undimensionalise" the geometric quantities, so that all masses, lengths and times are divided by $m_{eff}$ to obtain their value in dimensionless Planck units. In the example above, $m_{eff} = M/q = 0.1M$, so a geometrised quantity of $2M$ becomes $2M/m_{eff} = 20$ in dimensionless (and thus Planck) units. We are essentially using the freedom inherent in geometric units to choose the mass scaling which achieves a convenient simulation unit. Converting from Planck units into scaled geometrised units is simply the reverse of this process.

\section{Lie derivatives - a brief description and notes on use}
\label{sec-appendix1B}

The concept of a Lie derivative is in many respects more fundamental than the concept of a covariant derivative since it does not require an affine connection to be defined on the manifold. A full discussion of Lie derivatives can be found in Schutz \citep{SchutzGeo}, or other standard texts on GR and differential geometry. Here we will summarise the key ideas and show how they are used in the derivation of the extrinsic curvature and in the BSSN equations. We assume a coordinate basis throughout. 

\subsection{Lie Derivatives}

\begin{figure}
\begin{center}
\includegraphics[width=.95\textwidth]{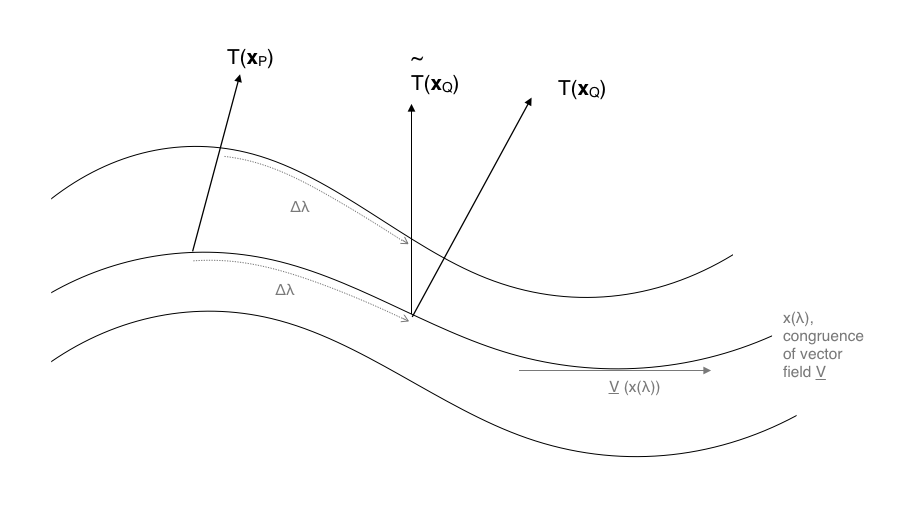}
\caption[The Lie Derivative]{The Lie Derivative is a measure of how much a tensor $T^a_b$ changes as one moves infinitesimally along the congruence of a vector field $\vec{V}$.}
\label{fig-LieDerivative}
\end{center}
\end{figure}

If one has a vector field $\vec{V}$ on a manifold, it is possible to define the \emph{integral curves} $\bf{x}(\lambda)$ by integrating the relation for the coordinates
\begin{equation}
\frac{dx^a}{d\lambda} = V^a ({\bf{x}}(\lambda)) \ ,
\end{equation}
which simply insists that that tangent to the curve at each point is the vector $\vec{V}$ at that point. This forms a \emph{congruence}, a family of such curves which fill the manifold, with affine parameter $\lambda$. The integral curves are thus like streamlines in an (ideal) fluid flow, with the vector $\vec{V}$ being the fluid velocity at each point; see figure \ref{fig-LieDerivative}.

The Lie derivative of a tensor $T^a_b$ with respect to $\vec{V}$, written as $\pounds_{\vec{V}} T^a_b$, is a measure of how much that tensor changes as one moves infinitesimally along the congruence. We have already emphasised in Chapter \ref{ch-Introduction} that comparing tensorial objects (like vectors) at different points on a manifold is an ambiguous concept. This is because the manifold may be curved and we must define how to transport the object from point to point.

When we define the covariant derivative, we introduce the affine connection and the concept of parallel transport in order to make a comparison. We require the covariant derivative to be metric compatible, i.e.
\begin{equation}
\nabla_c g_{ab} = 0 \, , 
\end{equation}
which means that it preserves the scalar product of two vectors as they are transported infinitesimally along the manifold. The covariant derivative can then be expressed in terms of the partial derivatives and the Christoffel symbols for a general tensor $T^a_b$ as
\begin{equation}
\nabla_c T^a_b = \partial_c T_a^b + T^d_b \Gamma^a_{cd} -  T^a_d \Gamma^d_{bc} \, . \label{covariant}
\end{equation}

The procedure for taking the Lie derivative is slightly different - again we want to somehow compare our tensors at two different points, but this time we use an infinitesimal coordinate transform, generated by the vector field, to drag the tensor at one point an infinitesimal distance along the congruence, and compare it to the ``actual'' value of the tensor at the new location. Consider the points $P$ and $Q$ on an integral curve of $\vec{V}$, separated by $\Delta \lambda$, as in figure \ref{fig-LieDerivative}. What we called the ``actual'' value of the tensor at $Q$ is given by Taylor expanding from the point P
\begin{equation}
 T^a_b ({\bf{x}}_Q) =  T^a_b ({\bf{x}}_P + \Delta \lambda ~ \vec{V})
 =  T^a_b ({\bf{x}}_P) + \Delta \lambda ~ V^c \partial_c T^a_b + O(\Delta \lambda^2) \ .
\end{equation}
We want to compare this to the tensor at $P$ which has been ``dragged along'' the congruence to Q using the infinitesimal coordinate transform
\begin{equation}
\frac{\partial x^{a'}}{\partial{x^b}} = \delta^a_b + \Delta \lambda ~ \partial_b V^a \ , \label{eqn:LieCoordTransform}
\end{equation}
where the primed indices relate to the new transformed coordinate system. We will call this object $\tilde{T}^{a'}_{b'} (\bf{x}_Q)$, and it is given by
\begin{equation}
\tilde{T}^{a'}_{b'} ({\bf{x}}_Q) = \frac{\partial x^{a'}}{\partial{x^c}} \frac{\partial x^{d}}{\partial{x^{b'}}} ~ T^c_d ({\bf{x}}_P) \ .
\end{equation}
We have the relation we need to transform the raised index via \eqn{eqn:LieCoordTransform}, and the second comes from inverting it
\begin{equation}
\frac{\partial x^a}{\partial{x^{b'}}} = \delta^a_b - \Delta \lambda ~ \partial_b V^a + O(\Delta \lambda^2) \ .
\end{equation}
We define the Lie derivative as 
\begin{equation}
\pounds_{\vec{V}} T^a_b \equiv \lim_{ \Delta \lambda \to 0} \left[ \frac{\tilde{T}^{a'}_{b'} ({\bf{x}}_Q) -  T^a_b ({\bf{x}}_Q)}{\Delta \lambda} \right] \ ,
\end{equation}
which using the above relations is
\begin{equation}
\pounds_{\vec{V}} T^a_b = V^c \partial_c T^a_b - T^c_b \partial_c V^a + T^a_c \partial_b V^c \ . \label{eqn:LieDeriv1}
\end{equation}
This generalises to tensors of higher orders in the usual way, with upper indices generating additional terms like the second, and lower indices generating additional terms like the third. Comparing this to the covariant derivative in \eqn{covariant}, we see that the Lie derivatives require the derivatives of $\vec{V}$ at each point, and it is this piece of additional structure that replaces the chosen connection (the Christoffel symbols) from the covariant derivative. The result is that the Lie derivative does not require a connection and thus (in the case of the Levi-Civita connection) it is independent of the metric. 

Note that one can show that the partial derivatives in \eqn{eqn:LieDeriv1} can be replaced by covariant ones with the same result, that is
\begin{equation}
\pounds_{\vec{V}} T^a_b = V^c \nabla_c T^a_b - T^c_b \nabla_c V^a + T^a_c \nabla_b V^c \, . \label{eqn:LieDeriv}
\end{equation}
We will use this result in the following section.

\subsection{Lie Derivatives and the extrinsic curvature}

We now use \eqn{eqn:LieDeriv} to show that the two definitions of the extrinsic curvature $K_{ab}$ are equivalent using the results above, as was stated in \ref{sec-ADMTheory}. Starting from the definition of $K_{ab}$ in terms of the Lie derivative along the normal direction of the spatial metric, per \eqn{eqn:Kabdefinition2}
\begin{equation}
K_{ab} \equiv - \onehalf \pounds_{\vec{n}} \gamma_{ab} = -\onehalf \left( n^c \nabla_c \gamma_{ab} + \gamma_{ac} \nabla_b n^c + \gamma_{cb} \nabla_a n^c \right) \, .
\end{equation}
Expanding out the spatial metric as $g_{ab} + n_a n_b$ and using the fact that the normal vector is orthgonal to its gradient $n_a \nabla_b n^a = 0$ this becomes
\begin{equation}
K_{ab} = -\onehalf \left( n^c n_a \nabla_c n_b + n^c n_b \nabla_c n_a + \nabla_a n_b + \nabla_b n_a \right) \, .
\end{equation}
Reversing the trick to replace the normal vectors with $n^a n_b = \gamma^a_b - g^a_b$ and using the fact that the 4-metric commutes with the covariant derivative we find
\begin{equation}
K_{ab} = -\onehalf \left( \gamma^c_a \nabla_c n_b + \gamma^c_b \nabla_c n_a \right) = - P^c_a \nabla_c n_b \ ,
\end{equation}
which is the alternative definition as per eqn \eqn{eqn:Kabdefinition1}.

\subsection{Lie derivatives and the evolution equations}
\label{sec-appendix1LieShift}

\subsubsection{ADM equations}

The ADM evolution equations are derived by finding the change of each tensor quantity $T$ along the normal direction to the spatial hyperslice, which is in a sense the most ``natural'' direction to consider. This must then be re-expressed as the change with respect to the coordinate time in terms of the gauge variables $\alpha$ and $\beta^i$, i.e.
\begin{equation}
\pounds_{\vec{n}} T = \frac{1}{\alpha} \pounds_{\alpha \vec{n}} T =  \frac{1}{\alpha} \left( \pounds_{\vec{t}} T - \pounds_{\vec{\beta}} T \right)  \ ,
\end{equation}
which is rearranged to give the evolution in coordinate time (recognising that the Lie derivative along $\vec{t}$ reduces to the partial derivative with respect to the coordinate time $t$) such that
\begin{equation}
\partial_t T = \alpha \pounds_{\vec{n}} T + \pounds_{\vec{\beta}} T  \ .
\end{equation}
The first term will be some combination of the evolution variables per the derivation of the change along the normal direction, to which one then adds the Lie derivative of $T$ along the shift vector. This is what gives rise to the terms like $\beta^i \partial_i T$ in each of the expanded ADM equations.

\subsubsection{BSSN equations}

In the BSSN evolution equations, the decomposition of the evolution variables into conformal quantities means that they are no longer tensors but \emph{tensor densities}. A tensor density $\tilde{T}$ of ``weight'' $w$ is a tensor $T$ multiplied by the determinant of the spatial metric $\gamma$ to the power $w/2$, i.e. 
\begin{equation}
\tilde{T} = \gamma^{w/2} T  \ .
\end{equation}
The Lie derivative then becomes
\begin{equation}
\pounds_{\vec{V}} \tilde{T} = \left[ \pounds_{\vec{V}} \tilde{T} \right]_{w=0} + w \tilde{T} \partial_i V^i  \ ,
\end{equation}
where the first term is the expression arising from \eqn{eqn:LieDeriv} as if $\tilde{T}$ were a normal tensor, and the second is the correction for the non zero tensor density.
The conformal factor $\chi$ as defined as $\gamma_{ij}=\frac{1}{\chi^2}\,\tilde\gamma_{ij}$ has weight $-1/3$, and the conformal metric and the traceless part of the extrinsic curvature have weight $-2/3$ according to their definitions. $K$ is a normal tensor. This leads to additional terms in the BSSN evolution equations, compared to the ADM versions. 

The evolution of $\tilde{\Gamma}^i$ is further complicated by the fact that it is not a true vector density either. This is clear from the fact that the Christoffel symbols $\Gamma^i_{jk}$ are not tensors, therefore neither is their contraction. One thus obtains second derivatives of the shift in the Lie derivative, in addition to the term to account for the tensor density weight of $2/3$, as can be seen in the second line of \eqn{eqn:dtgamma2} . 

%% file: Appendix2/appendix2.tex

\chapter{Summary of Equations}
\label{sec-appendix2}

\ifpdf
    \graphicspath{{Appendix2/Figs/Raster/}{Appendix2/Figs/PDF/}{Appendix2/Figs/}}
\else
    \graphicspath{{Appendix2/Figs/Vector/}{Appendix2/Figs/}}
\fi

\pagebreak

\section{Summary of the ADM equations}
\label{sec-appendix1C}

\begin{tcolorbox}

For the standard  $3+1$ ADM decomposition per York of the spacetime metric
\begin{equation}
ds^2=-\alpha^2\,dt^2+\gamma_{ij}(dx^i + \beta^i\,dt)(dx^j + \beta^j\,dt)
\end{equation}

$\gamma_{ij}$ is the induced metric on the spatial slices with timelike unit normal
\begin{equation}
n^\mu = \frac{1}{\alpha}\left(\partial_t^\mu - \beta^i\,\partial_i^\mu\right)
\end{equation} 

The extrinsic curvature is defined as
\begin{equation}
K_{ij} = -\frac{1}{2}\,(\pounds_{\vec{n}}\gamma_{ij})
\end{equation}

The Hamiltonian constraint
\begin{equation}
\mathcal{H} = R + K^2-K_{ij}K^{ij}-16\pi \rho 
\end{equation}

The Momentum constraint
\begin{equation}
\mathcal{M}_i = D^j (\gamma_{ij} K - K_{ij}) - 8\pi S_i
\end{equation}

The definition of the extrinsic curvature (and evolution equation for $\gamma_{ij}$)
\begin{equation}
\partial_t \gamma_{ij} = - 2\alpha K_{ij} + D_i \beta_j + D_j \beta_i 
\end{equation}

The evolution equation for $K_{ij}$
\begin{multline}
\partial_t K_{ij} = \beta^k \partial_k K_{ij} + K_{ki} \partial_j \beta^k + K_{kj} \partial_i \beta^k - D_i D_j \alpha \\ 
+ \alpha \left({}^{(3)}R_{ij} + K K_{ij} - 2 K_{ik} K^k_j \right) + 4 \pi \alpha \left( \gamma_{ij}(S - \rho) - 2 S_{ij} \right)
\end{multline}

Where the various components of the matter stress tensor are defined as
\begin{equation}
\rho = n_a\,n_b\,T^{ab}\,,\quad S_i = -\gamma_{ia}\,n_b\,T^{ab}\,,\quad S_{ij} = \gamma_{ia}\,\gamma_{jb}\,T^{ab}\,,\quad S = \gamma^{ij}\,S_{ij} 
\label{eq:Mattereqns}
\end{equation}

\end{tcolorbox}

\section{Summary of the BSSN equations}
\label{sec-appendix1D}

\begin{tcolorbox}
In the BSSN formalism used in our $\grchombo$ simulations, the induced metric is decomposed as 
\begin{equation}
\gamma_{ij}=\frac{1}{\chi^2}\,\tilde\gamma_{ij} \quad \det\tilde\gamma_{ij}=1 \quad \chi = \left(\det\gamma_{ij}\right)^{-\frac{1}{6}} 
\end{equation}
The extrinsic curvature is decomposed into its trace, $K=\gamma^{ij}\,K_{ij}$, and its traceless part $\tilde\gamma^{ij}\,\tilde A_{ij}=0$ as
\begin{equation}
K_{ij}=\frac{1}{\chi^2}\left(\tilde A_{ij} + \frac{1}{3}\,K\,\tilde\gamma_{ij}\right)
\end{equation}
The conformal connections $\tilde\Gamma^i=\tilde\gamma^{jk}\,\tilde\Gamma^i_{~jk}$ where $\tilde\Gamma^i_{~jk}$ are the Christoffel symbols associated with the conformal metric $\tilde\gamma_{ij}$.

The evolution equations for BSSN are
\begin{align}
&\partial_t\chi=\frac{1}{3}\,\alpha\,\chi\, K - \frac{1}{3}\,\chi \,\partial_k \beta^k + \beta^k\,\partial_k \chi\, \label{eqn:dtchi2}\\
&\partial_t\tilde\gamma_{ij} =-2\,\alpha\, \tA_{ij}+\tgamma_{ik}\,\partial_j\beta^k+\tgamma_{jk}\,\partial_i\beta^k-\frac{2}{3}\,\tgamma_{ij}\,\partial_k\beta^k +\beta^k\,\partial_k \tgamma_{ij}\, \label{eqn:dttgamma2} \\
&\partial_t K = -\gamma^{ij}D_i D_j \alpha + \alpha\left(\tilde{A}_{ij} \tilde{A}^{ij} + \frac{1}{3} K^2 \right) + \beta^i\partial_iK + 4\pi\,\alpha(\rho + S) \label{eqn:dtK2} \\
&\partial_t\tilde A_{ij} = \chi^2\left[-D_iD_j \alpha + \alpha\left( R_{ij} - 8\pi\,\alpha \,S_{ij}\right)\right]^\textrm{TF} + \alpha (K \tA_{ij} - 2 \tA_{il}\,\tA^l{}_j)  \nonumber \\
&\hspace{1.3cm} + \tA_{ik}\, \partial_j\beta^k + \tA_{jk}\,\partial_i\beta^k-\frac{2}{3}\,\tA_{ij}\,\partial_k\beta^k+\beta^k\,\partial_k \tA_{ij}\,   \label{eqn:dtAij2} \\
&\partial_t \tilde \Gamma^i=-2\,\tA^{ij}\,\partial_j \alpha +2\,\alpha\left(\tilde\Gamma^i_{jk}\,\tA^{jk}-\frac{2}{3}\,\tilde\gamma^{ij}\partial_j K - 3\,\tA^{ij}\frac{\partial_j \chi}{\chi}\right) \nonumber \\
&\hspace{1.3cm} +\beta^k\partial_k \tilde\Gamma^{i} +\tilde\gamma^{jk}\partial_j\partial_k \beta^i +\frac{1}{3}\,\tilde\gamma^{ij}\partial_j \partial_k\beta^k \nonumber \\
&\hspace{1.3cm} + \frac{2}{3}\,\tilde\Gamma^i\,\partial_k \beta^k -\tilde\Gamma^k\partial_k \beta^i - 16\pi\,\alpha\,\tilde\gamma^{ij}\,S_j\, \label{eqn:dtgamma2}
\end{align} 

The covariant derivative of the lapse in the term $D_iD_j \alpha$ is calculated with reference to the full spatial metric, and not the covariant one, ie
\begin{equation}
D_iD_j \alpha = \partial_i \partial_j \alpha - \Gamma^k_{ij} \partial_k \alpha \label{eqn:D2alpha}
\end{equation}

\end{tcolorbox}

\begin{tcolorbox}
where
\begin{equation}
\Gamma^k_{ij} = \tGamma^k_{ij} - \frac{1}{\chi} \left(  \delta^k_i \partial_j \chi + \delta^k_j \partial_i \chi - \tgamma_{ij} \tgamma^{kl} \partial_l \chi  \right) \label{eqn:D2alpha2}
\end{equation}

The Ricci tensor $R_{ij}$ is split into conformal and non conformal parts $R_{ij} = \tilde R_{ij} + R^\chi_{ij}$ which are calculated as
\begin{equation}
\tilde R_{ij} = -\frac{1}{2}\tgamma^{lm}\partial_m\partial_l\tgamma_{ij}+\tGamma^k\tGamma_{(ij)k}+\tgamma^{lm}(2\tGamma^k_{l(i}\tGamma_{j)km}+\tGamma^k_{im}\tGamma_{klj}) \label{eqn:conformalR2}
\end{equation}
and
\begin{equation}
R^{\chi}_{ij}=\frac{1}{\chi}(\tD_i\tD_j\chi + \tgamma_{ij}\tD^l\tD_l \chi)-\frac{2}{\chi^2}\tgamma_{ij}\tD^l\chi \tD_l\chi. \label{eqn:Rchi2}
\end{equation}

The scalar field matter evolution equations are
\begin{equation}
\partial_t \phi = \alpha \Pi_M +\beta^i\partial_i \phi \label{eqn:dtphi2}
\end{equation}
\begin{equation}
\partial_t \Pi_M=\beta^i\partial_i \Pi_M + \alpha\partial_i\partial^i \phi + \partial_i \phi\partial^i \alpha+\alpha\left(K\Pi_M-\gamma^{ij}\Gamma^k_{ij}\partial_k \phi+\frac{dV}{d\phi}\right) \label{eqn:dtphiM2} 
\end{equation} 

\end{tcolorbox}